\documentclass[11pt]{article}
\pdfoutput=1
\usepackage[margin=1in]{geometry}
\usepackage{amsmath, amssymb, braket, dsfont,authblk,etoolbox}
\usepackage[mathscr]{euscript}
\usepackage{graphicx}
\usepackage{rotating, mathtools}
\usepackage[matrix,arrow]{xy}
\usepackage[numbers]{natbib}
\usepackage{hyperref}

\usepackage{color}
\definecolor{hanpurple}{rgb}{0.32, 0.09, 0.98}
\definecolor{bluecolor}{rgb}{0.2, 0.5, 0.85}
\definecolor{dkgreen}{rgb}{0,0.5,0}
\definecolor{ltgreen}{rgb}{0.1,.59,.43}

\usepackage{caption}
\newcommand{\setmuskip}[2]{#1=#2\relax}      

\newcommand{\kett}[1]{ {|{#1}\rangle\mkern-2mu\rangle} }
\newcommand{\bigkett}[1]{ {\big|{#1}\big\rangle\!\big\rangle} }
\newcommand{\bigket}[1]{ {\big|{#1}\big\rangle} }

\newcommand{\defineas}{\mathrel{\overset{\textrm{def}}{=}}}
\newcommand{\Msixj}[2]{\begin{bmatrix} #1 \\ #2 \end{bmatrix}}
\newcommand{\Wsixj}[2]{\begin{Bmatrix} #1 \\ #2 \end{Bmatrix}_{\!q}}
\newcommand{\smallM}[2]{\left[\!\begin{smallmatrix} #1 \\ #2 \end{smallmatrix}\!\right]}
\newcommand{\Tr}{\operatorname{Tr}}
\newcommand{\qTr}{\operatorname{\widetilde{Tr}}}
\newcommand{\Vdim}{\operatorname{dim}}
\newcommand{\SU}[1]{{\widehat{\mathfrak{su}}(#1)}}		

\newcommand{\I}{{\mathds{1}}}
\newcommand{\D}{{\mathcal{D}}}
\newcommand{\modT}{{\mathbf{T}}}

\usepackage{xifthen}

\newcommand{\APotts}{\mathord{\vcenter{\hbox{\includegraphics[scale=0.8]{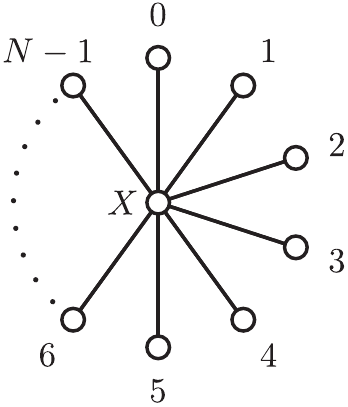}}}}}
\newcommand{\Ak}{\mathord{\vcenter{\hbox{\includegraphics[scale=1]{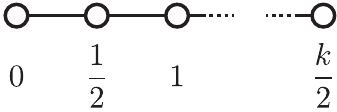}}}}}

\newcommand{\AintHalfInta}{\mathord{\vcenter{\hbox{\includegraphics[scale=1]{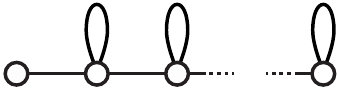}}}}}
\newcommand{\AintInta}{\mathord{\vcenter{\hbox{\includegraphics[scale=1]{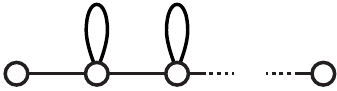}}}}}
\newcommand{\AintIntb}{\mathord{\vcenter{\hbox{\includegraphics[scale=1]{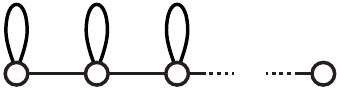}}}}}
\newcommand{\AintHanfIntb}{\mathord{\vcenter{\hbox{\includegraphics[scale=1]{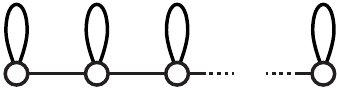}}}}}

\newcommand{\Txxx}{\mathord{\vcenter{\hbox{\includegraphics[scale=1]{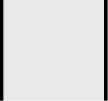}}}}}
\newcommand{\TTadb}{\mathord{\vcenter{\hbox{\includegraphics[scale=1]{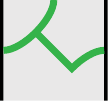}}}}}
\newcommand{\TTada}{\mathord{\vcenter{\hbox{\includegraphics[scale=1]{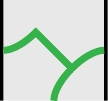}}}}}
\newcommand{\TxGG}{\mathord{\vcenter{\hbox{\includegraphics[scale=1]{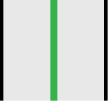}}}}}
\newcommand{\TGGx}{\mathord{\vcenter{\hbox{\includegraphics[scale=1]{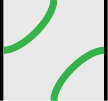}}}}}
\newcommand{\TGGG}{\mathord{\vcenter{\hbox{\includegraphics[scale=1]{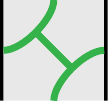}}}}}
\newcommand{\TGZG}{\mathord{\vcenter{\hbox{\includegraphics[scale=1]{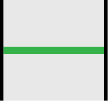}}}}}

\newcommand{\ShadowNotation}{\mathord{\vcenter{\hbox{\includegraphics[scale=1.3]{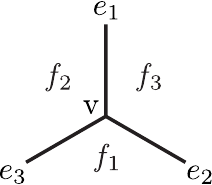}}}}}

\newcommand{\ZDisc}{\mathord{\vcenter{\hbox{\includegraphics[scale=1]{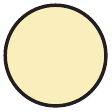}}}}}
\newcommand{\ZDiscG}{\mathord{\vcenter{\hbox{\includegraphics[scale=1]{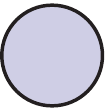}}}}}
\newcommand{\ZDiscGLoop}{\mathord{\vcenter{\hbox{\includegraphics[scale=1]{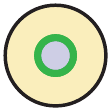}}}}}
\newcommand{\ZDiscGDual}{\mathord{\vcenter{\hbox{\includegraphics[scale=1]{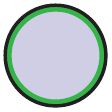}}}}}

\newcommand{\TransferSymmb}{\mathord{\vcenter{\hbox{\includegraphics[scale=1]{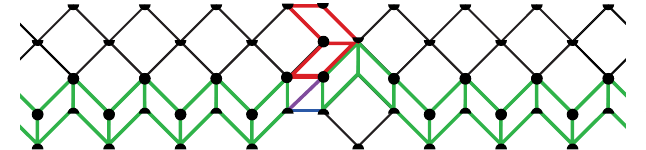}}}}}
\newcommand{\TransferSymma}{\mathord{\vcenter{\hbox{\includegraphics[scale=1]{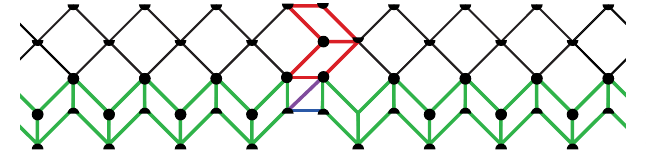}}}}}
\newcommand{\TransferSymmc}{\mathord{\vcenter{\hbox{\includegraphics[scale=1]{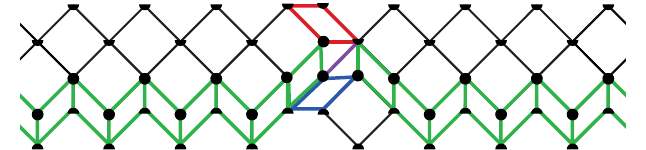}}}}}
\newcommand{\TransferSymmd}{\mathord{\vcenter{\hbox{\includegraphics[scale=1]{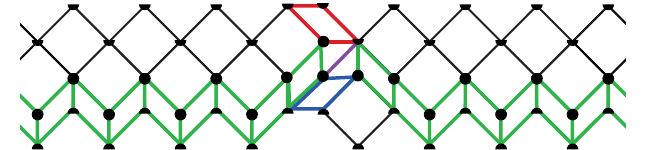}}}}}
\newcommand{\TransferSymme}{\mathord{\vcenter{\hbox{\includegraphics[scale=1]{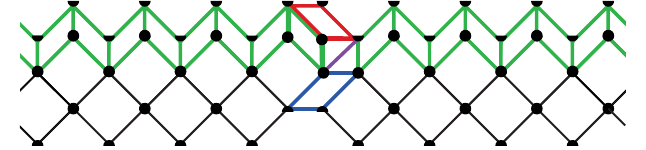}}}}}
\newcommand{\TransferSymmf}{\mathord{\vcenter{\hbox{\includegraphics[scale=1]{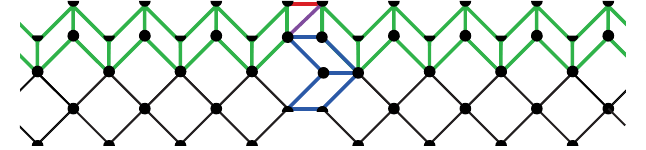}}}}}

\newcommand{\TransfMxIsingDots}{\mathord{\vcenter{\hbox{\includegraphics[scale=1.25]{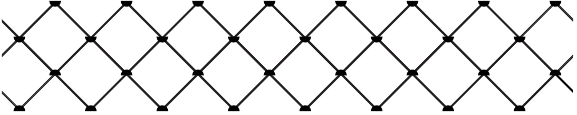}}}}}
\newcommand{\TopologicalSymmetryprime}{\mathord{\vcenter{\hbox{\includegraphics[scale=1.25]{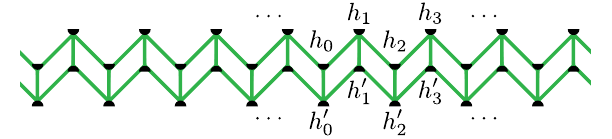}}}}}

\newcommand{\Topologicalsymmetrya}{\mathord{\vcenter{\hbox{\includegraphics[scale=1.25]{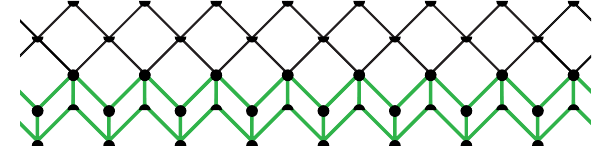}}}}}
\newcommand{\TopologicalSymmetryb}{\mathord{\vcenter{\hbox{\includegraphics[scale=1.25]{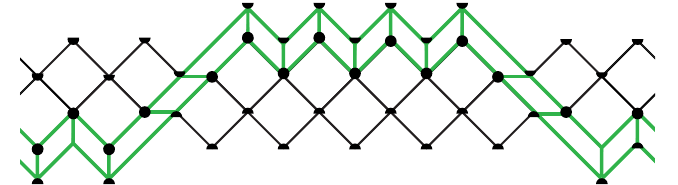}}}}}
\newcommand{\TopologicalSymmetryc}{\mathord{\vcenter{\hbox{\includegraphics[scale=1.25]{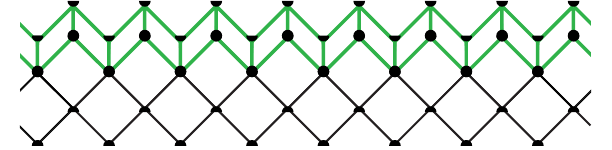}}}}}

\newcommand{\BWbare}{\mathord{\vcenter{\hbox{\includegraphics[scale=1.4]{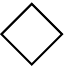}}}}}
\newcommand{\BWTLbare}{\mathord{\vcenter{\hbox{\includegraphics[scale=1.4]{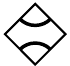}}}}}
\newcommand{\BWIdbare}{\mathord{\vcenter{\hbox{\includegraphics[scale=1.4]{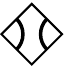}}}}}
\newcommand{\BWsmall}{\mathord{\vcenter{\hbox{\includegraphics[scale=0.9]{figures/BW.pdf}}}}}

\newcommand{\BWloopdefectG}{\mathord{\vcenter{\hbox{\includegraphics[scale=1.4]{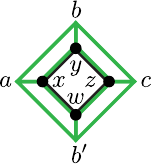}}}}}
\newcommand{\BWLoopDefectRemoved}{\mathord{\vcenter{\hbox{\includegraphics[scale=1.6]{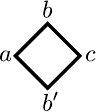}}}}}

\newcommand{\RhombiiTilingBoundaryRandom}{\mathord{\vcenter{\hbox{\includegraphics[scale=1.0]{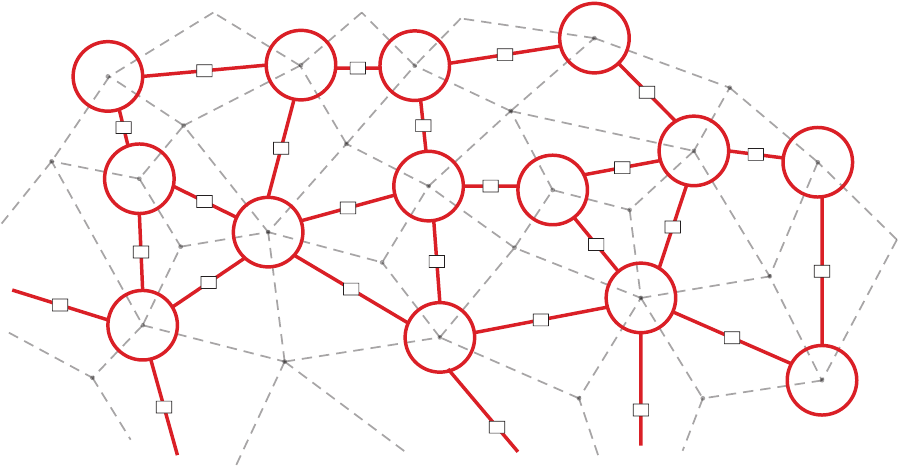}}}}}

\newcommand{\TVa}{\mathord{\vcenter{\hbox{\includegraphics[scale=1.0]{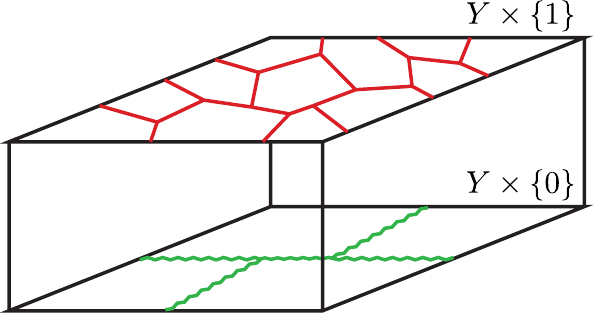}}}}}
\newcommand{\TVb}{\mathord{\vcenter{\hbox{\includegraphics[scale=0.8]{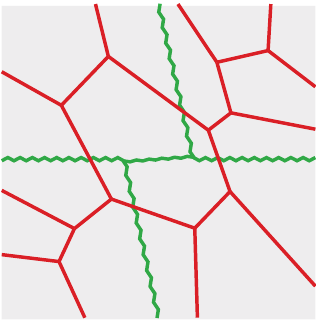}}}}}
\newcommand{\TVup}{\mathord{\vcenter{\hbox{\includegraphics[scale=1.0]{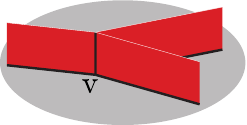}}}}}
\newcommand{\TVdown}{\mathord{\vcenter{\hbox{\includegraphics[scale=1.0]{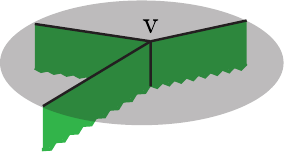}}}}}
\newcommand{\TVcross}{\mathord{\vcenter{\hbox{\includegraphics[scale=1.0]{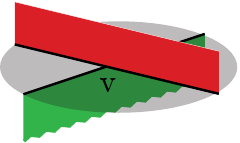}}}}}

\newcommand{\TVvertup}{\mathord{\vcenter{\hbox{\includegraphics[scale=1.0]{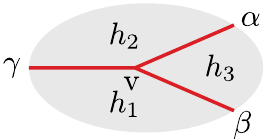}}}}}
\newcommand{\TVvertdown}{\mathord{\vcenter{\hbox{\includegraphics[scale=1.0]{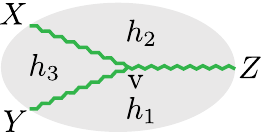}}}}}
\newcommand{\TVvertcross}{\mathord{\vcenter{\hbox{\includegraphics[scale=1.0]{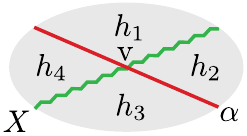}}}}}

\newcommand{\FExampleSpinZeroOne}{\mathord{\vcenter{\hbox{\includegraphics[scale=1.2]{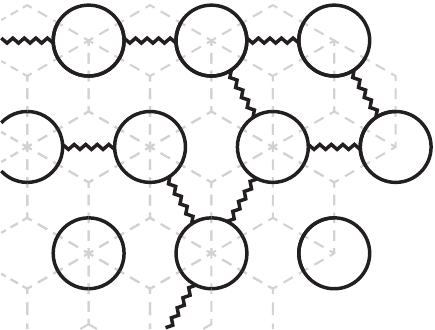}}}}}

\newcommand{\Ef}{\mathord{\vcenter{\hbox{\includegraphics[scale=0.65]{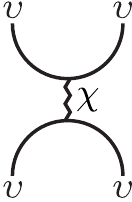}}}}}
\newcommand{\EfRhombii}{\mathord{\vcenter{\hbox{\includegraphics[scale=0.9]{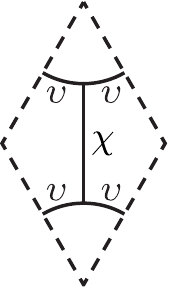}}}}}

\newcommand{\RVOne}{\mathord{\vcenter{\hbox{\includegraphics[scale=0.65]{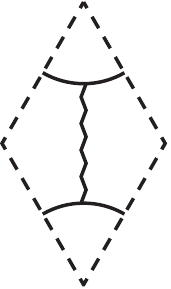}}}}}
\newcommand{\RVZero}{\mathord{\vcenter{\hbox{\includegraphics[scale=0.65]{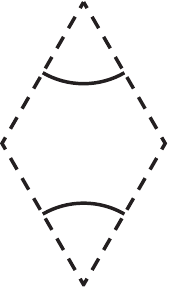}}}}}
\newcommand{\RHOne}{\mathord{\vcenter{\hbox{\includegraphics[scale=0.65]{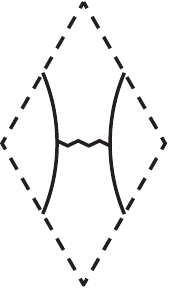}}}}}
\newcommand{\RHZero}{\mathord{\vcenter{\hbox{\includegraphics[scale=0.65]{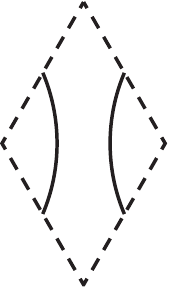}}}}}

\newcommand{\RVoD}{\mathord{\vcenter{\hbox{\includegraphics[scale=0.65]{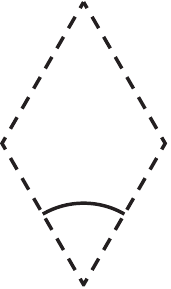}}}}}
\newcommand{\RHoL}{\mathord{\vcenter{\hbox{\includegraphics[scale=0.65]{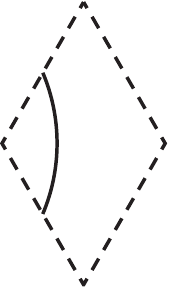}}}}}
\newcommand{\RVtwist}{\mathord{\vcenter{\hbox{\includegraphics[scale=0.65]{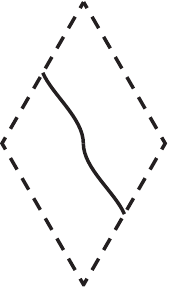}}}}}
\newcommand{\Rblank}{\mathord{\vcenter{\hbox{\includegraphics[scale=0.65]{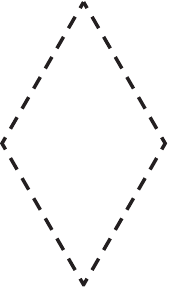}}}}}

\newcommand{\fBoxprime}{\mathord{\vcenter{\hbox{\includegraphics[scale=1.2]{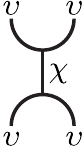}}}}}
\newcommand{\fboxa}{\mathord{\vcenter{\hbox{\includegraphics[scale=1.2]{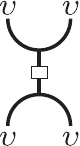}}}}}

\newcommand{\EftildeRhombii}{\mathord{\vcenter{\hbox{\includegraphics[scale=0.9]{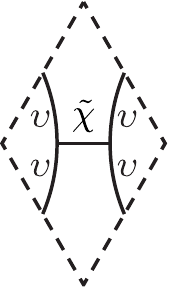}}}}}
\newcommand{\QD}{\mathord{\vcenter{\hbox{\includegraphics[scale=0.9]{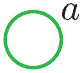}}}}}
\newcommand{\recouplinga}{\mathord{\vcenter{\hbox{\includegraphics[scale=0.9]{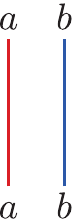}}}}}
\newcommand{\recouplingb}{\mathord{\vcenter{\hbox{\includegraphics[scale=0.9]{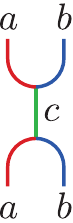}}}}}
\newcommand{\bubblea}{\mathord{\vcenter{\hbox{\includegraphics[scale=0.9]{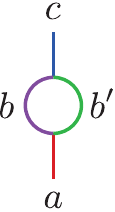}}}}}
\newcommand{\bubbleb}{\mathord{\vcenter{\hbox{\includegraphics[scale=0.9]{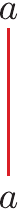}}}}}

\newcommand{\FSL}{\mathord{\vcenter{\hbox{\includegraphics[scale=0.9]{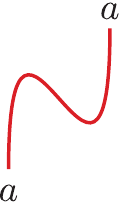}}}}}
\newcommand{\FSR}{\mathord{\vcenter{\hbox{\includegraphics[scale=0.9]{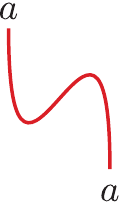}}}}}

\newcommand{\Epm}{\mathord{\vcenter{\hbox{\includegraphics[scale=0.9]{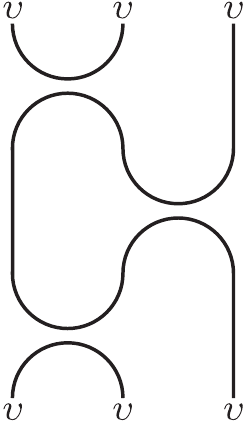}}}}}
\newcommand{\Epmb}{\mathord{\vcenter{\hbox{\includegraphics[scale=0.9]{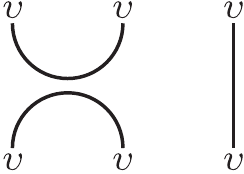}}}}}
\newcommand{\Esquared}{\mathord{\vcenter{\hbox{\includegraphics[scale=0.9]{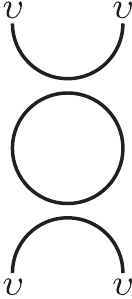}}}}}
\newcommand{\Eprime}{\mathord{\vcenter{\hbox{\includegraphics[scale=0.9]{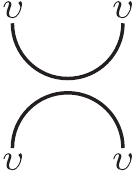}}}}}

\newcommand{\DefCommSolprimea}{\mathord{\vcenter{\hbox{\includegraphics[scale=0.9]{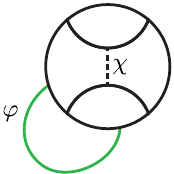}}}}}  
\newcommand{\DefCommSolprimeb}{\mathord{\vcenter{\hbox{\includegraphics[scale=0.9]{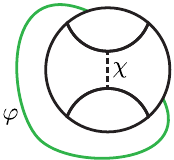}}}}}  
\newcommand{\DefCommSolprimec}{\mathord{\vcenter{\hbox{\includegraphics[scale=0.9]{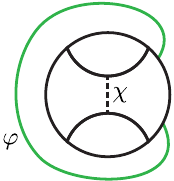}}}}}  
\newcommand{\DefCommSolprimed}{\mathord{\vcenter{\hbox{\includegraphics[scale=0.9]{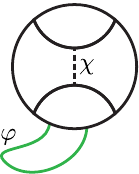}}}}}  

\newcommand{\tubeDefect}{\mathord{\vcenter{\hbox{\includegraphics[scale=0.7]{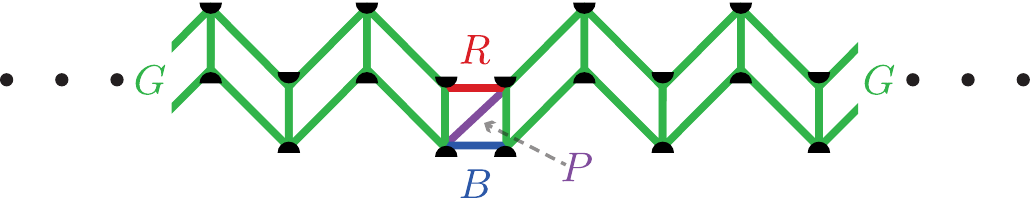}}}}}  

\newcommand{\eva}{\mathord{\vcenter{\hbox{\includegraphics[scale=0.7]{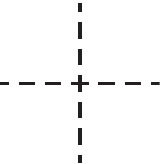}}}}}  
\newcommand{\evb}{\mathord{\vcenter{\hbox{\includegraphics[scale=0.7]{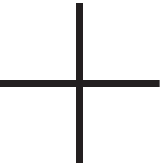}}}}}  
\newcommand{\evc}{\mathord{\vcenter{\hbox{\includegraphics[scale=0.7]{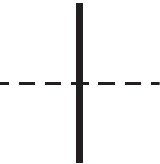}}}}}  
\newcommand{\evd}{\mathord{\vcenter{\hbox{\includegraphics[scale=0.7]{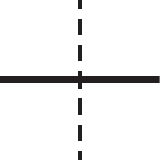}}}}}  
\newcommand{\eve}{\mathord{\vcenter{\hbox{\includegraphics[scale=0.7]{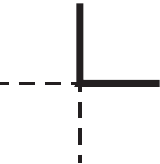}}}}}  
\newcommand{\evf}{\mathord{\vcenter{\hbox{\includegraphics[scale=0.7]{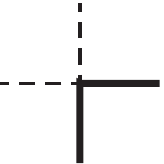}}}}}  
\newcommand{\evg}{\mathord{\vcenter{\hbox{\includegraphics[scale=0.7]{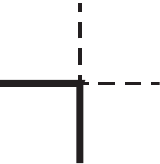}}}}}  
\newcommand{\evh}{\mathord{\vcenter{\hbox{\includegraphics[scale=0.7]{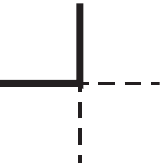}}}}}

\newcommand{\DefCommSol}{\mathord{\vcenter{\hbox{\includegraphics[scale=1.2]{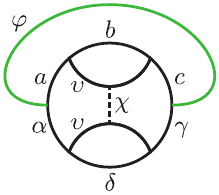}}}}}
\newcommand{\DefCommSolLa}{\mathord{\vcenter{\hbox{\includegraphics[scale=1.2]{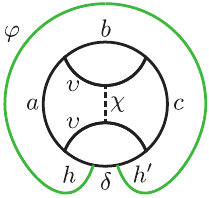}}}}}
\newcommand{\DefCommSolLb}{\mathord{\vcenter{\hbox{\includegraphics[scale=1.2]{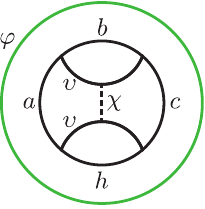}}}}}
\newcommand{\DefCommSolRa}{\mathord{\vcenter{\hbox{\includegraphics[scale=1.2]{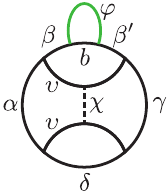}}}}}
\newcommand{\DefCommSolRb}{\mathord{\vcenter{\hbox{\includegraphics[scale=1.2]{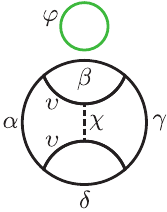}}}}}

\newcommand{\LineDefectInserta}{\mathord{\vcenter{\hbox{\includegraphics[scale=1.2]{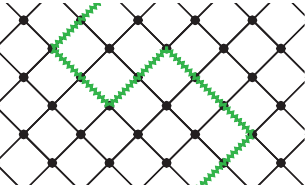}}}}}
\newcommand{\LineDefectInsertb}{\mathord{\vcenter{\hbox{\includegraphics[scale=1.2]{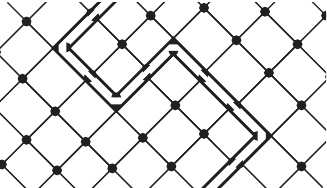}}}}}
\newcommand{\LineDefectInsertc}{\mathord{\vcenter{\hbox{\includegraphics[scale=1.2]{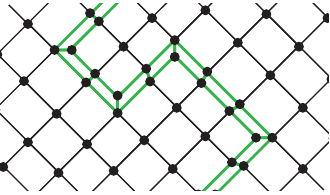}}}}}

\newcommand{\TrivalentComm}{\mathord{\vcenter{\hbox{\includegraphics[scale=1.2]{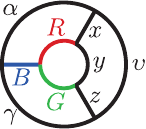}}}}}
\newcommand{\TrivalentCommL}{\mathord{\vcenter{\hbox{\includegraphics[scale=1.2]{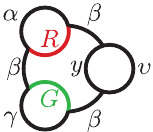}}}}}
\newcommand{\TrivalentCommR}{\mathord{\vcenter{\hbox{\includegraphics[scale=1.2]{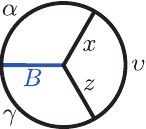}}}}}

\newcommand{\BubbleProof}{\mathord{\vcenter{\hbox{\includegraphics[scale=1.2]{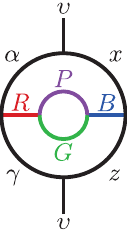}}}}}
\newcommand{\BubbleProofL}{\mathord{\vcenter{\hbox{\includegraphics[scale=1.2]{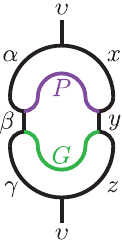}}}}}
\newcommand{\BubbleProofR}{\mathord{\vcenter{\hbox{\includegraphics[scale=1.2]{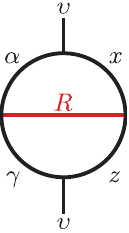}}}}}

\newcommand{\BWfour}[5]{\mathord{ \raisebox{0.2ex}{$\scriptstyle{#2}$}\mkern2mu\overset{#3}{\underset{#4}{#1}}\mkern2mu\raisebox{0.2ex}{$\scriptstyle{#5}$} }}

\newcommand{\BWbarex}[1]{{\mathord{\ooalign{ \vphantom{$\Big|^2$}\cr\hidewidth\ensuremath{#1}\hidewidth\cr$\vcenter{\hbox{\includegraphics[scale=1.4]{figures/BW.pdf}}}$\cr\vphantom{$\Big|_q$} }}}}		
\newcommand{\BWx}[5][]{\mathord{ \raisebox{0.2ex}{$\scriptstyle{#2}$}\mkern2mu\overset{#3}{\underset{#4}%
	{\ifthenelse{\isempty{#1}}{\BWbare}{\BWbarex{#1}}}%
	}\mkern2mu\raisebox{0.2ex}{$\scriptstyle{#5}$} }}
\newcommand{\BWabcd}{\BWfour{\BWbare}{a}{b}{b'}{c}}
\newcommand{\BWabcdx}[1]{\BWfour{\BWbarex{#1}}{a}{b}{b'}{c}}
\newcommand{\BWTL}{\mathord{\raisebox{0.2ex}{$\scriptstyle{a}$}\mkern2mu\overset{b}{\underset{b'}{\BWTLbare}}\mkern2mu\raisebox{0.2ex}{$\scriptstyle{c}$}}}
\newcommand{\BWId}{\mathord{\raisebox{0.2ex}{$\scriptstyle{a}$}\mkern2mu\overset{b}{\underset{b'}{\BWIdbare}}\mkern2mu\raisebox{0.2ex}{$\scriptstyle{c}$}}}

\newcommand{\BWbareDotsx}[1]{{\mathord{\ooalign{ \vphantom{$\Big|^2$}
\cr\hidewidth\ensuremath{#1}\hidewidth\cr$\vcenter{\hbox{\includegraphics[scale=1.4]{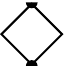}}}$\cr\vphantom{$\Big|_q$} }}}}		
\newcommand{\BWabcdDotsx}[1]{\BWfour{\BWbareDotsx{#1}}{a}{b}{b'}{c}}

\newcommand{\DefectSquare}{\mathord{\vcenter{\hbox{\includegraphics[scale=1.5]{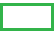}}}}}

\newcommand{\DefectSquarex}[4]{\mathord{\sideset{^{#1}_{#3}}{^{#2}_{#4}}{\mathop{\DefectSquare}}}}

\newcommand{\DefectCommuteDprime}{\mathord{\vcenter{\hbox{\includegraphics[scale=1.3]{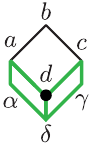}}}}}
\newcommand{\DefectCommuteDprimeprime}{\mathord{\vcenter{\hbox{\includegraphics[scale=1.3]{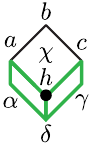}}}}}
\newcommand{\DefectCommuteUprime}{\mathord{\vcenter{\hbox{\includegraphics[scale=1.3]{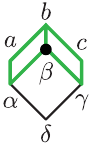}}}}}
\newcommand{\DefectCommuteUprimeprime}{\mathord{\vcenter{\hbox{\includegraphics[scale=1.3]{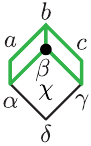}}}}}

\newcommand{\DefectCommuteSprimeprime}{\mathord{\vcenter{\hbox{\includegraphics[scale=1.3]{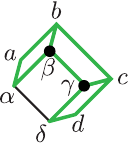}}}}}
\newcommand{\DefectCommuteNSprimeprime}{\mathord{\vcenter{\hbox{\includegraphics[scale=1.3]{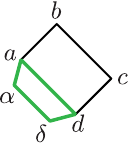}}}}}

\newcommand{\Trivalent}{\mathord{\vcenter{\hbox{\includegraphics[scale=0.3]{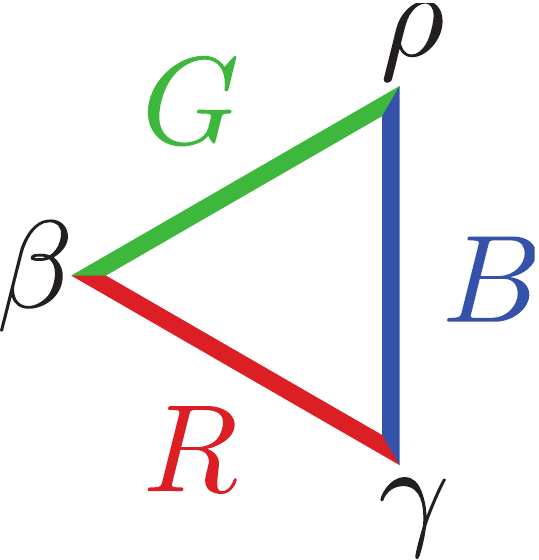}}}}}

\newcommand{\BubbleDiagram}{\mathord{\vcenter{\hbox{\includegraphics[scale=1.3]{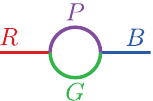}}}}}
\newcommand{\BlueLine}{\mathord{{\hbox{\includegraphics[scale=1.3]{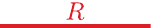}}}}}

\newcommand{\BubbleMacroLeft}{\mathord{\vcenter{\hbox{\includegraphics[scale=1.5]{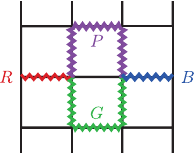}}}}}
\newcommand{\BubbleMacroRight}{\mathord{\vcenter{\hbox{\includegraphics[scale=1.5]{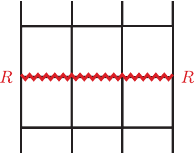}}}}}

\newcommand{\FmoveMacroLeft}{\mathord{\vcenter{\hbox{\includegraphics[scale=1.3]{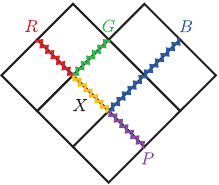}}}}}
\newcommand{\FmoveMacroRight}{\mathord{\vcenter{\hbox{\includegraphics[scale=1.3]{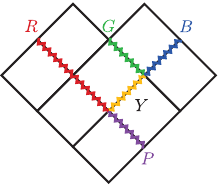}}}}}

\newcommand{\IsingBubbleMicroLeft}{\mathord{\vcenter{\hbox{\includegraphics[scale=.6]{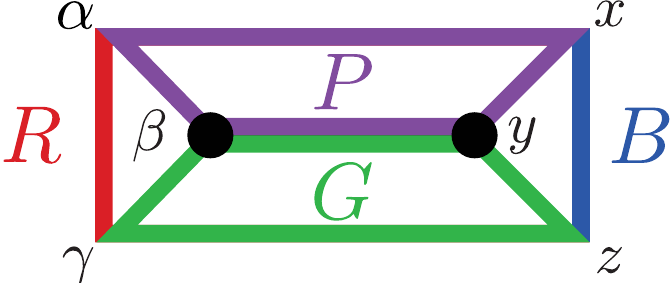}}}}}
\newcommand{\IsingBubbleMicroRight}{\mathord{\vcenter{\hbox{\includegraphics[scale=.6]{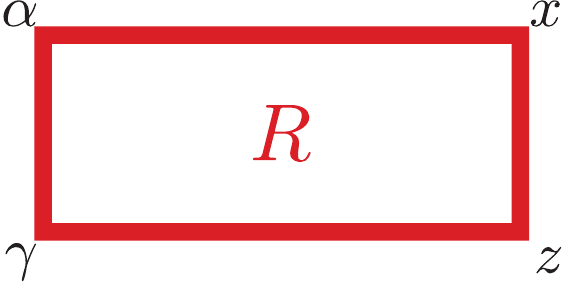}}}}}

\newcommand{\halfdot}{\mathord{\vcenter{\hbox{\includegraphics[scale=1]{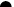}}}}}
\newcommand{\fulldot}{\mathord{\vcenter{\hbox{\includegraphics[scale=1]{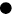}}}}}

\newcommand{\psia}{\mathord{\vcenter{\hbox{\includegraphics[scale=1]{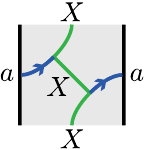}}}}}

\newcommand{\TubeObja}{\mathord{\vcenter{\hbox{\includegraphics[scale=1.5]{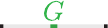}}}}}
\newcommand{\TubeObjb}{\mathord{\vcenter{\hbox{\includegraphics[scale=1.5]{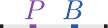}}}}}
\newcommand{\TubeBPRG}{\mathord{\vcenter{\hbox{\includegraphics[scale=1]{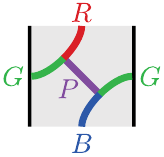}}}}}
\newcommand{\TubeMor}{\mathord{\vcenter{\hbox{\includegraphics[scale=1]{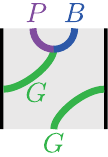}}}}}

\newcommand{\FLeftProof}{\mathord{\vcenter{\hbox{\includegraphics[scale=1.2]{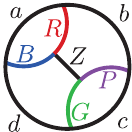}}}}}
\newcommand{\FRightProof}{\mathord{\vcenter{\hbox{\includegraphics[scale=1.2]{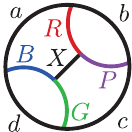}}}}}

\newcommand{\FLeft}{\mathord{\vcenter{\hbox{\includegraphics[scale=2.0]{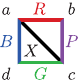}}}}}
\newcommand{\FRight}{\mathord{\vcenter{\hbox{\includegraphics[scale=2.0]{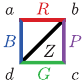}}}}}

\newcommand{\FusionTree}{\mathord{\vcenter{\hbox{\includegraphics[scale=1]{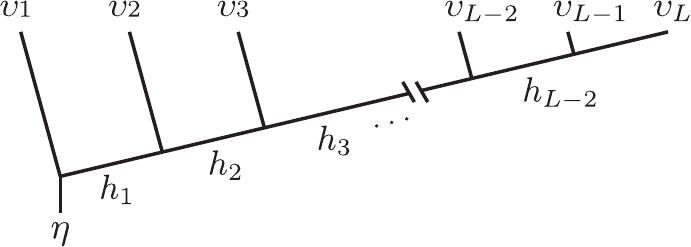}}}}}
\newcommand{\Destroyer}{\mathord{\vcenter{\hbox{\includegraphics[scale=1]{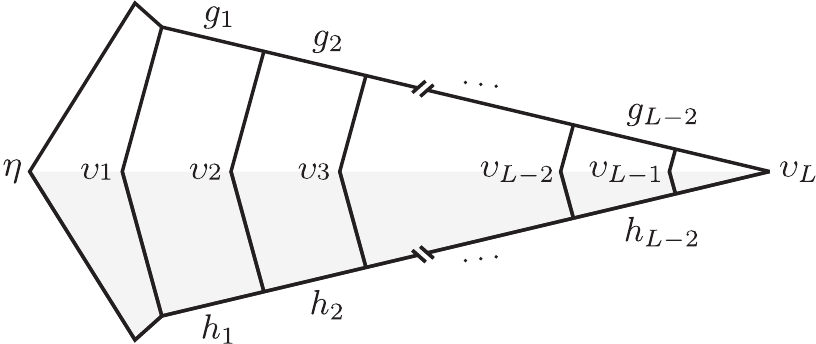}}}}}

\newcommand{\HilbOpen}{\mathord{\vcenter{\hbox{\includegraphics[scale=1]{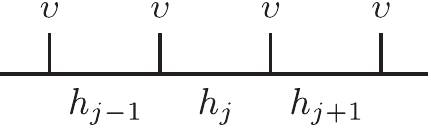}}}}}

\newcommand{\Hilbopenprime}{\mathord{\vcenter{\hbox{\includegraphics[scale=1.5]{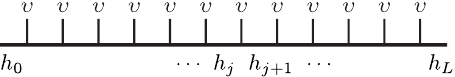}}}}}
\newcommand{\HilbOpenZprime}{\mathord{\vcenter{\hbox{\includegraphics[scale=1.5]{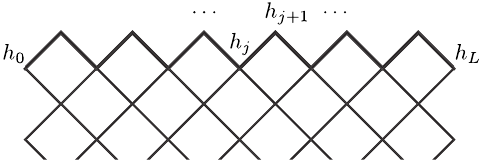}}}}}

\newcommand{\VminusIsing}{\mathord{\vcenter{\hbox{\includegraphics[scale=1]{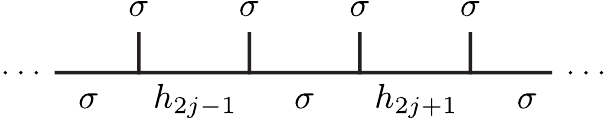}}}}}
\newcommand{\VplusIsing}{\mathord{\vcenter{\hbox{\includegraphics[scale=1]{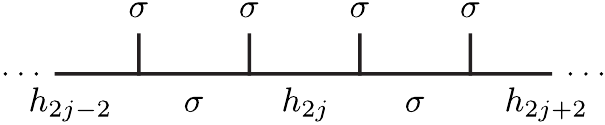}}}}}

\newcommand{\HilbTwisted}{\mathord{\vcenter{\hbox{\includegraphics[scale=1.5]{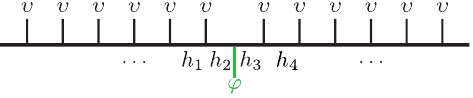}}}}}
\newcommand{\HilbFromZTwisted}{\mathord{\vcenter{\hbox{\includegraphics[scale=1.5]{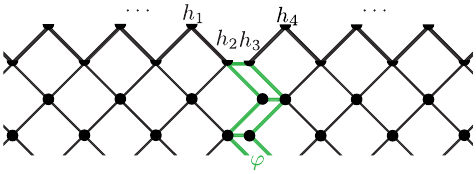}}}}}

\newcommand{\TSymmprime}{\mathord{\vcenter{\hbox{\includegraphics[scale=1]{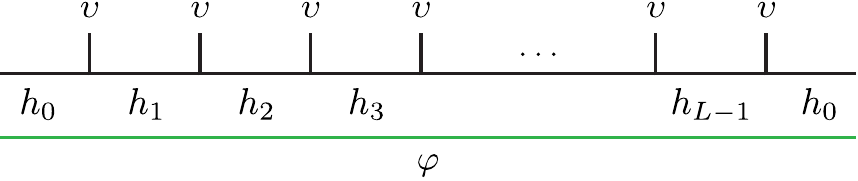}}}}}
\newcommand{\DomainTransfermatrixUpsilon}{\mathord{\vcenter{\hbox{\includegraphics[scale=1]{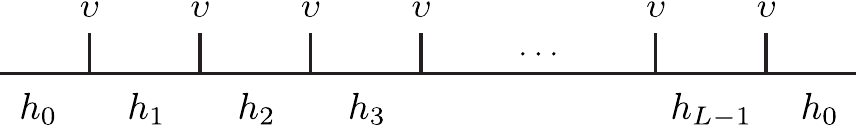}}}}}

\newcommand{\TSymmbprime}{\mathord{\vcenter{\hbox{\includegraphics[scale=1]{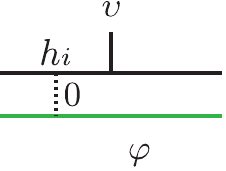}}}}}
\newcommand{\TSymmcaprime}{\mathord{\vcenter{\hbox{\includegraphics[scale=1]{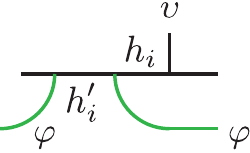}}}}}
\newcommand{\TSymmcprime}{\mathord{\vcenter{\hbox{\includegraphics[scale=1]{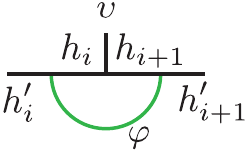}}}}}
\newcommand{\TSymmdprime}{\mathord{\vcenter{\hbox{\includegraphics[scale=1]{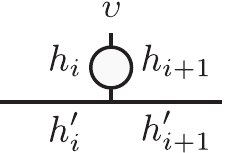}}}}}
\newcommand{\TSymmeprime}{\mathord{\vcenter{\hbox{\includegraphics[scale=1]{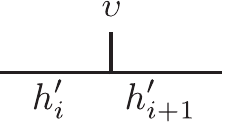}}}}}

\newcommand{\FmoveLsmall}{\mathord{\vcenter{\hbox{\includegraphics[scale=1.1]{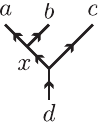}}}}}
\newcommand{\FmoveRsmall}{\mathord{\vcenter{\hbox{\includegraphics[scale=1.1]{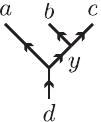}}}}}
\newcommand{\FmoveC}{\mathord{\vcenter{\hbox{\includegraphics[scale=1.3]{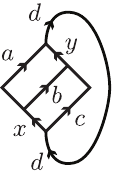}}}}}

\newcommand{\FmoveLNA}{\mathord{\vcenter{\hbox{\includegraphics[scale=1.6]{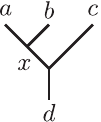}}}}}
\newcommand{\FmoveRNA}{\mathord{\vcenter{\hbox{\includegraphics[scale=1.6]{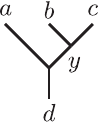}}}}}

\newcommand{\FmoveHprime}{\mathord{\vcenter{\hbox{\includegraphics[scale=1.6]{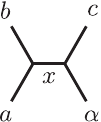}}}}}
\newcommand{\FmoveVprime}{\mathord{\vcenter{\hbox{\includegraphics[scale=1.6]{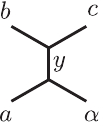}}}}}

\newcommand{\FmoveLsmallNA}{\mathord{\vcenter{\hbox{\includegraphics[scale=1.0]{figures/Fmove_left.pdf}}}}}
\newcommand{\FmoveRsmallNA}{\mathord{\vcenter{\hbox{\includegraphics[scale=1.0]{figures/Fmove_right.pdf}}}}}
\newcommand{\FmoveCNA}{\mathord{\vcenter{\hbox{\includegraphics[scale=1.3]{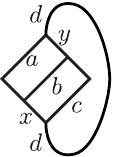}}}}}
\newcommand{\FSquared}{\mathord{\vcenter{\hbox{\includegraphics[scale=1.3]{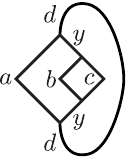}}}}}

\newcommand{\tube}{\operatorname{Tube}}
\newcommand{\mcc}{\mathcal{C}}
\newcommand{\mcd}{\mathcal{D}}

\newcommand{\mct}{\mathcal{T}}
\newcommand{\mcz}{\mathcal{Z}}
\newcommand{\mcm}{\mathcal{M}}

\newcommand{\annulusx}{\mathord{\vcenter{\hbox{\includegraphics[scale=0.8]{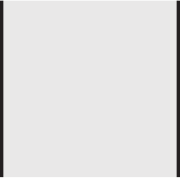}}}}}
\newcommand{\annulusy}{\mathord{\vcenter{\hbox{\includegraphics[scale=0.8]{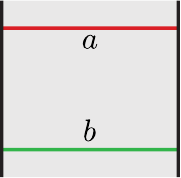}}}}}
\newcommand{\annulusz}{\mathord{\vcenter{\hbox{\includegraphics[scale=0.8]{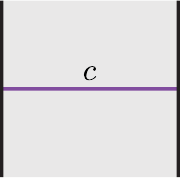}}}}}

\newcommand{\AnnulusHa}{\mathord{\vcenter{\hbox{\includegraphics[scale=1]{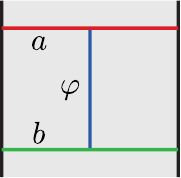}}}}}

\newcommand{\annulusphi}{\mathord{\vcenter{\hbox{\includegraphics[scale=1]{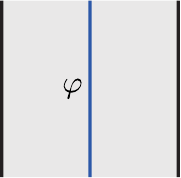}}}}}
\newcommand{\annulusaH}{\mathord{\vcenter{\hbox{\includegraphics[scale=1]{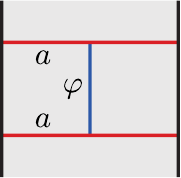}}}}}
\newcommand{\annulusaloopprime}{\mathord{\vcenter{\hbox{\includegraphics[scale=1]{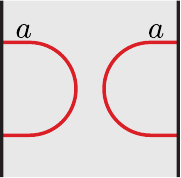}}}}}
\newcommand{\annulusaloop}{\mathord{\vcenter{\hbox{\includegraphics[scale=1]{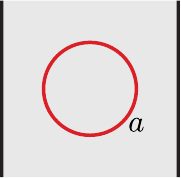}}}}}

\newcommand{\bcc}{\mathord{\vcenter{\hbox{\includegraphics[scale=1]{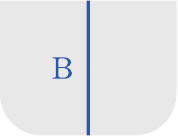}}}}}
\newcommand{\bcb}{\mathord{\vcenter{\hbox{\includegraphics[scale=1]{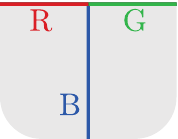}}}}}
\newcommand{\bca}{\mathord{\vcenter{\hbox{\includegraphics[scale=1]{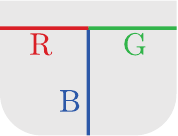}}}}}

\newcommand{\tetradiagramNA}{\mathord{\vcenter{\hbox{\includegraphics[scale=1.5]{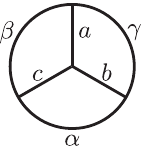}}}}}
\newcommand{\tetradiagram}{\mathord{\vcenter{\hbox{\includegraphics[scale=1.5]{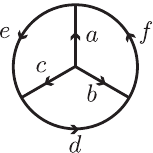}}}}}

\newcommand{\circlespentagonNA}{\mathord{\vcenter{\hbox{\includegraphics[scale=1.6]{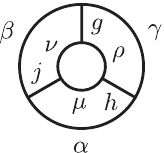}}}}}
\newcommand{\circlespentagon}{\mathord{\vcenter{\hbox{\includegraphics[scale=1.6]{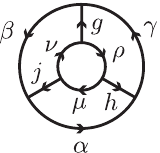}}}}}
\newcommand{\vertexabc}{\mathord{\vcenter{\hbox{\includegraphics[scale=1.1]{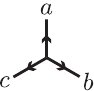}}}}}
\newcommand{\vertexabcC}{\mathord{\vcenter{\hbox{\includegraphics[scale=1.1]{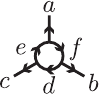}}}}}

\newcommand{\vertexabcNA}{\mathord{\vcenter{\hbox{\includegraphics[scale=1.6]{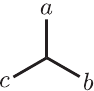}}}}}
\newcommand{\vertexabcCNA}{\mathord{\vcenter{\hbox{\includegraphics[scale=1.6]{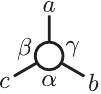}}}}}

\newcommand{\omegaf}{\mathord{\vcenter{\hbox{\includegraphics[scale=1.4]{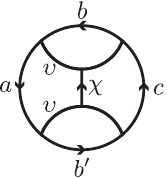}}}}}
\newcommand{\omegafNA}{\mathord{\vcenter{\hbox{\includegraphics[scale=1.4]{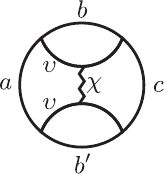}}}}}

\newcommand{\Trivalenta}{\mathord{\vcenter{\hbox{\includegraphics[scale=.7]{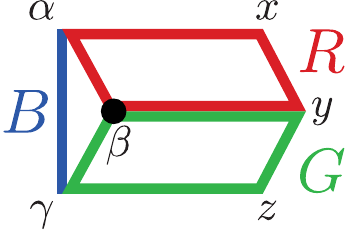}}}}}
\newcommand{\Trivalentb}{\mathord{\vcenter{\hbox{\includegraphics[scale=.7]{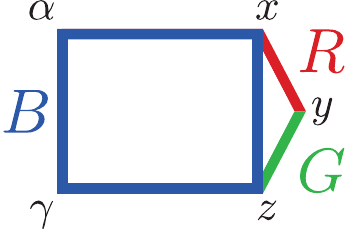}}}}}

\newcommand{\idremovRHS}{\mathord{\vcenter{\hbox{\includegraphics[scale=1]{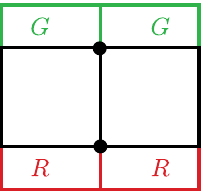}}}}}
\newcommand{\idremovLHS}{\mathord{\vcenter{\hbox{\includegraphics[scale=1]{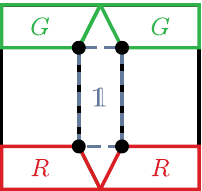}}}}}

\newcommand{\Yabc}{\mathord{\vcenter{\hbox{\includegraphics[scale=1]{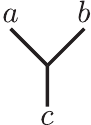}}}}}
\newcommand{\YabcX}{\mathord{\vcenter{\hbox{\includegraphics[scale=1]{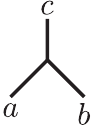}}}}}

\newcommand{\IsingFusionRuleSigmaPsiSigma}{\mathord{\vcenter{\hbox{\includegraphics[scale=1]{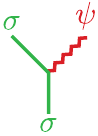}}}}}
\newcommand{\IsingFusionRulePsiPsiId}{\mathord{\vcenter{\hbox{\includegraphics[scale=1]{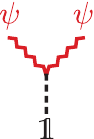}}}}}
\newcommand{\IsingFusionRuleSigmaSigmaId}{\mathord{\vcenter{\hbox{\includegraphics[scale=1]{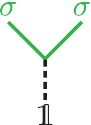}}}}}

\newcommand{\IsingFusionRuleSigmaSigmaSigma}{\mathord{\vcenter{\hbox{\includegraphics[scale=1]{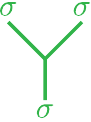}}}}}
\newcommand{\IsingFusionRuleIdIdPsi}{\mathord{\vcenter{\hbox{\includegraphics[scale=1]{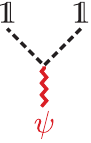}}}}}
\newcommand{\IsingFusionDiagram}{\mathord{\vcenter{\hbox{\includegraphics[scale=1.4]{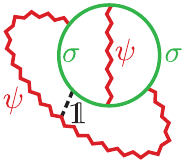}}}}}
\newcommand{\IsingFusionRulePsiPsiPsi}{\mathord{\vcenter{\hbox{\includegraphics[scale=1]{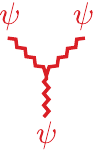}}}}}

\newcommand{\UaaSmall}{\mathord{\vcenter{\hbox{\includegraphics[scale=0.7]{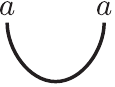}}}}}

\newcommand{\ALD}{\mathord{\vcenter{\hbox{\includegraphics[scale=1]{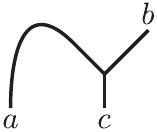}}}}}
\newcommand{\ARD}{\mathord{\vcenter{\hbox{\includegraphics[scale=1]{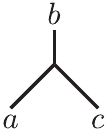}}}}}
\newcommand{\BLD}{\mathord{\vcenter{\hbox{\includegraphics[scale=1]{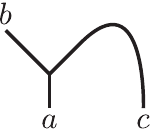}}}}}
\newcommand{\BRD}{\mathord{\vcenter{\hbox{\includegraphics[scale=1]{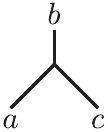}}}}}

\newcommand{\FusionDiagram}{\mathord{\vcenter{\hbox{\includegraphics[scale=1]{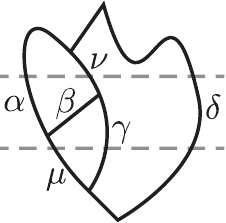}}}}}
\newcommand{\FusionDiagramVu}{\mathord{\vcenter{\hbox{\includegraphics[scale=1]{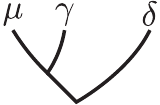}}}}}
\newcommand{\FusionDiagramOpVu}{\mathord{\vcenter{\hbox{\includegraphics[scale=0.9]{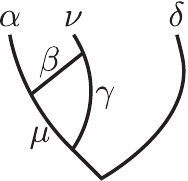}}}}}
\newcommand{\FusionDiagramVuFlipped}{\mathord{\vcenter{\hbox{\includegraphics[scale=1]{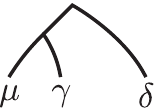}}}}}
\newcommand{\FusionDiagramVd}{\mathord{\vcenter{\hbox{\includegraphics[scale=1]{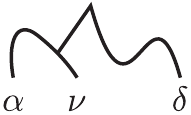}}}}}
\newcommand{\FusionDiagramOp}{\mathord{\vcenter{\hbox{\includegraphics[scale=1]{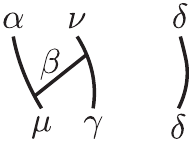}}}}}

\newcommand{\FusionDiagrama}{\mathord{\vcenter{\hbox{\includegraphics[scale=1]{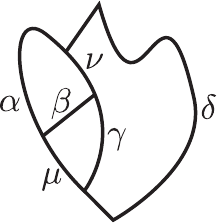}}}}}
\newcommand{\FusionDiagramb}{\mathord{\vcenter{\hbox{\includegraphics[scale=1]{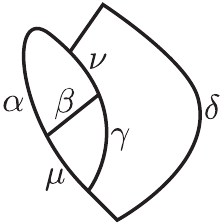}}}}}
\newcommand{\FusionDiagramc}{\mathord{\vcenter{\hbox{\includegraphics[scale=1]{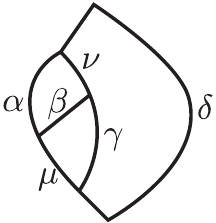}}}}}
\newcommand{\FusionDiagramd}{\mathord{\vcenter{\hbox{\includegraphics[scale=1]{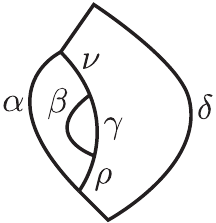}}}}}
\newcommand{\FusionDiagrame}{\mathord{\vcenter{\hbox{\includegraphics[scale=1]{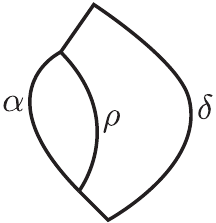}}}}}

\newcommand{\TetvTBWSS}{\mathord{\vcenter{\hbox{\includegraphics[scale=.4]{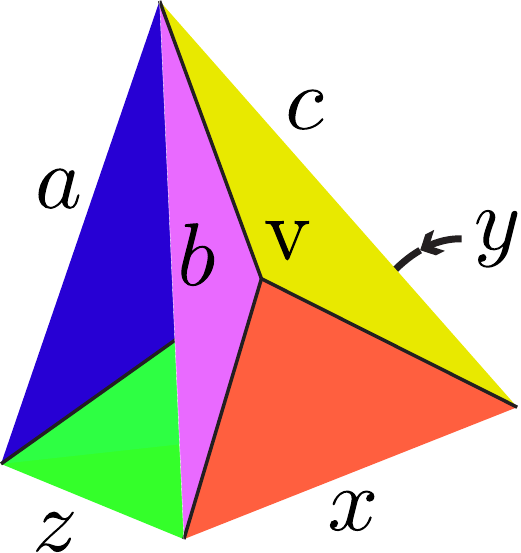}}}}}

\newcommand{\IdentityTerm}{\mathord{\vcenter{\hbox{\includegraphics[scale=1.2]{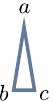}}}}}

\newcommand{\TrivalentConsistencya}{\mathord{\vcenter{\hbox{\includegraphics[scale=1.3]{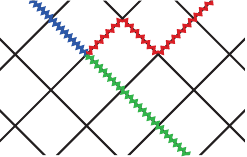}}}}}
\newcommand{\TrivalentConsistencyb}{\mathord{\vcenter{\hbox{\includegraphics[scale=1.3]{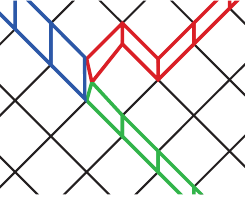}}}}}
\newcommand{\TrivalentConsistencyc}{\mathord{\vcenter{\hbox{\includegraphics[scale=1.3]{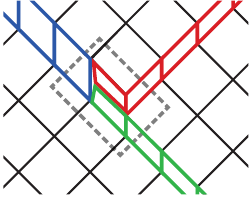}}}}}
\newcommand{\TrivalentConsistencyd}{\mathord{\vcenter{\hbox{\includegraphics[scale=1.3]{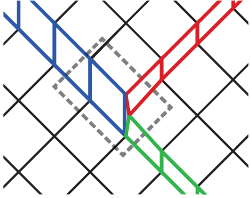}}}}}
\newcommand{\TrivalentConsistencye}{\mathord{\vcenter{\hbox{\includegraphics[scale=1.3]{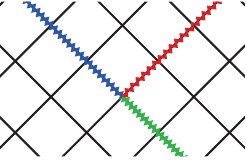}}}}}

\newcommand{\TrivalentJunctionInserta}{\mathord{\vcenter{\hbox{\includegraphics[scale=1.7]{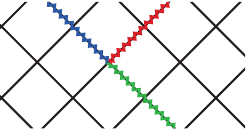}}}}}
\newcommand{\TrivalentJunctionInsertb}{\mathord{\vcenter{\hbox{\includegraphics[scale=1.7]{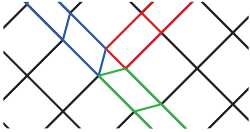}}}}}

\newcommand{\IsingBubbleRightId}{\mathord{\vcenter{\hbox{\includegraphics[scale=.3]{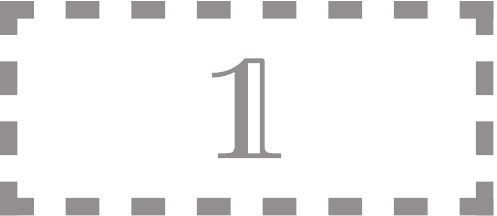}}}}}

\newcommand{\Hcsgsbasis}{\mathord{\vcenter{\hbox{\includegraphics[scale=1]{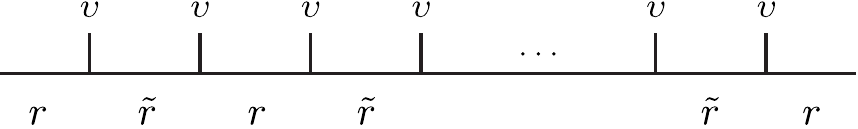}}}}}
\newcommand{\Hcsgs}{\mathord{\vcenter{\hbox{\includegraphics[scale=1]{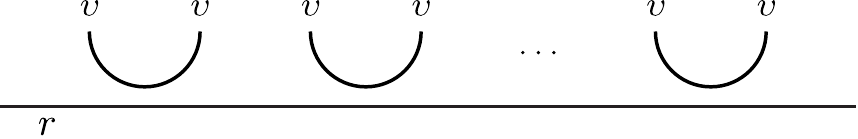}}}}}
\newcommand{\kinkrwx}{\mathord{\vcenter{\hbox{\includegraphics[scale=1]{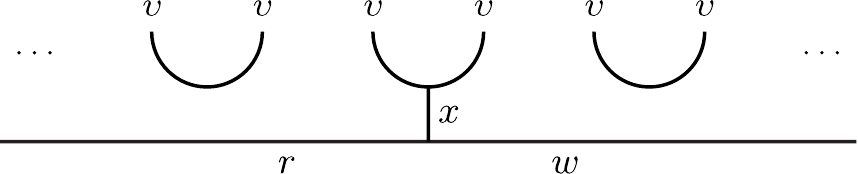}}}}}

\newcommand{\kinkconfigr}{\mathord{\vcenter{\hbox{\includegraphics[scale=1]{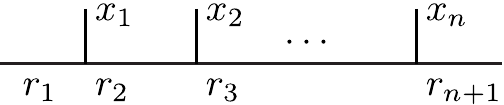}}}}}
\newcommand{\kinkconfigw}{\mathord{\vcenter{\hbox{\includegraphics[scale=1]{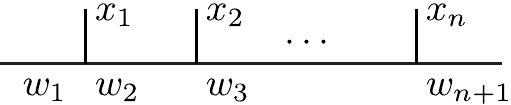}}}}}

\definecolor{purple}{rgb}{0.5,0,0.5}

\definecolor{orange}{rgb}{1,0.5,0}
\definecolor{slategray}{RGB}{112,128,144}
\newcommand{\comment}[1]{{}}

\numberwithin{equation}{section}		
\allowdisplaybreaks     

\title{\vspace{-1.5cm}Topological Defects on the Lattice: Dualities and Degeneracies\vspace{.3cm}}

\author{David Aasen$^{1}$, Paul Fendley$^2$, Roger S. K. Mong$^{3}$}

\affil{\vspace{.4cm}$^1$Microsoft Quantum, Microsoft Station Q, and Kavli Institute for Theoretical Physics, \\University of California, Santa Barbara, CA 93106-6105, USA}
\affil{
$^2$ All Souls College and Rudolf Peierls Centre for Theoretical Physics, \\ University of Oxford, Clarendon Laboratory, Oxford OX1 3NP, UK}
\affil{$^3$Department of Physics and Astronomy and Pittsburgh Quantum Institute, \\ University of Pittsburgh, Pittsburgh, PA 15260, USA  \vspace{-.1cm}}

\begin{document}

\maketitle

\begin{abstract}
We construct topological defects in two-dimensional classical lattice models and quantum chains.  
The defects satisfy local commutation relations guaranteeing that the partition function is independent of their path.  
These relations and their solutions are extended to allow defect lines to fuse, branch and satisfy all the properties of a fusion category.  We show how the two-dimensional classical lattice models and their topological defects are naturally described by boundary conditions of a Turaev-Viro-Barrett-Westbury partition function.
These defects allow Kramers-Wannier duality to be generalized to a large class of models, explaining exact degeneracies between non-symmetry-related ground states as well as in the low-energy spectrum. 
They give a precise and general notion of twisted boundary conditions and the universal behavior under Dehn twists. 
Gluing a topological defect to a boundary yields linear identities between partition functions with different boundary conditions, allowing ratios of the universal $g$-factor to be computed exactly on the lattice.  
We develop this construction in detail in a variety of examples, including the Potts, parafermion and height models. 
\vspace{0.3cm}
\end{abstract}

%
%
%
%
%

\tableofcontents


\section{Introduction}

The renormalization group provides a method for understanding how a field theory arises as the continuum limit of a many-body lattice model. It gives both quantitative and qualitative information on which microscopic degrees of freedom and couplings remain relevant, and so enter into the effective long-distance description. In its half-century of existence, it has transformed the way theorists think about both statistical mechanics and field theory.

It is important to remember, however, that the renormalization group is not the only method for understanding how continuum arises from lattice. In this paper we go deeper than effective field theory, as we exploit the remarkable confluence between integrable lattice models, conformal field theory, and mathematics. In particular, we utilise the profound connection between knot and link invariants \cite{Jones1987} and lattice statistical mechanics \cite{Jones1989,Wadati1989,Wu1992}, a line of development that had started well before \cite{Temperley1971,Fortuin1971,Baxter1976}. 
Since this classic work, substantial progress has been made in developing the mathematics involved. In particular, much more is known about {\em fusion categories}. Among other things, fusion categories give a precise way of manipulating graphs while preserving their topological invariants  \cite{MSReview89,barrett1999,kup_spider,blob_paper,jones_pa,Kitaev2006}. Our aim is to show how the topological framework not only provides an elegant setup for lattice statistical mechanics, but gives a perspective both qualitatively and quantitatively useful. 

We use the inherent topological structure of a wide class of two-dimensional lattice models to define {\em topological defect lines} with remarkable properties. The partition function of a lattice model defined using a fusion category is independent of deformations of the paths of any topological defects present. Moreover, the construction results in a host of exact linear identities relating partition functions with different defect configurations. The reason the topological approach is so powerful is that many constraints are necessary to enforce consistency. The solutions of these constraints are simple to express in terms of topological data, whereas a brute-force approach typically is horrendously complicated.  

Universal results can be obtained simply and exactly from lattice computations.
One main result is our demonstration how lattice topological defects implement {\em dualities}, extending the work of Kramers and Wannier \cite{Kramers1941} well beyond the Ising and Potts models where it is familiar.
We hope to convince that these defects provide the best way of understanding why and how dualities work for two-dimensional classical lattice models and for quantum spin chains. The key observation is that distinct models can be defined using the same fusion category. The dualities then arise in a very natural fashion, as the topological defects built using the category separate regions described by the different models. When the model is the same on both sides, the self-duality places strong constraints on it, allowing us to compute universal quantities.

The topological defects we construct inherit properties from the fusion category used to define them, providing a set of linear identities relating partition functions involving different defect configurations.  In essence, the microscopic definitions of the lattice model and its Boltzmann weights result in macroscopic constraints on the partition functions.  
A basic example is
\begin{align}
 \BubbleMacroLeft \; =\; \delta_{RB} \sqrt{\frac{d_P d_G}{d_R}} \; \BubbleMacroRight 
\label{bubble}
\end{align}
where $R,G,P,B$ label the different types of defects. The numbers $d_a$ are given by the category, and are known as {quantum dimensions}. Roughly speaking, the weight of a single defect line with label $a$ is $d_a$. 
A more general set of relations are known as $F$-moves.  They describe a change of basis, relating partition functions with distinct defect configurations as
\begin{align}
\FmoveMacroLeft \ =\  \sum_{Y} \big[ F_{P}^{RGB} \big]_{XY}\FmoveMacroRight\quad .
\label{MacroF}
\end{align}
The coefficients are called $F$-symbols, and are a glorified version of the $6j$ symbols familiar from quantum mechanics.
Finding a consistent set of $F$-symbols satisfying \eqref{MacroF} results in a huge number of non-linear relations.
The category hands us the solution.

Relations like \eqref{bubble} and \eqref{MacroF} are exact on the lattice. When a lattice model we study has a continuum limit, such partition function identities become exact relations for the corresponding field-theory quantities. These relations allow us to compute exact universal quantities in terms of topological invariants. When the lattice model is critical and ensuing field theory is conformally invariant, these quantities can be compared to those computed using  conformal field theory (CFT) \cite{BPZ:CFT:84}. One type of universal quantity we derive exactly and rigorously on the lattice is the ratio of $g$-factors for conformal boundary conditions \cite{Cardy1989,AffleckLudwig}. Another is the shift in momentum arising from twisting the boundary conditions using a topological defect. This quantity is related to the eigenvalues of a Dehn twist, and hence the conformal spin in the corresponding CFT \cite{Cardy1986}. The relations \eqref{bubble} and \eqref{MacroF} themselves correspond nicely to those arising directly in CFT \cite{Petkova2001,Frohlich:2006,Frohlich2009}. 

The connection of our approach to CFT brings the story full circle. Fusion categories were prefigured in the statistical mechanics of models having CFTs describing their continuum limit \cite{Temperley1971,Fortuin1971,Baxter1976}.  
Canonical examples of such lattice models are the Ising and Potts models  \cite{Baxter1982,Wu1992} and their ``parafermion'' generalizations \cite{Fradkin1980,Fateev82}, along with the integrable height models of Andrews, Baxter and Forrester \cite{Andrews1984}.  Characterizing the topological properties of fields in rational CFT provided a crucial impetus for the introduction of modular tensor categories (fusion categories with more data, including braiding)  \cite{MSReview89}. Further connections of CFT to topology and knot invariants came through the Chern-Simons approach to the Jones polynomial \cite{Witten:1989}. Our work exploits the subsequent progress in the understanding of fusion categories to derive new properties of the very same lattice models.  
We emphasize, however, that even though integrable models and conformal field theory are deeply embedded in the story, our method yields many results for non-integrable and non-critical models as well, duality being one prominent example.

Despite this paper being a sequel, it is self-contained. In \cite{Aasen2016} (henceforth referred to as Part~I), we defined the Ising model in traditional fashion, and explained how topological defects had some rather magical properties. Underlying all these results were fusion categories. The magic remains in the much more general results described here, but now we display our tricks.

We are well aware that many readers may be unfamiliar with the topological approach to lattice models. We therefore have written three review sections describing the setup. In section \ref{sec:Fusion_Theory} we give a concise introduction to fusion categories, focusing on how to compute topological invariants. In section \ref{sec:statmech}, we explain how to build lattice height models using these categories. The Boltzmann weights of Potts, parafermion and ABF height models, among others, then can be related to topological invariants. In section \ref{modules} we explain the mathematics underlying our construction of lattice topological defects, the Turaev-Viro-Barrett-Westbury state sum.

Sections \ref{sec:topological_defect_lines} and \ref{sec:trivalent} contain our core results. The definition of lattice topological defect lines and the proof they satisfy the defect commutation relations is given in section \ref{sec:topological_defect_lines}. In section \ref{sec:trivalent}, we define topological junctions of defects, and prove that defect lines themselves satisfy the rules of the fusion category. In our opinion, the brevity of these two sections demonstrates the substantial value of the categorical approach.

We then give applications. In section \ref{sec:duality} we implement Kramers-Wannier duality using topological defects, and generalize it to a much larger class of models. Section \ref{sec:selfduality} explores a particular consequence of self-duality, enabling the derivation of degeneracies in the ground and low-lying states of certain non-critical models. We show in section \ref{sec:bdry} how our setup provides a natural way of describing boundary conditions in terms of states. In the critical case, it gives a way of computing universal ratios of $g$-factors easily on the lattice. In section \ref{sec:twists}, we define twisted boundary conditions in terms using topological defects and derive identities for the resulting partition functions. By understanding the behavior under Dehn twists, we explain how to extract exact results for conformal scaling dimensions as well. The conclusions are given in section \ref{sec:conclusion}.


\section{Fusion categories and diagrammatics}
\label{sec:Fusion_Theory}

In this section we describe the foundational tool in our analysis, a {\em fusion category}.  Many good reviews cover this subject and much more, including \cite{MSReview89,barrett1999,Bakalov2001,etingof2005,
Kitaev2006}. 
Fusion categories are well known to play important roles in representation theory, in conformal field theory, in the properties of anyons, and in topological quantum field theory.
They play an equally important role in the lattice models studied in this paper, essential both in the construction and in the analysis.
We aim here to provide a practical and computationally useful introduction.

\begin{figure}[h]
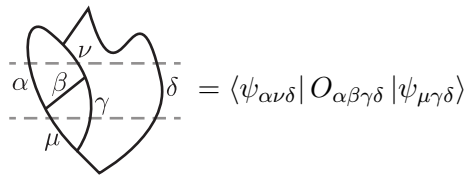

	\[\FusionDiagram \; = \bra{\psi_{\alpha \nu \delta}} O_{\alpha \beta \gamma \delta} \ket{\psi_{\mu \gamma \delta}}\]
	\caption{A fusion diagram, i.e., a labeled planar graph. For simplicity we leave out vertex labels and edge orientations. Using the standard notation of quantum mechanics, the diagram here can be viewed as an inner product between bra labeled $\mu, \gamma, \delta$, a ket labeled $\alpha, \nu, \delta$, and an operator with labels $\alpha, \beta, \gamma, \delta$.	 }	
	\label{fig:Fusion-diagram}
\end{figure}

One particularly important aspect of a fusion category is that it allows one to associate a number with a ``fusion diagram'', an example of which is shown in Fig.~\ref{fig:Fusion-diagram}. 
This association is called ``evaluating'' the diagram, and the resulting number is independent of any local deformations of the diagram.
The fusion category plays two roles in this evaluation. 
It provides identities between the evaluations of different diagrams that can be applied repeatedly and judiciously to yield a sum over simple diagrams.  Just as importantly, the fusion category also guarantees that the evaluation of a given diagram is invariant under topology-preserving manipulations (after a few simplifying assumptions).
For this reason it is a central tool in knot theory, where a topological invariant is associated with a knot or link by evaluating a corresponding sum over fusion diagrams. 

\subsection{Fusion rules}
\label{sec:fusionrules}

To begin gently, we first discuss the fusion rules associated with the simple objects of a fusion category. The simple objects $a,b,c$ are the labels on the diagrams to be evaluated. The fusion rules are specified by non-negative integers $N_{ab}^c$ which govern how two simple objects labelled by $a$ and $b$ fuse into $c$:
\begin{align}
	a \otimes b  = \bigoplus_c N_{ab}^c \, c.
\label{Nabcdef}	
\end{align}
The tensor product $\otimes$ is the fusing, and the sum is over all simple objects.
We say $a$ and $b$ fuse to $c$ when $N_{ab}^c> 0$, with this integer giving the number of \emph{fusion channels}.
In category language, $N_{ab}^c = \text{dim} \; \text{mor}( a \otimes b \rightarrow c) $. 
We refer to the $N_{\ast\ast}^\ast$ as the \emph{fusion symbols}.

A familiar example of fusion rules come from taking tensor products of representations of algebras or groups, where each irreducible representation $a,b,c,\dots...$ corresponds to a simple object of the category. The fusion rules specify how to write the tensor product $a\otimes b$ as sums over irreps labelled by $c$.  
A physical example arises when simple objects correspond to operators in a field theory,
with the fusion describing their behaviour under the operator product expansion: $N_{ab}^c$ is the number of times the operator $c$ appears in the operator product expansion of operators $a$ and $b$.
This example has played a particularly important role in conformal field theory in two \cite{MooreSeiberg:89,MSReview89} and higher \cite{Poland2018} spacetime dimensions. 
Another physical manifestation arises from the study of anyons in topologically ordered systems.
Each (isomorphism class of) simple object is identified with a different species of anyon, and the equality in \eqref{Nabcdef} means that at long distances the right- and left-hand sides cannot be distinguished \cite{Kitaev2006}.

The simple objects and the corresponding fusion coefficients $N_{ab}^c$ in a fusion category must satisfy a variety of conditions, including \cite{etingof2015,etingof2005}
\begin{enumerate}
\item	 A fusion category has a finite number of simple objects. 
\item
Among the objects there must be a unique ``identity'', labelled $\I$ or $0$, such that $\I \otimes a = a \otimes \I = a$ for all $a$. 
This implies that $N_{a \I}^b = N_{\I a}^b = \delta_{ab}$. 
\item
For every object $a$ there exists a unique `dual' object $\bar{a}$ such that $\I$ appears in their fusion product.
For the most part in this paper, we assume that all objects are self-dual, so that in the anyon picture every particle is its own antiparticle.
This implies that $a \otimes a = \I \oplus \cdots$, with the fusion multiplicity for the identity channel equal to one.
Equivalently $N^\I_{ab} = \delta_{ab}$.
\item
The fusion rules must be associative:
\begin{align}
	a \otimes(b \otimes c) = (a\otimes b) \otimes c \quad
	\Rightarrow \quad \sum_{x} N_{ax}^z N_{bc}^x = \sum_{y} N_{ab}^y N_{yc}^z \,.
\label{associativity}
\end{align}
\item	A pivotal structure of the fusion category implies that if $a \otimes b = c \oplus \cdots$, then $c \otimes b^\ast = a \oplus \cdots$.
In terms of fusion symbols,
\begin{align}
	N_{ab}^c = N_{b \bar{c}}^{\bar{a}} = N_{\bar{c} a} ^{\bar{b}} = N_{\bar{b} \bar{a}} ^{\bar{c}} \,.
	\label{eq:pivotal_N}
\end{align}
Combined with the self-dual condition, the fusion symbol $N_{ab}^c$ are completely symmetric under permutations of the three objects.
\end{enumerate}

{\em Quantum dimensions} are important quantities associated with a fusion category.  
We first define a set of matrices $N_a$ based on the fusion coefficients, each of which has entries $(N_a)^c_b=N^c_{ab}$.
For each simple object, we define its quantum dimension to be
\begin{align}
	d_a \defineas \operatorname{maximal\ eigenvalue}\big[ N_a \big] .
	\label{eq:QD_def}
\end{align}
This also known as the Frobenius-Perron dimension.
As the entries of the matrix $N_a$ are non-negative and the sum of each column is positive ($\sum_c N_{ab}^c > 0$), $d_a$ is unique, real and positive.
As a consequence of the associativity condition~\eqref{associativity}, the quantum dimensions satisfy
\begin{align}
	d_a d_b = \sum_c N_{ab}^c d_c \,.
	\label{eq:ddNd}
\end{align}
As we will explain, the quantum dimension provides a measure of the Hilbert space dimension associated to a simple object.

We here give a few examples of fusion rules.

\paragraph{Ising.}
One of the most important examples is given by the Ising fusion rules, which arise for example in the operator products of the primary fields in the Ising conformal field theory \cite{BPZ:CFT:84}. There are three elements denoted $\I, \sigma$, and $\psi$. The non-trivial fusion rules are given by:
\begin{align}
	\psi \otimes \psi = \I, \quad \sigma \otimes \sigma = \I \oplus \psi, \quad \psi \otimes \sigma = \sigma.
	\label{eq:Ising_fusion_rules}
\end{align}
The remaining fusion rules can be inferred from the rules above, giving for example $N_{\psi \psi}^\I = N_{\sigma \sigma}^\I = N_{\sigma \sigma}^\psi = N_{\psi \sigma}^\sigma = N_{\sigma \psi}^\sigma = 1$.
The quantum dimensions of these basis elements are given by $d_\I = d_\psi = 1$ and $d_\sigma=\sqrt{2}$.

\paragraph{Fibonacci.}
Another common example has only two elements, $\I$ and $\tau$, with fusion rules
\begin{align} 
	\I \otimes \tau &= \tau \otimes \I = \tau,
	&	\tau \otimes \tau &= \I \oplus \tau.
\label{fusionFib}
\end{align}
The reason for the name Fibonacci is apparent from the non-trivial fusion rule in (\ref{fusionFib}); continually fusing $\tau$ elements together gives multiplicities of the Fibonacci numbers.
Here we have $d_\I=1$ and $d_\tau=\frac{1+\sqrt{5}}{2}$ is the golden ratio.

\paragraph{$\SU2$.}  The most familiar example of fusion rules come from the representations of the group or Lie algebra $\SU2$.
The elements are the irreducible representations, which we label using physics conventions by their spin, the non-negative integers and half-integers $0,\,\frac12,\,\dots$, with $0$ being the identity object.
Combining two spins gives a direct sum of representations, e.g.\ $\frac12 \otimes \frac12 = 0 \oplus 1$.
In general, fusion symbols for $\SU2$ are
\begin{align}
	N^a_{bc} \;=\; \bigg\{ \begin{tabular}{l @{\qquad} l}
		1	&	$a+b \geq c$,\; $b+c \geq a$,\; $c+a \geq b$,\; and $a+b+c \in \mathbb{Z}$,
	\\[0.2ex]	0	&	otherwise.
	\end{tabular}
	\label{eq:def_SU2_N}
\end{align}
The quantum dimensions coincides with the irrep dimensions: $d_a = 2a+1$.

However, there is no $\SU2$ fusion category, as there are infinite number of irreducible representations. Instead, we consider the

\paragraph{$\mathcal{A}$-series.}
The fusion category associated to $\mathcal{A}_{k+1}$, with $k$ a non-negative integer, has $k+1$ simple objects labeled $0,\frac12,\dots,\frac{k}{2}$.  
These fusion rules are very similar to those of $\SU2$ in \eqref{eq:def_SU2_N} but require also the ``truncation'' $a+b+c \leq k$, giving
\begin{align}
	N^a_{bc} \;=\; \bigg\{ \begin{tabular}{l @{\qquad} l}
		1	&	$a+b \geq c$,\; $b+c \geq a$,\; $c+a \geq b$,\; $a+b+c \in \mathbb{Z}$,\; and $a+b+c \leq k$,
	\\[0.2ex]	0	&	otherwise.
	\end{tabular}
	\label{eq:def_Ak_N}
\end{align}
The quantum dimensions of these objects are
\begin{align}
	d_h = \frac{ \sin\frac{\pi(2h+1)}{k+2} }{ \sin\frac{\pi}{k+2} } = \frac{q^{h+\frac12} - q^{-(h+\frac12)}}{q^{\frac12}-q^{-\frac12}} ,\qquad\quad
		\textrm{with } q & = \exp\frac{2\pi i}{k+2}\ .
	\label{eq:def_QD}
\end{align}
These fusion rules for $\mathcal{A}_{k+1}$ are identical to those of another fusion category $\SU2_k$, familiar from the study of Wess-Zumino-Witten models in CFT \cite{KZ}. Although closely related, the two categories are not identical, as the additional data described in section \ref{sec:category} differs. 
This  ``$\mathcal{A}$-series'' encompasses the first two examples:
The Ising fusion algebra is equivalent to that of $\mathcal{A}_{3}$ with the identification  ($\I,\sigma,\psi \rightarrow 0,\frac12,1$), while Fibonacci is the subcategory of $\mathcal{A}_4$ restricted to integer elements with ($  \I,\tau\rightarrow 0, 1$).
In general, as we explain below, they are associated with the height models introduced by Andrews, Baxter and Forrester (ABF) \cite{Andrews1984}, and their ``fused'' generalizations.

\subsection{Fusion category data}
\label{sec:category} 

The following data completely defines a fusion category:
\begin{itemize}
\setlength{\itemsep}{0ex}
\setlength{\parskip}{0.1ex}
\item	list of simple objects,
\item	fusion symbols $N^\ast_{\ast\ast}$,
and \item	the $F$-symbols.
\end{itemize}
The $F$-symbols, defined in \eqref{eq:Fmove_def} below, describe linear relations among diagrams. They provide the means to evaluate an arbitrary labeled planar diagram.
Those involving the identity object are written in terms of quantum dimensions $d_\ast$, and so determine them as well.
Of course,  these quantities are not independent. As we detail, a number of self-consistency equations must be satisfied.

To give vast simplifications to the computations and diagrammatics, we impose several restrictions on fusion categories utilized.  We emphasize however that the ideas and methods presented in this paper are applicable to a broader class. The simplifying assumptions are:
\begin{itemize}
\item	We usually assume that simple objects are self-dual: $a \cong a^\ast$.
This condition allows us to omit arrows in our diagrams.
\item There is a non-degenerate trace that can be used to define an inner product; see the discussion around \eqref{dagdef} for details. 
This condition allows us to treat vector spaces as Hilbert spaces, and ensures that the quantum dimensions are always positive.
\item	The fusion category is multiplicity-free, i.e., $N_{ab}^c = 0$ or $1$ for all simple objects $a,b,c$.
\item	The Frobenius-Schur indicators are all $+1$.
\item	We require that there exist a gauge (i.e., basis of splitting operators) such that all the $A$- and $B$-symbols be $+1$ whenever $N_{ac}^b \neq 0$ (see \eqref{ABsymbols}).
\end{itemize}

A fusion category can be presented in many different ways, and also may be extended to contain additional structures such as braiding.
In this paragraph, we state in a more formal language the type of categories we will be utilizing.
We are interested in the data associated with a $\mathbb{C}$-linear semi-simple spherical pivotal fusion category. 
The data are given by a collection of objects built from a finite list of simple objects $\{a, b, c, \cdots \}$;
	a list of morphisms $\operatorname{mor}(a \to b)$ for each pair of objects $a,b \in \mcc$;
	as well as two kinds of multiplications: vertical composition $\circ$ of morphisms and a tensor product $\otimes$ for horizontal composition of morphisms and objects.
The pivotal%
	\footnote{A pivotal fusion category means that there are isomorphisms $V_c^{ab} \xrightarrow{A} V_{\bar{a}c}^b$ and $V_c^{ab} \xrightarrow{B} V_{c\bar{b}}^a$.
		Diagrammatically, this allows fusion diagrams to be rotated.  A consequence is Eq.~\eqref{eq:pivotal_N}.}
 and spherical%
	\footnote{A spherical fusion category is one where the left and right traces agree. In terms of the fusion diagrams we discuss below [e.g.\ see \eqref{eq:def_Tetra2}], the trace corresponds to evaluating the closed loop formed by connecting a line labelled by the simple object $a$ with itself.
		The left and right traces differ by the two ways one can close this strand into a loop on the plane.
		For the categories we are interested in we have $\Tr a = d_a$.}
	structures are additional constraints on the fusion and morphism spaces.

\subsection{Fusion diagrams}
\label{sec:fusiondiagrams}

An exceptionally useful tool in our analysis comes from the relation of a fusion category to the calculus of fusion diagrams.
A fusion diagram is built from a labeled planar graph with no crossings, only trivalent vertices.
Each edge in the planar graph is labelled by an object in the fusion category, and the 
vertices by morphisms of the surrounding labeled edges to the vacuum.
At any trivalent vertex, the labels  $a,b,c$ of the edges touching the vertex must obey $N_{ab}^c=1$. 
The number of allowed labelings for a given graph and fusion category is computable via a ``crossing'' symmetry \cite{Simon2012}.
In our setting, trivalent vertices will be fundamental, and can be used to build any fusion diagram. Hence morphisms in higher-valence vertices can always be split up into compositions of morphisms that can be represented as trivalent vertices.

\begin{figure}
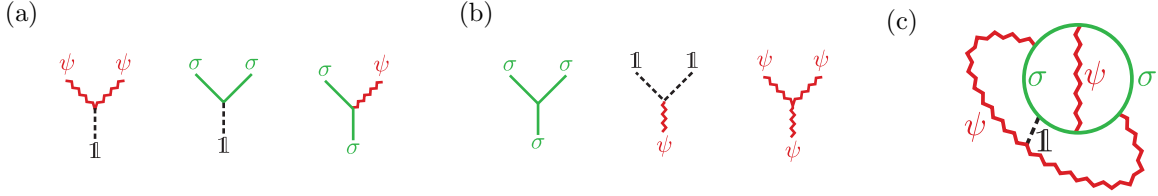

	\center{
	\raisebox{10mm}{\small(a)} \begin{minipage}{47mm}
		\begin{align*} \IsingFusionRulePsiPsiId \quad\quad \IsingFusionRuleSigmaSigmaId \quad\quad \IsingFusionRuleSigmaPsiSigma \end{align*}
	\end{minipage}
	\hspace{5mm}
	\raisebox{10mm}{\small(b)} \begin{minipage}{31mm}
		\begin{align*} \IsingFusionRuleSigmaSigmaSigma \quad\quad \IsingFusionRuleIdIdPsi \quad \quad \IsingFusionRulePsiPsiPsi \end{align*}
	\end{minipage}
	\hspace{5mm}
	\raisebox{9mm}{\small(c)} \begin{minipage}{35mm}
		\begin{align*}\IsingFusionDiagram\end{align*}
	\end{minipage}
	}
	\\
	\caption{
		(a) Possible vertices permitted by the fusion rules Eq.~\eqref{eq:Ising_fusion_rules} in the Ising fusion category.
		(b) Examples of vertices forbidden by the Ising fusion rules.
		(c) An example of a fusion diagram in the Ising fusion category, with elements $\I$, $\sigma$, and $\psi$.
	}
	\label{fig:Isingfusiondiagram}
\end{figure}

We will explain in the next subsection that when the diagram has no ends (boundaries), it is associated with a complex number, the ``evaluation'' of the diagram, and is invariant under smooth deformations which preserve the topology. When the diagram has ends, we associate it with a vector or wave function.
We are free to add/remove lines with trivial labels (which depending on context we write as $0$ or $\I$) to a diagram without changing its evaluation or the associated vector space. The allowed vertex labelings for the Ising fusion rules are illustrated in Fig.~\ref{fig:Isingfusiondiagram}(a); in \ref{fig:Isingfusiondiagram}(b) we show vertices that are {\em not} allowed here.
An example of a diagram without ends for the Ising fusion category is given in Fig.~\ref{fig:Isingfusiondiagram}(c).

We often work with operators acting on vector spaces, as lattice models in statistical mechanics are convenient to analyze using transfer matrices and Hamiltonians. A vector space is associated to a diagram with ends. The simplest are $V^{a_1 a_2 \cdots a_n}$ and $V_{a_1 a_2 \cdots a_n}$, the vector spaces spanned by diagrams with open legs $a_1, \dots, a_n$ emanating upward and downwards respectively. 
For example, the bottom and top pieces of the fusion diagram in figure \ref{fig:Fusion-diagram} live in the vector spaces
\begin{align}
	\FusionDiagramVu &\in V^{\mu \gamma \delta} ,
&	\FusionDiagramVd &\in V_{\alpha\nu\delta} \,.
\end{align}
When the labels on the external legs match, we can glue the two together to give a closed diagram that can be evaluated by the techniques given below. This defines an inner product on the vector spaces, where 
$V^{a_1 a_2 \cdots a_n}$ and  $V_{a_1 a_2 \cdots a_n}$  are dual to each other. The inner product of two diagrams with mismatched legs is simply defined to be zero.
These vector spaces are canonically isomorphic via conjugation; we can convert vectors between $V_{a_1a_2\cdots}$ and $V^{a_1a_2\cdots}$ by simply flipping the figure vertically.
In the context of our example:
\begin{align}
	\FusionDiagramVuFlipped = \left( \FusionDiagramVu \right)^\dag .
\label{dagdef}	
\end{align}
This definition yields a Hermitian inner product structure that allows us to treat them as Hilbert spaces. We often utilize the  convenient quantum-mechanical notation of bras and kets to represent the elements of these vector spaces.

Vector spaces with upper and lower indices are defined in the obvious way. The elements of such spaces can be thought of as linear operators: a diagram with legs $a_1,a_2,\dots$ above and $b_1,b_2,\dots$ below maps the vector space $V^{b_1b_2\cdots}$ to $V^{a_1a_2\cdots}$. The vector space spanned by these operators is denoted $V^{a_1\cdots}_{b_1\cdots}$.
A diagram can be split into many pieces, and evaluated as a product of its constituents. 
The topmost part becomes the bra, the bottommost the ket, and the intervening slices operators, for example
\begin{align}
	\FusionDiagram \;=\; \Braket{\FusionDiagramVd | \FusionDiagramOp | \FusionDiagramVu} .
	\label{eq:FusionDiagramCut}
\end{align}
For example, the middle part of Eq.~\eqref{eq:FusionDiagramCut} is an operator
\begin{align}
	\FusionDiagramOp \;\;\in V_{\mu\gamma\delta}^{\alpha\nu\delta}
\end{align}
which takes a wavefunction with legs $\nu,\gamma,\delta$ to one with legs $\alpha,\nu,\delta$:
\begin{align}
	\left[\FusionDiagramOp\right] \Ket{\FusionDiagramVu} \;\;=\;\; \Bigg|\,\FusionDiagramOpVu\,\Bigg\rangle \,.
\end{align}

The building blocks for operators are the splitting and fusion vertices, along with ``identity'' operators:
\begin{align}
	\Yabc &\in V^{ab}_c \,,
&	\YabcX &\in V_{ab}^c \,,
&	\bigg|_a \in V^a_a \,.
\end{align}
One also encounters diagrams such as $\UaaSmall \in V^{aa}_0$, but this is simply a special case of the splitting operator.
The dimensionality of these vector spaces are given by
\begin{align}
	\Vdim V^{ab}_c = \Vdim V_{ab}^c = N_{ab}^c  \,,
	\qquad	\Vdim V^a_b = \delta_{ab}  \,.
\end{align}
The physical interpretation of these equations is as follows.
The former equation says that the number of linearly independent ways to fuse $a \otimes b \rightarrow c$ is precisely given by $N_{ab}^c$.
The latter equation ensures that simple objects cannot ``mutate'' into another.

The lattice models we study are conveniently defined in terms of a \emph{fusion tree}.
For any sequence $\upsilon_1,\upsilon_2,\dots$ and a fusion channel $\eta$, it is a diagram of the form
\begin{align}
	\FusionTree \quad\in V^{\upsilon_1 \upsilon_2 \cdots \upsilon_L}_\eta .
	\label{eq:RFusionTree}
\end{align}
The diagram illustrated is a right-branching fusion tree, in that each $\upsilon_i$ sprouts from the main branch with labels $\eta,h_1,h_2,\dots$.
The labels must satisfy the fusion rules, i.e., $N^{h_{i-1}}_{\upsilon_i h_i} = 1$ for each $i$.
Here we let $h_0 = \eta$ and $h_{L-1} = \upsilon_L$ to simplify notation.
The key statement is as follows.
\begin{align}
	\text{The set of right-branching fusion trees forms an orthogonal basis for $V^{\upsilon_1 \upsilon_2 \cdots \upsilon_L}_\eta$}.
\end{align}
In other words, the basis of states for $V^{\upsilon_1 \cdots \upsilon_L}_\eta$ are enumerated by the set of $h_1,\dots,h_{L-2}$ satisfying the appropriate fusion constraints.
(These states have non-trivial normalization factors, which we define below.)
Equivalently, we may define these vector spaces recursively as
\begin{align}
	V^{\upsilon_1 \upsilon_2 \cdots \upsilon_L}_\eta &\cong \bigoplus_{h_1} V^{\upsilon_1 h_1}_\eta \otimes V^{\upsilon_2 \upsilon_3 \cdots \upsilon_L}_{h_1} .
	\label{eq:V_recur_def}
\end{align}
In general, for an arbitrary operator space,
\begin{align}
	V^{a_1a_2\cdots}_{b_1b_2\cdots} &\cong \bigoplus_x V^{a_1a_2\cdots}_x \otimes V^x_{b_1b_2\cdots} .
	\label{eq:Vop_decomp}
\end{align}

Consider a fusion tree with large $L$ and all $\upsilon_j=\upsilon$ the same.
The asymptotic growth of the dimension of this vector space depends on the quantum dimension $d_\upsilon$ as
\begin{align}
	\Vdim V^{\overbrace{\scriptstyle\upsilon\upsilon\upsilon\cdots\upsilon\upsilon\upsilon}^{L\text{ indices}}}_\eta \sim (d_\upsilon)^L \text{ as } L \rightarrow \infty\ .
	\label{eq:QD_asymp}
\end{align}
This follows from the definition (\ref{eq:QD_def}) via the recurrence relation \eqref{eq:V_recur_def}. For example, in the Fibonacci category, it is easy to see that these dimensions are related to Fibonacci numbers, which indeed grow with $L$ as $\big(\frac{1+\sqrt{5}}{2}\big)^L$.

\subsection{Evaluating fusion diagrams}
\label{sec:evaluation}

We now turn to the method of evaluation, the process for associating a number with any fusion diagram without ends.
This is done iteratively by applying successive manipulations to the diagram.
In this section we introduce all the linear relations on diagrams necessary to evaluate any planar graph. 
To do so, we will need to introduce additional symbols in our description of a fusion category, namely $\varkappa$-, $A$-, $B$-, and $F$-symbols.
Many highly constraining self-consistency equations for these symbols are needed so that topologically equivalent graphs evaluate to the same number.

\paragraph{Isotopy moves.}
The simplest move corresponds to a process where we start with particle $a$, generate an additional pair out of the vacuum, and annihilate one of the newly produced particles with the original.
Since the vector space $V^a_a$ associated to this process is one-dimensional, the only possibility is to pick up a phase:
\begin{align}
	\FSR \;=\; \varkappa_a\, \bubbleb \;=\; \FSL\ ,
\label{FSdef}
\end{align}
where $\varkappa_a$ is called the Frobenious-Schur indicator of $a$ (note that we have assumed that $a$ is self dual), and takes on a value $\pm1$.
The Frobenius-Schur indicators are invariants of a fusion category, independent of the fusion symbols.

As we discuss the rules for planar calculus, we will also illustrate them by evaluating the diagram in Fig.~\ref{fig:Fusion-diagram}.
The first move is to remove the bend in the $\delta$ line using \eqref{FSdef}:
\begin{align}
	\FusionDiagrama \;=\; \varkappa_{\delta}\, \times\, \; \FusionDiagramb \,.
\end{align}
Next we look at the ``pivotal'' structure of the theory, which means there are isomorphisms between fusing and splitting spaces. With the restriction that all $N_{ab}^c=0,1$ these isomorphisms involve only 
a phase due to bending of the lines, namely
\begin{align}
	\ALD= A^{\bar{a} c}_b \ARD\ , \qquad\quad	 \BLD  = \varkappa_b B^{b \bar{c}}_a \BRD	\,.
	\label{ABsymbols}
\end{align}
These define the $A$- and $B$-symbols, which in general are also complex phase factors. As opposed to the Frobenius-Schur indicators, the $A$- and $B$-symbols are not gauge invariant, meaning they depend on the basis chosen.
Applying an ``$A$-move'' to our example gives
\begin{align}
	\FusionDiagramb \; =\; A^{\bar{\alpha} \nu}_\delta \times \; \FusionDiagramc \,.
\end{align}
Most of the categories we study have all $\varkappa_\delta=1$, and all $A$- and $B$-symbols $+1$ whenever $N_{ac}^b \neq 0$.
Along with requiring all particles to be self-dual, these two conditions allow us to make arbitrary deformations to a diagram without incurring any phases.

Lastly, we remark that non-simple objects may also label edges on a fusion diagram.
Semisimplicity of the category implies that any object can be written as a direct sum of simple objects. 
Correspondingly, to evaluate a diagram containing a non-simple object $x = a_1 \oplus a_2 \oplus \cdots$ decomposed in terms of simple objects $a_i$, one replaces the evaluation of a diagram with label $x$,
by a sum of evaluations of diagrams with edges labeled $a_i$.

\paragraph{$F$-moves.}
The $F$-move is the workhorse of the fusion category. It allows us to make non-trivial transformations on the diagrams that preserve the evaluation.  They arise as basis changes for the vector spaces defined via fusion diagrams.
There are two ways to write a basis for the operators in $V^{abc}_d$, either as left-branching or right-branching trees.
\begin{align}
	\FmoveLNA &\in \bigoplus_x V^{ab}_x\otimes V^{xc}_d \cong V^{abc}_d,
&	\FmoveRNA &\in \bigoplus_y V^{ay}_d \otimes V^{bc}_y \cong V^{abc}_d,
	\label{eq:Vabc_decomp}
\end{align}
The operators in the left-branching basis are enumerated by $x$ while those of the right-branching basis by $y$, the unitary map between the two bases is referred to as an $F$-move.
The coefficients for the transformation are provided by the $F$-symbols, which carries six indices: four for the external labels and two internal ones that label orthogonal vectors in each space.
Diagrammatically it is defined via
\begin{align}
	\FmoveLNA = \sum_y \big[ F^{abc}_d \big]_{xy} \FmoveRNA\ . 
	\label{eq:Fmove_def}
\end{align}
The transformation $\big[ F^{abc}_d \big]_{xy}$ is unitary in the indices $x,y$, evident from the fact that it provides the isomorphism between the left/right side of Eq.~\eqref{eq:Vabc_decomp}.

An $F$-move can be applied to any pair of adjacent trivalent vertices in any diagram; adjacent vertices are connected by one (or more) lines. Applying the $F$-move then modifies the diagram. In our ongoing example, this results in
\begin{align}
	\FusionDiagramc \; =\; \sum_\rho \big[ F^{\alpha \beta \gamma}_{\bar{\delta}} \big]_{\mu \rho}  \times \; \FusionDiagramd\ .
\label{ongoingexample}	
\end{align}

Considering vector spaces with more indices leads to a variety 
of consistency conditions that guarantee different sequences of $F$-moves produce identical results \cite{MooreSeiberg:89}.
The pentagon equation, given below in (\ref{pentagon}), is an important one.
Fusion categories by definition satisfy these conditions, and many solutions are known and classified.
A general classification does not yet exist, although
all modular tensor categories (fusion categories that also have a braiding among other constraints) with up to five objects are known \cite{Rowell, Bruillard2016}.
The consistency conditions for a given fusion algebra do not have a unique solution; $\mathcal{A}_{k+1}$ and $\SU2_k$ have the same fusion rules but different $F$-symbols.
The main differences are that the latter category has non-trivial Froebenius-Schur indicators, while all quantum dimensions are positive.
We utilize $\mathcal{A}_{k+1}$ here; its $F$-symbols are written out in Appendix~\ref{app:TetraSym}.

\paragraph{Loop and bubble removal}

Once one has solved for the $F$-symbols then every planar diagram can then be reduced to one that only has bubbles left in it, with each bubble having at most two lines attached to it. A bubble with no lines attached evaluates to the quantum dimension of the label, while one with two lines attached can be reduced to a line with no bubble:
\begin{align}
	\QD = d_a \,, \qquad\qquad \bubblea = \delta_{ac}\sqrt{\frac{d_b d_{b'}}{d_a}}\ \bubbleb \;.
	\label{eq:loop_removal}
\end{align}
The former is a special case of the latter with $a=c=0$. Note also that any bubble with only one line attached to it (not including identity labels) evaluates to zero. 
One can use these to find a special case of the $F$-symbol:
\begin{align}
	\recouplinga \;=\; \sum_c \sqrt{\frac{d_c}{d_a d_b}}\ \recouplingb
	\qquad\Rightarrow\qquad \big[F^{\bar{a}ab}_b\big]_{0c} = \sqrt{\frac{d_c}{d_a d_b}} \,.
\label{F0c}
\end{align}
This relation on the left-hand side follows by joining the $a$ and $b$ lines in two places using the $F$-move, and then using \eqref{eq:loop_removal} to remove the resulting bubble.

Using these relations we can finish evaluating Fig.~\ref{fig:Fusion-diagram}. Repeatedly applying \eqref{eq:loop_removal} to \eqref{ongoingexample} gives
\begin{align}
	&\FusionDiagramd \; = \;  \delta_{\nu \rho} \sqrt{\frac{d_\beta d_\gamma}{d_\rho}} \; \FusionDiagrame
	\; = \; \delta_{\nu \rho} \sqrt{\frac{d_\beta d_\gamma}{d_\rho}} \sqrt{\frac{d_\alpha d_\rho}{d_\delta}} d_\delta \ ,	\cr
	&\implies \qquad \FusionDiagrama \;=\; \big[ F^{\alpha\beta\gamma}_\delta \big]_{\mu\nu} \sqrt{d_\alpha d_\beta d_\gamma d_\delta} \,.
\end{align}
We could have evaluated this diagram by using a different initial $F$-move to derive one of the many consistency conditions on $F$-symbols.  For example, the evaluation is also $[F^{\alpha\mu\gamma}_\nu]_{\beta\delta}\sqrt{d_\alpha d_\gamma d_\mu d_\nu}$,

We now can compute the inner product for the fusion trees in Eq.~\eqref{eq:RFusionTree}.
The inner product of two right-branching trees in $V^{\upsilon_1\cdots\upsilon_L}_\eta$ is 
\begin{align}
	\Destroyer  \;\;=\;\; \sqrt{d_\eta} \prod_{j=1}^L \sqrt{d_{\upsilon_j}} \prod_{j=1}^{L-2} \delta_{h_j g_j}  \,.
	\label{eq:destroyer}
\end{align}
This diagram has been evaluated by repeated application of Eq.~\eqref{eq:loop_removal}.
Notably, the two right-branching fusion trees are orthogonal unless the inner indices $\{h\}$ and $\{g\}$ agree; this justifies our earlier claim that the states form an orthogonal basis.

As an example of the power of the $F$-symbols, we can use them to derive another identity for the quantum dimensions:
\begin{align}
\sum_{a,b \in {\cal C}} d_a d_b N_{ab}^z = d_z \mathscr{D}^2\ , \qquad \hbox{where } \mathscr{D}^2\equiv{\sum_{c \in {\cal C}} d_c^2}\ . 
\label{ddNdD}
\end{align}
The sums are over all simple objects in the category, and $\mathscr{D}$ is called the {\em total quantum dimension}. 
To derive this relation, let $\text{cl}(a)$ denote a closed planar loop labeled by $a$, which of course evaluates to $d_a$. Using \eqref{eq:loop_removal} and \eqref{F0c}, we have 
\[\mathscr{D}^2 d_z = \sum_{a \in {\cal C}} d_a \,\text{cl}(a) \text{cl}(z) = \sum_{a \in {\cal C}} d_a\, \text{cl}(a\otimes z) = \sum_{a,b \in{\cal C}} d_a N_{az}^b\, \text{cl}(b) =  \sum_{a,b \in{\cal C}} d_ad_b N_{az}^b\ .\] 
By using the symmetries of the $N_{\ast\ast}^\ast$ symbols described in section~\ref{sec:fusionrules}, \eqref{ddNdD} follows.

\subsection{Tetrahedral symbols}
\label{sec:tetra}
The $F$-symbols satisfy a number of useful symmetry properties. These are conveniently displayed by utilising \emph{tetrahedral symbols}, which are proportional to the $F$-symbols. We define the tetrahedral symbols in terms of the diagram:
\begin{align}
	\Msixj{a & b & c}{\alpha & \beta & \gamma}
	& \;\defineas\;
	\frac{1}{\sqrt{d_a d_b d_c d_\alpha d_\beta d_\gamma}} \tetradiagramNA .
	\label{eq:def_tetra_NA}
\end{align}
Recall from Eq.~\eqref{eq:Fmove_def} that the $F$-symbols give the transformation between two fusion trees in $V_d^{abc}$.
We can thus extract any matrix element of $F_d^{abc}$ via
\begin{align}
	\big[ F_d^{abc} \big]_{x,y} = 
	\frac{\qTr\left[ \left(\,\FmoveRsmallNA\,\right)^{\!\!\dag} \;\; \FmoveLsmallNA \right]}
	     {\qTr\left[ \left(\,\FmoveRsmallNA\,\right)^{\!\!\dag} \;\; \FmoveRsmallNA \right]}
	= \frac{\FmoveCNA}{\FSquared} \;.
	\label{eq:def_Tetra2}
\end{align}
Here $\qTr$ denotes the ``quantum trace'', which simply takes two diagrams with identical external legs joins them together.%
	\footnote{
	Here we have shown the right trace.
	The left trace is given by connecting the line labelled by $d$ on the left side of the picture; 
	the assumption of a spherical category means that both traces agree.}
The numerator is identical (up to a relabeling) to the diagram in Eq.~\eqref{eq:def_tetra_NA}, and the denominator is the normalization of the tree diagram given in Eq.~\eqref{eq:destroyer}.
Therefore, the tetrahedral symbols are related to the $F$-symbols via
\begin{align}
	\big[ F_\alpha^{abc} \big]_{x,y} &= \sqrt{d_x d_y} \Msixj{a & b & x}{c & \alpha & y} .
	\label{Ftosixj}
\end{align}
In context of tetrahedral symbols, the $F$-move is thus
\begin{align}
	\FmoveHprime = \sum_y \sqrt{d_x d_y} \Msixj{a & b & x}{c & \alpha & y} \FmoveVprime .
	\label{eq:tetra_Fmove}
\end{align}
The definition (\ref{eq:def_tetra_NA}) thus gives a nice way of seeing which $F$-symbols vanish.
\begin{align}
\hbox{if }	N^{{a}}_{bc} N^\alpha_{\beta c} N^\beta_{\gamma a} N^\gamma_{\alpha b} = 0\
	\;\Rightarrow\;\ \Msixj{a & b & c}{\alpha & \beta & \gamma} = 0 \;.
\label{vanishingF}
\end{align}
A special case follows from \eqref{F0c}:
\begin{align}
	\Msixj{{a} & a & 0}{b & b & c}
	= \Msixj{a & b & {c}}{{b} & a & 0}
	= \frac{N^c_{ab}}{\sqrt{d_ad_b}}\ .
\label{tetra0}
\end{align}
Using an $F$-move and then removing the bubble gives another useful identity
\begin{align}
	\vertexabcCNA = \sqrt{d_ \alpha d_\beta d_\gamma} \Msixj{a & b & c}{\alpha & \beta & \gamma} \vertexabcNA\ .
	\label{eq:Trivalent_Bubble}
\end{align}

\paragraph{Properties.}
The tetrahedral symbols are convenient as they possess all the symmetries of a tetrahedron.
The symbols are invariant under permutation of the columns:
\begin{align}
	\begin{array}{ccccccc @{} l}
		\Msixj{a & b & c}{\alpha & \beta & \gamma}
	&	=	&	\Msixj{{a} & {c} & {b}}{\alpha & \gamma & \beta}
	&	=	&	\Msixj{{c} & {b} & {a}}{\gamma & \beta & \alpha}
	&	=	&	\Msixj{{b} & {a} & {c}}{\beta & \alpha & \gamma}
	&	.
	\\[-1.6ex]
	&&	\hspace{3ex} \underbracket{\hspace{4ex}}_{\text{exchange}} \hspace{0.7ex}
	&&	\underbracket{\hspace{7ex}}_{\text{exchange}} \hspace{0.7ex}
	&&	\underbracket{\hspace{4ex}}_{\text{exchange}} \hspace{4ex}
	\end{array}
	\label{eq:tetrasym_permute_cols}
\end{align}
They are also invariant under interchanging the rows of any two columns:
\begin{align}
	\begin{array}{ccccccc @{} l}
		\Msixj{a & b & c}{\alpha & \beta & \gamma}
	&	=	&	\Msixj{{a} & \beta & {\gamma}}{{\alpha} & b & {c}}
	&	=	&	\Msixj{{\alpha} & {b} & \gamma}{{a} & {\beta} & c}
	&	=	&	\Msixj{\alpha & {\beta} & {c}}{a & {b} & {\gamma}}
	&	.
	\\[-1.6ex]
	&&	\hspace{3ex}\underbracket{\hspace{4ex}}_{\text{swap rows}}
	&&	\underbracket{\hspace{7ex}}_{\text{swap rows}}
	&&	\underbracket{\hspace{4ex}}_{\text{swap rows}}\hspace{3ex}
	\end{array}
	\label{eq:tetrasym_swap_rows}
\end{align}
Exchanging the top and bottom row is \emph{not} a symmetry: $\smallM{a&b&c}{d&e&f} \neq \smallM{d&e&f}{a&b&c}$ in general.

\paragraph{Orthogonality and pentagon equations.}
The tetrahedral symbols inherit all the properties of the $F$-symbols. 
The unitarity of the $F$-symbols implies
\begin{align}
	\sum_y d_y \Msixj{a & b & x}{c & d & y} \Msixj{a & b & x'}{c & d & y}
	= \frac{ \delta_{xx'} N^{{x}}_{ab} N^{x}_{c{d}} }{d_x} .
	\label{Orthogonality6j}
\end{align}
The pentagon equation
\begin{align}
	\sum_x d_x
		\Msixj{\rho & \nu & g}{\beta & \gamma & x}
		\Msixj{\mu & \rho & h}{\gamma & \alpha & x}
		\Msixj{\nu & \mu & j}{\alpha & \beta & x}
	=	\Msixj{g & h & j}{\alpha & \beta & \gamma}
		\Msixj{g & h & j}{\mu & \nu & \rho} .
		\label{pentagon}
\end{align}
is a consequence of evaluating the diagram
\begin{align}
	\circlespentagonNA.
	\label{pentagon_wheel}
\end{align}
in two different ways. 
The right-hand side follows from \eqref{eq:Trivalent_Bubble} and \eqref{eq:def_tetra_NA}, while the left-hand side follows from three applications of \eqref{eq:tetra_Fmove}.


\section{Statistical mechanics from fusion categories}
\label{sec:statmech}

Here we explain how to define two classes of statistical-mechanical models using the data of a fusion category described in section \ref{sec:Fusion_Theory}. The degrees of freedom in these two classes are respectively {\em geometric objects}, and {\em heights}. 
A key advantage of the latter class is that the Boltzmann weights are {\em local}. If the weights are positive as well, any field theory resulting from taking the continuum limit must necessarily be unitary. Thus many of the height models are well known, the most prominent being the Ising model analyzed in depth in Part~I.
While the connection of the fusion categories to the geometric models is fairly obvious, we explain how many widely studied height models (and some not-so-widely studied) are also best described in terms of category data.
Although virtually all of the results in this section can be found in various places in the literature (see e.g.\ Refs.~\cite{Wadati1989,Wu1992} for reviews), many of the lessons learned have been neglected in the statistical-mechanical world.

A key feature of the formulation here is that, as opposed to most of these studies, we are \emph{not} restricted to the study of integrable models.
When the Boltzmann weights are chosen to be uniform in space, some of the simpler models we study are integrable (and often critical).
Examples include the loop models \cite{Baxter1976,Nienhuis1990} coming from the Temperley-Lieb algebra \cite{Temperley1971} and the related ABF height models \cite{Andrews1984}.
However, beyond Temperley-Lieb, integrability requires even further fine-tuning.
Moreover, when the Boltzmann weights are staggered or disordered, the models are typically neither critical nor integrable, even when the uniform case is. 
Thus the tools we develop are among the few reliable analytical ways of understanding their physics.

The lattice models we study are defined by the following data:
\begin{enumerate}
\item A graph $G$ that gives the location of the degrees of freedom. In completely packed geometric models, they live on faces of $G$, while in height models the live on the vertices of $G$. 
\item	A fusion category $\mathcal{C}$ with the appropriate assumptions laid out in Sec.~\ref{sec:Fusion_Theory}.
\item	A distinguished object $\upsilon$ from $\mathcal{C}$.
\item	A choice of amplitudes $A(\chi)$ for each element $\chi$ in the fusion product of $\upsilon$ with itself: $\chi\in \upsilon \otimes \upsilon$.
\end{enumerate}
This structure allows us to define two types of closely related models, where the degrees of freedom are written in terms of geometrical objects (Sec.~\ref{sec:loopmodels}) and heights (Sec.~\ref{sec:heights}) respectively.
In both cases we give the most general possible Boltzmann weights for this configuration space that can be written as the evaluation of diagrams from the fusion category. These are necessarily consistent with the diagrammatic tools developed.
We will finish the section by giving several examples of well known models recast in this language, as well as some lesser-known ones.

\subsection{Loop and geometric models}
\label{sec:loopmodels}
We first describe how to to define {\em geometric models} using a fusion category. In these models the degrees of freedom are written most naturally in terms of geometric objects such as loops or nets, the former being the best known.
The physical space on which these models live is defined in terms of a graph $G$ built of quadrilaterals, such as the square lattice. In this paper we focus on planar quadrilateral graphs, but the methods can be generalized to any cell decomposition of a surface. The degrees of freedom in the geometric models live on the edges and quadrilaterals of $G$, whereas in the height models discussed later in this section they live on the vertices. 

The role of the fusion category ${\cal C}$ in the geometric models is fundamental, giving the framework to define both the degrees of freedom and the most interesting part of the Boltzmann weights. The distinguished object $\upsilon$ in ${\cal C}$ governs the degrees of freedom on the edges of $G$,
while its fusion channels govern degrees of freedom on each quadrilateral, and each configuration in a geometric model corresponds to a fusion diagram. The topological part of the Boltzmann weights then are defined utilising the evaluation of fusion diagrams. When these elements are chosen appropriately, one recovers many of the well-known examples we discuss below.

Precisely, we associate to each edge of $G$ a strand labeled by $\upsilon$, the distinguished object in the category. Each such strand goes between two adjacent quadrilaterals, so that each quadrilateral has four entering strands and a single fusion channel $\chi\in \upsilon \otimes \upsilon$, pictured as
\begin{align} 
\EfRhombii\ .
\label{loopfusion}
\end{align}
We assemble all the strands and fusion channels into a fusion diagram $\mathcal{F}$, for example \begin{align}
\RhombiiTilingBoundaryRandom
\label{Zfusion}
\end{align}
A ``boundary'' edge of $G$ belongs to only one quadrilateral.  For simplicity, here we imposed a boundary condition by joining the strands through boundary edges to one of their neighbors; we later discuss a variety of possible boundary conditions. 
The strands of the fusion diagram labeled by a box are those associated with each quadrilateral, whereas those without a box all have labels $\upsilon$. Here the objects $\upsilon$ and $\chi$ are self-dual; for non self-dual objects one must include an arrow.

For simplicity, we first describe in more detail  the case where $\upsilon$ is a simple object of ${\cal C}$. 
The resulting geometric model is called {\em completely packed}, because there are effectively no degrees of freedom on the edges; all are fixed to $\upsilon$. 
The non-trivial degrees of freedom thus live inside each quadrilateral $Q$ of $G$, and correspond to the possible fusion channels $\chi \in \upsilon \otimes \upsilon$.
In other words, the allowed $\chi$ are the basis vectors of the $V^{\upsilon \upsilon}_{\upsilon \upsilon}$, and each configuration corresponds to specifying one of these per box in (\ref{Zfusion}).
The number of degrees of freedom per quadrilateral is therefore simply the dimension of this vector space.
For example, for the geometric models built on ${\cal A}_{k+1}$ with $\upsilon=\tfrac12$, there are two states per quadrilateral labeled $\chi=0,1$.

The Boltzmann weight for each configuration is a product of both locally and non-locally determined pieces. 
The local weights are most conveniently given by specifying a vector $A_Q \in V^{\upsilon \upsilon}_{\upsilon \upsilon}$ for each $Q$.
It is a non-simple object, and so can be expanded in terms of the basis vectors $\chi$ as
\begin{align}
	A_Q\equiv \fboxa =\sum_{\chi \in \upsilon \otimes \upsilon} A_Q(\chi) \frac{\sqrt{d_\chi}}{d_\upsilon} \; \fBoxprime \ .
	\label{AQbasis}
\end{align}
This local weight amounts to weighting each appearance of $\chi$ on a given face $Q$ with the amplitude $A_{Q}(\chi)$, with the extra quantum dimensions put in to make the amplitude multiply a projection operator (see \eqref{Pf} below).
We have included the subscript to emphasize that even when the amplitudes are spatially dependent, all our results for topological defects still apply.

The non-local part is more interesting. Labelling each box by one of the allowed $\chi$ gives a fusion diagram ${\cal F}$.
A natural choice of non-local Boltzmann weight is then the evaluation of ${\cal F}$ using the category ${\cal C}$, which we denote as eval${}_{\cal C}[{\cal F}]$.
The procedure to compute the evaluation is described in section \ref{sec:evaluation}.
The result depends only on the topology of the diagram, and is central to the power of our approach. 
The amplitude for each ${\cal F}$, its full Boltzmann weight, and the partition function for the completely packed model are therefore
\begin{align}
	A({\cal F}) &= \prod_Q \frac{\sqrt{d_{\chi_Q}}}{d_\upsilon} \, A_{Q}(\chi_Q) \,,
&	W({\cal F}) &= A({\cal F}) \operatorname{eval}_{\mcc}[{\cal F}] \,,
&	Z_{\rm geo}(G) &= \sum_{\cal F} W({\cal F}) \,.
	\label{ZW}
\end{align}
The sum over ${\cal F}$ in the completely packed case is the sum over all allowed labelings $\chi_Q$ for each $Q$.

In the picture (\ref{Zfusion}), we have made a particular choice of basis on each quadrilateral in order to draw the fusion diagram ${\cal F}$.
There are two possible ways of drawing this line, i.e., in (\ref{loopfusion}) the fusion can be either in the vertical or horizontal directions. The two bases are related by the $F$-move
\begin{align}
\EfRhombii  = \sum_{\widetilde{\chi}}  \Msixj{\upsilon & \upsilon & \chi}{\upsilon & \upsilon & \widetilde{\chi}} \sqrt{d_\chi d_{\widetilde{\chi}}}  \;  \EftildeRhombii
\label{Frhombus}
\end{align}
described in (\ref{eq:Fmove_def}).
Letting ${\cal F}_{\widetilde{\chi}^{}_Q}$ be the fusion diagram given by replacing some $\chi^{}_Q$ in ${\cal F}$ by $\widetilde{\chi}^{}_Q$ in the other direction,  the corresponding evaluations are related via
\begin{align}
	\operatorname{eval}_{\mcc}[{\cal F}]= \sum_{\widetilde{\chi}} [F^{\upsilon\upsilon\upsilon}_{\upsilon}]_{\chi \,\widetilde{\chi}} \operatorname{eval}_{\mcc}[{\cal F}_{\widetilde{\chi}^{}_Q}]\ .
\end{align}
Defining
\begin{align}
	A_{Q}(\widetilde{\chi}) = \sum_{\chi} d_\chi \Msixj{\upsilon & \upsilon & \chi}{\upsilon & \upsilon & \widetilde{\chi}} A_{Q}(\chi) \,;
	 \qquad
	\widetilde{W}({\cal F}_{\widetilde{\chi}^{}_Q })= \operatorname{eval}_{\mcc}[{\cal F}_{\widetilde{\chi}^{}_Q}] \frac{\sqrt{d_{\widetilde{\chi}}}}{d_\upsilon} A_{Q}(\widetilde{\chi}_Q)\prod_{Q'\ne Q } \frac{\sqrt{d_\chi} (\chi_{Q'})}{d_\upsilon}A_{Q'}
\label{relateweights}
\end{align}
and noting that the matrix $F^{\upsilon\upsilon\upsilon}_{\upsilon}$ is invertible gives
\begin{align}
	Z_{\rm geo}(G) =\sum_{\cal F} W({\cal F})= \sum_{{\cal F}_{\widetilde{\chi}^{}_Q}} \widetilde{W}({\cal F}_{Q, \widetilde{\chi}})\ .
	\label{relateweights2}
\end{align}
Thus the partition function is independent of choices of direction on any quadrilateral $Q$, as long as the local part of the Boltzmann weights is related by (\ref{relateweights}).

More general geometric models come from relaxing the complete-packing constraint and allowing $\upsilon$ to be a non-simple object. Decomposing it into a direct sum of simple objects $\upsilon^{(l)}$ as
\begin{align}
\upsilon = \bigoplus_l^{} B_l \upsilon^{(l)}\ ,
\label{Bupsilon}
\end{align}
the coefficients $B_l$ give a weight for each choice of $\upsilon_l$ on a given edge. They can be taken to be $0$ or $1$, since any other values can be absorbed into a redefinition of the amplitudes. Each configuration can still be described by a fusion diagram like in (\ref{Zfusion}), 
except now each strand $s$ without a box carries a label  $\upsilon^{(l_s)}$ instead of $\upsilon$.
The labels allowed for $\chi_Q$ are then the allowed fusion channels for the labels  
on the four strands entering $Q$, i.e., the basis elements of the vector space $V^{\upsilon^{(l_3)}\upsilon^{(l_4)}}_{\upsilon^{(l_1)}\upsilon^{(l_2)}}$.
The Boltzmann weights and partition function remain of the form (\ref{ZW}), but here the local weights now depend on all four of these labels as well as $\chi$.  In the special case where $\upsilon=0 \oplus \upsilon'$, each edge is either empty or covered by a strand. Such models are called ``dilute".

More complicated geometric models are defined using fusion categories not having the simplifying assumptions outlined in Section \ref{sec:Fusion_Theory}. 
Relaxing the requirement that all simple objects be self-dual yields geometric models where the strands have arrows. Fusion categories with $N_{ab}^c>1$ results in models with extra degrees of freedom. 
We could also define a model on any triangulation by splitting each quadrilateral into two triangles and assigning a splitting or fusion operator to each triangle, or on any cell decomposition by choosing an object for each edge, and an element in the fusion space of the edge labels for every face. However, the height models discussed below are naturally defined on the quadrilaterals; on a triangulation or generic cell decomposition, the Boltzmann weights cannot be guaranteed to be positive in an obvious way.

In mathematical language, 
\begin{align}
A_Q \in \text{mor}(\mathds{1} \rightarrow  \otimes_{e \in Q} \upsilon),
\end{align}
where the $e \in Q$ runs over all the edges adjacent to the face $Q$. 
A basis for $\text{mor}(\mathds{1} \rightarrow  \otimes_{e \in Q} \upsilon)$ can be decomposed in terms of trivalent fusion spaces. 
For example if all $B_l=1$, and there are $r$ edges adjacent to face $Q$ then we have $A_Q \in \text{mor}(\mathds{1} \rightarrow  \otimes_{e \in Q} \upsilon) \cong \bigoplus_{l_1, \cdots, l_r} V^{l_1 \cdots l_r}$.

\subsection{Examples of geometric models }
\label{sec:geometricexamples}

\paragraph{Completely packed self-avoiding loop models.}

The best-known class of geometric models are \emph{loop models}, where the degrees of freedom are closed loops.
Although avoiding branching seems somewhat antithetical to our approach, we show how $F$-moves make it quite simple to describe loop models in terms of fusion categories.
One payoff is that even though the Boltzmann weights are non-local in either approach, the $F$-moves give \emph{local} relations between partition functions.

Loop models are built using the ${\cal A}_{k+1}$ fusion category. 
In a {\em completely packed} loop model, each edge of ${\cal Q}$ has an $\upsilon = \tfrac12$ line passing through it. 
This object (corresponding to the ``fundamental'' spin-$\tfrac12$ representation in the related quantum-group algebra \cite{Pasquier1990}) obeys the fusion rule $\tfrac12\otimes \tfrac12=0\oplus 1$. 
There are two degrees of freedom per quadrilateral corresponding to the two fusion channels $\chi = 0$ or $1$.
This has a very convenient graphical representation. 
When $\chi=0$, we leave that bit of the fusion diagram as empty, since adding or removing the identity line never changes the evaluation of the diagram.
When $\chi=1$, we draw this as a zig-zag line. Thus a typical fusion diagram for the geometrical model coming from the ${\cal A}_{k+1}$ fusion category with $\upsilon=\tfrac12$ can be drawn as
\begin{align} \FExampleSpinZeroOne\ .
\label{spinzeroone}
\end{align}
It is easy to check using the $F$-moves (\ref{relateweights}) that the evaluation vanishes for any graph that contains a ``tadpole'', a $\upsilon=\tfrac12$ loop with a single zig-zag line attached to it.

$F$-moves make it easy to see why this geometric model is equivalent to the completely packed self-avoiding loop model in its traditional definition.  
Here they relate the two fusion channels $0,1$ on a given quadrilateral $Q$ to those $\tilde{0},\tilde{1}$ in the other orientation. 
Written as a matrix where the rows and columns are labeled by $0,1$, the needed $F$-matrix elements are
\begin{align}
	\Big[ F^{\frac12 \frac12 \frac12}_{\frac12} \Big] = \frac{1}{d_{\frac12}} \begin{pmatrix} 1 &\sqrt{d_1}\cr \sqrt{d_1} &-1 \end{pmatrix} \,.
\end{align}
The quantum dimensions are given by \eqref{eq:def_QD}.
As (\ref{relateweights2}) indicates, the evaluation and the partition function are invariant if on any $Q$ we replace
\begin{align}
\RVZero\; \;  &= \; \; \frac{1}{d_{\frac12}} \; \; \RHZero\; \; + \; \; \frac{\sqrt{d_1}}{d_{\frac12}} \; \; \RHOne \;.
\label{TLF0} \\[5pt]
\RVOne \; \;  &=\; \; \frac{\sqrt{d_1}}{d_{\frac12}} \; \; \RHZero \; \; -\;\; \frac{1}{d_{\frac12}} \; \; \RHOne \;.
\label{TLF1}
\end{align}
Any two of the four possible fusion channels $0,1,\tilde{0},\tilde{1}$ are linearly independent, and so form a basis.
Thus each of the zig-zag lines in the fusion diagrams like (\ref{spinzeroone}) can be replaced with a linear combination of lines that avoid each other, namely
\begin{align}
\RVOne \; \;  &=\; \; \frac{d_{\frac12}}{\sqrt{d_1}} \; \; \RHZero \; \; -\;\; \frac{1}{\sqrt{d_{1}}} \; \; \RVZero \;.
\label{avoidcrossing}
\end{align}
All the configurations in this geometric model can be built in terms of diagrams using only ``avoided crossings" labeled $0$ and $\tilde{0}$ on each quadrilateral $Q$
on each quadrilateral. If we chose a boundary condition where no $\upsilon$ lines end, all the fusion diagrams can be rewritten as linear combinations of fusion diagrams ${\cal F}$ of completely packed self-avoiding loops.

The evaluation of a diagram consisting of self-avoiding loops is easy.
Since each loop is labelled by the object $\upsilon=\tfrac12$ here, it contributes a factor $d_{\frac12} = 2\cos \tfrac{\pi}{k+2}$ to the evaluation.
Thus when ${\cal F}$ has $N_\text{loops}$ closed loops, the partition function for completely packed loops is simply 
\begin{align}
	Z_{\rm CPL} 
		= \sum_{\cal F} \big(d_{\frac12}\big)^{N_\text{loops} - N_\text{quadr}} \prod_Q A_{Q} (\chi_Q)
	\label{ZCPL}
\end{align}
where $\chi_Q=0$ or $\tilde 0$ depending on how the two loops avoid on face $Q$. The local weights $A_Q(0)$ and $A_Q({\tilde 0})$ can be related to those $A_{Q}(0)$, and $A_Q(1)$ by using (\ref{relateweights}). 

This partition function of completely packed loops has long been studied, for example in the Fortuin-Kasteleyn expansion for the Q-state Potts model \cite{Fortuin1971}, where Q=$\big(d_{\frac12}\big)^2$.
While expressing the states in terms of loops is quite elegant and useful, it obscures the fact that there are local manipulations such as $F$-moves that leave (\ref{ZCPL}) invariant. The number of loops is inherently a non-local quantity; one needs to follow the entire path of the loop to know which segments comprise it.
Being able to do local manipulations on the configurations without changing the partition function is essential to our analysis. Thus even when there are seemingly simpler rewritings, we usually write the partition function as the sum over evaluations of fusion diagrams, rather than expressions like (\ref{ZCPL}).
Using such a basis also has distinct advantages when studying quantum loop and net models, for example providing a natural inner product for the quantum model \cite{Fendley2008}. 

\paragraph{Dilute loop models.}

A well-known example without the complete packing is the ``dilute'' loop model, where $\upsilon = 0\oplus \tfrac12$ in the ${\cal A}_{k+1}$ category, corresponding to unoccupied or occupied edges respectively \cite{Nienhuis1990}.
Sometimes this model known as the $O(n)$ model, because of its relation to models of magnets when the loop weight is an integer \cite{BN1989}. Since all possible fusion rules preserve the number of strands mod $2$, there must be an even number of strands entering each quadrilateral. This results in nine allowed configurations on each quadrilateral, which can be chosen to be
\begin{align} \begin{aligned}
	&\Rblank \,,
	&&\RVZero \,,
	&&\RVoD \,,
	&&\reflectbox{\rotatebox[origin=c]{180}{$\RVoD$}} \,,
	&&\RHoL \,,
	&&\reflectbox{{$\RHoL$}} \,,
	&&\RVtwist \,,
	&&\reflectbox{\rotatebox[origin=c]{180}{$\RVtwist$}} \,,
	&&\RVOne \,.
\label{diluteloops}
\end{aligned} \end{align}
As with the completely packed model, one can use \eqref{avoidcrossing} to get rid of the zig-zag line and replace it with self-avoiding loops.
A particular choice of local Boltzmann weights $A_{Q}$ for these nine configurations results in an interesting integrable model \cite{Nienhuis1990}, to which we return in section \ref{sec:gmin}.

\paragraph{Higher-spin models.}
The completely packed loop models defined above are closely related to the Kac-Moody algebra $SU(2)_k$ and the corresponding quantum-group deformation of the $SU(2)$ algebra, with the simple object $\upsilon=\tfrac12$ corresponding to the spin-$\tfrac12$ representation of the algebras.
An obvious generalization is to take $\upsilon$ to be a higher-spin value.
For $\upsilon>\tfrac12$ there is no way to rewrite the configurations in terms of loop configurations in the plane, so utilising fusion diagrams is the only sensible way of defining such geometric models. 

For example, when $\upsilon=1$, its fusion algebra is $1\otimes1 = 0\oplus1\oplus2$, so that three numbers $A_Q(0)$, $A_Q(1)$, and $A_Q(2)$ are required to specify the partition function.
Graphs can be manipulated by using the $F$-matrix
\begin{align}
	\big[ F^{111}_{1} \big] = \frac{1}{d_1}
		\begin{pmatrix} 1 &\sqrt{d_1}&\sqrt{d_2}
		\\ \sqrt{d_1}& \frac{d_1(d_1-2)}{d_1-1}&-\frac{\sqrt{d_1d_2}}{d_1-1}
		\\ \sqrt{d_2} & -\frac{\sqrt{d_1d_2}}{d_1-1}&\frac{1}{d_1-1}
		\end{pmatrix}
	\label{Fmat1}
\end{align}
with the rows and columns are labeled by the $0,1,2$ channels.
One can generalize (\ref{TLF0},\,\ref{TLF1}) to write three linear relations relating the six possible graphs on each $Q$.
We see that indeed one cannot reduce configurations to a combination of self-avoiding loops. 

For any $\upsilon$ there exist integrable Boltzmann weights \cite{Kulish:1981}, but as in the dilute case, they require fine-tuning.
One example for $\upsilon=1$ comes from rewriting the ``19-vertex model'' solution of the Yang-Baxter equation \cite{Zamolodchikov1980} in terms in terms of topological invariants \cite{Wadati1989}, giving
\begin{align}
	A_Q(0) &= \frac{\sin(\gamma+u)}{\sin(\gamma-u)} \,,
&	A_Q(1)&= 1 \,,
&	A_Q(2) &= \frac{\sin(2\gamma-u)}{\sin(2\gamma+u)} \,.
\label{spinoneYBE}
\end{align}
where $\gamma = \frac{\pi}{k+2}$, while $u$ can vary. 
A more unusual solution for $\upsilon=1$ comes from the Izergin-Korepin or $A_2^{(2)}$ solution of the Yang-Baxter equation \cite{Izergin:1980,Jimbo:1985}, where the amplitudes obey
\begin{align}
	A_Q(0) &= \frac{\cos(3\gamma+u)}{\cos(3\gamma-u)} \,,
&	A_Q(1) &= \frac{\sin(2\gamma+u)}{\sin(2\gamma-u)} \,,
&	A_Q(2) &= 1 \,.
\label{IKYBE}
\end{align}
We emphasize again that our results for topological defects do not require tuning the weights to these special integrable cases. 

\paragraph{BMW models.} 
Not surprisingly, there are fusion categories associated with the quantum-group deformations of Lie algebras more general than $SU(2)$.
An algebra generalizing the Temperley-Lieb algebra is known as the Birman-Murakami-Wenzl (BMW) algebra \cite{birman1989,murakami1990}.
It allows the definition of categories labeled by Lie algebras and an integer parameter $k$.
The corresponding geometric models amount to having $\upsilon$ be the vector representation $V$ of the (quantum-group deformed \cite{Gomez1996}) algebra.
In the $SO(n)$ and $Sp(n)$ cases for $k\ge 2$, it obeys the fusion algebra
\begin{align}
V\otimes V = \I \oplus A\oplus S \,,
\label{BMWmult}
\end{align}
giving the trivial, antisymmetric and symmetric representations respectively.
In the $SO(3) \sim SU(2)$ case, both the vector and antisymmetric representations are spin 1, because of the existence of the completely antisymmetric invariant tensor $\epsilon_{abc}$.

By choosing  $\upsilon=V$, a completely packed geometric model can be found for these BMW algebras. As follows from \eqref{BMWmult}, each $\chi_Q$ can be in any of three channels.
An associated solution of the Yang-Baxter equation can be extracted from \cite{Jimbo:1987}, and can be generalized to the dilute case $\upsilon=0\oplus V$  \cite{Grimm1994}.
More general integrable models exist where multiple strands are allowed on each edge, and so effectively there are different layers  \cite{Grimm1995}.
In such a case, $\upsilon$ then takes values in some direct product of categories $\mcc^{\text{layer}\,1}  \times \mcc^{\text{layer}\,2}$.

\subsection{Heights and shadows}
\label{sec:shadow}

The role of fusion categories in statistical mechanics goes well beyond geometric models.
Much less obviously, they play a central role in understanding an important class of lattice models with {\em local} Boltzmann weights.
Examples of this connection long predate the invention of fusion categories.  For example, the Fortuin-Kasteleyn cluster expansion relates the $Q$-state Potts model, a model with nearest-neighbor Boltzmann weights, to a cluster/loop expansion, as reviewed e.g.\ in Ref.~\cite{Baxter1982}.
We explain in this subsection how in general, fusion diagrams can be evaluated in terms of local data by using a technique known in the mathematical literature as the \emph{shadow world} \cite{Reshetikhin1988,turaev1990,turaev1992,kauffman1993,turaev2016quantum}.
Physical explanations of the correspondence in terms of braiding and knot invariants can be found in the review articles \cite{Wadati1989,Wu1992}, and in terms of Chern-Simons theory in \cite{Witten:1989}. 

The shadow-world construction gives the evaluation of a fusion diagram embedded in a sphere  by associating to it a sum over degrees of freedom living on its faces \cite{Reshetikhin1988,TuraevViro1992,Barrett1996}. These degrees of freedom are called {\em heights}, and can be used to define a statistical-mechanical model with many elegant properties. Terms in the sum are weighted by quantum dimensions and tetrahedral symbols of these degrees of freedom, and from these weights the corresponding height model can be constructed. 
This general class of models is often known as RSOS (restricted solid-on-solid) or IRF (interaction round a face) models, but we simply call them {\em height models}. Thus even though the Boltzmann weights are local in the height models, their partition functions can still be expressed in terms of fusion diagrams as in \eqref{ZW}. Although that seems to be making matters more complicated, the fusion-diagram form makes it much easier to find topological defects. We introduced this picture for the Ising model in Part~I, and found many useful applications.

Before describing the construction of topological defects though, we state the shadow-world formula and use it to define the Boltzmann weights of the height models. In the next section \ref{modules} we explain the origin of this formula, and how it can be generalized to define topological defects.  

As we have explained, a fusion diagram ${\cal F}$ is a labelled trivalent graph $\mathcal{G}$ with a collection of vertices $V$, edges $E$, and faces $F$. Each edge $e \in E$ is labelled by an object in the fusion category ${\cal C}$.
Since ${\cal G}$ is trivalent, the faces of the dual graph $\widehat{\cal G}$ are triangles centered on the vertices of $\cal{G}$. The first step in evaluating the fusion diagram ${\cal F}$ using the shadow world is to assign a ``height'' $h_f \in {\cal C}$ to each face $f \in F$ (or equivalently each vertex of $\widehat{\cal G}$ or the original graph $G$).  The allowed heights are thus in one-to-one correspondence with the simple objects. Thus for the ${\cal A}_{k+1}$ category, the heights take on values $0,\tfrac12,1,\dots \tfrac{k}{2}$. 

The next step is to assign a weight $w(\{h_f\};{\cal F})$ to each configuration of heights and edge labels. The evaluation of the fusion diagram using the shadow world is then simply given by summing the weight \eqref{shadoweval2} over all the heights:
\begin{align}
	\operatorname{eval}_{\cal C}[{\cal F}] = \sum_{\{h_f\}} w(\{h_f\}; {\cal F}) \,,
\label{shadoweval1}
\end{align}
The weight is associated with the vertices, edges, and faces of ${\cal G}$.
\begin{figure}[h]
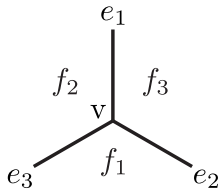

	\begin{align*} \ShadowNotation \end{align*}
	\caption{Notation for each vertex ${\rm v}$ and adjacent edges and faces.}
	\label{fig:ShadowNotation}
\end{figure}
The vertex weights are the non-trivial part.
Consider a single vertex $\rm v$ of ${\cal F}$ with adjacent edges $e_1$, $e_2$, $e_3$, and faces $f_1$, $f_2$, $f_3$ as shown in Fig.~\ref{fig:ShadowNotation}.
The associated weight is in terms of the tetrahedral symbol depending on the labels on the adjacent edges $\epsilon_i(e_i)$ and those on the neighboring faces $h_i(f_i)$:
\begin{align}
	\operatorname{Tet}({\rm v}) \;=\; \Msixj{\epsilon_1 & \epsilon_2 & \epsilon_3}{h_{1} & h_{2} & h_{3}} \,,
\end{align}
This tetrahedral symbol enforces the adjacency rule $\epsilon_1 \otimes \epsilon_2 = \epsilon_3+\dots$, as necessary for a fusion diagram to have a non-zero evaluation.
Additional requirements from the tetrahedral symbol such as $N_{h_{1} \epsilon_3}^{h_{2}} > 0$ means that the heights on two adjacent faces must differ by the fusion of the object on the edge they share.
The locally defined weights are then \cite{Reshetikhin1988}
\begin{align}
	w\big(\{h_f\}; {\cal F}\big) = \frac{1}{\sum_{a \in {\cal C}} d_a^2}
		\prod_{f \in F} d_{h_f} \times \prod_{e \in E} \sqrt{d_{\epsilon_e}} \times
		\prod_{{\rm v} \in V} \operatorname{Tet}({\rm v})~.
	\label{shadoweval2}
\end{align}
The constant prefactor arises because we include the exterior of the graph as a face, and comes from the useful identity \eqref{ddNdD}.
Thus as opposed to evaluating fusion diagrams by e.g.\ resolving intersections and counting loops as one does in the Jones polynomial, evaluating using (\ref{shadoweval1}, \ref{shadoweval2}) gives entirely a product of local weights.\footnote{The formula~\eqref{shadoweval2} requires that each face must be simply connected on $S^2$, and that each edge terminates at a pair of (possibly the same) vertices.  If not, $\I$ lines can be inserted to break up closed loops and noncontractible faces; i.e., turn the diagram into a non-empty-connected-graph for which the formula~\eqref{shadoweval2} is applicable.}

\subsection{Height models}
\label{sec:heights}

The shadow-world evaluation (\ref{shadoweval1}, \ref{shadoweval2}) makes it obvious how to define a model with the same partition function as a geometric model. The degrees of freedom are the heights, which are simple objects in the category. While the height description partially hides the elegant topological properties, a key advantage is the Boltzmann weights of the height model are local. 

As with the geometric models, the height models are defined using the three criteria listed at the beginning of section \ref{sec:statmech}: a fusion category ${\cal C}$, a distinguished object $\upsilon$ from ${\cal C}$, and a choice of amplitudes $A_{Q}$ for each quadrilateral.
In the shadow world the heights live on the {\rm faces} of the fusion diagram ${\cal F}$.
They therefore live on the {\em vertices} $v$ of ${G}$, the graph whose faces are all quadrilaterals.
The allowed heights are the elements of $\mathcal{C}$, so that each configuration corresponds to labelling each vertex of $G$ by some element of $\mathcal{C}$ subject to an adjacency rule.
Namely, two heights $h$ and $h'$ are only allowed to be on vertices connected by an edge of $G$ if $h$ appears in the fusion of $h'$ and $\upsilon$, i.e., $N_{h\upsilon}^{h'}\ne 0$.
The pivotal property \eqref{eq:pivotal_N} we are assuming means that $N_{h'\upsilon}^h$ must be non-vanishing as well. 
If $\upsilon$ is not self-dual, then one needs to include arrows on the edge connecting the adjacent heights.

The Boltzmann weights in the height model are given by a product of weights involving four-site interactions around each quadrilateral \cite{Baxter1982}. 
We use the notation where a labelled picture denotes a number depending on the heights on its vertices), so that each $\BWsmall$ is a Boltzmann weight which includes the interactions between all four spins $a$, $b$, $c$ and $b'$ around a quadrilateral.
The partition function thus is of the form
\begin{align}
	Z \quad = \sum_{\substack{\text{height}\\\text{configurations}}} \left( \prod_{{v}\in {G}} d_{h_{v}} \right)\, \prod_{Q} \BWabcd \,.
	\label{eq:Zheightdef}
\end{align}
where the factors $d_{h_v}$ are extracted for later convenience.
The adjacency rule requires that the Boltzmann weights on each quadrilateral obeys
\begin{align}
	\BWabcd \;\propto\; N_{ab'}^\upsilon N_{b'c}^\upsilon N_{cb}^\upsilon N_{ba}^\upsilon \,.
	\label{eq:BWproptoNNNN}	
\end{align}

The shadow-world construction relates the evaluation of a single fusion diagram to a sum over heights, where each term in the sum is a product over local weights.
However, the partition function of a geometric model is a sum over different fusion diagrams, given by expanding in fusion channels for each quadrilateral, as displayed in \eqref{AQbasis}.
This  expansion is over $\chi\in V^{\upsilon \upsilon}_{\upsilon \upsilon}$, i.e., $\chi$ with $N_{\upsilon\upsilon}^\chi>0$.
We draw the analogous expansion in terms of height-model weights as
\begin{align}
\BWabcd \;&=\; \sum_\chi A_Q(\chi) \; \BWabcdx{\chi} \,.
\label{Wexp}
\end{align}
The coefficients $A_Q(\chi)$ in \eqref{Wexp} are arbitrary,  but their precise expression depends on the basis for the space $V^{\upsilon \upsilon}_{\upsilon \upsilon}$ of fusion channels, i.e.\ which pairs of the $\upsilon$ we fuse together to get $\chi$. The two most natural bases have fusion diagrams with the $\chi$ line running horizontally or vertically, and \eqref{Frhombus} shows how the corresponding evaluations are related. As in the geometric model, the partition function is invariant if the corresponding $A_Q(\chi)$ are related by  \eqref{relateweights}. This relation between weights is sometimes referred to in the statistical-mechanical literature as the ``crossing'' property, by analogy with the properties of Feynman diagrams in field theory.
Inserting \eqref{Wexp} into the height-model partition function \eqref{eq:Zheightdef} then gives it as a sum over both fusion diagrams and heights:
\begin{align}
	Z =  \sum_{\text{heights}} \left[ \sum_{\cal F} \prod_{Q} A_Q(\chi_Q) \; \BWabcdx{\chi_{Q}} 
 \;\times \prod_{\rm v} d_{h_{\rm v}} \right] .
 \label{Zheightfusion}
\end{align}
The product over vertices remains over those in $G$, where the heights are defined.

We now use the shadow world weight \eqref{shadoweval2} to find Boltzmann weights of the height model that yield the same partition function as the corresponding geometric model. To this end, we change the order of sums in \eqref{Zheightfusion} 
and consider a particular fusion diagram. If we choose 
\begin{align}
	\BWabcdx{\chi} & 
	\;\defineas\; d_\chi \Msixj{\upsilon & \upsilon & \chi}{a & c & b} \Msixj{\upsilon & \upsilon & \chi}{a & c & b'}
	\;=\; \frac{\sqrt{d_\chi}}{d_\upsilon^2} \frac{1}{\sqrt{d_a d_b d_{b'} d_c}} \; \omegafNA \,,
	\label{eq:def_chi}
\end{align}
then summing over all allowed heights for this particular diagram gives the correct evaluation. 
Namely, using \eqref{eq:def_chi} in  \eqref{Zheightfusion} and comparing with (\ref{shadoweval1},\,\ref{shadoweval2}) gives
\begin{align}
	Z = \sum_{\cal F} A({\cal F}) \operatorname{eval}_{\mcc}[{\cal F}]
	= \sum_{\cal F} \operatorname{eval}_{\mcc}[{\cal F}] \prod_Q \frac{\sqrt{d_\chi}}{d_\upsilon} A_{Q}(\chi)\ .
\label{ZW2}
\end{align}
By definition \eqref{ZW}, the partition function of the completely packed geometric model is then the same, provided the boundary conditions are chosen appropriately.
Namely, for the fusion diagram in the geometric model to evaluate to a number, none of the strands can end on the boundary; they must close as e.g.\ in \eqref{spinzeroone}.
In the shadow world, each height lives on a face of the fusion diagram.
Therefore, for the evaluation in  \eqref{ZW2} to give the same answer as  \eqref{ZW}, all heights outside the fusion diagram (i.e., on the face at ``infinity'') must take on the same value. 
With this choice
\begin{align}
Z=Z_{\rm geo} \,.
\end{align}
Weights found from the shadow-world construction are not automatically positive, but are in the many examples we describe below.

The relation between height and geometric models applies to more general cases such as the dilute models, where $\upsilon$ is not necessarily a simple object. As explained after (\ref{Bupsilon}), one expands out $\upsilon$ in terms of the simple objects, and then generalizes (\ref{Zheightfusion},\ref{eq:def_chi}) to include labels on different strands. One then finds dilute height models, for example the integrable ones discussed in \cite{Warnaar1992}. 

Lastly we briefly remark on how to define the partition function on a graph with boundary, with a more detailed discussion given in section \ref{sec:bdry}. In geometric models, one can simply sew together adjacent strands, as in \eqref{Zfusion}, or do more complicated sewing by for example embedding $G$ and ${\cal F}$ on a cylinder or torus. In the height models, 
a fixed boundary condition is given by specifying one height per vertex along the boundary of $G$.
Of course, adjacent heights $h$ and $h'$ must contain $\upsilon$ in their fusion product, otherwise the Boltzmann weight which shares those heights will evaluate to zero.
One can design ``free" boundary conditions by taking various linear combinations of the fixed boundary condition. 
In the next subsection we investigate the transfer matrix, which is given by a special boundary condition on the cylinder.

\subsection{The transfer matrix for height models}

\begin{figure}[ht]
	\begin{align*}\xymatrix @!0 @M=2mm @R=32mm{
		&\Hilbopenprime \\
		&\HilbOpenZprime \ar[u]		
	}\end{align*}
\vspace{-0.3cm}
	\caption{%
		Defining the vector space of the height models in the transfer matrix/1+1D quantum picture.
		The heights on the 2D lattice are now written as labels on the trunk of the fusion tree.	\label{fig:PartitionCut}}
\end{figure}
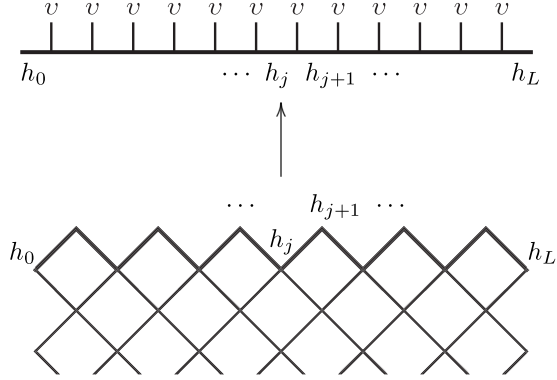

When the graph $G$ of quadrilaterals is regular, the transfer matrix gives a convenient way to parametrize the partition function.
The transfer matrix builds up the partition function one row at a time; roughly speaking, it implements Euclidean ``time'' evolution. 
One can then write the partition function in terms of an operator acting on a vector space ${\mathcal V}$ whose basis elements are the different height configurations. Since the nearest-neighbour heights are restricted to be related by fusion with $\upsilon$, one can represent the heights as labels in a fusion tree, as displayed in Figure \ref{fig:PartitionCut}.  With $\upsilon$ simple, the fusion tree has the main trunk with labels $h_j$ for the heights, and branches all labeled $\upsilon$ to enforce $N_{h_j h_{j+1}}^\upsilon = 1$.  For example, for the ABF models each fusion tree is labelled by requiring each $h_j=0,\tfrac12,1,\dots k/2$ with $|h_j-h_{j+1}|=\tfrac12$ for all $i$.

Fixed boundary conditions correspond to fixing edge heights $h_0$ and $h_{L}$ at the ends of the fusion tree, giving the basis elements as
\begin{align}
	\ket{h_0\cdots h_{j-1} h_j h_{j+1} \cdots h_{L}} \;\;=\;\; \dots \HilbOpen \dots \;\;\in\;\; \bigoplus_\eta V^{\cdots \upsilon \upsilon \upsilon \upsilon \cdots}_{h_0 h_{L}} \equiv \mathcal{V}_{h_0h_L}.
	\label{eq:fusiontreedef}
\end{align}
We are allowed to draw the fusion tree horizontally, since the diagrams are trivially pivotal, and all objects have trivial Frobenius-Schur indicators. 
For periodic boundary conditions, we set $h_{j+L}\equiv h_j$. A complete basis for the states in the corresponding vector space $\mathcal{V}$ is then given by the fusion trees
\begin{align}
\mathcal{V}=\bigoplus_{h_L} \mathcal{V}_{h_L h_L} \,.
\label{vper}
\end{align}
For either fixed or periodic boundary conditions, the dimension of ${\mathcal V}$ grows asymptotically as $(d_\upsilon)^L$  in (\ref{eq:QD_asymp}).
The dimension of the ABF Hilbert space therefore grows as $\big(2\cos\frac{\pi}{k+2}\big)^L$.

The transfer matrix $T$ is a linear operator taking ${\mathcal V}$ to $\mathcal{V}$. 
Since the Boltzmann weights are local in the height model, it can be built in terms of local operators $W_j$.
The operator $W_j$ is non-diagonal only on site $j$, i.e.,
\begin{align}
	W_j  : 
	\ket{h_1\dots h_{j-1}h_jh_{j+1}\dots h_L}\quad\to\quad  \ket{h_1\dots h_{j-1}h_j'h_{j+1}\dots h_L} .
\end{align}
The matrix elements of $W_j$  are labelled in terms of the heights around the quadrilateral, as in (\ref{eq:Zheightdef}), and so depend non-trivially only on the heights on sites $j-1$, $j$ and $j+1$:
\begin{align}
(W_j)_{\{h_j\},\{h'_{j}\}}^{}=\BWfour{\BWbareDotsx{}}{h_{j-1}}{h_j}{h'_j}{h_{j+1}}\   \defineas \ 
\BWfour{\BWbare}{h_{j-1}}{h_j}{h'_j}{h_{j+1}}\,\sqrt{d_b d_{b'}}
\label{semidotdef}
\end{align}
where the explicit expression is given in \eqref{eq:def_chi}.
The half-dots correspond to multiplying by $\sqrt{d_{h_{\rm v}}}$, in order to absorb the vertex factors $d_{h_{\rm v}}$ in (\ref{eq:Zheightdef}) into the $W_j$. 
For the square lattice, $T$ is built from the Boltzmann weights for a zig-zag block of faces, and is of the form
\begin{align}
T=\left(\prod_{j\ {\rm odd}} W_j\right)\left(\prod_{j\ {\rm even}} W_j\right) .
\label{Transfermatrixdef}
\end{align}
It can be illustrated by
\begin{align}
T = \TransfMxIsingDots\ .
\label{Tmatrixdef2}
\end{align}
The partition function on a torus for periodic boundary conditions is then $ Z = \Tr T^{R}$ for $R$ rows of the form (\ref{Transfermatrixdef}).

A convenient basis for the Boltzmann weights allows has generators that are orthonormal under stacking.  
Each operator $W_j$ is identified with a vector in the fusion space $V^{\upsilon \upsilon}_{\upsilon \upsilon}$, which according to Eq.~\eqref{eq:Vop_decomp} decomposes into $\bigoplus_\chi V^{\upsilon \upsilon}_\chi \otimes V^\chi_{\upsilon \upsilon}$. 
A complete basis for the vector space $V^{\upsilon \upsilon}_{\upsilon \upsilon}$ is given in pictorial language by 
\begin{align}
	P^{(\chi)} = \frac{\sqrt{d_\chi}}{d_\upsilon} \Ef \in V^{\upsilon \upsilon}_{\upsilon \upsilon} \,.
	\label{Pf}
\end{align}
We have picked the normalization so that 
\begin{align}
P^{(\chi)}P^{(\chi')} = \delta_{\chi\chi'}\,P^{(\chi)} \,,
\label{Pchisq}
\end{align}
as can be verified by using \eqref{eq:loop_removal}.
Since each operator is a projector, it can be thought of as projecting onto a fixed ``charge'' or ``channel'' $\chi \in \upsilon \otimes \upsilon$.
Using \eqref{eq:def_chi} gives a convenient basis
\begin{align}
	P^{(\chi)}  \bigket{v_{ac}^{(\chi)}}= N^a_{c\chi} \bigket{v_{ac}^{(\chi)}}\ ,\qquad\hbox{where }\ 
	\bigket{v_{ac}^{(\chi)}} = \sum_b \sqrt{d_b d_\chi} \Msixj{\upsilon & \upsilon & \chi}{a & c & b} \ket{a} \otimes \ket{b} \otimes \ket{c}\ .
\end{align}

The matrix elements of the projectors \eqref{Pf} acting on the full fusion tree can be related to the Boltzmann weights by using  \eqref{eq:Trivalent_Bubble}, \eqref{eq:tetra_Fmove} and \eqref{eq:def_chi}, giving
\begin{align}
	P_j^{(\chi)} \bigket{\cdots h_{j-1}h_jh_{j+1} \cdots}  = \sum_{h_j'}\, \BWfour{\BWbareDotsx{\chi}}{h_{j-1}}{h_j}{h'_j}{h_{j+1}} \, \bigket{\cdots   h_{j-1}h_j'h_{j+1} \cdots} \,.
\label{eq:Pf}
\end{align}
As one expects, the projectors for each $j$ sum to the identity operator:
\begin{align}
	\I = \sum_\chi P^{(\chi)}_j
\qquad	\Longleftrightarrow\qquad
	 \delta_{bb'} &= \sum_\chi \BWabcdDotsx{\chi}\ ,
\label{sumproj}
\end{align}
using (\ref{Pf}) and the resolution of the identity operator (\ref{F0c}). Alternatively, it can derived directly using (\ref{eq:Pf}) and the orthogonality relation (\ref{Orthogonality6j}).

The Boltzmann weights for the height models \eqref{Wexp} therefore are specified by an arbitrary linear combination of the projectors for each $Q$ 
\begin{align}
	W_j &= \sum_\chi A_Q(\chi) P_j^{(\chi)}
	&&\Longleftrightarrow
	&	\BWabcd &= \sum_\chi A_Q(\chi)  \; \BWabcdx{\chi}\ .
\label{WAP}
\end{align}
An important point is that these Boltzmann weights are {\em not} the most general for a height model with such adjacency restrictions. We are imposing much stronger conditions, in that they determined by evaluating some graph in a fusion category, and depend non-trivially on at most three successive heights. These are the most general weights with these constraints. Moreover, the conditions are in general less restrictive than needed for integrability. For example, in the ``spin-1'' height model found from the ${\mathcal A}_{k+1}$ category with $\upsilon=1$, integrability requires tuning the $A_Q(\chi)$ to be those in (\ref{spinoneYBE}) or (\ref{IKYBE}). In addition, integrability typically requires the $A_Q(\chi)$ be uniform in space (i.e.\ independent of $j$), while the defect lines we study will not require imposing any such restriction.

One can define an associated quantum chain by taking an appropriate limit of the Boltzmann weights.
As there is a one-to-one correspondence between fusion trees and heights along a row of the lattice, the Hilbert space for the open quantum chain remains $\mathcal{V}_{h_0 h_L}$. Choosing each $A_j(\chi_j)= 1 -\epsilon \mathcal{A}_j$ and expanding the transfer matrix in powers of $\epsilon$ gives for the leading nontrivival piece 
\begin{align} 
H = \sum_j \sum_{\chi_j} \mathcal{A}_j(\chi_j) P^{(\chi_j)}_j\ ,
\label{HPdef}
\end{align}
This $H$ is a sum over local operators acting on the Hilbert space displayed in \eqref{eq:fusiontreedef}, and so defines a Hamiltonian with nice properties. In particular, this Hamiltonian will commute with the topological defect creation operators defined below. 
Such Hamiltonians were renamed as ``anyon spin chains'', due to their relation with the restricted Hilbert space enjoyed by a linear array of anyons formed as the gapped excitations in a two-dimensional topological phase \cite{Feiguin2007}. 

A feature of treating these transfer matrices and Hamiltonians in a topological fashion is that it makes apparent how their spectrum splits into distinct sectors. In the models we consider, the $F$-matrix is unitary. Thus fusing the $\upsilon$ together in  \eqref{eq:fusiontreedef} by an $F$-move is a unitary transformation on $\mathcal{V}_{h_0 h_L}$. Repeating, we can make $h_0$ and $h_L$ closer and closer together, eventually leaving them with only a single object $C$ in between. This unitary transformation thus divides the spectrum into symmetry sectors labeled by $C\in h_0\otimes h_L$, giving intuition into why all sectors of the minimal conformal field theories were found in the lattice models by specifying three heights \cite{Saleur1989}. These sectors are reminiscent of those found by diagonalizing the action of a symmetry on $\mathcal{V}_{h_0 h_L}$, but as we will explain, they are not associated with any conventional symmetry. Rather, they can be understood as arising by the action of topological defect creation operators (a method much more practical than implementing this unitary transformation). Such defects are the main topic of this paper, and we define them in section \ref{sec:topological_defect_lines}.

\subsection{Examples of height models} 
\label{sec:heightexamples}

\paragraph{Andrews-Baxter-Forrester (ABF) models.}

The quintessential series of height models was introduced by Andrews, Baxter and Forrester (ABF) \cite{Andrews1984}.    The models we describe here are the trigonometric case of those defined in Ref.~\cite{Andrews1984}. The more general models involve Boltzmann weights defined in terms of elliptic functions, and cannot be written in terms of fusion category data (at least not in any known fashion).
We do however allow for general graphs $G$ and for non-uniform couplings $u_Q$, as opposed to square-lattice and uniform (and integrable) case considered in Ref.~\cite{Andrews1984}.

The ABF models are built from the ${\mathcal A}_{k+1}$ category,  as defined in (\ref{eq:def_Ak_N}), by taking $\upsilon = \frac12$. The $k+1$ basic objects give $k+1$ possible heights labeled  $\{ 0, \frac{1}{2}, 1, \cdots, \frac{k}{2} \}$ to highlight the similarity with the spins of $SU(2)$ and the quantum-group algebra $U_q(sl_2)$. (In the statistical-mechanics literature they are typically numbered $1,\dots,k+1$.)
Since fusing $h$ with $\upsilon=\frac{1}{2}$ gives either $h+\frac{1}{2}$ or $h-\frac{1}{2}$ (for $\tfrac{k}{2}>h>0$), adjacent heights can differ only by $\pm \frac{1}{2}$. These rules are conveniently encoded in an {\em adjacency graph}, where the  
vertices are the allowed heights, with any allowed pair of nearest-neighbour heights are connected by an edge. The adjacency graph for the ABF models is therefore the Dynkin diagram of the $A_{k+1}$ Lie algebra
\begin{align}\Ak\ .
	\label{Akadjacency}
\end{align}

The Boltzmann weights follow from \eqref{eq:def_chi}, and are related to projectors by using \eqref{eq:Pf}. For the projector onto $\chi\,$=$\,0$, the diagram is easily evaluated using bubble removal, and the projector onto $\chi\,$=$\,1$ immediately follows because the two must sum to the identity operator. Thus
\begin{align}
\BWfour{\BWbareDotsx{0}}{a}{b}{b'}{c }= \delta_{ac} \frac{1}{d_{\frac12}} \sqrt{\frac{d_{b}d_{b'}}{d_ad_c} }\ ,\qquad
\BWfour{\BWbareDotsx{1}}{a}{b}{b'}{c }= \delta_{bb'} - \BWfour{\BWbareDotsx{0}}{a}{b}{b'}{c }\ .
\label{ABFweights}
\end{align}
It is easy to verify these are projectors using the quantum dimensions (\ref{eq:def_QD}), which obey
$d_{\frac12}d_h = d_{h+\frac12} + d_{h-\frac12}$ (for $h \geq \tfrac12$). 
Since there are two channels, ignoring an overall constant the Boltzmann weight for each quadrilateral can be written in terms of a single parameter.
A convenient rewriting of the amplitudes is in terms of the ``spectral'' parameter $u_Q$ defined by
\begin{align}
	\BWabcd & \;=\; {\sin(\lambda-u_Q)} \delta_{bb'}\dfrac{1}{\sqrt{d_bd_{b'}}}
	\;+\; {\sin u_Q}\ \delta_{ac}\dfrac{1}{\sqrt{d_ad_c}} \,,
	\label{uQdef}
\end{align}
where $\lambda=\pi/(k+2)$, so that  $d_{\frac12} = 2 \cos\lambda$. This expression \eqref{uQdef} applies for the heights on any graph $G$ as defined above. In computing the partition function one must remember to also include the extra weight $d_{h_{\rm v}}$ for each vertex ${\rm v}$.
The relation of the trigonometric ABF models to the geometric picture from section \eqref{sec:geometricexamples} has long been known \cite{Pasquier:1987}. The two (non-orthogonal) channels on each quadrilateral from \eqref{uQdef} (called $0$ and $\tilde{0}$ near (\ref{avoidcrossing})) correspond to 
\begin{align}
	\BWId & \;=\; \sum_\chi \BWabcdx{\chi} = \frac{\delta_{bb'}}{d_b} \,,
&	\BWTL & \;=\; d_{\frac12} \times \BWabcdx{0} =\frac{\delta_{ac}}{d_a} \,.
\label{TLgeo}
\end{align}
Both \eqref{TLgeo} and  \eqref{uQdef} make apparent that the weights obey a ``crossing'' symmetry: sending $u_Q\to \lambda-u_Q$ is equivalent to rotating the square by 90 degrees.

These projection operators have some famous properties.  Defining $e_j$ by
\begin{align}
	e_j = d_{\frac{1}{2}}P^{(0)}\ ,
\label{TLproj}
\end{align}
these generators obey the {\em Temperley-Lieb algebra}  \cite{Temperley1971}.
\begin{align}
e_j^2 = d_{\frac12} e_j\,,\qquad
e_{j}e_{j\pm1}e_j = e_j\,, \qquad
e_j e_{j'} = e_{j'}e_j \ \text{ for } |j-j'| >1.  
\label{TLalgdef}
\end{align}  
By construction, these are all consistent with the rules in section \ref{sec:evaluation}, as using \eqref{TLgeo} gives
\begin{align}
\Esquared = d_\upsilon \Eprime\ ,\qquad\qquad \Epm = \Epmb 
\end{align}
The first relation is loop removal \eqref{eq:loop_removal}, while the second reflects isotopy invariance.

An integrable trigonometric ABF model is given by making $G$ the square lattice and the spectral parameter uniform in space: $u_Q=u$ for all $Q$. Integrability means that appropriately defined transfer matrices commute for different $u$.  Here, the model is critical, with its long-distance behavior depending on the sign of $u$. In the ferromagnetic case $0<u<\lambda$, the continuum limits are described by the $A$-series of CFT minimal models with central charge $c = 1 - \frac{6}{(k+1)(k+2)}$ \cite{BPZ:CFT:84, FQS84, Huse1984}; see section \ref{sec:gmin} below for a few more details. The CFTs describing the antiferromagnetic case $0>u>-\lambda$ are the $\mathbb{Z}_k$ parafermion conformal field theories with $c = 2\frac{k-1}{k+2}$ \cite{ZamoFateev:Parafermion:85}. 
For $k=2$ the CFTs are the same, as the square-lattice Ising ferromagnets and antiferromagnets are equivalent. 
As we discuss in more depth below, the $\mathcal{A}_{k+1}$ category used to define the lattice model is {\em not} the category describing the topological properties of the primary fields in CFT, but rather is a subcategory. 

One can define a critical quantum Hamiltonian $H$ commuting with the transfer matrix by taking the $u\to 0$ limit of \eqref{Transfermatrixdef}
\begin{align}
H_{\rm crit}= |u|^{-1} (1-T) =\pm \sum_j e_j\ .
\label{HTL}
\end{align}
This Hamiltonian is of the form \eqref{HPdef}, with next-nearest-neighbor interactions.
In section \ref{sec:selfduality} we consider staggering the terms in \eqref{HTL}, leading to a very interesting gapped theory.

\paragraph{Ising model.}
We showed at great length in Part~I how expressing the Ising model in terms of a fusion category yields a great deal of insight into the construction of topological defects and their consequences. 
The Ising category is equivalent to the $\mathcal{A}_3$ category, where the standard labels $\I$, $\sigma$ and $\psi$ for the objects in the former are $0,\,\tfrac12,\,1$ in the latter. Fusion with $\sigma$ using \eqref{eq:Ising_fusion_rules} indeed yields the adjacency graph (\ref{Akadjacency}) with $k=2$.

The Ising model is thus the ABF model with $k=2$ \cite{Andrews1984}, but the equivalence is not immediately obvious.
The conventional definition of the Ising model has only two different degrees of freedom on each site, whereas there are three allowed heights in the $\mathcal{A}_3$ description obeying the adjacency rules. Moreover, by construction the heights live on a bipartite graph, c.f.\ (\ref{Zfusion}).
Given these constraints, the only allowed height on half of the sites is $\sigma$, while on the other half the heights are either $\I$ or $\psi$.
The height model on the latter subgraph corresponds to the Ising model, with the $\I$ and $\psi$ corresponding to Ising ``spin'' degrees of freedom $h_n=0$ and $1$ respectively.

The interactions in this Ising model are the standard nearest-neighbor ones, since there only two vertices on each quadrilateral occupied by $h_n=0,1$.
To see this from the height construction, first note that the tetrahedral symbols for the theory are
\begin{align}
	\Msixj{\sigma & \sigma & a}{\sigma & \sigma & b}  = \frac{(-1)^{ab}}{\sqrt{2}} \,,
	\qquad \qquad  \Msixj{a & b & c}{\sigma & \sigma & \sigma}  = N_{ab}^c \,,
\label{Isingtetra}	
\end{align}
where $a,b,c = 0$ or $1$. 
The basis for the Boltzmann weights is therefore
\begin{align}
	\BWfour{\BWbareDotsx{\chi}}{\sigma}{a}{b}{\sigma} = \frac{1}{2}(-1)^{\chi(a-b)} \,,
	\qquad \BWfour{\BWbareDotsx{\chi}}{\; \; a}{\sigma}{\sigma}{b \; \;  } =  N_{\chi b}^a=\delta_{\chi,|a-b|} \,.
\label{Isingbasisweights}	
\end{align}
where $\chi$, $a$, $b$ are $0$ or $1$. The relation to the standard Ising Boltzmann weights is given in Part~I and in (\ref{Jxyu},\,\ref{Jxyu2}) below.

\paragraph{Higher spin ${\mathcal A}_{k+1}$ models}

The ``higher-spin'' or ``fused'' models  generalize the ABF models by using the same ${\mathcal A}_{k+1}$ fusion category, but with $\upsilon>\tfrac12$.  They  have different adjacency rules, and a height can be adjacent to itself when $\upsilon$ is an integer. Whenever $\upsilon$ is an integer, the adjacency graph splits into two pieces.
For example, for $\upsilon=1$,  
\begin{flalign} \qquad & \begin{array}{r @{\quad\quad } c @{\quad \quad \quad \quad } c }
 &\text{integer heights}&\text{half-integer heights}\\[2ex]
k\hbox{ odd }&\AintHalfInta&\AintIntb   \\[4ex]
k\hbox{ even }&\AintInta&  \AintHanfIntb
		\end{array} & 
\label{spin1heights}		
\end{flalign}
There are effectively two height models in such cases, one comprised of integer heights, and the other half-integer. This splitting into two theories occurs in a more general setting, where two subcategories $M,M' \subset \mathcal{C}$ and an $\upsilon \in \mathcal{C}$ such that $\upsilon \otimes M \subset M$ and $\upsilon \otimes M' \subset M'$. We
show in section \ref{sec:generalized} how such models are dual to each other by displaying an appropriate topological defect. Namely, if there exists an object $b \in \mathcal{C}$ such that $b \otimes M \subset M'$ and $b \otimes M' \subset M$, the defect labeled by $b$ maps between models with labels in $M$ and labels in $M'$.

The Boltzmann weights for higher-spin ${\mathcal A}_{k+1}$ models are specified as always by fixing the coefficients $A_Q(\chi)$ in each of the $2\upsilon+1$ channels. The projectors are computed using \eqref{eq:def_chi} and \eqref{eq:Pf}, with the appropriate associated tetrahedral symbols needed to define these are given explicitly in Appendix~\ref{app:TetraSym}.  Integrable cases exist for any $\upsilon$ \cite{Kulish:1981,Date:1987}. For example, in the spin-1 case they have amplitudes (\ref{spinoneYBE}) or (\ref{IKYBE}) with $u$ uniform in space and with the projectors written in terms of the heights. The former case with $-\gamma<u<0$ has continuum limit described by the minimal supersymmetric conformal field theories with $c = \frac32 - \frac{12}{k(k+2)}$ \cite{Date:1987}.

\paragraph{$N$-state clock and Potts models.}
Here we show how to cast the $N$-state clock and Potts models in terms of the Tambara-Yamagami fusion category based on $\mathbb{Z}_N$~\cite{Tambara1998}.
Not only are these models of fundamental importance in statistical mechanics, but recently they and their parafermionic operators also have been studied intensively in the search for platforms for topological quantum computation \cite{Alicea16}.
Moreover, they provide a nice example that generalizes the models above to include non-self-dual objects, labeled by arrows to distinguish an element from its conjugate.

The $\mathbb{Z}_N$ Tambara-Yamagami fusion category 
has $N+1$ simple objects, and
$N$ of them obey Abelian fusion rules. Labeling the Abelian objects by an integer $a$, $b$ modulo $N$, they obey
\begin{align}
	a\otimes b = (a+b) \,.
\label{clockabelian}
\end{align}
with the right-hand side interpreted mod $N$.
The remaining non-Abelian object we denote $X$, and its fusion rules are
\begin{align}
	 a \otimes X = X \otimes a = X, \qquad X \otimes X = \bigoplus_{a= 0}^{N-1} a\ .
\label{clockX}
\end{align}
As apparent from these rules, the quantum dimensions of these objects are $d_{a} = 1$ and $d_X = \sqrt{N}$.
The abelian objects have conjugates given by $\bar{a} = (N-a)$, while $X$ is self-conjugate. When writing fusion diagrams, we therefore must keep track of arrows. The non-trivial $F$-symbols are
\begin{align}
	\left[F^{aXb}_X\right]_{XX} =\left[F^{XaX}_b\right]_{XX} = \omega^{ab}, \qquad \left[ F^{XXX}_X\right]_{ab} = \frac{1}{d_X}\omega^{-ab}\ ,
	\label{FPotts}
\end{align}
where $\omega=e^{2\pi i/N}$. More details on this category are given in Appendix~\ref{app:Potts}.

Since the objects here are not self-conjugate, the formalism of sections \ref{sec:Fusion_Theory} and \ref{sec:statmech}  requires generalization.
We give some detail in appendix \ref{app:notselfdual} on how this works.
The formula \eqref{eq:def_chi} for the Boltzmann weights in each fusion channel generalizes to 
\begin{align}
	\BWabcdx{\chi} & \;\defineas\; \frac{\sqrt{d_\chi}}{d_\upsilon^2 \sqrt{d_ad_bd_{b'}d_c} }
		\; \omegaf
	\;=\; d_\chi \Msixj{\upsilon & \upsilon & \chi}{a & c & b} \Msixj{\upsilon & \upsilon & \chi}{a & c & b'}^\ast \,.
	\label{weightsdef}
\end{align}
The rotated version generalizing \eqref{Frhombus} is found from an $F$-move, giving
\begin{align}
	\BWabcdx{\raisebox{-0.4ex}{\begin{sideways}$\widetilde{\chi}$\end{sideways}}}
	\;=\; d_{\widetilde{\chi}} \Msixj{\upsilon & \upsilon & \widetilde{\chi}}{d & b & a} \Msixj{\upsilon & \upsilon & \widetilde{\chi}}{d & b & c} 
	\;=\; d_{\widetilde{\chi}} \sum_{\chi}
		\Msixj{\upsilon & \upsilon & \widetilde{\chi}}{\upsilon & \upsilon & \bar{\chi}}
		\BWabcdx{\chi} \;.
\end{align}

The lattice models built from the $\mathbb{Z}_N$ Tambara-Yamagami with $\upsilon = X$ are known as {\em clock} models. The adjacency graph generalizes that of the $N=2$ Ising case to
\begin{align}
	\APotts\ .
\end{align}
The Boltzmann weights are built from the projectors $P^{(\chi)}$, where $\chi=0\dots, n-1$, as $N_{XX}^X=0$.
Using \eqref{weightsdef} we find
\begin{align}
	\BWfour{\BWbareDotsx{\chi}}{X}{a}{b}{X} &= \frac{\omega^{\chi(a-b)}}{d_X^2},
	\qquad \BWfour{\BWbareDotsx{\chi}}{\; \; a}{X}{X}{b \; \;  } =  N_{\chi b}^a = \delta_{\chi,(a-b)\,{\rm mod}\ n} \,,
\label{clockprojectors}	
\end{align}
The amplitudes and weights defined by \eqref{WAP} have a manifest $\mathbb{Z}_N$ symmetry under a cyclic shift of the $N$ states. Since the graph $G$ is bipartite, all heights on one sublattice will be $X$, while all the degrees of freedom $a=0\dots N-1$ live on the other sublattice.
The clock models thus provide a natural generalization of the Ising model to a $N$-state system where the interactions are effectively nearest-neighbor.

A famous special case of the clock model is known as the Potts model. Here the $\mathbb{Z}_N$ symmetry is enhanced to the permutation group $S_N$, which requires that the couplings obey 
\begin{align}
\hbox{Potts: }\quad A_Q(\chi)=A_Q(\chi')\ \hbox{ for all }\chi,\chi'\ne 0\ .
\end{align}
The Boltzmann weights then can be parametrized as in the $\mathcal{A}_k$ case \eqref{uQdef}, where here $d_X=2\cos\lambda=\sqrt{N}$.
Thus for $N>4$, $\lambda$ is imaginary and we must take $u$ imaginary to keep the Boltzmann weights real.
The Potts model is integrable only when the couplings are uniform in space (i.e., $A_Q$ is independent of $Q$ well) and $G$ is one of certain lattices, including the square \cite{Baxter1982}.
The Potts models also can be written in terms of generators satisfying the Temperley-Lieb algebra \eqref{TLalgdef}.
The generators are defined as $e_j=d_X P_j^{(\mathds{1})}$, so that their matrix elements are
\begin{align}
(e_j)_{\{h\},\{h'\}}= \begin{cases}
 \sqrt{N} \delta_{h_{j-1}h_{j+1}}\,\prod_{l} \delta_{h_l\,h'_l}\quad & \hbox{ for }h_j=X\ ,\\
\tfrac{1}{\sqrt{N}}\, \prod_{l\ne j} \delta_{h_l\,h'_l} & \hbox{ for } h_{j-1}=h_{j+1}=X.
\end{cases}
\label{TLPotts}
\end{align}
This form enabled the original derivation that Potts models are critical on the square lattice only for $u$ uniform and $N\le 4$  \cite{Temperley1971}. 
Another interesting case is the ``parafermion'' lattice model \cite{Fateev82}, which is integrable on the square lattice when $u_Q=u$ is uniform in space and the $A_Q$ obey
\begin{align}
\hbox{integrable parafermions: }\qquad\  \frac{A_Q(\chi)}{A_Q(\chi-1)}\ =\  
\frac{1-e^{iu}\omega^{\chi+\frac12}}{e^{iu}-\omega^{\chi+\frac12}}\ .\qquad
	\label{paraweights}
\end{align}

These integrable clock models are critical for all $N$, and their critical limit is described by the $\mathbb{Z}_N$-invariant parafermion conformal field theory \cite{ZamoFateev:Parafermion:85}.
These are the same $c=\frac{2(N-1)}{N+2}$ CFTs that describe the critical antiferromagnetic ABF models, but the lattice models are very different for $N>2$.
Here the $\mathbb{Z}_N$ symmetry is an internal one, shifting the spins on each site.
For $N=2,3$, the integrable parafermion and Potts models are the same, but for $N=4$ they are distinct (although both critical). 
As with the ABF case, the Tambara-Yamagami category used to define the lattice models differs from that governing the fusion of the primary fields in corresponding CFT. Here it is not even a subcategory: the field corresponding to the object $X$ does not appear in the modular tensor category. Rather, both categories have an abelian subcategory involving the objects $a=0\dots n-1$.

\paragraph{Birman-Murakami-Wenzl (BMW) models. }
The Boltzmann weights for the integrable BMW height models are given in Ref.~\cite{Jimbo:1987}, and explicit expressions for the projectors can be reverse-engineered from these.
The projectors also can be found in Ref.~\cite{Reshetikhin1988b}.
The tetrahedral symbols of the fusion categories associated with the BMW algebra can be inferred from the Racah coefficients derived in Ref.~\cite{Dai2001}.

\paragraph{Fusion multiplicites.}

For theories which have multiple fusion channels, i.e., integers $N_{ab}^c>1$, there are additional degrees of freedom on each link to signify how the adjacent heights fuse together. 
Our results generalize in an obvious way.


\section{Lattice models as defects in the Turaev-Viro-Barrett-Westbury state sum}
\label{modules}

The equivalence of the geometric and height lattice models relied on the shadow-world construction.
In this section we explain the origin of this construction. Moreover, we show how the same setup can be used to 
define lattice topological defects with a host of elegant properties. We make use of the Turaev-Viro-Barrett-Westbury state sum (TVBW state sum)~\cite{TuraevViro1992,Barrett1996} to arrive at these results, and to 
derive further topological properties of the two-dimensional height and geometric models.
The key formula \eqref{shadoweval2} relating the geometric models to the height models arises directly from evaluating the TVBW state sum on a certain 3-manifold with a particular boundary condition. Even more strikingly, we show how lattice topological defects naturally arise as a consequence of a more complicated boundary condition for the TVBW state sum. In sections~\ref{sec:topological_defect_lines} and \ref{sec:trivalent} we analyze these defects and prove their topological invariance using a more direct approach.  As such the reader should feel free to skip this section on a first reading.

\subsection{Turaev-Viro-Barrett-Westbury state sum}
\label{TVBW state sumwBdry}
We breifly review the Turaev-Viro-Barrett-Westbury state sum~\cite{TuraevViro1992,Barrett1996}.
The TVBW state sum describes a 2+1D topological quantum field theory (TQFT). 
It takes a fusion category $\mcc$ as an input, 
and assigns a number to every closed 3-manifold and a vector space to every 3-manifold with boundary. 
Picking a particular boundary condition, as we do below, again results in a number. 

The TVBW state sum can be viewed as a statistical partition function built on a cell decomposition of a 3-manifold $\mcm$. 
For now we take $\mcm$ to be closed. 
For convenience we require the cell decomposition to be dual to a triangulation; 
each 1-cell should belong to the boundary of three 2-cells,
and each 0-cell should belong to the boundary of six 2-cells and four 1-cells. 
The states in the TVBW state sum are given by assigning a simple object $x_f \in \mcc$ to each 2-cell $f$.
The TVBW state sum is given by
\begin{align}
	\label{TV0}
	Z_\text{TVBW}(\mcm) = \mathscr{D}^{-2N_{\text{3-cells}}} \sum_{\{ x_f \}} \,
		\prod_{\text{2-cells }f} \!\! d_{x_f} \prod_{\text{0-cells }{\rm v}} \!\! \operatorname{Tet}({\rm v})
\end{align}
where $N_{\text{3-cells}}$ is the number of 3-cells, $\mathscr{D}^2 = \sum_{x \in \mcc} d_x^2$,
	and $\operatorname{Tet}({\rm v})$ is a 6j-symbol whose labels are determined by the six 2-cells which meet at ${\rm v}$.
The 6j-symbol assigned to a vertex ${\rm v}$ depends on the adjacent 2-cell labeling as
\begin{align}
\operatorname{Tet}({\rm v} ) =  \Msixj{a & b & c}{x & y & z} \ ,\qquad\qquad
\TetvTBWSS \;.
\end{align}
The face labelled with height $y$ is behind all the other visible faces.

A few comments are in order. 
If any triple of two cell labels meeting at a 1-cell cannot fuse to vacuum the Boltzmann weight for that labeling vanishes (see \eqref{vanishingF}). 
This partition function evaluates to $\mathscr{D}^{-2}$ on the three-sphere.
The partition function is independent of choice of cell decomposition, and as such is describing a topological quantum field theory (TQFT). 
The TQFT is known as the {\em Drinfeld center} of the fusion category $\mcc$, and is denoted $\mathcal{Z}(\mcc)$.
The Drinfeld center is a {\em modular tensor category}, a fusion category possessing more structure, namely braiding and a non-degenerate modular $S$-matrix ($\det{\bf S } \neq 0$). 
Many topological properties of the lattice models we describe in this paper can be understood by viewing them as special boundary conditions to the TVBW state sum as we describe in the next subsection.

We now describe how to incorporate boundaries into the TVBW state sum~\cite{Balsam2010,Kirillov2011}.
The boundary of an open 3-manifold is a closed 2-manifold.
The fusion category $\mcc$ associates a vector space to every closed 2-manifold: $X \mapsto V(X,\mcc)$.%
	\footnote{The dimension of $V(X,\mcc)$ is equal to the number of ground states of $\mathcal{Z}(\mcc)$ on the manifold $X$.  For example, $\Vdim V(S^2,\mcc) = 1$ and $\Vdim V(T^2,\mcc)$ is the number of simple objects of $\mathcal{Z}(\mcc)$.
	}
As such, the TVBW topological quantum field theory assigns a vector to $\mcm$ living on $V(\partial\mcm,\mcc)$.
Furthermore, gluing two manifolds which share a common boundary together to form a closed manifold is equivalent to taking an inner product of their respective vectors, and results in the state sum given by \eqref{TV0}.

Fusion diagrams on 2-manifolds are dual vectors in $V^\ast(\partial\mcm,\mcc)$, that is, they provide a linear map $V(\partial\mcm,\mcc) \to \mathbb{C}$.
Consequently, the TVBW state sum assigns a number to any fusion diagram $\mathcal{F}$ on the boundary of a manifold $\mcm$.
To evaluate this number we again pick a cell decomposition of $\mcm$ dual to a triangulation that is compatible with $\mathcal{F}$: the 1-cells (edges) of $\mathcal{F}$ lie at the boundary of 2-cells of $\mcm$, and the 0-cells (vertices) of $\mathcal{F}$ are in the boundary of 1-cells of $\mcm$.
Again, to each 2-cell of $f\in\mcm$ we assign a simple object $x_f$.
For the 2-cells $f$ intersecting the boundary, $x_f$ is that $(f \cap \partial\mcm) \in \mathcal{F}$ determined by the boundary diagram $\mathcal{F}$.
From the cell decomposition,
\begin{align}
	\label{TV1}
	Z_\text{TVBW}(\mcm; \mathcal{F}) =
		\mathscr{D}^{-2N^\text{bulk}_\text{3-cells}}
		\prod_{\substack{ \text{boundary}\\\text{1-cells } e\in\mathcal{F} }} \mkern-20mu\sqrt{d_{x_e}}
		\; \sum_{\{ x_f \}} \,
		\prod_{\substack{ \text{bulk}\\\text{2-cells } f }} \!\! d_{x_f}
		\prod_{\substack{ \text{0-cells } {\rm v} }} \!\! \operatorname{Tet}({\rm v})  \;.
\end{align}
We sum over the $x_f$ assigned to the bulk 2-cells (those which do not have support on $\partial \mcm$).

We remark that there is some arbitrariness in defining $Z_\text{TVBW}(\mcm; \mathcal{F})$. 
The bulk weights are fixed by \eqref{TV0}.
The weights associated with boundary 0- and 1-cells are chosen so that the state sum is invariant under $F$-moves of the fusion diagram $\mathcal{F}$.
We pick the overall normalization so that that the ``identity diagram'' ($\mathcal{F}$ with only $\I$ lines) evaluates to one on $S^2$.

\subsection{Lattice models as defects in the Turaev-Viro-Barrett-Westbury state sum}

The partition functions of the geometric and height models with and without topological defects are naturally described using TVBW state sums, with the lattice configurations entering into the boundary conditions on the 3d space. 
We first pick a closed 2-manifold $Y$, and embed in it the graph $G$ used to build the lattice models. 
We let $\mcm = Y \times I$, and choose boundary conditions on $Y\times \{1 \} $ and $Y \times \{ 0 \}$.  Each boundary condition is a fusion diagram, which we label as $\bra{\mathcal{F}}$ and $\ket{D}$ for $Y\times \{1 \} $ and $Y \times \{ 0 \}$ respectively. 
Doing the TVBW state sum over $\mcm$ with these boundary conditions produces a number, which we write as $ \braket{\mathcal{F}|D}$. In a picture,
\begin{align}
\braket{\mathcal{F}|D} =Z_\text{TVBW}(Y \times I;\mathcal{F} \cup D) =  \TVa
\label{FDdef}
\end{align}
where for clarity we have suppressed the diagram labels, and drawn only a patch of the 3-manifold $Y\times I$.
Even though we adopt the same physics bra-ket notation, this TQFT inner product is distinct from that defined in \eqref{eq:destroyer}, as here the vector space is spanned by two-dimensional net configurations.

The degrees of freedom in the geometric lattice models are defined in terms of fusion diagrams $\mathcal{F}$ living on a graph $G$ embedded in $Y\times\{1\}$, as described in section~\ref{sec:loopmodels}. We define the state $\bra{\Psi}$ as a linear combination of all such diagrams with
\begin{align}
\label{netconfig}
\bra{\Psi} = 
\sum_{\mathcal{F}} A\big(\mathcal{F}\big) \big\langle \mathcal{F}\big|
\end{align}
where $A\big(\mathcal{F}\big)$ is the local part of the Boltzmann weights  \eqref{ZW}, the product of amplitudes $A_Q$. The partition function of the geometric model in the absence of defect lines is defined by 
\begin{align}
Z_{\rm geo} = \big\langle\Psi \big| 0\big\rangle = \sum_{\mathcal{F}} A\big(\mathcal{F}\big) \big\langle\mathcal{F}|0\big\rangle
\label{Z0}
\end{align}
where $|0\rangle$ is diagram with no lines. This expression may be used to define more general lattice models having different amplitudes at each vertex of $\mathcal{F}$, a $G$ not made of quadrilaterals, and/or $\upsilon$ not uniform in space.

\begin{figure}[tbp]
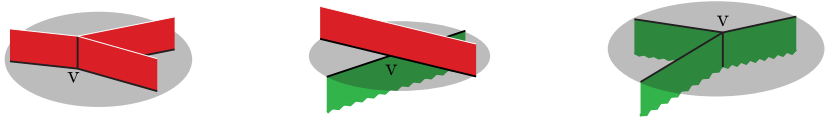

\begin{center}
$\TVup \quad \quad \quad \quad \TVcross \quad \quad \quad \quad \TVdown$
\caption{The three types of 0-cells appearing in the TVBW state sum for $\langle \mathcal{F}|D\rangle$.
The three 0-cells are indicated with ${\rm v}$.
The grey regions are 2-cells which sit inside the surface $Y \times \{ \frac{1}{2} \}$.
The red (green) 2-cells are found by extending the nets living on $Y \times \{1 \}$ ($Y \times \{0\}$) into the bulk until they meet $Y \times \{ \frac{1}{2} \}$.
}
\label{TVcells}
\end{center}
\end{figure}

The state-sum setup makes it easy to to define a partition function in the presence of defects. A defect configuration $\ket{D} \in V^*(Y,\mcc)$ is also given by a $\mathbb{C}$-linear combination of fusion diagrams, but now on $Y\times \{0\}$. The inner product  $\langle \Psi |D \rangle =Z_\text{TVBW}(Y \times I;\Psi \cup D)$ is given by evaluating the TVBW state sum on $Y \times I$, subject to the boundary condition $\bra{\Psi}$ on $Y \times \{ 1 \} $, and $\ket{D}$ on $Y \times \{ 0 \}$. The presence of the defects does not affect the amplitudes of the Boltzmann on each quadriateral $Q$ of $G$, so the partition function in the presence of defects is
\begin{align}
\big\langle\Psi \big| D\big\rangle = \sum_{\mathcal{F}} A\big(\mathcal{F}\big) \big\langle\mathcal{F}|D\big\rangle\ .
\label{ZD}
\end{align}
To compute $\braket{\mathcal{F}|D}$ using the state sum $\eqref{TV0}$, we choose a cell decomposition where the 1-cells appearing in the fusion diagrams for $\mathcal{F}$ and $D$ are boundaries of 2-cells of $Y \times I$,  and where those 2-cells terminate on the surface $Y \times \{ \frac{1}{2} \}$.
The 2-cells meeting the surface $Y \times \{ \frac{1}{2} \}$ can do so in the three ways shown in Figure~\ref{TVcells}.
An easy way to keep track of these data is by projecting the inner product \eqref{FDdef} into the plane: 
\begin{align}
\label{TVproj}
\braket{ \mathcal{F} | D } =\TVb \ .
\end{align}
The green strands represent 2-cells extending from $Y \times \{ 0 \}$ to $Y \times \{ \frac{1}{2} \}$ labeled with $\ket{D}$, while
the red strands represent 2-cells extending from $Y \times \{ 1 \}$ to $Y \times \{ \frac{1}{2} \}$ labeled by $\ket{\Psi}$. 
The gray shaded region makes up the remaining 2-cells, the faces of $Y \times \{\frac{1}{2}\}$.
Each of the latter 2-cells $f$ is labeled by a simple object of $\mathcal{C}$, denoted $h_f$.
The explicit formula for the inner product in \eqref{TV1} is then
\begin{align}
	\label{TVSS}
	\langle \mathcal{F} |D \rangle  \quad=\;\;
		\prod_{\substack{\text{red line}\\e}} \mkern-8mu \sqrt{d_e}
		\prod_{\substack{\text{green line}\\e}} \mkern-15mu \sqrt{d_e}
		\; \sum_{\{h_f\}} \;
		\prod_{\substack{\text{gray face}\\f}} \!\!\! d_{h_f}
		\;\prod_{\substack{\text{vertex}\\\rm v}} \! \operatorname{Tet}({\rm v}) \,,
\end{align}
where the sum is over all possible labelings of the gray 2-cells, and the products are over all 2-cells $f$, and vertices labeled $\rm v$. As in Eq.~\eqref{shadoweval2}, all 2-cells must be contractible.
We comment that the sum in \eqref{TVSS} for the ``identity graph'' (on both sides) evaluates to $\sum_a d_a^{2 \chi}$ where $\chi$ is the Euler characteristic of $Y$.
Explicitly writing out the tetrahedral symbols that appear from each type of 2-cell termination appearing in Fig.~\ref{TVcells} and in the projection \eqref{TVproj} gives
\begin{subequations} \begin{align}
	\TVvertup\;\;\; &= \operatorname{Tet}({\rm v}) = \Msixj{\alpha & \beta & \gamma}{h_1 & h_2 & h_3} ,
	\label{TV1a}\\
	\TVvertcross\;\;\; &= \operatorname{Tet}({\rm v})=\Msixj{X& h_1 & h_2}{\alpha & h_3 &h_4} ,
	\label{TV2}\\
	\TVvertdown\; &= \operatorname{Tet}({\rm v}) = \Msixj{X & Y & Z}{h_1 & h_2 & h_3} .
	\label{TV3}
\end{align} \end{subequations}

The key advantage of defining the lattice models in terms of the TVBW state sum  is that the TQFT inner product is invariant under local deformations of the defect configuration $\ket{D}$. 
Different configurations are related by the re-triangulation invariance of the TVBW state sum. 
The local weights in $\bra{\Psi}$ mean however that $\langle \Psi |D \rangle$ is not topologically invariant but instead a more interesting partition function.\footnote{We note that there are choices of $\ket{\Psi}$ which do lead to topologically invariant partition functions. 
These are parametrized by Frobenius algebra objects in $\mathcal{C}$ and referred to as topological boundary conditions.} The shadow-world formula \eqref{shadoweval2} follows from  \eqref{TV1a} (up to an unimportant overall factor of $\mathscr{D}^{-2}$) by setting $\ket{D}=\ket{0}$, the configuration with no defects. The heights are simply the labels $h_f$ on the faces.  Defects with very nice topological properties come from defining the defect weights using \eqref{TV2}, giving \eqref{eq:def_LDefect} below. By construction, such defects are topological, as they can be moved around at will without changing $\braket{ \Psi | D }$.  In equation \eqref{DefComm1} we show how to express this topological invariance in terms of local relations we name the defect commutation relations, and prove  directly that these weights satisfy them. Topological trivalent junctions of defects arise by defining the triangle weight \eqref{eq:def_trivalent}	using \eqref{TV3}. We again verify directly that they satisfy the appropriate commutation relations, namely \eqref{eq:trivalent_pentagon}.

We explore some further consequences of the Drinfeld center in Sec.~\ref{sec:twists} by relating the behaviour of the lattice models under Dehn twists to idempotents in the tube category, which correspond to objects in $\mcz(\mcc)$. In addition, we note that vertex operators in the lattice models are naturally labeled by objects in $\mcz(\mcc)$, and can be viewed as terminations of bulk Wilson lines on the boundary. 


\section{Topological defect lines}
\label{sec:topological_defect_lines}

We now focus on central topic of our paper, the construction of topological defect lines in lattice statistical-mechanical models and quantum spin chains. In the presence of a topological defect, the partition function is independent of any continuous deformation of the path of this defect.

The requirement that a defect be topological results in a huge number of constraints, prohibitively difficult to solve by brute force. We show here that the way to make progress is to rewrite
lattice models in terms of fusion categories, as described in Sections~\ref{sec:Fusion_Theory}, \ref{sec:statmech} and \ref{modules}.
The rewriting gives a simple construction of topological defects in an enormous set of models.
Our method is to use \eqref{TV2} to define the defect weights, and give a direct proof that such defects are topological.
Our work thus substantially generalizes and systematizes a construction applicable to certain integrable and critical lattice models \cite{Chui2001,Chui2002,Chui2003}.
Not only does our method allow extensions to many more critical and off-critical models, but it also gives a construction of topological defects that branch and fuse. This additional structure is instrumental in allowing the exact computation of universal quantities.

\subsection{The definition}

\begin{figure}[t] \centerline{%
\xymatrix @!0 @M=1mm @R=24mm @C=55mm{
		\LineDefectInserta \ar[r] & \LineDefectInsertb \ar[r] & \LineDefectInsertc \\	
	}}
\caption{
To insert a defect line we choose a path along the edges of the lattice, shown by the zig-zag line in the leftmost diagram.
We then split open the lattice along this path, as in the middle diagram. 
Finally we ``glue" the lattice back together with a line of defect weights inserted along the path to find the diagram on the right.
The resulting partition function is written in \eqref{partition_function_defect}.}
\label{fig:LineDefect}
\end{figure}
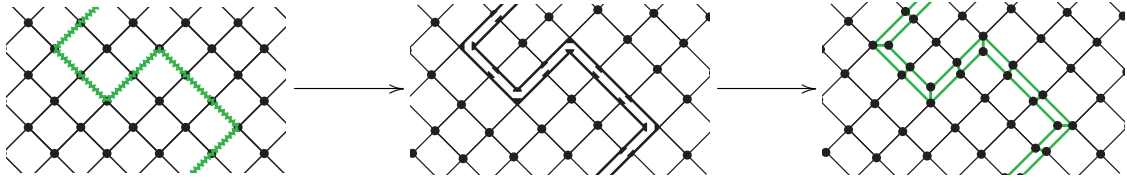

The defects we describe are defined by choosing a one-dimensional path along the edges of the quadrilaterals  comprising $G$ and insert a set of defect Boltzmann weights, as depicted in Figure~\ref{fig:LineDefect}.
Each vertex along the path is split into two, thus creating a string of defect quadrilaterals. 
We use the same notation for the defect weights as the Boltzmann weights, namely each defect weight is a number depending on the labels of the four surrounding vertices:
\begin{align}
\DefectSquarex{a}{b}{\alpha}{\beta},
\label{defectsquare}
\end{align}
The heights $a$,$b$ live on one side of the defect's path and the heights $\alpha$, $\beta$ the other.

The partition function in the presence of non-intersecting defect lines is now easy to define. Diagrammatically it is given by far right of Figure~\ref{fig:LineDefect}, with the convention that we sum over all heights. To be more precise, let $P$ be a path along the edges in the lattice, and $l \in P$ denote the edges in this path.
To each edge of $P$ we assign one defect weight \eqref{defectsquare}.
The partition function is
\begin{align}
	\mathcal{Z}_{D,P} = \sum_{\text{heights}} \left( \prod_{\text{faces}} \BWabcd \;\times \; \prod_{l \in P} \DefectSquarex{a}{b}{\alpha}{\beta} \times  \prod_{\text{vertices}} d_v \right). 
	\label{partition_function_defect}
\end{align}

\subsection{Defect commutation relations}
\label{sec:defects}

For a defect weight to be topological, the partition function defined in \eqref{partition_function_defect} must obey $\mathcal{Z}_{D,P} = \mathcal{Z}_{D,P'}$ for any two paths $P$ and $P'$ that can be continuously deformed into one another.
A necessary condition for a topological defect is that its weight (\ref{defectsquare}) and the Boltzmann weights obey the {\em defect commutation relations} defined in Part~I \cite{Aasen2016}:
\begin{align}
	\sum_d \DefectCommuteDprime \;=\; \sum_\beta \DefectCommuteUprime\ \ , \qquad\quad \sum_{b,c} \DefectCommuteSprimeprime=\DefectCommuteNSprimeprime\ .
	\label{DefComm1}
\end{align}
This set of equations holds for any choice of fixed heights around the outside, while the summations are over all allowed internal heights. The equations coming from rotating the pictures  must hold as well. 
The first of (\ref{DefComm1}) resembles the Yang-Baxter equation, but is a less demanding constraint. For example, it places no requirement that the Boltzmann weights be uniform in space, since it involves only one non-defect weight. Indeed, we will find topological defects for non-uniform and hence non-integrable models.
If the statistical-mechanical model were instead defined on a generic cell decomposition one would need to solve the analogous relations for every 2-cell.

Solving these equations by brute force is quite difficult, even for the Ising model analysed in Part~I.  However, when the Boltzmann weights are written as evaluations of diagrams in an underlying fusion category as described in Section~\ref{sec:statmech}, solving the defect commutation relations is straightforward.
In Section~\ref{modules} we showed how to define topological defect weights from the data associated with the Drinfeld center of the fusion category used to define the model. Here we prove directly that these weights obey the defect commutation relations, without appealing to the TVBW state sum. 
We find one solution for each simple object  $\varphi$ of the fusion category $\mathcal{C}$. A direct consequence of such a solution is that the partition function can only depend on the topology of the defect lines, for example which cycle(s) of a torus they wrap around.

The topological defect weight is nonzero only if the heights labeling its vertices satisfy fusion constraints similar to (\ref{eq:BWproptoNNNN}), but here involving $\varphi$ as well as $\upsilon$.
Namely, a height $a$ on one side of the defect path is related to the adjacent height $\alpha$ on the other side of the defect path by fusion with $\varphi$, i.e., $N_{a\varphi}^\alpha\ne 0$.
Adjacent heights on the same side of the defect still must have $\upsilon$ appearing in their fusion product as before.
Thus in analogy with (\ref{eq:BWproptoNNNN}) we have
\begin{align} 
\DefectSquarex{a}{b}{\alpha}{\beta} \propto N_{ab}^{\upsilon}N_{\alpha \beta}^{\upsilon}N_{a \alpha}^\varphi N_{b \beta}^{\varphi} \,,
\label{DWproptoNNNN}
\end{align}
for a topological defect specified by simple object $\varphi$.
The topological defect weights themselves turn out to have a remarkably simple form. 
They are proportional to an $F$-symbol, but are more naturally written using the tetrahedral symbols
\begin{align}
		\DefectSquarex{a}{b}{\alpha}{\beta} =
			\Msixj{\varphi & a & \alpha}{\upsilon & \beta & b} .
		\label{eq:def_LDefect}
\end{align}
{}From (\ref{vanishingF}) this symbol vanishes automatically unless $N_{a\alpha}^\varphi N^{\upsilon}_{ab} N_{\varphi b}^\beta  N^b_{a\upsilon} \ne 0$, so we do not need to impose the adjacency conditions in \eqref{DWproptoNNNN} separately. 
When the defect type is trivial ($\varphi = \mathds{1}$), the symbol reduces to
\begin{align}
	\Msixj{\mathds{1} & a & \alpha}{\upsilon & \beta & b}
	= \frac{N_{ab}^{\upsilon} \delta_{a\alpha}\delta_{b\beta}}{\sqrt{d_a d_b}} \label{eq:kronecker}\ 
\end{align}
so that the heights across this ``identity defect'' are equal. As described above in (\ref{semidotdef}) and represented by the dot on each vertex, the partition function includes a factor of $d_h$ for each vertex with height $h$ on it. 
These cancel the factors of $\sqrt{d_ad_b}$ coming from (\ref{eq:kronecker}), so including an identity defect line indeed changes absolutely nothing.

The corresponding defect weights for Ising are given in Part~I. Another simple example comes from the Fibonacci fusion category.
It is given by the subcategory of $\mathcal{A}_4$ consisting of two objects $\{0, 1 \}$, written as $\{0, \tau \}$ in the language common to the study of anyons \cite{Feiguin2007}.
The corresponding model is the $k=3$ ABF model where we label heights $\{0,\frac{3}{2}\}$ by $0$ and the heights $\{\frac12,1\}$ by $\tau$; because $G$ is bipartite we can restore the original heights if desired.
(In the language introduced in section~\ref{sec:duality} below, we consider one of the models $M^+$, $M^-$.)
We take $\varphi=\upsilon=\tau$, so that the only non-obvious fusion rule is $\tau\otimes\tau=0\oplus\tau$.
The non-zero defect weights are given by
\begin{align}
\DefectSquarex{\tau}{\tau}{\tau}{\tau} = -\frac{1}{d_{\tau}^2}\ , \qquad \DefectSquarex{0}{\tau}{\tau}{0} = \DefectSquarex{\tau}{0}{0}{\tau} = \DefectSquarex{0}{\tau}{\tau}{\tau}=\DefectSquarex{\tau}{0}{\tau}{\tau} = \DefectSquarex{\tau}{\tau}{0}{\tau} = \DefectSquarex{\tau}{\tau}{\tau}{0} = \frac{1}{d_{\tau}} \,,
\label{defectFib}
\end{align}
where $d_\tau=2\cos(\pi/5)=\phi=(1+\sqrt{5})/2$. 
It is quite tedious to check by brute force that they satisfy the defect commutation relations. Luckily, there is a better way.

The diagrammatic techniques developed in Section \ref{sec:Fusion_Theory} provide an efficient means to proving that (\ref{eq:def_LDefect}) indeed provides a solution to the defect commutation relations for all fusion categories. The key simplifying feature is that since all the weights are defined using quantities coming from evaluations in the fusion category, we can use $F$-moves to relate various combinations of them. 
For example, the series of steps to show that \eqref{eq:def_LDefect} solves 
\begin{align}
\sum_h \DefectCommuteDprimeprime\quad =\quad \sum_\beta\; \DefectCommuteUprimeprime
\end{align}
can be summarized by the following diagram:
\begin{align}
	\vcenter{\xymatrix @!0 @M=2mm @R=39mm @C=57mm{
		 \DefCommSolLa \ar[d] &\DefCommSol \ar[r] \ar[l]& \DefCommSolRa \ar[d] \\
		\DefCommSolLb & & \DefCommSolRb	
	}}
\label{defcommdiagram}
\end{align} 
The bottom two pictures each evaluate to a Boltzmann weight (recall \eqref{eq:def_chi}) with a single closed defect line present, with the two related by dragging the line through.  The arrows correspond to doing $F$-moves and other manipulations to relate various evaluations to each other. 

To write out these manipulations explicitly, we start at the middle-top diagram of \eqref{defcommdiagram} and follow the path left and downward.
(The arrows in \eqref{defcommdiagram} are drawn to show the following steps, but all steps are reversible, as $F$-matrices are invertible.)
Putting in the explicit $F$-symbols yields
\begin{align}
	\nonumber
\DefCommSol &= \sum_{h, h'} \Msixj{\varphi & a & \alpha}{\upsilon & \delta & h} \Msixj{\varphi & c & \gamma}{\upsilon & \delta & h'} \sqrt{d_{\alpha h \gamma h'}} \; \; \DefCommSolLa  \\
	\nonumber
		&=  \sum_{h} \Msixj{\varphi & a & \alpha}{\upsilon & \delta & h} \Msixj{\varphi & c & \gamma}{\upsilon & \delta & h} \sqrt{d_{\alpha h \gamma h}} \sqrt{\frac{ d_\delta}{d_{\varphi h}}} \; \; \DefCommSolLb \\
	\nonumber
		&=  \sum_{h} \Msixj{\varphi & a & \alpha}{\upsilon & \delta & h} \Msixj{\varphi & c & \gamma}{\upsilon & \delta & h} \sqrt{\frac{ d_{\alpha\gamma h\delta} }{ d_{\varphi} }} \times d_\varphi \times \frac{d_\upsilon^2}{\sqrt{d_\chi}} \sqrt{d_{abhc}} \;  \BWfour{\BWbarex{\chi}}{a}{b}{h}{c}\\
	\nonumber
		& = \frac{d_\upsilon^2}{\sqrt{d_\chi}}  \sqrt{d_{abc \alpha \gamma \delta \varphi}} \,  \sum_{h} d_h\Msixj{\varphi & a & \alpha}{\upsilon & \delta & h} \Msixj{\varphi & c & \gamma}{\upsilon & \delta & h}  \,  \BWfour{\BWbarex{\chi}}{a}{b}{h}{c} \\ 
		&=  \frac{d_\upsilon^2}{\sqrt{d_\chi}}  \sqrt{d_{abc \alpha \gamma \delta \varphi}} \,  \sum_h \DefectCommuteDprimeprime \;.
\label{Defcomm1proof}				
\end{align}
We have used a compact notation for the product of quantum dimensions $d_{abc\cdots} \defineas d_a d_b d_c \cdots$. 
Starting from the middle-top of~\eqref{defcommdiagram} and traversing the diagram right and downward results in
\begin{align}
	\nonumber
	\DefCommSol \; &=  \frac{d_\upsilon^2}{\sqrt{d_\chi}}  \sqrt{d_{abc \alpha \gamma \delta \varphi}} \sum_{\beta} d_\beta \Msixj{\varphi & \alpha & a}{\upsilon & b & \beta} \Msixj{\varphi & \gamma & c}{\upsilon & b & \beta}  \; \BWfour{\BWbarex{\chi}}{\alpha}{\beta}{\delta}{\gamma}
	\\[-4ex]
	& =  \frac{d_\upsilon^2}{\sqrt{d_\chi}}  \sqrt{d_{abc \alpha \gamma \delta \varphi}} \sum_\beta \DefectCommuteUprimeprime \;.
	\label{Defcomm2proof}
\end{align}
Together these imply the first set of the defect commutation relations shown in (\ref{DefComm1}). The rotated version of these commutation relations can be found in an identical way.
In fact, since a complete basis of fusion channels also occurs for $\chi$ going horizontally instead of vertically, one can literally rotate the diagrams and use the same proof.

The second set of commutation relations in \eqref{DefComm1}  follows from
\begin{align}
	\vcenter{\xymatrix @!0 @M=3mm @R=0mm @C=35mm{
	\DefCommSolprimed &	 \DefCommSolprimea \ar[l] \ar[r] &\DefCommSolprimeb \ar[r] & \DefCommSolprimec
	}} \,.
	\label{three-to-one-defcomm}
\end{align}
For clarity, we left out all the height and $\upsilon$ labels, keeping only the defect label $\varphi$ and the channel label $\chi$.
The proof of (\ref{three-to-one-defcomm}) comes from evaluating the fusion diagram shown second from the left in two ways.
Following the arrow directed to the left yields one tetrahedral symbol multiplied by the Boltzmann weight.
Following the arrows directed to the right yields three tetrahedral symbols multiplied by the Boltzmann weight, along with appropriate summations. The remaining equations in \eqref{DefComm1} are then proved by translating the pictures back into quadrilaterals and including the explicit expressions as in (\ref{Defcomm1proof}) and (\ref{Defcomm2proof}).

One nice feature of giving proofs in this fashion is that it makes it straightforward to extend the results of this paper to models on more general lattices and graphs. In fact, these methods give solutions to the defect commutation relations for a partition function defined on {\em any} cell decomposition in a similar way, as long as all Boltzmann weights are written as evaluations of diagrams in a fusion category.

\subsection{The defect-line creation operator}
\label{sec:defectcreationop}

Many applications of topological defects are easiest to see in the transfer-matrix formalism.  In this set-up, a $\varphi$-defect in the horizontal direction is created by acting with an operator ${\mathcal D}_\varphi$ on the same vector space $\mathcal{V}$ on which the transfer matrix acts. Since the path of any topological defect can be deformed by using the defect commutation relations, we show here how $\mathcal{D}_\varphi$ must {\em commute with the transfer matrix}. The ensuing symmetries and dualities are described in Section~\ref{sec:duality}. 

For simplicity we here consider a square lattice with transfer matrix given by \eqref{Tmatrixdef2}
and work in terms of height models, although are results are easily extended to geometric models and to other graphs. The defect-creation operators are easiest to define with periodic boundary conditions, so they act on the $\mathcal{V}$ described in (\ref{eq:fusiontreedef},\,\ref{vper}). This space is spanned by all configurations of heights that satisfy the fusion rules.
Labeling by $Z_\varphi$ the toroidal partition function in the presence of a $\varphi$ defect line wrapped around this cycle, the defect-creation operator ${\mathcal D}_\varphi$ is then defined so that
\begin{align}
Z_\varphi= \hbox{tr}\,\mathcal{D}_{\varphi}\left(T_L \right)^M\ .
\label{Zphidef}
\end{align}
where $T_L$ is the transfer matrix acting on $L$ sites with periodic boundary conditions. Since $Z_\varphi$ remains invariant under deformations of the path of the defect, this definition does not determine ${\mathcal D}_\varphi$ uniquely. We fix the ambiguity by requiring $\mathcal{D}_\varphi$ to create a straight horizontal defect.  Its matrix elements for taking a state $\ket{h'_0 h'_1 \cdots h_{L-1}'}$ to  $\ket{ h_0 h_1 \cdots h_{L-1}}$ are then
\begin{align}\
({\mathcal D}_\varphi)_{\{h\},\{h'\}}
	& \;=\; \cdots \TopologicalSymmetryprime \cdots
	\notag
	\\ &	\;=\;	\prod_{n=0}^{L-1} \Msixj{\varphi & h_n & h'_n}{\upsilon & h'_{n+1} & h_{n+1}} \sqrt{d_{h_n} d_{h'_n}}\ .
	\label{eq:BW_duality}
\end{align}

Repeatedly applying the defect commutation relations \eqref{DefComm1} makes it easy to show that {\em any} topological defect can be commuted through the transfer matrix, and that a stronger statement holds:
\begin{align} 
P^{(\chi)}_j \mathcal{D}_\varphi = \mathcal{D}_\varphi P^{(\chi)}_j\ ,
\label{PDDP}
\end{align}
where $j$ denotes the spatial location of the projector $P^{(\chi)}$ defined in \eqref{eq:Pf}.
To prove \eqref{PDDP}, we note that the defect commutation relations give for any choice of external labels
\begin{align}
\nonumber
T(\{A_Q\}) \mathcal{D}_\varphi  &= \Topologicalsymmetrya \\
&=   \TopologicalSymmetryb
\label{TDDT}\\
\nonumber
&=\TopologicalSymmetryc\qquad =\quad  \mathcal{D}_\varphi T(\{A_Q\})\ .
\end{align}
We here and henceforth adopt the convention that in such pictures, the internal heights are summed over, while external are fixed. The commutation \eqref{TDDT} holds for any choice of the amplitudes $A_Q$ of the projectors in each Boltzmann weight, even when they are not uniform in space and when they are not integrable ones. Because we can choose all the $A_Q$ to give the identity  away from $j$, \eqref{PDDP} holds as well.

The commutation relation (\ref{TDDT}) makes the defect-creation operator $\mathcal{D}_\varphi$ look like a symmetry generator: It is a non-trivial operator that commutes with the transfer matrix. When the quantum dimension $d_\varphi=1$, the eigenvalues of $\mcd_\varphi$ are of unit norm and it is unitary.  It can indeed generate a symmetry, as we describe in section \ref{sec:symm}. In general, however, $\mcd_\varphi$ is not unitary and instead generates a duality, as described in section \ref{sec:duality}. It is worth noting that one always can split ${\cal V}$ into sectors labelled by distinct eigenvalues of ${\mathcal D}_\varphi$. Using the nomenclature coming from systems with topological order, these eigenvalues label a global ``flux'' \cite{Feiguin2007}.

A nice feature of the topological defect-creation operators is that they satisfy the same fusion rules as the objects that label them, namely
\begin{align}
\mcd_A \mcd_B = \sum_C N_{AB}^C \mcd_C
\label{DDeqD}
\end{align}
This identity can be proved by brute force,  by manipulating diagrams as in \eqref{defcommdiagram}, or by 
showing that the defect lines themselves satisfy \eqref{eq:loop_removal} and \eqref{F0c}. Here we present the brute-force proof. 
By definition,
\begin{align}
({\mcd}_A \mcd_B)_{\{h\},\{h'\}} \;=\;	\sum_{\{ x \}} \prod_{n=0}^{L-1} d_{x_n}\Msixj{A & h_n & x_n}{\upsilon & x_{n+1} & h_{n+1}}
	 \Msixj{B & x_n & h'_n}{\upsilon & h'_{n+1} & x_{n+1}} \sqrt{d_{h_n} d_{h'_n}}\ ,
	 \label{DADB}
\end{align}
where the sum is over all all heights $x_n$. We expand each pair of tetrahedral symbols into three by using the pentagon equation \eqref{pentagon}, giving
\begin{align*}
	\big({\mcd}_A \mcd_B\big)_{\{h\},\{h'\}} = \! \sum_{\{ C\},\{ x \}} \prod_{n=0}^{L-1} 
	d_{C_n} \Msixj{C_n & h_n & h_n'}{\upsilon & h_{n+1}' & h_{n+1}}
	d_{x_n}  \Msixj{A & B & C_n}{h_n' & h_n & x_n}\Msixj{A & B & C_{n-1}}{h_n'& h_n & x_n} \sqrt{d_{h_n} d_{h'_n}}.
\end{align*}
Conveniently, the sum over all $x_n$  now can be performed using the orthogonality relation \eqref{Orthogonality6j}.
All $C_n$ then must be equal while the factors of $d_{C_n}$ cancel, leaving us with (\ref{DDeqD}): 
\begin{align} 
({\mcd}_A \mcd_B)_{\{h\},\{h'\}} \; = \; \sum_{C} N_{AB}^C \prod_{n=0}^{L-1} \Msixj{C & h_n & h_n'}{\upsilon & h_{n+1}' & h_{n+1}}  \sqrt{d_{h_n} d_{h'_n}} = \sum_C N_{AB}^C (\mcd_C)_{\{h\},\{h'\}}\ .
\end{align}
A more elegant  and general proof uses the trivalent junctions we describe in section \ref{sec:trivalent}. We prove that all defect loops fuse as do their labels, no matter what their path. 

An illuminating way of viewing the defect-creation operators is in terms of the diagrammatic techniques used to prove the defect commutation relations.
This picture was utilized in the study of spin chains with Hamiltonian \eqref{HPdef}, where the fact that $[\mcd_\varphi, H]=0$ was derived and termed ``topological symmetry'' \cite{Feiguin2007}.
In this picture, acting with $\mcd_\varphi$ corresponds to fusing a closed loop $\varphi$ wrapped around a cycle into the ``spine" of the fusion tree \eqref{eq:fusiontreedef} with periodic boundary conditions  \eqref{vper}, starting with
\begin{align}
\nonumber
\mathcal{D}_{\varphi}&\left( \DomainTransfermatrixUpsilon \right)& \\
&\qquad = \; \TSymmprime&
\label{TsymmGoldenChain} \ .
\end{align}
We then fuse the defect line into the fusion tree using \eqref{F0c}, i.e.\
\begin{equation}
 \TSymmbprime  = \sum_{h_i'} N_{\varphi \mkern1mu h_i}^{h_i'} \sqrt{\frac{d_{h_i'}}{{d_{\varphi}d_{h_i}}}} \TSymmcaprime \ .
 \label{tsymmbprime}
\end{equation}
Doing this fusion on every horizontal edge of the tree gives for each $\upsilon$ vertex a picture
\begin{align}
\TSymmcprime &= \sqrt{d_{\varphi} d_{\upsilon} } \Msixj{\varphi & h_i & h_{i+1}}{\upsilon & h_{i+1}' & h_{i}'} \TSymmdprime
\notag\\
&= \sqrt{d_{\varphi} d_{\upsilon} } \Msixj{\varphi & h_i & h_{i+1}}{\upsilon & h_{i+1}' & h_{i}'}  \sqrt{\frac{d_{h_i} d_{h_{i+1}}}{d_\upsilon}} \, \TSymmeprime \,.
\label{downstrand}
\end{align}
Utilizing \eqref{tsymmbprime} and \eqref{downstrand} for every edge and vertex shows that the definitions  \eqref{TsymmGoldenChain} and \eqref{eq:BW_duality} are equivalent.
In this picture the fusion of defects \eqref{DDeqD} is obvious, and it also makes apparent how the eigenvalues of $\mcd_\varphi$ can be thought of as a flux through the cylinder or torus.


\section{Microscopic to Macroscopic}
\label{sec:trivalent}

Our approach has enabled the explicit construction of topological defects in a large class of lattice models.
We here show that these defects possess some remarkable properties.  We saw already in \eqref{DDeqD} that defect lines wrapped around a cycle obey the same fusion algebra as the corresponding objects do. In this section we go much further and show that the topological defects themselves obey the rules of a fusion category described in Section \ref{sec:Fusion_Theory}, the same rules used in the construction of the lattice model. The microscopic definitions thus lead to macroscopic constraints on the partition functions.

To this end, we define junctions of defects in terms of a triangle weight by using \eqref{TV3}, guaranteeing their topological invariance by deriving commutation relations generalizing \eqref{DefComm1}. We then establish how defects describe a diagrammatic calculus with fusion rules, $F$-symbols and so on. We see that the defects belong to the same fusion category used in constructing the model, and so all the quantum dimensions and $F$-symbols are the same. 

\subsection{Closed defect loops}

The simplest rule in a fusion category is that evaluating a closed loop labeled by $\varphi$ gives $d_\varphi$. The analogous statement for topological defect relates the partition functions in the presence and absence of a single defect loop. The derivation of the defect commutation relations is also useful here. Following \eqref{defcommdiagram} from the bottom-left-most diagram to the bottom-right-most diagram gives
\begin{align}
\sum_{x,y,z,w}\BWloopdefectG\ =\ d_\varphi\ \BWLoopDefectRemoved\ .
\label{removeloop}
\end{align}
The green defect weights carry label $\varphi$ and so removing the loop results in multiplying the partition function by the quantum dimension $d_\varphi$. 
The defect commutation relations ensure local deformations of the defect path leave the partition function invariant, so the loop around the quadrilateral in \eqref{removeloop} can be deformed to have any topologically trivial path. The corresponding partition function $Z_\varphi$ is therefore simply related to the partition function $Z$ in the absence of the loop:
\begin{align}
Z_\varphi = d_\varphi\,Z\ .
\label{ZdZ}
\end{align}
If a defect loop wraps around a cycle, it cannot be removed so easily, but we show below how the defect $F$-moves derived in this section still yield non-trivial identities between partition functions.

\subsection{Topological trivalent junctions}
\label{junction_solution}

\begin{figure}[h] \vspace{-1cm}
\centerline{%
 \xymatrix @!0 @M=9mm @C=65mm{
 \TrivalentJunctionInserta\ar[r] & \TrivalentJunctionInsertb
 }}
 \vspace{-.5cm}
\caption{
On the left, a schematic view of a trivalent junction occurring at the termination of three topological defects. 
On the right, the cracked-open lattice with a triangular face at the junction. 
}
\label{fig:Trivalent}
\end{figure}
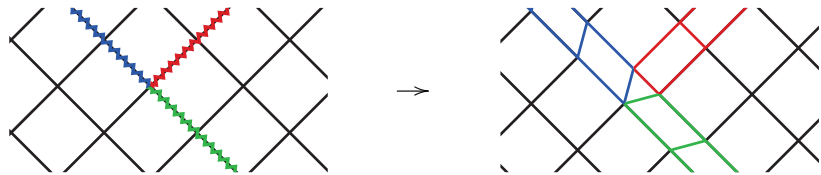

The basic defining feature of the fusion category is the trivalent vertex, which illustrates the fusion rules. Three defect lines meeting at a point is illustrated on the left in Figure~\ref{fig:Trivalent}, with the lattice cracked open on the right.
As apparent, a triangular face occurs at the junction. This triangle defect is labeled by six objects, three coming from the topological defects that it fuses and three from the height labels at the surrounding vertices.

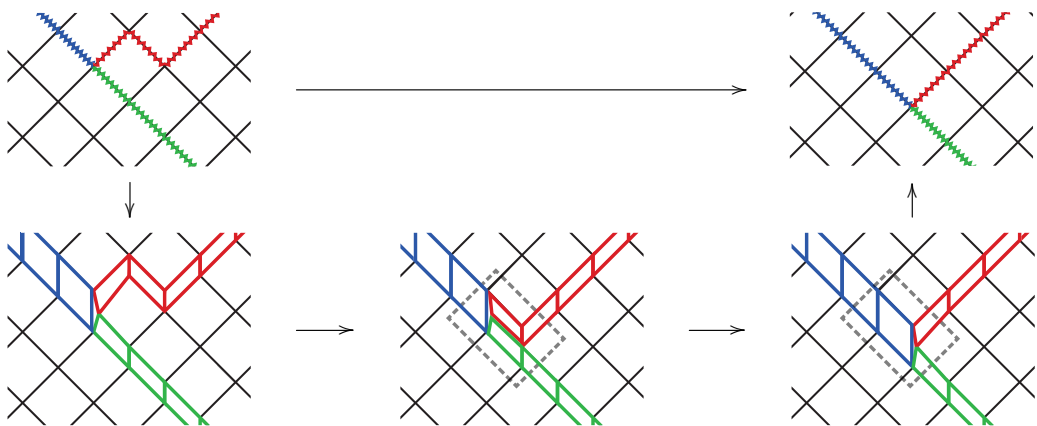
\begin{figure}[hb] \centerline{%
 \xymatrix @!0 @M=2mm @R=32mm @C=52mm{
 {\quad \TrivalentConsistencya \quad } \ar[rr]\ar[d]& & {\quad \TrivalentConsistencye \quad}  \\
  {\quad \TrivalentConsistencyb \quad } \ar[r]&{\quad  \TrivalentConsistencyc \quad} \ar[r]       & {\quad \TrivalentConsistencyd \ar[u] \quad}
}}
\caption{
Moving a trivalent junction on the lattice is schematically illustrated in the top row. The precise definition requires two steps, shown in the bottom row. The first step uses the defect commutation relations to move the red defect line around a face. 		The second step uses the triangle defect commutation relation (\ref{eq:trivalent_pentagon}) inside the boxed region. The dots are omitted for clarity. }
		\label{fig:TrivalentConsistency}
\end{figure}

The next step in showing that the defects obey the rules of the fusion category is to find triangle weights that make the trivalent junction topological, so that the partition function is independent of its location.  Namely, we need to impose that Figure~\ref{fig:TrivalentConsistency} is commutative. Along the bottom path of the diagram, a triangle defect weight is moved across two defect Boltzmann weights. 
Thus in addition to the defect commutation relations, we also need the consistency conditions
\begin{equation}
\sum_\beta \Trivalenta\quad =\quad \Trivalentb\ ,
\label{eq:trivalent_pentagon}
\end{equation}
where the internal height is summed over and external ones fixed.  

As with the defect commutation relations, the triangle commutation relation results in a huge number of constraints. Solving them by brute force is rather difficult, even for the Ising model treated in Part~I \cite{Aasen2016}. 
Luckily, the shadow-world analysis hands us the solution in \eqref{TV3}, suggesting we define
the triangle weight as
\begin{align}
	\Trivalent = \left( d_{R}d_G d_B\right)^{\frac{1}{4}}
		\Msixj{R & G & B}{\rho & \gamma & \beta}\ ,
 \label{eq:def_trivalent}	
\end{align}
where $\rho$, $\beta$, $\gamma$ are heights at the corners of the triangle.  We chose the overall constant so that the defects satisfy versions of \eqref{eq:loop_removal} and \eqref{F0c}, as we show below. There are as many different kinds of trivalent junctions as there are non-zero fusion rules $N_{RG}^{B}$.

Proving directly that triangle defects from \eqref{eq:def_trivalent} are topological follows from evaluating the center diagram in 
\begin{align}
	\vcenter{\xymatrix @!0 @M=2mm @R=24mm @C=42mm{
		\TrivalentCommL &\TrivalentComm \ar[r] \ar[l]& \TrivalentCommR\\
	}}
	\label{TrivalentProof}
\end{align}
in two different ways, identical to the derivation of the pentagon equation \eqref{pentagon} from the diagram \eqref{pentagon_wheel}.
To get to the left-hand side of (\ref{TrivalentProof}), we use three $F$-moves, and then evaluate it using the bubble removal \eqref{eq:loop_removal}. The ensuing expression is proportional to the left-hand side of \eqref{eq:trivalent_pentagon}:
\begin{align}
\TrivalentComm &= \sqrt{d_{xBz}}\sum_\beta d_\beta^{\frac{3}{2}} \Msixj{R & y & x}{\upsilon & \alpha & \beta} \Msixj{B & R & G}{\beta & \gamma & \alpha} \Msixj{G & y & z}{\upsilon & \gamma & \beta}\;  \TrivalentCommL \cr
& = \sqrt{d_{xyz RBG \alpha \gamma \upsilon}} \sum_\beta d_\beta \Msixj{R & y & x}{\upsilon & \alpha & \beta} \Msixj{B & R & G}{\beta & \gamma & \alpha} \Msixj{G & y & z}{\upsilon & \gamma & \beta} \cr
& =   \sqrt{d_{xyz RBG \alpha \gamma \upsilon} }\sum_\beta \Trivalenta\ .
\end{align}
To get to the right-hand side of (\ref{TrivalentProof}), we do a single $F$-move and then use \eqref{eq:loop_removal}, giving 
\begin{align}
\TrivalentComm &= \sqrt{d_{RGy}}  \Msixj{B & x & z}{y & G & R} \;  \TrivalentCommR \cr
		           & =  \sqrt{d_{xyz RBG \alpha \gamma \upsilon} }  \Msixj{B & x & z}{y & G & R}  \Msixj{B & x & z}{\upsilon & \gamma & \alpha} \cr
		           & =  \sqrt{d_{xyz RBG \alpha \gamma \upsilon} } \;  \Trivalentb
\end{align}
Equating the two shows that \eqref{eq:def_trivalent} indeed solves the trivalent defect commutation relation \eqref{eq:trivalent_pentagon}.

The topological defects thus obey the same fusion algebra as the objects in the category.  For a self-dual category, the $N_{RG}^B$ do not change under permutation of indices, so it does not matter which two we fuse to get the third, as is obvious from the pictures. To generalize to a non-self-dual category such as Tambara-Yamagami, one needs to keep track of the arrows.

\subsection{Topological defects as a fusion category}

Using the topological trivalent junction defined by \eqref{eq:def_trivalent},  we can define the partition function with an arbitrary trivalent network of defects, independent of deformations of the defects and their junctions. Here we show the defects have even deeper properties: they satisfy all the properties of a fusion category outlined in Section~\ref{sec:Fusion_Theory}.
In particular, we show that the defect lines obey \eqref{eq:loop_removal} and \eqref{eq:Fmove_def}, giving linear identities among partition functions with different defect configurations.

We first show that identity defect lines can be removed freely.
For $\varphi=\mathds{1}$, \eqref{eq:kronecker} and  \eqref{eq:def_trivalent} give
\begin{equation}
\mathord{\sideset{^{a}_{\alpha}}{^{b}_{\beta}}{\mathop{\IsingBubbleRightId}}}\ = \Msixj{\mathds{1} & a & \alpha}{\upsilon & \beta & b} = \  \frac{1}{\sqrt{d_a d_b}}\,\delta_{a\alpha}\delta_{b\beta}\  ,\qquad
\IdentityTerm\; = \; \delta_{bc} N_{ab}^\upsilon N_{ac}^\upsilon \frac{1}{\sqrt{d_b}}\ .
\end{equation}
\begin{align}
\idremovLHS\ =\ \idremovRHS\ .
\end{align}
The symmetry of the tetrahedral symbols means this relation holds for any rotated version of this diagram, so we can always remove identity strands.

Next we generalize the defect-loop removal \eqref{removeloop} relation to bubbles, deriving \eqref{bubble}. The only data needed involve the defect lines themselves, so we can write this diagram more abstractly as
\begin{align}
	\BubbleDiagram \;=\; \delta_{RB} \sqrt{\frac{d_{P}d_G}{d_R}}  \; \BlueLine \; ,
\label{bubbleremoval}
\end{align}
where $R,P, G,$ and $B$ label the lines.
This relation is exactly the same as \eqref{eq:loop_removal}, hence generalizing the microscopic definitions to the macroscopic defect lines. In terms of the defect weights \eqref{eq:def_trivalent} and \eqref{eq:def_LDefect}, \eqref{bubbleremoval} is
\begin{align}
\sum_{\beta, y } \; \IsingBubbleMicroLeft \; = \delta_{RB} \sqrt{\frac{d_{P}d_{G}}{d_R}} \;\IsingBubbleMicroRight\ ,
\label{bubble_removal}
\end{align}
The elegant method of proving \eqref{bubble_removal} is to evaluate a single diagram in two different ways, as with the proof of the triangle commutation relation. The diagram here is
\begin{align}
	\vcenter{\xymatrix @!0 @M=3mm @R=24mm @C=42mm{
	\BubbleProofL &\BubbleProof \ar[r] \ar[l]& \BubbleProofR
	}}\ .
	\label{BubbleProof}
\end{align}
Following the arrow to the left results in the left-hand side of \eqref{bubble_removal}, while following the arrow to the right results in the right-hand side of \eqref{bubble_removal}.

We finish off our demonstration that the defects obey the rules of the fusion category by deriving their $F$-moves. 
Just as in \eqref{eq:Fmove_def}, the $F$-symbol tells us how to relate the different ways of fusing four defect lines together. 
Four defect lines meet at two trivalent junctions. Using \eqref{eq:trivalent_pentagon} and \eqref{DefComm1}, the junctions can always be moved together to give
\begin{align} 
\FRight \quad \quad \text{or} \quad \quad \FLeft\ .
\label{fboxLR}
\end{align}
Either side of \eqref{fboxLR} provides a complete basis for the junctions of four defect lines. 
The two bases are related by noting that \eqref{eq:tetra_Fmove} gives
\begin{align}
\FLeftProof \; = \sum_X 
\left[F^{BRP}_G \right]_{ZX} \;\FRightProof
\end{align}
and that the evaluations of these two diagrams result in the same weights appearing in the two expressions \eqref{fboxLR}, up to an identical overall constant. We thus obtain 
\begin{align}
	\FRight \;=\; \sum_X \left[F^{BRP}_G \right]_{ZX} \FLeft \ .
	\label{eq:quad_move}
\end{align}
The tetrahedral symbol in \eqref{eq:quad_move} enforces the fusion rules in each of the four triangles.  The two triangle defects can then be moved back away from each other, giving \eqref{MacroF}.
We thus have derived the remarkable fact that the defect lines themselves obey the same $F$-moves as the fusion diagrams used to defined the Boltzmann weights for the microscopic degrees of freedom.

The trivalent vertex for the defects and the ensuing bubble removal and $F$-moves are the main results of this section. Together they show that diagrammatic calculus explained in Section~\ref{sec:Fusion_Theory} applies to the topological line defects themselves.
We have thus defined topological defects and their junctions in a host of lattice models, and established they satisfy a variety of remarkable properties. Some of these properties are summarized in the Table \ref{summarytable}. In the rest of the paper (and in a sequel or two) we exploit these properties. 

\vspace{0.5cm}

\begin{table}[ht]
	\fbox{\begin{minipage}{\textwidth}
	\vspace{3mm}\begin{center}{\begin{minipage}{158mm}
		{ \setlength{\parskip}{-1.2ex} \begin{itemize}
      \setlength{\itemsep}{0pt}
      \setlength{\parskip}{0.7ex}
      \setlength{\parsep}{0ex}
			\item[--]	$N^\ast_{\ast\ast}$ are the fusion symbols, defined in \eqref{Nabcdef}.
			\item[--]	$d_\ast$ are the quantum dimensions defined in \eqref{eq:QD_def}, and $d_{x_1 \cdots x_n} = d_{x_1} \cdots d_{x_n}$.
			\item[--]	The $F$-symbols relate diagrams via \eqref{eq:Fmove_def} and 	\eqref{eq:quad_move}.
			\item[--]	The tetrahedral symbols \eqref{eq:def_tetra_NA} relate to the $F$-symbols via $\big[ F_d^{abc} \big]_{xy} = \sqrt{d_{xy}} \smallM{a & b & x}{c & d & y}$.
   	\end{itemize}}
		\vspace{-0.8cm}
\begin{align*}\begin{array}{llc}
	\text{Partition function:}
	&	Z = \displaystyle \sum_{\text{heights}} \Bigg( \prod_{\text{vertices}}\! d_v \;\times \prod_{\text{elements}}\!\! W \Bigg)
\\[4ex]	\text{Plaquette weight:}
	&	\displaystyle \BWbare \;=\; \sum_\chi A(\chi) \, \BWabcdx{\chi} 
		\\[5pt]&\displaystyle\mkern52.6mu =\; \sum_\chi A(\chi) \, d_\chi \Msixj{\upsilon & \upsilon & \chi}{a & c & b} \Msixj{\upsilon & \upsilon & \chi}{a & c & b'}
	&	\eqref{eq:def_chi}
\\[4ex]	\text{Vertex dot:}
	&	\displaystyle \text{(full dot)~~} \overset{h}{\fulldot} \;=\; d_h  \qquad\qquad  \text{(half dot)~~} \overset{h}{\halfdot} = \sqrt{d_h}
	&	\eqref{semidotdef}
\\[3ex]	\text{Defect line weight:}
	&	\mathord{\ooalign{$\sideset{^a_\alpha}{^b_\beta}{\mathop{\DefectSquare}}$\cr\hidewidth$\varphi$\hidewidth}}
				 \;=\; \Msixj{\varphi & a & \alpha}{\upsilon & \beta & b} 
	&	\eqref{eq:def_LDefect}
\\[3ex]	\text{Trivalent junction:}
	&	\Trivalent  \;=\; (d_R d_G d_B)^{\frac14} \Msixj{R & G & B}{\rho & \gamma & \beta}
	&	\eqref{eq:def_trivalent}
\\[5ex]	\text{Transfer matrix:}
	&	T =\; \TransfMxIsingDots
	&	\eqref{Tmatrixdef2}
\\[4ex]	\text{Defect-line creation:}
	&	{\mathcal D}_\varphi \;=\;  \TopologicalSymmetryprime 
	&	\eqref{eq:BW_duality} 
\end{array}\end{align*}
	\end{minipage}}\end{center}
	\end{minipage}}
	\caption{Summary of how fusion categories appear in height models.}
\label{summarytable}	
\end{table}


\section{Duality}
\label{sec:duality}
In the preceding sections we proved that the defect lines defined by \eqref{eq:def_LDefect} are topological, in that the partition function in their presence is invariant under locally deforming them. The properties of the fusion category, in particular the invariance of evaluations under $F$-moves,  made it simple to solve the potentially huge number of equations arising from the defect commutation relations appearing in Sections~\ref{sec:topological_defect_lines} and \ref{sec:trivalent}.

Here we start developing applications of these results. In this section we show how any defect made from an object $\varphi$ with quantum dimension $d_\varphi>1$ yields a {duality}.  The key observation is in equation \eqref{TDDT}, showing that the operator creating a defect line commutes with the transfer matrix. When this operator is unitary, acting with it implements a symmetry transformation.
However, typically 
$\mcd_{\varphi}$ is not unitary, as is obvious from the fusion rules \eqref{DDeqD}: whenever $\mcd_{\varphi}$ fuses to more than one operator, its eigenvalues cannot be of unit norm. Thus whenever the quantum dimension $d_\varphi>1$, the corresponding $\mcd_{\varphi}$ cannot be unitary.  The conventional quantum-mechanical definition of symmetry requires a unitary transformation of the Hilbert space, and so we do not call acting with $\mcd_\varphi$ a symmetry transformation. Instead, we say it implements a {duality}. 

One of our central results is that dualities occur in {\em any} statistical-mechanical model defined in terms of a fusion category as above. Such dualities encompass the Kramers-Wannier duality of the Ising \cite{Kramers1941} and clock \cite{Wu1982} models, but our approach gives many non-trivial generalizations. Such a duality relates different couplings in the same model, as with Kramers-Wannier duality, or sometimes it relates two (seemingly) distinct models. We also describe self-dualities, where defects provide a non-trivial map of a model to itself.

\subsection{Grading and splitting}
\label{sec:splitting}

In order to understand how topological lattice defects gives symmetries and dualities, we must first describe how the configuration space of many of the models of interest split into sectors.
Seemingly different theories can originate from the same category even for a fixed $\upsilon$.
Not surprisingly, the resulting theories are closely related, and we explain subsequently how the generalized duality described here provides a way of mapping between them.

\paragraph{Grading.}

Many fusion categories can be {\em graded} by a finite group $G$~\cite{Etingof_2010,Barkeshli2019}.
The resulting sectors $\mcc_g$ are indexed by the elements of $G$, while the full fusion category is given by $\mcc = \bigoplus_g \mcc_g$. The fusion rules must obey the $G$ grading: the objects $a_g\in \mcc_g$ and $b_h\in \mcc_h$ obey $a_g \otimes b_h \in \mcc_{gh}$. 
In this paper we analyze only $\mathbb{Z}_2$-graded fusion categories, but many interesting generalizations exist. 
In a $\mathbb{Z}_2$-graded fusion category, the objects split into ``even'' and ``odd'' sets, so that fusion of two even or two odd objects gives an even object, while fusion of even and odd gives odd. 
The $\mathcal{A}_k$ fusion categories are graded in this fashion, with the `even' set $\mcc_{0} = \{ 0,1,2,\dots \}$ and the `odd' set $\mcc_{1} = \{ \frac12, \frac32, \dots \}$.
The splitting of the spin-1 models as displayed in \eqref{spin1heights} is a manifestation of this grading.
As apparent from (\ref{clockabelian},\,\ref{clockX}), the Tambara-Yamagami fusion categories~\cite{Tambara1998} used to build the clock models also have a $\mathbb{Z}_2$ grading,  with $\mcc_0 = \{ 0,1,\cdots, n-1\}$ and $\mcc_1 = \{ X \}$.

\paragraph{Splitting.}
A $\mathbb{Z}_2$ grading allows us to split the partition function into two pieces: $Z=Z_0 + Z_1$.  
The structure of the resulting models depends on whether the object $\upsilon$ used to build the models is even or odd. When $\upsilon$ is even, fusing with it preserves the sectors $\mcc_0$ and $\mcc_1$, so the heights in a given configuration are either all even or all odd. In effect, our construction yields two separate models: a model $M_0$ comprised of entirely of even heights, and a model $M_1$ comprised entirely of odd heights, with partition functions $Z_0$ and $Z_1$ respectively. An example of such splitting is in the $\mathcal{A}_{k+1}$ models with $\upsilon=1$, where the incidence diagrams for $M_0$ and $M_1$ are given in \eqref{spin1heights}. 

When  $\upsilon$ is odd, the grading results in a different sort of splitting. Because the graph $G$ on which the heights live is bipartite, on one of the two sublattices, all the heights are even and the other are odd.  The ABF models have $\upsilon$ odd, and in their transfer-matrix formulation, the vector space $\mathcal{V}$ displayed in   \eqref{eq:fusiontreedef} splits into two disjoint parts $\mathcal{V}_0\oplus \mathcal{V}_1$, where the subscript is $(2h_1+1)\,$mod 2.  
For example, for Ising, basis elements in each are labelled by the fusion trees
\begin{subequations} \begin{align}
&\VplusIsing \in \mathcal{V}_+
\label{Vplusdef}\\
&\nonumber\\
&\VminusIsing \in \mathcal{V}_-
\label{Vminusdef}
\end{align} \end{subequations}
Acting with the transfer matrix preserves these subspaces, and so $T$ is block diagonal in the form 
\begin{equation}
T=\begin{pmatrix} T_0&0\cr 0&T_1 \end{pmatrix} .
\label{IsingTsplit}
\end{equation}
We again can think of two separate models $M_0$ and $M_1$ with partition functions $Z_0$ and $Z_1$.

To summarize, the lattice model built from any $\mathbb{Z}_2$ graded fusion category can be split as
\begin{align*}
	\upsilon\hbox{ even:} \quad&\begin{cases} M_0\ &\text{all even heights}, \\
	M_1\ &\text{all odd heights},
	\end{cases}\cr
	\upsilon\hbox{ odd:} \quad &\begin{cases} M_0\ 
	&\text{even heights on $\mathcal{V}_+$, odd on $\mathcal{V}_-$}, \\
	M_1\ &\text{even heights on $\mathcal{V}_-$, odd on $\mathcal{V}_+$}.
	\end{cases}
\end{align*}

\subsection{Dualities and Symmetries}
\label{sec:symm}

Crossing a defect line labeled by $\varphi$ requires fusing with the object $\varphi$. Thus when $\varphi$ is odd, the corresponding defect line separates a region of model $M_0$ from a region of model $M_1$. Since the defect lines are deformable, moving the line in effect gives a mapping from $M_0$ to $M_1$.
In the transfer matrix, \eqref{TDDT} requires 
\begin{align}
T_0\, \mathcal{D}_{\varphi}= \mathcal{D}_{\varphi}\, T_1 \,,\qquad  T_1\,\mathcal{D}_{\varphi}= \mathcal{D}_{\varphi} \,T_0\ ,\qquad\hbox{ for }\varphi\hbox{ odd.}
\label{TDDTpm}
\end{align}
The corresponding partition functions are thus related as well, and when $d_\varphi>1$, we say the defect implements a {\em duality}.  The map \eqref{TDDTpm} gives powerful identities for the corresponding partition functions,  as shown in Part~I for Ising and below more generally.

Crossing a defect with even $\varphi$ leaves one in the same model. Such a $\mcd_\varphi$ still commutes with the transfer matrix,  its action is quite non-trivial when $d_\varphi>1$,. We thus say such a defect implements a {\em self-duality}. It results in linear identities relating e.g.\ partition functions with different boundary conditions. Another application is given in section \ref{sec:selfduality}, where we derive non-symmetry-related degeneracies in the spectrum of off-critical models.

The $d_\varphi=1$ case is much simpler to understand.
As the corresponding defects obey $\mcd_\varphi\mcd_{\overline\varphi}=1$, duality is too glorious a name.
When $d_\varphi=1$ and $\varphi$ is odd, the map is not particularly interesting, as it merely shows $M_0$ and $M_1$ are equivalent by a relabeling.  The $\varphi$ odd case is more interesting, since as apparent in \eqref{TDDTpm}, such a $d_\varphi=1$ defect implements a {\em symmetry}.
For example, in the $\mathcal{A}_{k+1}$ models, $d_{\frac{k}2}=1$, as \eqref{eq:def_Ak_N} and  \eqref{DDeqD} give for all $h$
\begin{align}
\frac{k}2 \otimes h =\frac{k}{2}-h \qquad \implies\qquad \mcd_{\frac{k}{2}}\mcd_h=\mcd_{\frac{k}{2}-h}\,.
\label{heightflip}
\end{align}
Acting with $\mcd_{\frac{k}{2}}$ thus exchanges each height $h_j$ in $\mathcal{V}$ with the height $\tfrac{k}{2} -h_j$. For the ABF models, this exchange reflects the adjacency diagram \eqref{Akadjacency} around its midpoint. The fusion rules respect this exchange,  and it is straightforward to check from the explicit expressions in appendix \ref{app:TetraSym} the Boltzmann weights for any $\upsilon$ are also invariant. For $k$ even, sending each height $h$ to $\tfrac{k}{2}-h$ preserves $\mathcal{V}_0$ and $\mathcal{V}_1$, and so acting with $\mcd_{\frac{k}{2}}$ implements a symmetry. For example, in the Ising case, $\mcd_1=\mcd_\psi$ is the generator of spin-flip symmetry. For $k$ odd, $\mcd_{\frac{k}{2}}$ maps between $\mathcal{V}_0$ and $\mathcal{V}_1$ by a trivial relabeling of the heights, merely showing that the height models ${M}_0$ and ${M}_1$ are identical. 

Similarly, in the clock models, defects defined using the Abelian objects implement the $\mathbb{Z}_N$ symmetries. Namely, the operator $\mcd_a$ cyclically shifts all the spins in $\mathcal{V}$, with the fusion rule \eqref{clockabelian} giving the necessary
\begin{align}
\mcd_a\mcd_b = \mcd_{(a+b)\,{\rm mod}\ N}\ .
\end{align}
Another defect implements charge-conjugation symmetry in the clock models, sending each height $h\to N-h$. 
Including it in a fusion algebra with the objectes in the $\mathbb{Z}_N$ Tambara-Yamagami category allows the construction of a larger category,  a $\mathbb{Z}_2$ extension of the $\mathbb{Z}_N$ Tambara-Yamagami category. 
In the larger category, the subcategory of Abelian simple objects is given by $\text{Vec}_{D_N}$.
These ``$C$-disorder'' defects are described in the CFT setting in \cite{Zamolodchikov86}.

The action of a defect labeled by $\varphi$ in a $\mathbb{Z}_2$ graded theory thus depends on whether $\varphi$ is even or odd, and whether $d_\varphi=1$ or not. 
To summarize, the dualities and symmetries are
\begin{align*}
\varphi\hbox{ even (or no grading):} \quad&\begin{cases} \hbox{self-duality }\ M_0\to M_0,\ M_1\to M_1\quad&d_\varphi>1\quad\\
\hbox{symmetry}&d_\varphi=1
\end{cases}\cr
\varphi\hbox{ odd:} \quad&\begin{cases} \hbox{duality }\ M_0\to M_1,\ M_1\to M_0\qquad\quad&d_\varphi>1\\
\hbox{relabeling}&d_\varphi=1
\end{cases}
\end{align*}

\subsection{Kramers-Wannier duality defects}
\label{sec:KWduality}

We here describe how topological defects implement Kramers-Wannier (KW) duality, before moving on to the generalizations. As well as giving an elegant setting for the many known results, our framework allows new results to be derived, even in such a well-studied model as Ising. For example, we showed in Part~I how rigorous lattice calculations give precise constraints for critical exponents in any continuum limit for Ising, and we extend the results below in section \ref{sec:twistedbc}.

By a Kramers-Wannier duality we mean those where acting with the duality transformation twice yields a sum over symmetry transformations. Namely, a topological defect $\mcd_{X}$ implementing a KW duality obeys
\begin{align}
\left(\mcd_X \right)^2 =\mathds{1}+\sum_{a=1}^{N-1} \mcd_a\qquad\hbox{ with all }d_a=1\ .
\label{KWdef}
\end{align}
In Ising, the duality defect obeys $\mcd_\sigma^2 = \mathds{1} +\mcd_\psi$, where $\mcd_\psi$ is the spin-flip generator.  The definition \eqref{KWdef} therefore provides a natural generalization to the clock/Potts models. 

KW duality defects are not invertible. For example, for Ising the right-hand side of \eqref{KWdef} projects onto all states even under spin-flip. The traditional way of dealing with this fact is to use the symmetries to split the vector space $\mathcal{V}$ into sectors where the duality transformation can be inverted and those where it acts trivially. Although sometimes useful calculationally, this splitting obscures the general structure. Moreover, it makes such dualities seem quite special to the Ising/Potts models, where such a splitting is easy to do. 

The canonical example of Kramers-Wannier duality is in the Ising model. The lack of invertibility of $\mcd_\sigma$ manifests itself in the presence of two free-energy minima in the low-temperature phase (and so two ground states in the quantum-spin-chain limit), while the high-temperature phase has a unique minimum. 
The duality defects in the clock/Potts models arise from the operator $X$ in the $\mathbb{Z}_N$ Tambara-Yamagami category. The Ising model is the special case with $N=2$, and one of many nice features of our construction is that the generalization to Potts and clock models is very simple.  The fusion rules in  \eqref{clockabelian} and \eqref{clockX} make immediately apparent that $\mcd_X$ obeys \eqref{KWdef}, with the right-hand side the sum over all spin-shift operators. It follows that $(\mcd_X)^3 = N\,\mcd_X$, so that this duality defect is indeed not invertible for $N>1$. 

A special feature of the Ising/clock models is that all the fluctuating degrees of freedom live on one sublattice. One can think of ${M}_0$ and $M_1$ each as being nearest-neighbor models, the former on a graph $G^+$ and the latter on the dual graph $G^-=\widehat{G^+}$. Namely, because $G$ is bipartite, its vertices split into two sets, those of $G^\pm$. The edges in $G^+$ join any two vertices that share a quadrilateral in $G$, and likewise for $G^-$.  The interactions within each of $G^\pm$ are then indeed nearest neighbor. 

As the odd object $X$ labels the KW duality defect, an $X$ is always across from a fluctuating degree of freedom and vice versa. The duality defect therefore maps between a clock model $M_0$ on the graph $G^+$ and $M_1$ on the dual graph $G^-$.  The defect weights are given in terms of tetrahedral symbols by  \eqref{eq:def_LDefect}, with those for Ising given explicitly in Part~I.
These tetrahedral symbols are found from \eqref{ATY6j} with $A = \mathbb{Z}_N$ and $\chi(a,b) =  \exp\!\left[\frac{-2\pi i}{N}ab\right]$, yielding
\begin{align}
\label{Pottsdefect}
	\DefectSquarex{X}{\,a}{\,b}{X} &= \frac{1}{d_X} \exp\!\left[-\frac{2\pi i}{N}ab\right] \,,
&	\DefectSquarex{\,a}{X}{X}{\,b} &= \frac{1}{d_X} \exp\!\left[\frac{2\pi i}{N}ab\right] \,.
\end{align}
with $a,b=0,\dots, N-1$.
Their symmetry properties given in \eqref{colexchange}.

The models ${M}_0$ and ${M}_1$ have the same types of degrees of freedom, and are defined via the {same} couplings $A_Q$ for each quadrilateral $Q$ of $G$. The partition functions $Z_0$ and ${Z}_1$ thus depend on the same amplitudes. However, the duality effectively {\em changes} the couplings. We here explain why using more conventional definitions. The Potts coupling $J_Q$ for a given bond is defined by setting the weight to be $e^{J_Q}$ if the two spins at the end of the bond are the same, and $1$ if they are different. The basis weights given in \eqref{Isingbasisweights} and  \eqref{clockprojectors} depend on whether the fluctuating spins are on the top and bottom of $Q$, or on the sides.  To simplify the discussion, we work in the transfer matrix on the square lattice. Each Boltzmann weight depends only on four heights $h'_{n-1}, h'_{n}, h'_{n+1}, h_{n}$. In ${M}_0$ the spins all reside on vertices with even labels, so the Potts interaction couples the spins $h'_{n}$ and $h_n$ for $n$ even while it couples the spins $h'_{n-1}$ and $h'_{n+1}$ when $n$ is odd. In ${M}_1$, the reverse holds. To further simplify, we let the couplings depend spatially only on whether $n$ is even and odd respectively.  The couplings defined in \eqref{uQdef} then depend on $u_{\rm even}$ and $u_{\rm odd}$ respectively.
The subtlety comes in relating them to the traditional Potts couplings $J_x$ and $J_y$: which $u$ is horizontal and which is vertical depends on the model. Namely, for $M_0$, using \eqref{WAP} and \eqref{uQdef} gives
\begin{align}
	e^{J_x} = \cot \left(\lambda-u_{\rm odd} \right)\ , \qquad  
	e^{J_y}=	\frac{\sin(2\lambda-u_{\rm even})}{\sin u_{\rm even} }\qquad\hbox{ in }M_0.
\label{Jxyu}
\end{align}
Where for Potts $\lambda$ is defined by $2\cos\lambda=\sqrt{N}$, so that $\lambda$ and $\mu$ are imaginary for $N>4$. At $N=4$, one takes a limit $u\propto\lambda\to 0$. 
To get $M_1$, we must change the orientation on each $Q$, so
\begin{align}
	e^{J_x} =	\cot\left(\lambda-u_{\rm even} \right)\ , \qquad e^{J_y} =\frac{\sin(2\lambda-u_{\rm odd})}{\sin u_{\rm odd} } \qquad\hbox{ in }M_1.
\label{Jxyu2}
\end{align}
Thus while both $M_0$ and $M_1$ are both $N$-state Potts models, in general the two have different couplings and are defined on different graphs.

The ``self-dual'' line in the Potts model (and in an integrable clock model) is therefore described by $u_{\rm even}=u_{\rm odd}$ for every $Q$. We put self-dual in quotes because in our nomenclature we describe this mapping as a (Kramers-Wannier) duality, {\em not} a self-duality, even at the critical point. We justify this naming by the fact that even at $u_{\rm even}=u_{\rm odd}$, this map in general not only changes the model to one on a different lattice but, as detailed in \eqref{sec:bdry}, changes different boundary conditions even on the square lattice.  We reserve self-dual without quotes to mean dualities implemented by $\upsilon$ even, as we discuss next.

\subsection{Self-duality in the eight-vertex model and XXZ spin chain}
\label{sec:8v}

Here we illustrate self-duality in the eight-vertex model, another fundamental model of statistical mechanics. 
This self-duality  \cite{Baxter1978,Baxter1982,Wu1989} however comes from an even-$\upsilon$ defect, but is very similar to Kramers-Wannier duality.  We recover it by using topological defects, and make this similarity precise. This calculation provides a useful demonstration of the general approach. 

In our approach, the eight-vertex model arises from considering the $\mathcal{A}_5$ category with $\upsilon=1$, i.e.\ $k=4$.
The fusion rules for a spin-1 object here are
\begin{align}
	1\otimes1 &=0 \oplus1\oplus2 \,, &	1\otimes2 &= 1\,,
&	\frac12\otimes1 &= \frac12\oplus\frac32 = \frac32 \otimes 1\,, 
\label{A5fusion}
\end{align}
As apparent in \eqref{spin1heights}, with $\upsilon=1$ there are two distinct models, with $M_0$ comprised entirely of integer heights and $M_1$ half-integer.

The model $M_1$ built from \eqref{A5fusion} is the eight-vertex model.
At each site we have a two-state system with heights either $\tfrac12$ or $\tfrac32$ and no adjacency constraints.
The weights depend only on whether they are the same or different, because of the spin-flip symmetry.
We then can rewrite configurations in terms of the domain walls living on the edges of the dual graph $\widehat{G}$.
A domain wall is placed on such an edge if the neighboring heights on $G$ are different, i.e., one is $\tfrac12$ and the other $\tfrac32$, while the edge is left empty if the two heights are the same.
On the plane with free boundary conditions the map from heights to domain walls is therefore two to one.
With only two allowed heights at each vertex of $G$, there must be an even number of domain walls touching each vertex of $\widehat{G}$.%
	\footnote{The eight-vertex model is usually written in terms of arrows on each edge of $\widehat{G}$ instead of domain walls. Since our $G$ is bipartite, it is easy to define a map between the two.}
Since these domain walls cannot end, there must be an even number of them touching each vertex of $\widehat{G}$, and so eight possible configurations at each vertex (hence the name).
These are
\begin{align}
\underbrace{
\text{\rotatebox[origin=c]{45}{$\eva$}}
\;\; \text{\rotatebox[origin=c]{45}{$\evb$}}
}_{a_Q}
\quad\underbrace{
\text{\rotatebox[origin=c]{45}{$\evc$}}
\;\;\text{\rotatebox[origin=c]{45}{$\evd$}}
}_{b_Q}
\quad\underbrace{
\text{\rotatebox[origin=c]{45}{$\evh$}}
\;\; \text{\rotatebox[origin=c]{45}{$\evf$}}
}_{c_Q}
\quad\underbrace{
\text{\rotatebox[origin=c]{45}{$\evg$}}
\;\; \text{\rotatebox[origin=c]{45}{$\eve$}}
}_{d_Q}
\label{eightvertices}
\end{align} 
labeled by the corresponding Boltzmann weights $a_Q,b_Q,c_Q,d_Q$ \cite{Baxter1982}. 

Each vertex of $\widehat{G}$ is at the center of a quadrilateral $Q$, so these Boltzmann weights are related to the amplitudes $A_Q(\chi)$.  The projectors can be written in terms of the Pauli matrices $X_{j,j+1}$, $Y_{j,j+1}$ and $Z_{j,j+1}$ acting on the two-state system on the edge of $\widehat{G}$ separating the heights $j$ and $j+1$ in $\mathcal{V}_-$. Since $\upsilon =1$, the first fusion rule in \eqref{A5fusion} means there are three channels $\chi=0,1,2$ in \eqref{WAP}, and the tetrahedral symbols in \eqref{tetrahalf} give
\begin{subequations} \label{P012} \begin{align}
	P_Q(0) &\rightarrow \frac{1}{4}\left( X\otimes X - Y \otimes Y + \mathds{1}\otimes \mathds{1} + Z\otimes Z \right) \,,\\
	P_Q(1) &\rightarrow \frac{1}{2}\left( \mathds{1}\otimes \mathds{1} -X\otimes X  \right) \,,\\
	P_Q(2) &\rightarrow \frac{1}{4}\left( X\otimes X + Y \otimes Y + \mathds{1}\otimes \mathds{1} - Z\otimes Z \right) \,.
\end{align} \end{subequations}
Comparing to \eqref{eightvertices} gives
\begin{align} \begin{aligned}
2a_Q&= A_Q(1) + A_Q(2)\ ,\qquad\quad 2b_Q = A_Q(2)-A_Q(1)\ ,\cr 
2c_Q &= A_Q(0)+A_Q(1)\ ,\qquad\quad 2d_Q=A_Q(0)-A_Q(1)\ .
\end{aligned} \label{8vweights}
\end{align}
These projectors and Boltzmann weights obey a $\mathbb{Z}_2$ symmetry exchanging domain walls with empty edges (which is called the ``zero field'' condition in the eight-vertex language).  

With the fusion rules \eqref{A5fusion}, a spin-1 defect also preserves the integer and half-integer heights, and therefore preserves $M_0$ and $M_1$ individually. As its fusion rules from \eqref{A5fusion} are non-trivial (it has quantum dimension $d_1=2$), it must implement a {self-duality}.  The defect creation operator $\mcd_1$ commutes with the transfer matrix of any eight-vertex model whose Boltzmann weights obey \eqref{8vweights}. Such weights satisfy $a_Q-b_Q=c_Q-d_Q$, which indeed is a self-duality condition of the eight-vertex model \cite{Baxter1978,Wu1989}. While the eight-vertex model is integrable for any choice of $a,b,c,d$ independent of $Q$, this self-duality holds even in the non-integrable case where the weights dependent on $Q$.

Explicitly, the defect weights for $M_1$ are
\begin{align}
	\DefectSquarex{a}{\,b}{a'}{\,b'} \;&=\; \Msixj{1 & a &b}{1 & b' & a'}\; =\;
			\dfrac{1}{\sqrt{3}} \cos\!\left[ \dfrac{\pi(a+b+a'+b')}{2} + \dfrac{\pi}{3} \right] \quad\quad\hbox{for }\       a,a',b,b'\in\big\{\tfrac12,\,\tfrac32\big\} \,,\quad		
\end{align}
making apparent how the self-duality implemented by this defect is much more complicated than a symmetry transformation. The corresponding defect-creation operator is then
\begin{align}
\label{defopAfive}
\mathcal{D}_1 | \{ x_j \} \rangle &= \sum_{\{ y_j \} } \prod_{j=1}^N\sqrt{d_{x_j} d_{y_j}} \Msixj{1 & y_j &y_{j+1}}{1 & x_{j+1} & x_j}  | \{ y_j \} \rangle\cr
&=\sum_{\{ y_j \} } \prod_{j=1}^N\sqrt{d_{x_j} d_{y_j}} 
\dfrac{1}{\sqrt{3}} \cos\!\left[ \dfrac{\pi(x_j+y_j+x_{j+1}+y_{j+1})}{2} + \dfrac{\pi}{3} \right] | \{ y_j \} \rangle
\end{align}
for $x_j,y_j \in \{1/2, 3/2\}$.
The resulting self-duality is similar to Kramers-Wannier duality as a consequence of the spin-2 defect here generating spin-flip symmetry as in \eqref{heightflip}.
A shift makes the connection more apparent, as
\begin{align}
	\Big(\mcd_1 - \tfrac14 \big(\mathds{1} + \mcd_2\big)\Big)^2\ = \ \tfrac98\big(\mathds{1}+ \mcd_2\big) \,.
\end{align}
The right-hand side is of the same form as \eqref{KWdef} for Ising, projecting onto states even under spin-flip (i.e., here eigenvectors of $\mcd_2$ with eigenvalue $1$).
Thus in this sector, $\mcd_1 -\frac12$ acts in the same fashion as Ising Kramers-Wannier duality. However,  $(\mcd_1)^2-\mcd_1$ annihilates spins odd under spin-flip here, as opposed to the Ising $\mcd_\sigma^2=0$ acting on such states. 

This self-dual transfer matrix can be mapped on the famed XXZ spin chain by taking an appropriate limit. Namely, defining couplings $\Delta$ and $\epsilon$ by
\begin{align}
A_Q(0)=1-\epsilon\ , \qquad A_Q(1)=1-\Delta\epsilon\ ,\qquad A_Q(2) = 1 +\epsilon
\end{align}
for all $Q$, the transfer matrix can be written as $T=1-\epsilon H$ for small $\epsilon$, where
\begin{align}
H_{\rm XXZ}=\sum_j \Big({-}\Delta X_j X_{j+1} - Y_jY_{j+1} + Z_jZ_{j+1}\Big)\ .
\end{align}
After a unitary transformation to flip the signs of $X_{2n}$ and $Y_{2n}$, we recover the XXZ Hamiltonian (with $X$ and $Z$ exchanged from the standard conventions). Since our defects commute with the transfer matrix for any choice of the $A_Q(\chi)$, the defect creation operator $\mcd_1$ commutes with $H_{\rm XXZ}$ as well. 
This Hamiltonian obeys a $U(1)$ symmetry, and standard results \cite{Temperley1971,Baxter1982} show that it is critical when $|\Delta|\le 1$.

\subsection{Generalized dualities from defects}
\label{sec:generalized}

We showed that defects give an elegant way of implementing Kramers-Wannier duality in the Ising, Potts and clock models. The models $M_0$ and $M_1$ related by Kramers-Wannier duality have the same degrees of freedom, albeit on different lattices and with different boundary conditions. Here we give a few examples of dualities that are not Kramers-Wannier.

Using $\varphi=\tfrac12$ generalizes the Ising duality defect to all of the ABF models, mapping between $\mathcal{M}_0$ and $\mathcal{M}_1$. 
In the ABF models, ${M}_0$ and ${M}_1$ have the same degrees of freedom, but we must keep the $A_Q(\chi)$ (chosen with some orientation) fixed on each quadrilateral $Q$.  
These defects generalize those found using integrability \cite{Chui2001,Chui2002} to allow for non-uniform couplings and hence no integrability.  
The ensuing duality results in linear identities relating the partition functions, some of which we discuss below. 
The generalized duality in the $k=3$ ABF model \cite{Feiguin2007} has already been discussed in the Hamiltonian limit where the $k=3$ chain at the critical point was redubbed the ``golden chain'' and the self-duality called ``topological symmetry''.%
	\footnote{In \cite{Feiguin2007}, both $\upsilon$ and $\varphi$ are taken to be 1, not 1/2.  However, for $k=3$ the two are equivalent, up to relabeling heights on even and odd sites.}
The $\varphi=\frac12$ defect weights from \eqref{eq:def_LDefect} are
\begin{align}
	\DefectSquarex{a}{b}{a'}{b'} = \begin{bmatrix} \frac12& a & a' \\[3pt] \frac12 & b' & b \end{bmatrix}
	\label{ABFdefect}
\end{align}
Using \eqref{tetrahalf} gives them explicitly:
\begin{subequations} \begin{align}
	\DefectSquarex{\quad a}{\,a+\frac12}{a+\frac12}{\,a} &= \DefectSquarex{a+\frac12}{\,a}{\quad a}{\,a+\frac12}
	= \frac{ (-1)^{2a} }{ d_a d_{a+\frac12} } \,,\qquad
\\	
	\DefectSquarex{\quad a}{\,a+\frac12}{a+\frac12}{\,a+1}
	&= \DefectSquarex{a+\frac12}{\,a}{\,a+1}{\,a+\frac12}
	= \DefectSquarex{\,a+1}{\,a+\frac12}{a+\frac12}{\,a}
	= \DefectSquarex{a+\frac12}{\,a+1}{\quad a}{\,a+\frac12}
	= \frac{1}{d_{a+\frac12}} \,.
\end{align} \end{subequations}

An example where $M_0$ and $M_1$ look very different is given by the $\mathcal{A}_5$ category with $\upsilon=1$. As discussed in section \ref{sec:8v}, $M_1$ is the eight-vertex model. In the model $M_0$, the allowed heights $0$, $1$, $2$ are integer-valued. This model also is rather nice, generalizing the ``PXP" model and deformations \cite{Fendley2004} to a two-species hard-particle model. The heights $0$ and $2$ play the role of the two species, while any site with height $1$ is considered empty. The adjacency graph from \eqref{spin1heights} with $k=4$ ensures that the hard particles cannot be adjacent to themselves nor to each other, but only to an empty site.  (This adjacency graph alternatively is the $\widehat{D}_5$ Dynkin diagram \cite{Pasquier1987} if one gives different labels to the heights on $G^+$ and $G^-$.)
The projectors and transfer matrix/Hamiltonian follow from the $F$-matrix \eqref{Fmat1}.

Acting with $\mcd_{\frac12}$ on the eight-vertex model $M_1$ gives a duality to the two-species hard-square $M_0$. Using \eqref{eq:def_LDefect} and the tetrahedral symmetries allows us to use \eqref{tetrahalf} to give the explicit expressions for this duality defect:
\begin{align} \begin{split}
	\DefectSquarex{\mkern2mu 0}{\;1}{\frac12}{\,\frac12} = \DefectSquarex{\mkern2mu 0}{\;1}{\frac12}{\,\frac32} = \DefectSquarex{\mkern2mu 2}{\;1}{\frac32}{\,\frac12} = \DefectSquarex{\mkern2mu 2}{\;1}{\frac32}{\,\frac32} &= 2^{-\frac12}3^{-\frac14} \,,
\\	\DefectSquarex{\mkern2mu 1}{\;1}{\frac12}{\,\frac12} = -\Big(\DefectSquarex{\mkern2mu 1}{\;1}{\frac12}{\,\frac32}\Big) = \DefectSquarex{\mkern2mu 1}{\;1}{\frac32}{\,\frac32} &= 2^{-1} 3^{-\frac14}\,.
	\label{8vhardsquare}
\end{split} \end{align}
The remainder are found by reflecting horizontally.
The self-dual line of the eight-vertex model maps to a self-dual line in this two-species hard-square model.
Indeed, the self-duality ($\varphi=1$) of the latter is given by the defect weights
\begin{align}
	\DefectSquarex{a}{\,b}{a'}{\,b'} \;=\dfrac{1}{2} \sin\dfrac{\pi(ab'+a'b)}{2} \qquad\quad\hbox{for }  a,a',b,b'\in\{0,1,2\} \,.
\end{align}
A section of this integrable self-dual line (that dual to $|\Delta|<1$) is critical. The self-duality is exact on the lattice and the critical line has continuously varying exponents, as opposed to the deformed PXP model. Off the self-dual line, however, it is not guaranteed that the duality \eqref{8vhardsquare} between models persists, although it may. It thus would be interesting to analyze this mapping further, and to explore the phase diagram away from the self-dual line, i.e.\ by allowing for interactions not written in terms of the projectors $P_0$, $P_1$ and $P_2$.

As any object $\varphi$ with $d_\varphi>1$ in any fusion category results in a non-trivial duality or self-duality, we could give many more examples. The ensuing duality results in linear identities relating the partition functions, some of which we discuss below. Using $\varphi=1$ results in a self-duality in any of the $\mathcal{A}_{k+1}$ models. We use this self-duality in the next section \ref{sec:selfduality} to derive degeneracies in the ground state and the low-lying states of ABF models with Boltzmann weights staggered in space.  Self-duality has already been used to explore phase diagrams of $\mathcal{A}_{k+1}$ Hamiltonian with $\upsilon=\varphi=1$ for couplings uniform in space but with arbitrary amplitudes \cite{Trebst2008,Gils2013}.

\section{Degeneracies from self-duality}
\label{sec:selfduality}

One remarkable feature of our setup is that the dualities and self-dualities explained in section~\ref{sec:duality} occur in any lattice model built on a fusion category.
Whereas dualities help understand how different theories are equivalent to each other, self-dualities allow a given theory to be probed deeper.
One famed use of self-duality is to locate critical points.
However, although self-duality is enough information to fix the critical point in a few simple cases like Ising, it does not always.
For example, in Potts with $N\ge 4$, the self-dual point is gapped.
The $\mathcal{A}_{k+1}$ models with $\upsilon=1$ are not in general critical even when self-dual and uniform in space, instead having a rich phase diagram explored in \cite{Trebst2008,Gils2013,Vernier2017}.
We do emphasize however that if one breaks the duality by no longer defining the Boltzmann weights as a sum over the projectors of the fusion category, then typically the models are not critical.
Thus we believe it is fair to say that typically, critical points in height models are self-dual, but not the converse.

We show in this section one striking consequence of self-duality. Acting with duality defect operators explains the presence of unusual degeneracies in the ground and excited states of the spectrum of certain (non-integrable) gapped spin chains. These degeneracies are between states not related by any symmetry. Moreover, in some (perhaps all) such cases, the scaling limit of such chains turns out to be an integrable field theory. The presence of the self-duality provides intuition into why this is so. In an integrable theory, the scattering matrix of the low-energy excitations obeys special properties. We explain how the self-duality provides a simple way of constraining it further.

In this section we study the quantum spin-chain limit, because it is more transparent to probe the structure of the ground states than the minima of the free energy. To simplify further, we consider the case where the Hamiltonian \eqref{HPdef} involves only projectors onto the identity:
\begin{align}
H = \sum_{j=1}^L u_j P^{(0)}_j\ .
\label{HP}
\end{align}
This  Hamiltonian acts on the Hilbert spaces $\mathcal{V}_\pm$, where the inner product is defined by making height configurations orthonormal.
We take periodic boundary conditions, so that we can utilize the defect-creation operators defined in section~\ref{sec:defectcreationop}.
We make no assumptions about integrability; typically $H$ will be integrable only if the $u_j$ are uniform in space. As we hope will be obvious from our analysis, the methods will apply to more complicated sums of projectors as well. 

For the ABF or Potts models, this Hamiltonian (\ref{HP}) is a standard one. In these cases it is typically written in terms of the Temperley-Lieb generator $e_j=d_{\upsilon}P^{(0)}_j$, as shown for ABF in (\ref{TLproj}) and Potts in \eqref{TLPotts}. In this section we study mainly the staggered case, where $u_j \propto -1 +t(-1)^j$. 
The staggered Hamiltonian is then
\begin{align} 
H_{\rm stag} = 
H_{\rm crit}+H_{\rm alt} \,,\qquad H_{\rm crit} = -\sum_{j=1}^L e_j \,,\qquad H_{\rm alt} = t\sum_{j=1}^L(-1)^j e_j \,,
\label{Hcritpert}
\end{align}
with $L$ even. Here we put the sign in front of $H_{\rm crit}$ to make the models ferromagnetic. For Potts, one also needs $N\le 4$ to make the uniform case $t=0$ critical like ABF. This Hamiltonian \eqref{Hcritpert} can be obtained directly from the transfer matrix (\ref{Transfermatrixdef}) by taking $u_Q$ from \eqref{uQdef} to be small, as shown in \eqref{HTL} for the critical ABF case.
Although $H_{\rm alt}$ breaks the integrability, it has long been known from a variety of approaches that it is relevant and results in a gap.
The Coulomb-gas approach gives the scaling dimension of $H_{\rm alt}$ as $\frac{k+4}{2k+2}$ \cite{denNijs1979}, relevant for any $k$. This operator corresponds to $\Phi_{2,1}$ in the conformal field theory describing the continuum limit of the ABF model \cite{Pasquier:1987} (see section \ref{sec:gmin} for a more thorough explanation of the notation). Interestingly in light of the following, the perturbed conformal field theory is integrable even though this lattice model is not \cite{Smirnov1991,Chim1991}.

\subsection{Degenerate ground states}
\label{sec:degenerategs}

Diagonalizing $H$ to find its ground states obviously is an intractable problem in general.
Our strategy therefore is to first consider the completely staggered limit, where the model can be solved trivially.
We then use self-duality to 
show that the ground states remain degenerate, up to exponentially small finite-size corrections, throughout an entire phase.

In the completely staggered case $t=1$, the Hamiltonian \eqref{Hcritpert} becomes a sum of commuting operators, as no $e_{2a}$ appear.
Each term thus can be diagonalized independently and all eigenvalues of $H_{\rm cs}$ computed.
Since $e_{j}^2=d_{\upsilon} e_j$ by construction, it has only eigenvalues $d_\upsilon$ and 0.
These ground states are easily constructed explicitly, since the matrix elements of $e_{2a-1}$ vanish unless $h_{2a-2}=h_{2a}$. 
The ground states are thus labeled by a fixed height $h_{2a-2}=h_{2a}=r$, and are given by
\begin{align}
	\kett{r} = \ket{ r\tilde{r}r\tilde{r}r\tilde{r}\cdots } ,\qquad 
	\ket{\tilde r} \defineas \frac{1}{\sqrt{d_rd_\upsilon}}\sum_{x}N_{r \upsilon}^x \sqrt{d_x} \ket{x} ,
	\label{staggeredgs}
\end{align}
Graphically these states can be written in the fusion-tree basis defined in \eqref{eq:fusiontreedef} as
\begin{align} 
	\kett{r} \;=\; \Hcsgsbasis\ ,
\label{fusiontreebasis}
\end{align}
where as always the vertical lines are in the $\upsilon$ representation here.
Using \eqref{F0c}, this particular tree can be rewritten as
\begin{align} 
	\kett{r} \;=\; \Hcsgs \ ,
\label{csgscupbasis}
\end{align}
since the cups are invariant under $e_{2a-1}$. 

The expression \eqref{csgscupbasis} gives the ground states for the more general completely staggered models, where  $e_{j}$ in \eqref{Hcritpert} is replaced with $P_{j}^{(0)}$ as in \eqref{HP}.
The ground states of all these completely staggered models therefore obey
\begin{align}
	H_{\rm cs} = \sum_{a}^{L/2} P^{(0)}_{2j-1} \qquad\implies\qquad
	H_{\rm cs} \kett{r} = -d_{\upsilon}L\, \kett{r}
	\label{energycs}
\end{align}
for even $L$.
For ABF, $\upsilon=\tfrac12$, while for Potts $\upsilon=X$.

The number of such ground states is simply the number of allowed heights $r$. For the ABF model with $k$ odd, there are $(k+1)/2$ ground states in each of $M_0$ and in $M_1$, corresponding to setting  $r$ integer and half-integer respectively.
For $k$ even, however, the models $M_0$ and $M_1$ are not equivalent, but rather dual to each other via $\mcd_{\frac12}$.
Indeed, there are $k/2+1$ ground states in $M_0$, with $r=0,1,\dots,k/2$, but one less ground state in $M_1$, with $r=1/2,3/2,\dots,(k-1)/2$.  
Since the number of ground states is not invariant under this duality, $\mcd_{\frac12}$ is not invertible for $k$ even. 

The operators $\mcd_\varphi$ commute with the Hamiltonian for any staggering. Their action therefore must therefore take ground states to ground states. We find in general how $\mcd_{\varphi}$ acts on on any of the ground states by using the fusion algebra.
This is most easily seen by first applying $\mcd_\varphi$ on the state given in \eqref{csgscupbasis}, and then changing basis.  This amounts to fusing an `$\varphi$' strand with an `$r$' stand, yielding the strikingly simple formula
\begin{align}
	\mcd_\varphi \kett{r} = \sum_{s} N_{\varphi r}^s \kett{s} .
	\label{defectongs}
\end{align}
In the ABF models, applying $\mcd_{1/2}$ repeatedly to $\kett{0}$ gives all the ground states in the completely staggered limit.

We can easily verify \eqref{defectongs} in the Potts models by using the defect weights in the first line of \eqref{Pottsdefect}, yielding \begin{align}
\mcd_{X} \kett{a} = \kett{X} , \qquad\quad 
\mcd_{X} \kett{X} = \sum_{b=0}^{N-1} \kett{b}
\label{DXa}
\end{align}
for any $a=0,\,\dots,\,N-1$.
The states $\kett{a}$ have all fluctuating spins fixed to be $a$, while $\kett{X}$ is the equal-amplitude sum over all states.
The former states live in the model $M_0$, while the latter live in $M_1$.
Thus we recover the usual picture: the $N$ ground states  $\kett{a}$ in $M_0$ means that it is in the ordered phase, while the unique ground state $\kett{X}$ in $M_1$ indicates it is in the disordered phase. The defect-creation operator still implements Kramers-Wannier duality even in this extreme limit.

The ground states form a multiplet under the action of the defect creation operators. Since for $d_a>1$ the transformation is not unitary, no symmetry relates different values of $r$ in general. Although a local (in terms of heights) order parameter distinguishes the ground states, the degeneracies are {\em not} due to spontaneously breaking any symmetry. Sometimes this phenomenon is referred to as ``topological symmetry breaking'', as occurs in two-dimensional quantum systems with topological order; see e.g.\  \cite{Bais2009}. 
In the related 3d Chern-Simons topological field theory, Wilson loops are akin to defect lines, since they have no energy \cite{Nayak2008}. In the non-Abelian case, inserting Wilson loops around cycles thus gives degenerate states unrelated by any symmetry.

The states $\kett{r}$ are exact ground states only in the completely staggered limit $t=1$.
Nonetheless, the degeneracy must hold away from this limit, up to splitting exponentially small in the size of the system.
This fact follows from doing perturbation theory around $t=1$, where different states $\kett{r}$ and $\kett{r'}$ mix only at $\big(\frac{L}{2}\big)$\textsuperscript{th} order.
Indeed, since the perturbation is a sum over terms with next-nearest neighbor interactions, to get a non-vanishing matrix element, the perturbation must change heights $h_{2a}=r$ to $h_{2a}=r'$ for all $a$. 
Letting $E_r$ be the energy of the eigenstate $\kett{r}$ deformed away from $t$, this perturbative argument shows that the energy splitting $E_r-E_{r'}\propto (1-t)^{L/2}$ is exponentially small for $0<t<1$.
One thus expects the degeneracies remain throughout the region $0<t\le 1$, and only at the critical point $t=0$ does the splitting finally become proportional to $1/L$.
Moreover, even though the precise expression of $\kett{r}$ will change, the action 
(\ref{defectongs}) of $\mcd_a$ on these degenerate ground states will remain throughout the phase.

\begin{figure}[htbp]
   \centering
   \includegraphics[scale=1.5]{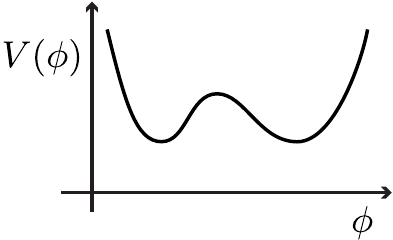} 
   \caption{A sketch of the effective Landau-Ginzburg potential for the staggered Fibonacci model, i.e.\ ferromagnetic ABF with $k=3$.}
   \label{fig:TwoMinimaPlot}
\end{figure}

The Landau-Ginzburg approach gives a  nice heuristic picture for these degenerate ground states \cite{Lassig1990}. Here the ground states are characterized as minima in a potential for some field $\phi$. For $k=3$, there are two such minima.  However, since the two ground states are not related by any symmetry, the potential must be drawn asymmetrically, as in Figure \ref{fig:TwoMinimaPlot}.

These ground-state degeneracies not required by symmetries resemble those found arising in two-dimensional gapped systems with  non-Abelian topological order \cite{Nayak2008}.
The degeneracies have the identical mathematical structure, with the self-duality in our models playing the role of topological order.
This correspondence of course is not a coincidence, as the same fusion categories describing our topological defects also govern the properties of anyons in a theory with topological order.  
The precise correspondence between degeneracies in 1d and 2d models has already been exploited to build a 2d topological phases with Fibonacci anyons by coupling together quantum chains \cite{Mong2014}.
Indeed, the Fibonacci structure for the 3-state Potts CFT perturbed by $\Phi_{1,3}$ described below in section \eqref{sec:AFM} is precisely the structure exploited there. 
This suggests that a two-dimensional quantum version of the self-duality described here will be useful in understanding non-Abelian topological order. Indeed, such self-duality is found in the quantum net models described in \cite{Fendley2008}, and the results here hint that many generalizations of such are possible.

\subsection{Degenerate excited states}

Self-duality allows us to go beyond the ground states to make statements about degeneracies among the low-lying excited states. Acting with the defect operators shows that such degeneracies go hand-in-hand with the degenerate ground states described previously.
Namely, since the defect creation operators commute with the Hamiltonian, excited states form degenerate multiplets that we can find when some of the states are kinks.
Moreover, we can show directly and easily that there must exist states other than the kinks called {\em breathers}.
We then can provide a very simple explanation of degeneracies among the single-particle excitations in certain integrable field theories.

A kink is one kind of low-energy state occurring in theories with multiple ground states.
An eigenstate with a kink at location $2b-1$ looks locally like distinct ground states to the right and to the left of $2b-1$.
In the completely staggered case $t=1$,  a single kink corresponds to having $h_{2b-2}\ne h_{2b}$. 
Elsewhere, the configuration looks like \eqref{staggeredgs}.
The kinks $K^{r,r+1}$ and $K^{r+1,r}$ in the ABF models at $t=1$ necessarily have the height $h_{2b-1}=r+\frac12$, and are
{\setmuskip{\medmuskip}{0mu}
\begin{align}
	\bigkett{K^{r,r+1}_b} &= \bigket{\cdots \, r \, \tilde{r} \, r \, \tilde{r} \, r \underbracket{ (r+\tfrac12) }_{\mkern-12mu\text{site $2b-1$}\mkern-12mu} \, (r+1) \, \widetilde{r+1} \, (r+1) \, \widetilde{r+1} \cdots} \,,
\cr
	\bigkett{K^{r+1,r}_b} &= \bigket{\cdots \, \widetilde{r+1} \, (r+1) \, \widetilde{r+1} \, (r+1) \underbracket{(r+\tfrac12)}_{\mkern-12mu\text{site $2b-1$}\mkern-12mu} r \, \tilde{r} \, r \, \tilde{r} \, r \, \cdots} \,.
\label{Kdef}
\end{align}}%
For example, for $k=2$ or $k=3$ there are two ground states in $M_0$ and so two types of kinks: $K^{0,1}$ and $K^{1,0}$.
In the Ising case, these are the usual domain walls, while in the Fibonacci case $k=3$ these are much less obvious objects.
In the effective field theory with potential in figure \ref{fig:TwoMinimaPlot}, the kinks are field configurations that interpolate between the two minima (hence the name).

More generally, these excitations can be expressed as a projector acting on the ground state~\eqref{csgscupbasis}.
To make a kink on single site, we act with $P^{(x)}_{2b-1}$ from \eqref{Pf}, where $x\in \upsilon\otimes\upsilon\ne 0$.
Using~\eqref{eq:Trivalent_Bubble}, a kink between ground states $r$ and $w$ takes the form 
\begin{align}
	\bigkett{K^{rw ; x}_b } \;=\quad \kinkrwx \ ,
\label{kinkstate}
\end{align}
with $N_{\upsilon \upsilon}^x N_{rw}^x \neq 0$ and $x\ne 0$. 
The operator $-P^{(0)}_{2b-1}$ annihilates this kink state, while it gives $-d_\upsilon$ acting on the ground state. The kink state therefore has energy $d_\upsilon$ higher than the ground state. Similarly, a multi-kink state must satisfy global fusion constraints. When $n$ excitations are present they live in the vector space $\bigoplus_{\{x_j \in \upsilon \otimes \upsilon \}}V^{r_1 x_1}_{r_2} \otimes V^{r_2x_2}_{r_3} \otimes \cdots \otimes V^{r_{n} x_{n}}_{r_{n+1}}$, a basis for which is given by allowed labelings of the following fusion tree
\begin{align}
\kinkconfigr\ .
\end{align} 
For periodic boundary conditions, $r_{n+1} =r_1$. The energy of such an $n$-kink configuration relative to the ground state is $nd_\upsilon$ for any allowed choice of the $x_j$. 

One important consequence of the definition \eqref{kinkstate} is that it includes excitations where $r=w$, i.e.\ the ground states on the left and right are identical. Such excitations are typically called ``breathers'' instead of kinks. The name arises from classical field theory, where it is a solution of the equations of motion for a kink and antikink oscillating back and forth.  From the effective-potential point of view illustrated in figure  \ref{fig:TwoMinimaPlot}, the breather $B^r$ is a localized excitation where the field strength is oscillating around the minimum corresponding to $r$.  Although in such a classical picture, such modes occur for any energy, in the quantum chain the breather energies are quantized to specific values. In models such as sine-Gordon, an explicit semiclassical calculation can be done \cite{ZamZam}. 
One thing to note is that no breather $K^{0,0}$ occurs in the ABF models, as between two heights $0$ is $\tilde{0}=\frac12$, the ground state.  Likewise, there is no $K^{\frac{k}{2},\frac{k}{2}}$ either.  Thus for $k=3$, the breather $K^{1,1}$ only occurs around one of the minima of the potential in figure  \ref{fig:TwoMinimaPlot}, the shallower one.

Since multiple ground states persist in the gapped phase $t>0$, the kink states should remain a part of the spectrum as well. They qualitatively resemble the $t=1$ expressions \eqref{Kdef}, interpolating between two regions that locally look like ground states. Due to \eqref{eq:loop_removal}, each excitation has energy $d_\upsilon$ relative to the ground state at $t=1$, but the expression will vary with $t$, presumably becoming gapless as $t\to 0$. 
However, while all the kink states will remain in the spectrum,  in general symmetries alone do not guarantee that they all remain degenerate. Moreover, the argument that breathers remain stable and degenerate as $t<1$ no longer automatically holds, as it appears they can mix with other topologically trivial states in perturbation theory. 

Nevertheless, the degeneracies between all kinks and breathers do persist for $t<1$ as a simple consequence of self-duality. This degeneracy was originally observed numerically in the field theory for $k=3$  \cite{Lassig1990}, and generalized to all $k$ by using rather subtle arguments involving the integrability of the perturbed conformal field theory and the (inferred) quantum-group symmetry \cite{Lassig1990,Smirnov1991,Chim1991,Dorey2002a,Dorey2002b}. Our work provides a direct lattice derivation of this striking result.

As explained in section \ref{sec:generalized}, the operator $\mcd_1$ implements a self-duality in the ABF models. Acting with it on kink configurations gives other configurations with the same energy. Since $\mcd_1$ is non-local it is possible for it to change the type of kinks, or even map to non-kink configurations. Consider again the completely staggered case, where the kink states are ground states everywhere except at one height. Since acting with $\mcd_1$ takes ground states to a linear combination of ground states, it must do the same on kinks except in the region of that height. Since the defect creation operators are defined with periodic boundary conditions, we must consider two-kink configurations. For example, because $\mcd_1$ acting on any height $0$ must give the height $1$, we have
\begin{align} \begin{split}
	\mcd_1 \bigkett{K^{0,1}_b K^{1,0}_{b'}} &=
	\mcd_1 \ket{\cdots 0\tilde{0} 0\tilde{0} 1\tilde{1}1 \cdots 1 \tilde{1}1 \tilde{0} 0\tilde{0} 0 \cdots}
	\\&= \alpha_{0} \bigkett{ K^{1,0}_b K^{0,1}_{b'} } + \alpha_{1} \bigkett{ K^{1,1}_b K^{1,1}_{b'} } + \alpha_2 \bigkett{ K^{1,2}_b K^{2,1}_{b'} }
	\label{D1K01}
\end{split} \end{align}
for some coefficients $\alpha_j$.
The state $\bigkett{ K^{1,1}_b }$ is a breather. Explicitly, in the ABF models
\begin{align}
	\bigkett{ K^{rr}_b } = \ket{\cdots r \, \tilde{r} \, r \, \tilde{r} \, r \mkern1mu B^r r \, \tilde{r} \, r \cdots} \,,\qquad
	\ket{B^r} \defineas \frac{1}{\sqrt{d_{\frac12}d_{r}}}\left(\sqrt{d_{r+\frac12}}\, \Ket{r-\tfrac{1}{2}} - \sqrt{d_{r-\frac12}}\, \Ket{r+\tfrac{1}{2}}\right) .
\end{align}

As $\mcd_1$ commutes with the Hamiltonian, the breathers and the kinks are thus degenerate in the completely staggered case, even though the states look quite different.
The single-particle excited states, both kinks and breathers, thus can be thought of as forming a multiplet under the action of $\mcd_1$, in the same way that the ground states are.    
Moving away from complete staggering, the computation of the energies using perturbation theory is much more difficult than for the ground states, since the Hamiltonian mixes kink/breather configurations at different sites.

Luckily, we do not need to do this calculation to see that the degeneracies remain throughout the ordered phase.
All we need is the self-duality. Because the distinct degenerate ground states remain in the gapped case, the kinks remain low-lying excitations.
Repeatedly applying $\mcd_1$ to these states maps the different kinks among each other as well as to various breather modes.
Since $\mcd_1$ commutes with the Hamiltonian for any $t$, all the states obtained this way must be degenerate.
The degeneracies observed in perturbed conformal field theory exactly occur on the lattice.

Working out the precise action of a defect on kinks and breathers configuration takes a little work.
In the completely staggered theory for a general category, a kink can be labeled in terms of three objects as $\bigkett{ K^{rw;x}_b }$ (see \eqref{kinkstate}), where $r= h_{2a}$ for $a<b$ and $w=h_{2a}$ for $a \geq b$ while $x$ labels the flavour of kink that sits between the $r$ and $w$ vacua.
For the state to exist we require $N_{rw}^x \neq 0$, with $x=1$ in the ABF case. 
Acting with $\mcd_\varphi$ on a basis element results in
\begin{align}
\mcd_{\varphi} \left( \kinkconfigr  \right) &\cr
= \sum_{w_j} \prod_j \sqrt{d_{r_j} d_{w_j}}& \Msixj{x_j & w_j & w_{j+1}}{\varphi &r_{j+1} & r_j} \kinkconfigw\qquad
\label{defectkink}
\end{align}
where we have assumed periodic boundary conditions so that $r_{n+1} = r_1$ and $w_{n+1} = w_1$.
For the ABF kink in \eqref{D1K01} we can use \eqref{F0c} to get $\alpha_{w'}=\sqrt{d_{w'}/d_1^2}$.

The self-duality of the (nonintegrable for $t\ne 1$) lattice model gives even more information about the long-distance behavior, providing a strong indication that the corresponding perturbed CFT is integrable. The argument comes from studying the scattering of two-particle states. The scattering matrix must commute with $\mcd_\varphi$, as it describes the asymptotic behavior of the states under time evolution by the Hamiltonian. Knowing the coefficients from \eqref{defectkink} provides a strong constraint on its form. Since in the staggered ABF case the kinks and breather form a degenerate spin-1 multiplet, the two-particle scattering matrices themselves can be written as a sum over projectors (\ref{AQbasis}). The simplest (and only nice) solution of this constraint is for two particles always scatter into two. Having no particle production is the hallmark of an integrable field theory, as a consequence of the additional conserved charges \cite{ZamZam}. Scattering matrices then can be non-diagonal only when there are exact degeneracies of the type we derived using the defect-creation operators. Although our argument goes no further, the much-more-detailed analysis using perturbed conformal field theory does show the model is integrable \cite{Smirnov1991,Chim1991}. Integrability then constrains further the relative coefficients of the projectors  \cite{Smirnov1991,Chim1991,Fendley2002}, and by analyzing the bound-state structure one finds the overall prefactor  \cite{Dorey2002a,Dorey2002b}. The coefficients turn out to be the Izergin-Korepin solution of the Yang-Baxter equation \eqref{IKYBE}.

\subsection{The antiferromagnetic case}
\label{sec:AFM}

The upshot of our analysis is that the self-duality of the ABF models requires that the allowed low-energy {excitations} of the staggered Hamiltonian form a degenerate multiplet. The general form \eqref{defectkink} indicates that the same degeneracies occur in many other theories.  Indeed, any time some completely staggered limit of \eqref{HP} can be found that admits multiple ground states, any $\mcd_\varphi$ preserving $M^\pm$ will map between them. The above arguments suggest that if there is a long-distance perturbed conformal field theory describing this limit, it should be integrable. 

One simple but striking generalization is to consider the antiferromagnetic version of \eqref{Hcritpert} for the ABF models. 
For these models, translation symmetry in the continuum model corresponds on the lattice to translation by $k$ sites \cite{Huse1984}.
Translation by a single site becomes a $\mathbb{Z}_k$ internal symmetry of the parafermion CFT describing the continuum limit.
We should then stagger the couplings on every $k$\textsuperscript{th} site instead of every other, so that the effective field theory describing the perturbation preserves the same translation symmetry. We consider the simplest such possibility
\begin{align}
	H^{\rm AF} &= H^{\rm AF}_{\rm crit} - t_{\rm AF}\sum_{a=1}^{L/k} e_{ka} \,,
&	H^{\rm AF}_{\rm crit} &= -H_{\rm crit} = \sum_{j\in\mathbb{Z}} e_j \,,
&	H^{\rm AF}_{\rm cs} &= \lim_{t_{\rm AF}\to 1}H^{\rm AF} = \sum_{\frac{j}{k}\notin \mathbb{Z}} e_j
	\label{HAF}
\end{align}
with $L$ a multiple of $2k$. The staggering breaks the $\mathbb{Z}_k$ symmetry of the critical theory.

First consider the completely staggered limit $t_{\rm AF}\to 1$, with Hamiltonian $H_{\rm AF,cs}$ given in \eqref{HAF}. A zero-energy ground state is given by any state annihilated by 
all $e_j$ with $j$ not a multiple of $k$. One such state in $\mathcal{V}_+$ is 
\begin{align}
	\kett{0}_{\rm AF} \defineas \Ket{ \tfrac12\,1\,\tfrac32\,\cdots \tfrac{k-1}{2}\,\tfrac{k}{2}\, \tfrac{k-1}{2}\,\cdots\,\tfrac12\,0\,\tfrac12\,1\cdots \tfrac12\,0 }, \quad\text{ i.e., }
h_{nk+a}=
\begin{cases}
 \tfrac{a}{2}\quad &\hbox{ for } n \hbox{ odd,}\cr
 \tfrac{k-a}{2}\quad &\hbox{ for } n \hbox{ even}
 \end{cases}
\label{AFgs}
\end{align}
for $a=0,\dots,k-1$.
As apparent from \eqref{ABFweights},  $e_j\ket{\cdots h_{j-1}h_jh_{j+1}\cdots}=0$ when $h_{j-1}\ne h_{j+1}$.
Thus by construction, $e_j \kett{0}_{\rm AF}=0$ for all  $j\ne nk$, and so $H^{\rm AF}_{\rm cs}$ annihilates $\kett{0}_{\rm AF}$.

One can construct other ground states by brute force, but there is no need: we simply act with $\mcd_1$ and can generate the others.
For example, for the Fibonacci case $k=3$, we use  \eqref{defectFib} for the defect weights, rewriting them in terms $0,\frac{3}{2}\to 0$ and $\frac12,1\to \tau$.
Along with $\kett{0}_{\rm AF} = \ket{0\,\tau\,\tau\,0\,\tau\,\tau\,0\,\cdots}$, we have the ground state
\begin{align}
	\kett{1}_{\rm AF} \defineas \mcd_1\kett{0}_{AF}\propto \ket{ {-\!-} \, \tau \, {-\!-} \, \tau \, {-\!-} \, \tau \, \cdots } ,
	\qquad\text{where } \ket{-\!-} \defineas \frac{ \ket{0\tau} + \ket{\tau0} }{\phi} - \frac{ \ket{\tau\tau} }{\phi^{3/2}} \,,
	\label{AFgs1}
\end{align}
where $\phi=d_\tau=2\cos(\pi/5) = \tfrac{1}{2}\big(1+\sqrt{5}\big)$, the golden mean.
Using \eqref{ABFweights} it is straightforward to verify that each term in $H_{\rm AF,cs}$ annihilates $\kett{1}_{\rm AF}$ as well.

Acting with $\mcd_1$ thus gives a degenerate multiplet of ground states, as \eqref{defectongs} requires. What is remarkable here is that because $\mcd_1$ is the same for ferromagnetic and antiferromagnetic ABF models, the ground-state structure is the same as well! We find the same number of ground states and the same action of $\mcd_1$, even though the Hamiltonians, the explicit expressions, and the critical CFTs are completely different. We therefore can run the same argument showing that the ground states will remain stable when $t_{\rm AF}<1$, and expect that they persist until the critical point at $t_{\rm AF}=0$. We also can rerun the same arguments to show that the kinks, breathers and scattering matrices also behave in the same fashion. The low-energy spectra should be identical, although possibly the overall function in front of the scattering matrix could be different. This structure was indeed inferred from quantum-group symmetries in \cite{Fateev1990}, but our approach gives a simple way of seeing it directly and intuitively. 

The antiferromagnetic models have another interesting facet: as apparent from \eqref{HAF}, positive and negative $t_{\rm AF}$ result in distinct theories.
We have showed that the positive $t_{\rm AF}$ theory is gapped, but as the completely staggered limit $t_{\rm AF}=-1$ is rather trivial, it is possible that the model is gapless for  $t_{\rm AF}$ negative.
Field-theory arguments  indicate that for $t_{\rm AF}$ negative, the model is indeed gapless with a flow to another CFT \cite{Fateev1991}.
The fact that all defect lines commute with $H_{\rm AF}$ throughout the flow places an enormous constraint on any CFT describing the endpoint of the flow, as it must have the same duality properties (and e.g.\ the $g$ factors we discuss below).
The obvious candidate for the endpoint CFT is therefore the $k$\textsuperscript{th} minimal CFT describing the ferromagnetic point.
Indeed, the much more detailed field-theory arguments  \cite{Fateev1991} indicate that the flow from the $\mathbb{Z}_k$ parafermion theory indeed is to the $k$\textsuperscript{th} minimal CFT.
For example, for the Fibonacci case $k=3$ the flow is from the three-state Potts CFT with $c=4/5$ to the tricritical Ising CFT with $c=7/10$.
Our arguments do not prove this flow occurs, but do make this flow from antiferromagnet to ferromagnet rather natural.


\section{Boundary states, conformal and otherwise}
\label{sec:bdry}

The topological defects constructed in this paper have several fundamental properties. The basic one is that the partition function remains invariant under their deformation. Even more profoundly, the properties derived in section \eqref{sec:trivalent} allow partition functions with different defect configurations to be related. This simplest such property $Z_\varphi = d_\varphi\,Z$ from (\ref{removeloop},\ref{ZdZ}) relates partition functions with and without a loop. More intricate but still incredibly useful are the invariances under  $F$-moves. 

In this section we use these invariances to derive a variety of properties of boundary conditions. Our results provide a natural lattice version of the boundary-state formalism in CFT \cite{Cardy1989}, and generalize it to off-critical models as well. The precise definition of linear combinations of boundary conditions allows us to define linear operators acting on them, including the defect-creation operators.  The ensuing equivalences between partition functions with different boundary conditions allows us to derive exact values of certain universal quantities. 
In particular, we give an exact and rigorous lattice derivation of ratios of the universal ``$g$-factors'' \cite{AffleckLudwig} characterizing conformal boundary conditions.  The only assumption required is the standard physicist one that a lattice model becomes a CFT in the continuum limit. As such, it provides a strong constraint on any such identification. Our method gives a precise way of finding $g$-factors not only in CFT, but in non-critical theories as well.

\subsection{Gluing defects to the boundary}
\label{sec:gluing}

We start our study of boundaries by considering height models on a planar graph $G$ embedded in a disc, so that the $\upsilon$ lines (which live on edges of the dual graph) dangle across the single boundary.  These dangling ends and the heights between them form a periodic fusion tree \eqref{eq:fusiontreedef}.  The space of all boundary conditions we study is therefore given by $\mathcal{V}$ from \eqref{vper}, with each basis element $\ket{\mathcal{B}}$ labeled by a particular height configuration around the boundary.
For $L_b$ heights in this tree, the dimension of $\mathcal{V}$ grows for large $L_b$ as $d_\upsilon^{L_b}$.
In standard language, each $\ket{\mathcal{B}}$ amounts to taking a {\em fixed} boundary condition.
All other boundary conditions we study are linear combinations $\ket{B} = \sum_{\mathcal{B}} \alpha_{\mathcal{B}} \ket{\mathcal B}$ for some coefficients $\alpha_{\mathcal B}$. 

The most convenient definition of the Boltzmann weights is to consider some fixed boundary condition with heights $h_j$ for $j=1,\dots,L_b$ along the boundary, and then take 
\begin{align}
e^{-\beta H(\{ h_{\rm v} \})} = \prod_{p} \BWx{\alpha_p}{\beta_p}{\delta_p}{\gamma_p}\, \times \prod_{{\rm v}\ {\rm in}\ \rm {bulk}}d_{h_{\rm v}} \times \prod_{j=1}^{L_b} \sqrt{d_{h_j}} \,,
\label{discweight}
\end{align}
so that the heights along the boundary each receive a weight $\sqrt{d_{h_j}}$, the half-dot in the notation in \eqref{semidotdef}.
The disc partition functions for a fixed and arbitrary boundary conditions are then 
\begin{align}
Z\big(\ket{\mathcal{B}}\big) = \sum_{ \{ h_{\rm v}\}\ {\rm in}\ {\rm bulk}} e^{-\beta H( \{ h_{\rm v} \}) } \,,\qquad Z\big(\ket{B}\big) =\sum_{\alpha_{\mathcal{B}}} \alpha_{\mathcal{B}} Z\big(\ket{\mathcal{B}}\big) \equiv\  \ZDisc \Bigg|_{\ket{B}} \,.
\label{Zbdry}
\end{align}
We have defined a convenient pictorial notation here. The value of the partition function will of course depend on the precise graph $G$ used to define the model, but we suppress this label.
One nice thing about the definitions (\ref{Zbdry}) is that the partition function is a linear function on $\mathcal{V}$:
\begin{align}
\ZDisc \Bigg|_{\alpha \ket{A}+\beta \ket{B}}  =\ \alpha \times \ZDisc \Bigg|_{ \ket{A}}+\beta \times \ZDisc \Bigg|_{\ket{B}} \,.
\label{Zlin}
\end{align} 

We therefore can define linear operators acting on the boundary state. A particularly useful set are the defect creation operators, defined as in \eqref{TsymmGoldenChain}, with the local weights given in \eqref{eq:BW_duality}. These by construction are linear operators in $\mathcal{V}$. Because of the defect commutation relations, it is rather simple to prove  all sorts of equivalences between partition functions with different boundary conditions. The easiest is to exploit
 (\ref{removeloop}), which in pictures yields
 \begin{align}
\ZDisc \Bigg|_{\ket{B}} =  \frac{1}{d_\varphi}\, \ZDiscGLoop \Bigg|_{\ket{B}} =\ \frac{1}{d_\varphi}\, \ZDiscGDual \Bigg|_{\ket{B}}  = \frac{1}{d_\varphi} \, \ZDiscG \Bigg|_{\D_\varphi |B\rangle} \,.
\label{Bdual}
\end{align}
The proof of \eqref{Bdual} amounts to explaining what the pictures mean. We ``nucleate'' a loop of type $\varphi$ in the bulk using  (\ref{removeloop}), The shading in the picture is a reminder that  the models on the inside and outside of the loop are different when $\varphi$ is odd. We then stretch the loop to the boundary using the defect commutation relations. For a fixed boundary condition $\ket{\mathcal{B}}$, the heights $h_j$ on the outside of the defect loop are those in the fusion tree of $\mathcal{B}$.  Given the definition \eqref{eq:BW_duality} of the defect line, the ensuing partition function is exactly the same as if we omit the defect line and instead have boundary state $\mcd_\varphi\ket{\mathcal{B}}$. Because this action is linear as in \eqref{Zlin}, this argument works for any boundary condition $\ket{B}$, yielding \eqref{Bdual}. Nucleating a defect loop and ``gluing'' it to the boundary thus gives a rigorous and exact relation between different partition functions. 

For example, applying  \eqref{Bdual} gives a simple way of understanding how Kramers-Wannier duality acts on boundary conditions.
In the clock models, the boundary state $\kett{a} \equiv \ket{aXaX\cdots}$ corresponds to the ``fixed-$a$" boundary condition, where all fluctuating spins on the boundary are fixed to be the same value $a=0,\dots,N-1$. (The length $L_b$ of the boundary must be even, as $G$ is bipartite.)
We have recycled the notation of section \ref{sec:degenerategs}, as these boundary states are identical to the ground states in the completely staggered limit. 
Using \eqref{Bdual} with a spin-shift defect gives then $Z\big(\kett{a}\big) = Z\big(\kett{b}\big)$ for all $a,b$.
The effect of gluing a duality defect $\mcd_X$ to $\kett{a}$ is already given in \eqref{DXa}:  $\mcd_X |A\rangle = \kett{X}$.
The boundary state $\kett{X}$ is the equal-amplitude sum over all fixed boundary states in $M_0$.
It corresponds to a ``free'' boundary condition in conventional language. The equivalence \eqref{Bdual} means
\begin{align}
	{Z}_1\big(\kett{X}\big) = \sqrt{N}\, Z_0(\kett{a}\big) \,.
	\label{Zfreefixed}
\end{align}
where the dual partition functions $Z_0$ and $Z_1$ are as described in section \ref{sec:duality}. Using  \eqref{KWdef} gives 
$\mcd_{X}\ket{{\rm free}} = \sum_{b}\ket{b}$, so no other types of boundary conditions are generated by gluing again.
In the $N=2$ Ising case, these relations reduce to the Ising case discussed in Part~I as they must. 

On the cylinder, a distinct fusion tree describes each boundary, and the definitions \eqref{discweight} and  \eqref{Zbdry} generalize in the obvious way.
The partition function is then a linear function on $\mathcal{V} \times \mathcal{V}$, so we label it as $Z\big(\ket{A},\ket{B}\big)$.
The defect creation operators act on each boundary individually, as they do for the disc. 
Since the defects can be moved through the bulk, we can unglue a defect loop from one boundary, move it across the cylinder, and glue it to the other boundary. Such loops cannot be removed as in \eqref{Bdual}, as they cannot be shrunk to surround a single quadrilateral. More generally, we can take advantage of the fusion rules for the defect lines \eqref{DDeqD} and fuse together defect operators coming from both boundaries. This fusion yields the remarkably simple identity between partition functions with distinct boundary conditions:
\begin{align}
\begin{split}
	\substack{\mathcal{D}_a |A\rangle\\
	\annulusx\\
	\mathcal{D}_b |B\rangle}\ =\  \substack{\ket{A} \\  \annulusy \\ \ket{B}}\ &=\  \sum_{c}N^c_{ab}\, \substack{\ket{A} \\
	 \annulusz \\
	  \ket{B}}  
	\\[6pt]
	\implies\quad 
	Z\big(\mcd_a|A\rangle,\mcd_b|B\rangle\big) &= \sum_c N_{ab}^c\, Z\big(\mcd_c|A\rangle,\ket{B}\big)
	\label{annulusfusion}
\end{split}
\end{align}
for any $A$, $B$, $a$ and $b$. 
We have drawn a cylinder by identifying the left and right edges of the gray square.
When the fusion ring is commutative, i.e., $a\otimes b \cong b\otimes a \Rightarrow N_{ab}^c = N_{ba}^c$, the identity extends to
$Z\big(\mcd_a|A\rangle,\mcd_b|B\rangle\big) = Z\big(\mcd_b|A\rangle,\mcd_a|B\rangle\big)$.

Applied to Potts/clock, \eqref{annulusfusion} yields
\begin{align}
	Z\big( \kett{a},\kett{b} \big) = Z\big( \kett{a{+}c} , \kett{b{+}c} \big) \,,\qquad
	Z_0\big( \kett{a} , \kett{X} \big) = Z_1\big( \kett{X},\kett{a} \big) \,.
\end{align}
Here we let $\kett{a+b} \defineas \kett{ (a+b)\bmod N }$.
While the first relation follows from the $\mathbb{Z}_N$ symmetry, those in the second are not so obvious, as dragging a duality defect from one boundary to the other changes the weights to their duals.
Even less obvious is the consequence of \eqref{KWdef}, which is
\begin{align}
	Z_0\big( \kett{X},\kett{X} \big) = \sum_{b=0}^{N-1} Z_1\big( \kett{a},\kett{b} \big)
	\label{Zff}
\end{align}
for any $a$. The dual of free boundary conditions on both boundaries of the cylinder is proportional to the sum over all fixed-$a$ boundary conditions on both sides. 
This linear relation is of course consistent with the corresponding conformal field theory results \cite{Cardy1986b}, but  applies to any Potts or clock couplings on any graph with the topology of a cylinder.

\subsection{Boundary states in and from conformal field theory}
\label{sec:conformal}

Changing boundary conditions by gluing defect lines provides a powerful method for understanding boundary states.  Here we show how the boundary-state formalism used to understand conformal boundary conditions \cite{Cardy1989} is intimately related to our setup. The payoff in linking the two approaches goes both ways: our work provides a precise lattice way of deriving identities for partition functions such as \eqref{annulusfusion}, and provide exact lattice calculations of universal quantities. In particular, we show how our exact results for the $g$-factor characterizing conformal boundary states match perfectly the CFT results.

Conformal invariance still holds in the presence of boundaries when the boundary conditions are chosen appropriately. Such a conformal boundary condition (CBC) can be found for each primary field in a rational conformal field theory  \cite{Ishibashi1988}. These ``Ishibashi states'' are constructed by a boundary-state formalism, where one treats the boundary as a Hilbert space on which the generators of conformal symmetry act. By clever arguments using modular invariance, Cardy showed how to relate take linear combinations of these boundary states to reproduce boundary conditions expected from taking the continuum limit of a lattice model \cite{Cardy1989}.  Both types of boundary states can be labeled using primary operators in (not necessarily the same) CFT.  We show here how the ``Cardy states" in CFT naturally correspond to the continuum limit of states built using defect-creation operators.

For many (perhaps all) of the statistical-mechanical models built from a fusion category, at least one choice of lattice and couplings $A_Q$ yields a critical model. Typically, a CFT is believed to describe the corresponding continuum limit, a fact rigorously shown for the Ising model \cite{Smirnov2012}.  In this event, it is natural to expect that the lattice topological defects found by our construction become conformal defects, with the partition function in their presence remaining conformally invariant. Then if a given lattice boundary condition $\ket{B}$ becomes a CBC in the continuum limit, we expect that the state $\D_\varphi|B\rangle$ found by gluing a topological defect to it will yield a CBC. 

In a rational conformal field theory on the cylinder with conformal boundary conditions labeled by primary fields $\mathscr{A}$ and $\mathscr{B}$, the partition function is of the form \cite{Cardy1989}
\begin{align}
Z_{\mathscr{AB}} = \sum_\mathscr{C} N^\mathscr{C}_{\mathscr{AB}}\chi^{}_\mathscr{C} \,,
\label{ZABCFT}
\end{align}
where $\mathscr{A}$ and $\mathscr{B}$ label the two boundary conditions, $N^\mathscr{C}_{\mathscr{AB}}$ is a non-negative integer, while $\chi_\mathscr{C}$ is the character corresponding to the primary field $\mathscr{C}$ in the CFT.  
In a rational conformal field theory, there are a finite number of such fields. Note we use subscripts to distinguish the CFT partition function from the lattice one.
The central tenet of the boundary-state formalism is that this partition function also can be written as 
\begin{align}
	Z_{\mathscr{AB}} = \Braket{\mathscr{A} | e^{-R H_L} | \mathscr{B} }
\label{ZAHB}
\end{align}
where $H_L$ is the CFT Hamiltonian on a circle of circumference $L$, $R$ the length of the cylinder, and $\ket{\mathscr{A}}$ and $\ket{\mathscr{B}}$ are the boundary states labeled by particular primary fields.
We expect that when the lattice couplings are tuned to make the model critical, there exist appropriate identifications of boundary states and conditions so that $Z\big(\ket{A},\ket{B}\big) \to Z_{\mathscr{AB}}$ in the continuum limit.

All three labels in \eqref{ZABCFT} correspond to primary fields in a rational CFT (RCFT).
They must therefore also correspond to objects in a fusion category~\cite{MooreSeiberg:89} associated with the RCFT. 
It is natural therefore to expect that the integers $N^\mathscr{C}_{\mathscr{AB}}$ correspond to the fusion coefficients in this category.
Cardy's boundary-state formalism \cite{Cardy1989} provides a method for understanding how and why. Here we show how similar considerations apply directly on the lattice, so that we can compute some universal quantities exactly without having to appeal to any CFT results. As we mentioned above and will discuss in more detail below, the fusion categories of the lattice model and the CFT typically are not the same (Ising providing one of the few examples where they are). Our work therefore provides powerful constraints on which CFT can describe the continuum limit of a given lattice model.

A valuable tool for characterizing boundary conditions is known as the ``$g$-factor'' \cite{AffleckLudwig}. It is something like a ground-state degeneracy, and is easiest to define when space is a cylinder. In the limit the cylinder is very long, i.e.\ $R\gg L$, the leading contribution to \eqref{ZAHB} comes from the lowest eigenvalue $E_0$ of $H_L$. Denoting by $\ket{0}_{\rm CFT}^{}$ the corresponding eigenstate, \eqref{ZAHB} reduces to 
\begin{align}
	Z_{\mathscr{AB}}  \rightarrow g_\mathscr{A}g_\mathscr{B} e^{-R E_0 } \,,\qquad g_\mathscr{B}\equiv \braket{\mathscr{B}|0}_{\rm CFT}^{}\,.
	\label{gCFT}
\end{align}
This number $g_\mathscr{B}$ associated with each boundary state is the $g$-factor.   Its logarithm can be thought of as the entropy associated with the boundary condition. For this reason, $g_\mathscr{A}g_\mathscr{B}$ is sometimes referred to as the ``ground-state degeneracy", but this degeneracy need not be an integer, as apparent from the definition \eqref{gCFT}.

In a CFT, the ground state is unique, and the $g$-factor is universal \cite{AffleckLudwig}.  Indeed, doing a modular transformation on the characters in \eqref{ZABCFT}, i.e.\ exchanging space and time, gives \cite{Cardy1989,AffleckLudwig}
\begin{align}
g_\mathscr{A} g_\mathscr{B} = \frac{1}{\mathscr{D}_{\rm CFT}^{}}\sum_\mathscr{C} N^\mathscr{C}_{\mathscr{AB}} d_\mathscr{C} \,,\qquad \mathscr{D}_{\rm CFT}^{}=\sqrt{\sum_{\mathscr{A}} d_\mathscr{A}^2}
\label{ggS}
\end{align}
where the sum in the total quantum dimension $\mathscr{D}_{\rm CFT}^{}$ from \eqref{ddNdD} is over all primary fields in the CFT, or equivalently, all the simple objects in the category governing the fusion and braiding in the CFT. The quantum dimensions arise from the explicit expression of the modular $S$ matrix entries involving the identity, see e.g.\ equation (225) in \cite{Kitaev2006}. When objects are not all self-dual one must again keep track of arrows, but the idea is the same. 
A simple exact expression for the $g$ factors applies when the boundary conditions correspond to the Cardy states associated with primary fields. Then $N^\mathscr{C}_{\mathscr{AB}}$ is the fusion symbol for the corresponding primary fields in the CFT \cite{Cardy1989}. Doing the sum in \eqref{ggS} by exploiting  \eqref{eq:ddNd}  yields 
\begin{align}
g_\mathscr{A} g_\mathscr{B} =  \frac{d_\mathscr{A}d_\mathscr{B}}{\mathscr{D}_{\rm CFT}^{}} \qquad \implies \qquad g_\mathscr{A}= \frac{d_\mathscr{A}}{\sqrt{\mathscr{D}_{\rm CFT}^{}}} \,.
\label{ggdd}
\end{align}

With topological defects, we can derive closely related equations exactly and rigorously on the lattice. We define the partition function on a $L_b\times R$ square lattice analogously to \eqref{ZAHB} as
\begin{align} 
	Z\big(\ket{A},\ket{B}\big) = \Braket{A | T^{R} | B }\,,
\end{align} 
where the transfer matrix $T$ acts on the usual vector space $\mathcal{V}$ defined by a periodic tree. In the long-cylinder limit where $R/L_b$ is large, the largest eigenvalue $\Lambda$ of the transfer matrix dominates. (In a quantum Hamiltonian limit like \eqref{ZAHB}, the lowest energy level $E_0\propto -\ln \Lambda$.)  When the corresponding eigenstate $\ket{\Lambda}$ is unique, we then define the lattice $g$-factor as
\begin{align}
Z\big(\ket{A},\ket{B}\big)  \sim g\big(\ket{A}\big)g\big(\ket{B}\big) \Lambda^R \,,\qquad g\big(\ket{B}\big)\equiv \braket{\Lambda| B}\,.
\label{gdef}
\end{align}
Nowhere in the lattice definition of $g$-factor did we utilize conformal invariance, so the definition \eqref{gdef} applies equally well for any lattice model and boundary condition. A test of this universality is that $g$ factors computed in various geometries give the same answer.

Non-universal subleading terms in the bulk part of the partition function complicate extracting the universal contribution from coming from the boundary. We therefore compute only {\em ratios} of $g$ factors, where the bulk contributions cancel. (In CFT, the $g$ factors can be computed directly, as bulk free energy is exactly zero and the leading subleading term universal \cite{Blote1986,Affleck1986}.) For any graph with topology of the disc,
\begin{align}
\frac{g\big(\ket{A}\big)}{g\big(\ket{B}\big)} = \lim_{L_b\to\infty}\frac{Z\big(\ket{A}\big)}{Z\big(\ket{B}\big)} \,.
\label{gratiodef}
\end{align}
The corresponding expression on the cylinder follows from \eqref{gdef}, with the ratio making the factors of $\Lambda$ cancel. A remarkable consequence of our work is that any time two boundary conditions are related by gluing a topological defect, the corresponding ratio of $g$ factors follows instantly. Namely, on a disc, \eqref{Bdual} gives for any $L_b$
\begin{align}
\frac{g\big(\mcd_\varphi \ket{B}\big)}{g\big(\ket{B}\big)} = d_\varphi \,.
\label{gdphi}
\end{align}
This exact relation holds for {\em any} boundary state $\ket{B}$ in any lattice model built using a fusion category: the quantum dimensions give exact ratios of $g$ factors!
The identity \eqref{annulusfusion} for cylinder partition functions gives a nice consistency check. Using it with the definition \eqref{gdef} gives
\begin{align}
g\big(\mcd_a|A\rangle\big)g\big(\mcd_b|B\rangle\big)=\sum_a N_{ab}^c\, g\big(\mcd_c |A\rangle\big) g\big(\ket{B}\big) \,.
\end{align}
Using \eqref{gdphi} then yields \eqref{eq:ddNd}.

Comparing \eqref{gdphi} with \eqref{ggdd} makes clear how our lattice result fits in beautifully with the CFT analysis. 
We expect that with appropriately chosen couplings and boundary states,
\begin{align}
\frac{g\big(\ket{A}\big)}{g\big(\ket{B}\big)}\quad\longrightarrow\quad \frac{g_\mathscr{A}}{g_\mathscr{B}}
\end{align}
in the continuum limit. These relations between $g$-factors and quantum dimensions had an earlier physical interpretation in a correspondence between thermodynamic entropy and topological entanglement entropy \cite{Fendley2006,Qi2012}. This correspondence provides a very strong justification for believing that the continuum limit of these lattice models at a critical point corresponds to a conformal field theory, and gives exact information to make the identification precise.

\subsection{From lattice to CFT}

We showed in section \ref{sec:conformal} the similarities between the CFT boundary-state formalism and the lattice approach of section \ref{sec:gluing}.
Here we deepen the correspondence by identifying specific lattice boundary states that yield
conformal boundary conditions labeled by primary fields in the rational CFT.

\subsubsection{Three-state Potts and parafermions}

We have already seen three distinct boundary conditions in Ising: fixed $+$, fixed $-$, and free, corresponding to the boundary states $\kett{0}$, $\kett{1}$ and $\kett{X}$ respectively. 
They are related by defect creation operators in \eqref{Zfreefixed} and \eqref{DXa}, so that \eqref{gdphi} requires 
\begin{align}
g\big(\kett{X}\big)=\sqrt{2} g\big(\kett{0}\big)=\sqrt{2} g\big(\kett{1}\big) \,.
\label{gIsing}
\end{align}
In the (chiral) Ising CFT and category, there are three primary fields $1,\psi$ and $\sigma$, with quantum dimensions $1,1,\sqrt{2}$ respectively, so that the total quantum dimension is $\mathscr{D}_{\rm CFT}^{}=2$. The detailed CFT analysis identifies the corresponding CBCs (Cardy states) with fixed $+$, fixed $-$ and free boundary conditions respectively \cite{Cardy1989}, so that $g_+=g_-=1/2$ and $g_{\rm free}=1/\sqrt{2}$. From \eqref{gIsing} it is apparent that our lattice analysis gives these ratios exactly, rigorously and directly. 

For Potts and clock models, \eqref{gIsing} for free and fixed boundary conditions generalize easily to
\begin{align}
g\big(\kett{X}\big)=\sqrt{N}\, g\big(\kett{a}\big)
\label{gfixedfreePotts}
\end{align}
for any fixed boundary condition with $a\,$=$\,0,\dots, N-1$. The connection with the corresponding CFTs here however is not as immediate as Ising. The free boundary condition does {\em not} correspond to a field in the Potts or parafermion CFT \cite{Cardy1989}. In category language, this distinction arises because the $\mathbb{Z}_N$ Tambara-Yamagami category used to define the lattice model is not a subcategory of the parafermion CFT category, as apparent from the CFT analysis \cite{ZamoFateev:Parafermion:85}. In CFT language, the field $X$ is one of the $C$-disorder operators described in \cite{Zamolodchikov86}. Our exact lattice result \eqref{gfixedfreePotts} requires that $g_{\rm free} = \sqrt{N} g_{a}$ in the CFT describing the continuum limit, whether or not the states corresponds to a primary field. In the $N=3$ case, it was argued  \cite{Affleck1998} that one can identify the appropriate field in the orbifold theory, and one indeed recovers the appropriate ratio $\sqrt{3}$. 

On the other hand, there are primary fields in the parafermion CFT not in the Tambara-Yamagami category. To find lattice analogs, we apply \eqref{gdphi} starting with some boundary condition not free or fixed. 
A classic paper \cite{Saleur1989} provides valuable guidance. By numerical analysis of the three-state Potts lattice model at its critical point, conformal boundary conditions corresponding to all six primary fields in the three-state Potts CFT are identified. The three fixed boundary conditions $\kett{a}$ with $a=0,1,2$ correspond to the three Cardy states for the identity and the two parafermion fields. The other three correspond to the ``not-$a$'' states $\bigkett{\overline{a}}$, where the state is an equal-amplitude sum over all basis states where the height $a$ does {\em not} appear, e.g.
\begin{align}
\bigkett{\overline{0}} = \Ket{\overline{0}\,X\,\overline{0}\,X\cdots} \,,\qquad    \Ket{\overline{0}} \defineas \frac{1}{\sqrt{2}}\big( \ket{1}+\ket{2}\big)\,.
\label{not0state}
\end{align}
Acting with a spin-shift defect on a not-$a$ states gives another one: $\mcd_b\bigkett{\overline{a}} = \bigkett{\overline{a+b}}$ where the sum is interpreted mod $3$. More intriguing is when we act with the duality defect on any of the not-$a$ states, giving an eighth conformal boundary condition \cite{Affleck1998}, corresponding to lattice boundary state $\bigkett{\overline{X}} \equiv D_X \bigkett{\overline{a}}$.
This boundary state (dubbed ``new'' in 1998) can be written out explicitly as a two-channel matrix-product state \cite{OBrien2019}. Applying \eqref{gdphi} then yields for the exact $g$-factor ratios
\begin{align}
	g\big(\bigkett{\overline{X}}\big)=\sqrt{3}\, g\big(\bigkett{\overline{a}}\big) \,.
\label{gnewPotts}
\end{align}
in agreement with the CFT analysis of the three-state Potts critical theory \cite{Saleur1989,Affleck1998}, where $g_{\rm new}=\sqrt{3}\,g_{\overline{a}}$. One can generalize \eqref{not0state} and \eqref{gnewPotts} in the obvious way to the $N$-state clock models, with the likely outcome that the analogous states will correspond to CBCs in the continuum $\mathbb{Z}_N$ parafermion theories.

While we can construct defect operators that map between $\kett{a}$ and $\bigkett{\overline{a}}$, such defects are not topological in the three-state Potts model, and so are not very useful. However,  lattice analogs of CFTs of course are not unique. Indeed, we have already discussed in section \ref{sec:AFM} how the antiferromagnetic integrable critical points in the ABF models also yield the parafermion CFTs in the continuum limit. The $\mathbb{Z}_k$ symmetry is generated by a lattice translation by two sites, while translation by $2k$ sites corresponds to translation symmetry in the continuum limit. The factors of two arise because translation by a single site merely maps between the models $M_0$ and ${M}_1$. 

With this identification, we can recover ratios of $g$-factors for the parafermion theories distinct from \eqref{gfixedfreePotts} and \eqref{gnewPotts}. Namely, from \eqref{gdphi} and \eqref{eq:def_QD} we have 
\begin{align}
\frac{g\big(\mcd_b|B\rangle\big)}{g\big(\ket{B}\big)} = d_b =\frac{\sin{\frac{(2b+1)\pi}{k+2}}}{\sin\frac{\pi}{k+2}}
\label{ggPotts}
\end{align}
for any boundary state $\ket{B}$. This ratio is agreement with the quantum dimensions of the primary fields in the parafermion CFT. Indeed, all such fields are found by taking a field of the $SU(2)_k$ Wess-Zumino-Witten CFT and modding out its $U(1)$ component, leaving the quantum dimension unchanged. Since the fusion algebra of the $\mathcal{A}_{k+1}$ category used to build the ABF models is the same as that of $SU(2)_k$, the quantum dimensions in the parafermion CFT are all given by \eqref{ggPotts} for some $b$. Note that in general, $\sqrt{N}$ does not appear in this list, confirming our earlier assertion that free boundary conditions are not associated with any primary field in the CFT. 

By considering two distinct lattice models, we thus have obtained all the ratios of $g$-factors corresponding to primary fields in the parafermion CFT, along with the free boundary condition and its dual. We can understand a little more about the explicit boundary states in 
the antiferromagnetic ABF models.
The state $\kett{0}_{\rm AF}$ defined in \eqref{AFgs} is the ground state of the completely staggered model, where the $\mathbb{Z}_k$ symmetry is explicitly broken, and a natural candidate for a state becoming a Cardy state in the conformal limit.
In this state, the heights obey $h_j=h_{j+2k}$, so that they correspond to a translation-invariant state in the continuum limit.
Moreover, this state is part of a multiplet of $k$ states, differing by translation by two sites.
Reinterpreting these states as boundary conditions, they provide the simplest candidates for the antiferromagnetic ABF analog of fixed boundary conditions.
Define $\mct$ as the translation operator by one site (to the left).
Its matrix elements are
\begin{align}
	{\mct}_{\{h\},\{h'\}} = \prod_{j=1}^{L} \delta_{h_j h'_{j+1}} \,,
\label{transdef}
\end{align}
we thus conjecture that in the continuum limit
\begin{align}
\mct^{2a} \kett{0}_{\rm AF} \to \ket{a}_{\rm CFT}^{} \,.
\label{Pottsconj1}
\end{align}
Further support comes from examining $\kett{1}_{\rm AF}=\mcd_1\kett{0}_{\rm AF}$, given explicitly in \eqref{AFgs1} for the $k=3$ case. For any $k$, it remains a product state, repeating every $2k$ sites. Thus it also is rather simple, suggesting it is also a conformal boundary condition. The not-$a$ states also are product states, and so we conjecture that in the continuum limit 
\begin{align}
\mct^{2a} \mcd_1\kett{0}_{\rm AF} \to \ket{\overline{a}}_{\rm CFT}^{} \,.
\label{Pottsconj2}
\end{align}
for $k=3$. The conjectures (\ref{Pottsconj1}, \ref{Pottsconj2}) are consistent with the ratio of $g$-factors known from the 3-state Potts CFT: $g_{\overline{a}}=\phi\,g_a$, where $\phi=2\cos\tfrac{\pi}{5} = (1+\sqrt{5})/2$ is the golden mean.

\subsubsection{Minimal models}
\label{sec:gmin}
 
Here we generalize the Ising results to find lattice analogs of all conformal boundary conditions in the unitary minimal models of CFT, the ``$A$-series'' \cite{BPZ:CFT:84,DiFrancesco1997}.
Our results are in harmony with the results of \cite{Behrend1998,Behrend2001,Chui2003}, but are obtained more directly and without exploiting integrability.

The connection between lattice models and minimal CFTs is more straightforward than in the parafermion models, as here the lattice category is a subcategory of the CFT category.
The CFTs in the $A$-series  
have central charge $c=1 - \frac{6}{(k+1)(k+2)}$ for integer $k\ge 2$.  The fusion category describing the topological properties of its fields can be written as $\mathcal{A}_k \times \mathcal{A}_{k+1} / \mathbb{Z}_2$.
The standard labeling of the CFT primary fields $\Phi_{r,s}$ is by a pair of integers $(r,s)$ in the range $1 \leq r \leq k$ and $1 \leq s \leq k+1$, and relates to ours by $r=2a+1$ for $a \in \mathcal{A}_{k}$, and $s = 2b+1$ for $b \in \mathcal{A}_{k+1}$. Modding out by the $\mathbb{Z}_2$ identifies the 
label $(r,s)$ with $(k+1-r,k+2-s)$, so there are only $\tfrac{k(k+1)}{2}$ primary fields in the $A$-series CFT.
The fields $\Phi_{2a+1,1}$ and $\Phi_{1,2b+1}$ obey the fusion rules of $\mathcal{A}_{k}$ and $\mathcal{A}_{k+1}$ respectively, and $\Phi_{r,s}$ (as an object in the fusion category) can be written as $\Phi_{r,1} \otimes \Phi_{1,s}$. The quantum dimensions of the primary fields are then simply the product of those in the two categories:
\begin{align}
	d_{r,s} = d_{\frac{r-1}2}^{(k)}d_{\frac{s-1}2}^{(k+1)} = \frac{ \sin\frac{r\pi}{k+1} \sin\frac{s\pi}{k+2} }{ \sin\frac{\pi}{k+1} \sin\frac{\pi}{k+2} } \,,
\label{drsdef}
\end{align}
where the superscripts label the category.
A thorough treatment of CBCs in this $A$-series was given in \cite{Behrend1999}. 
There are no ``extra'' boundary conditions here like in Potts, so we need consider only CBCs obeying \eqref{ggdd}.
All their $g$-factors are therefore given by \eqref{ggdd} and \eqref{drsdef} as
\begin{align}
	g_{r,s}=\frac{d_{r,s}}{\sqrt{\mathscr{D}_k}}\,,\qquad\quad
	\mathscr{D}_{k} = \frac{\sqrt{(k+1)(k+2)}}{\sqrt{8} \, \sin\frac{\pi}{k+1}\sin\frac{\pi}{k+2}} \,.
\label{gmin}
\end{align}

As with the parafermions, we analyze two distinct lattice models in the universality class of each $A$-series minimal model. One is the ferromagnetic $\mathcal{A}_{k+1}$ ABF model discussed at length above.
The other is the dilute-$\mathcal{A}_k$ height model \cite{Warnaar1992}.
The latter is described in our setup by  using the $\mathcal{A}_k$ category with $\upsilon=0\oplus \frac{1}{2}$, as opposed to the ABF $\upsilon=\frac12$.
Since $\upsilon$ is not a simple object, one must sum over configurations with either $0$ or $\frac12$ on each edge of $G$ (with an even number of each at each vertex), as displayed in \eqref{diluteloops}.
Using the shadow-world construction described in section \ref{sec:heights} gives a lattice height model in the same fashion as the ABF models, but here adjacent heights can be the same as well as differing by $\pm \frac12$.  

A special choice of weights in the dilute-$\mathcal{A}_k$ model gives a critical point described in the continuum limit as the same CFT coming from the ABF model built on the $\mathcal{A}_{k+1}$ category \cite{Nienhuis1990,Warnaar1992}.
Since a minimal CFT involves both $\mathcal{A}_k$ and $\mathcal{A}_{k+1}$ categories, we have the pleasant feature that different aspects of this CFT are described more naturally by the two different lattice models.
The simplest fixed boundary states in the dilute-$\mathcal{A}_{k}$ models are given by the completely ordered states
\begin{align}
\kett{a}_{\rm dilute} \equiv \ket{aaaa\cdots} = \mcd_a \kett{0}_{\rm dilute}
\label{totallyfixed}
\end{align}
for $a=0, \frac12,\dots,\frac{k-1}{2}$.
The ratios of $g$ factors for these states therefore obey
\begin{align}
\frac{g\big(\kett{a}_{\rm dilute})}{g\big(\kett{a'}_{\rm dilute})} = \frac{g\big(\kett{a}_{\rm dilute})/g\big(\kett{0}_{\rm dilute})}{g\big(\kett{a'}_{\rm dilute})/g\big(\kett{0}_{\rm dilute})} =\frac{d_a^{(k)}}{d_{a'}^{(k)}}  \,,
\label{diluteratio}
\end{align}
using \eqref{gdphi} as usual.
For the ABF models, the simplest boundary state is the completely ordered one $\kett{0} = \Ket{0\frac12 0\frac12\cdots}$.
Other nice ones $\kett{b}=\mcd_b\kett{0}$ defined by acting with the defect creation operators, and are given explicitly in various forms in \eqref{staggeredgs}, \eqref{fusiontreebasis} and \eqref{csgscupbasis}. For these, we have 
\begin{align}
\frac{g\big(\kett{b}\big)}{g\big(\kett{b'}\big)} = \frac{g\big(\kett{b})/g\big(\kett{0}\big)}{g\big(\kett{b'}\big)/g\big(\kett{0}\big)} =\frac{d_b^{(k+1)}}{d_{b'}^{(k+1)}} 
\label{ABFratio}
\end{align}
for $b,b'=0, \frac12,\dots,\frac{k}{2}$.
In \eqref{diluteratio} and \eqref{ABFratio} we indeed recover exact ratios of $g$-factors arising from the two distinct categories arising in the CFT.

To give intuition into how to identify such boundary states with CBCs, we analyze the Fibonacci case $k=3$ in detail. The best intuition into the corresponding CFT comes from a two-dimensional classical model that can {\em not} be written in terms of a fusion category. The Blume-Capel model is an Ising model with vacancies \cite{Blume1971}, so that each site is occupied by $+$, $-$ or is left empty. In addition to the usual Ising nearest-neighbor coupling, it includes a fugacity controlling the vacancy probability. The latter is an irrelevant perturbation of the Ising critical point, so a small fugacity preserves the critical phase transition between order and disorder. However, at large fugacity, vacancies dominate the disordered phase and make the transition first order. The critical and first-order lines meet at a tricritical point,  where all phases all coexist. The continuum limit of this Ising tricritical point is described by the $k=3$ minimal CFT with $c=7/10$. The phase diagram is nicely given by the Landau-Ginzburg theory. The tricritical point has a pure $\phi^6$ potential, while the potential along the first-order transition line has three degenerate minima corresponding to the two ordered and the one (vacancy-dominated) disordered states.

The three states $h=0,\frac12,1$ in the dilute-$\mathcal{A}_3$ model are naturally identified with  $+$, vacancy, and $-$ in the Blume-Capel model respectively, consistent with the $h\to 1-h$ symmetry. The main distinction between the two is that heights $0$ and $1$ are not allowed to be adjacent, as $\upsilon=0\oplus \tfrac12$. With appropriate tuning of Boltzmann weights it indeed has a critical point in the $c=7/10$ universality class \cite{Warnaar1992}.  The first-order transition line corresponds to coexistence among the corresponding ordered states $\kett{0}_{\rm dilute}$, $\bigkett{\tfrac12}_{\rm dilute}$, and
$\kett{1}_{\rm dilute}$.

The $\mathcal{A}_4$ ABF model is not so obviously similar to the Blume-Capel model. It has four possible heights with the rule that adjacent heights must differ by $\pm\tfrac12$. A useful piece of intuition comes from the fact that deforming the critical $\mathcal{A}_4$ ABF model along the first-order transition line preserves the integrability \cite{Andrews1984,Fendley2004}. The three coexisting ground states are the staggered ones: \cite{Huse1984}
\begin{align}
	\kett{0} = \Ket{0\,\tfrac{1}{2}\,0\,\tfrac{1}{2}\cdots} \,,\qquad \Ket{1\,\tfrac{1}{2}\,1\,\tfrac{1}{2}\cdots} \,,\qquad
	\mct\bigkett{\tfrac{3}{2}} = \mct\mcd_{\frac{3}{2}}\kett{0}=
	\Ket{1\,\tfrac{3}{2}\,1\tfrac{3}{2}\cdots} \,.
\label{firstordergs}
\end{align}
The translation operator $\mct$ is to preserve integer and half-integer heights on the sublattices. The first and last states were already encountered in e.g.\ \eqref{staggeredgs} and \eqref{ABFratio}.  Comparing \eqref{firstordergs} with the ground states in Blume-Capel or dilute-$\mathcal{A}_3$ shows that the analogs of the states $+$, vacancy and $-$ in the former are given by the nearest-neighbor pairs $0\tfrac12$, $1\tfrac12$ and $1\tfrac32$ in the latter.

There are six primary fields and hence six conformal boundary conditions in the tricritical Ising CFT. We can find lattice analogs of all of them in the $\mathcal{A}_4$ ABF model. The defect operators $\mcd_{\frac12}$ and $\mcd_1$ do not mix the states in \eqref{firstordergs}, but rather generate three more states
\begin{align}
	\mct\bigkett{\tfrac12} = \mct\mcd_{\frac12}\kett{0} \,,
	\qquad\mcd_1\Ket{1\,\tfrac{1}{2}\,1\,\tfrac{1}{2}\cdots} \,,\qquad \kett{1}=\mcd_1\kett{0} \,.
\label{Fibstates}
\end{align}
There are only six instead of twelve distinct states because $\mcd_{\frac32}\mcd_b=\mcd_{\frac32-b}$ here. Since $d_b\ge 1$ for all $b$, acting with a defect creation operator cannot decrease a state's $g$-factor.
We then identify those in \eqref{firstordergs} with the Cardy states $\Ket{\Phi_{r,1}}$ in the tricritical Ising CFT with $r=1,2,3$ respectively.
The remainder are identified by noting that 
for $k=3$ (c.f.\  \eqref{ABFratio})
\begin{align}
\frac{g\big(\mcd_{\frac12}\kett{B}\big)}{g\big(\kett{B}\big)} =
\frac{g\big(\mcd_1\kett{B}\big)}{g\big(\kett{B}\big)} =  \frac{1+\sqrt{5}}{2}= \frac{g_{r,2}}{g_{r,1}} \,,
\end{align}
strongly suggesting that each of the three states in \eqref{Fibstates} become the Cardy states $\Ket{\Phi_{r,2}}$. 

Going back to the dilute-$\mathcal{A}_3$ model, we already noted the three ordered states $\kett{0}_{\rm dilute}$, $\bigkett{\tfrac12}_{\rm dilute}$, and $\kett{1}_{\rm dilute}$ are analogous to the three in \eqref{firstordergs}, and so give a distinct lattice realization of the Cardy states $\Ket{\Phi_{r,1}}$.
As opposed to ABF, these states are related by defect-creation operators, and so we obtain from \eqref{diluteratio}
\begin{align}
\frac{g\big(\bigkett{\tfrac12}_{\rm dilute}\big)}{g\big(\kett{0}_{\rm dilute}\big)} =
\frac{ g\big(\bigkett{\tfrac12}_{\rm dilute}\big)}{g\big(\kett{1}_{\rm dilute}\big)}  
= \sqrt{2} = \frac{g_{2,1}}{g_{1,1}}=  \frac{g_{2,1}}{g_{3,1}} \,.
\end{align}
Thus we indeed obtain distinct exact $g$-factor ratios in the two lattice models.
Although the defects in the dilute-$\mathcal{A}_3$ model do not relate these to a state giving $\Ket{\Phi_{1,2}}$, the analogy with the ABF model suggests that it is a not-$0$ state.
Arguments using the Blume-Capel model and flows between conformal field theories \cite{Chim1995,Affleck2000} suggest that it is an equal-amplitude sum over all states not including height $0$ just like those in the Potts model. (Note, however, the analogous ABF state is a sum, the amplitudes are not equal.) Thus defining
\begin{align}
	\bigkett{\overline{0}}_{\rm dilute} \equiv \Ket{\overline{0}\,\overline{0}\,\overline{0}\cdots}  \,,\qquad    \Ket{\overline{0}} \equiv \tfrac{1}{\sqrt{2}}\Big( \Ket{\tfrac12}+\ket{1}\Big)\,,
\label{not0state2}
\end{align}
we conjecture that $\bigkett{\overline{0}}_{\rm dilute}$ becomes the Cardy state $\Ket{\Phi_{1,2}}$ in the continuum.
Likewise, $\bigkett{\overline{1}}_{\rm dilute}$ will become $\Ket{\Phi_{3,2}}$.
We then can identify $\mcd_{\frac12}\bigkett{\overline{0}}_{\rm dilute}=
\mcd_{\frac12}\bigkett{\overline{1}}_{\rm dilute}$ with the Cardy state $\Ket{\Phi_{2,2}}$, yielding the correct $g$-factor ratio $\sqrt{2}$.
These identifications agree with those made using the Blume-Capel model in \cite{Chim1995,Affleck2000}, and integrability in \cite{Behrend2001}. 

The upshot is that we conjectured lattice analogs of the Cardy states $\Ket{\Phi_{r,1}}$ in the ABF $\mathcal{A}_4$ model, and then acted with the defect creation operators to give the remaining  $\Ket{\Phi_{r,1}}$ states.
In the dilute-$\mathcal{A}_3$ model we conjectured lattice versions of the $\Ket{\Phi_{1,s}}$, and used defects to find the rest.
Any time two states are related by a $\mcd_a$, the corresponding lattice $g$-factor ratio is exact. The generalization to all $k$ is now apparent. In all the ferromagnetic ABF models, a first-order line with analogous degeneracies between $k$ states \cite{Andrews1984,Huse1984} suggests that we conjecture 
\begin{align}
	\mct^{r+s}\mcd_{\tfrac{s-1}{2}} \Ket{\tfrac{r-1}{2}\,\tfrac{r}{2}\,\tfrac{r-1}{2}\,\tfrac{r}{2}\,\cdots} \to \Ket{\Phi_{r,s}}_{\rm CFT}^{} \,.
\label{ABFconj3}
\end{align}
where the translation operator $\mct$ from \eqref{transdef} keeps even and odd heights on the appropriate sublattices, and so preserves the models $M_0$ and ${M}_1$.
This correspondence is consistent with the fact that $\mcd_{\frac{k}2}\kett{0} = \Ket{\frac{k}{2}\frac{k-1}{2}\frac{k}{2}\frac{k-1}{2}\cdots} = \bigkett{\frac{k}{2}}$ and that $\Phi_{r,s}$ and $\Phi_{k+1-r,k+2-s}$ are the same field.
It also gives the correct lattice $g$ factor ratios \eqref{gmin} for any $r$, $s$ and $s'$. In the dilute-$\mathcal{A}_k$ models, we have 
\begin{align}
	\mcd_{\frac{r-1}2}\kett{0}_{\rm dilute}=
	\bigkett{\tfrac{r-1}{2}}_{\rm dilute} \to\ \Ket{\Phi_{r,1}}_{\rm CFT}^{}
\label{diluteconj1}
\end{align}
for $r=1,\dots,k$.
This conjecture gives the correct ratios $g_{r,1}/g_{r',1}$ for any $r$ and $r'$.
Between the two lattice models, we then have exact lattice computations for all ratios of $g$ factors.

Various generalizations of these boundary states to higher-spin $\upsilon$ are straightforward to conjecture, as results from integrability \cite{Date:1987} give strong hints.  The same goes for the BMW models \cite{Jimbo:1987} and dilute BMW models \cite{Grimm1994}. It would be interesting to see if lattice analogs of all the boundary states could be conjectured as for the minimal CFTs.

\section{Twists and tubes}
\label{sec:twists}

Topological defects can be used to define {\em twisted boundary conditions} where the partition function is independent of the location of the twist. 
Here we introduce the rich structure describing such partition functions and compute universal quantities, the eigenvalues of the Dehn twist. When the continuum limit is a CFT, these yield exact results for conformal spins.

\subsection{Twisted boundary conditions}

\begin{figure}[hb] \centerline{%
 \xymatrix @!0 @M=2mm @R=30mm{
				&\HilbTwisted \\
				&\HilbFromZTwisted \ar[u]		
	}}
\caption{Ending a defect line in the boundary defines a twisted vector space $\mathcal{V}^{(\varphi)}$. }
\label{fig:PartitionCutTwisted}
\end{figure}
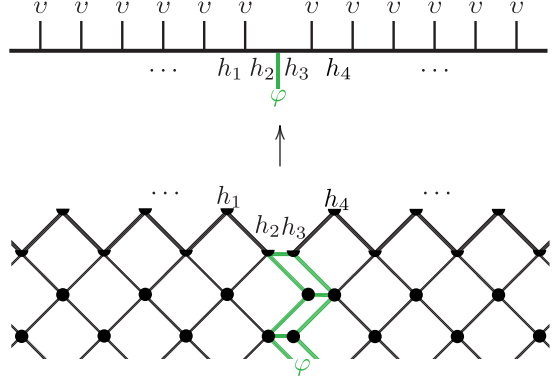

We consider twisted boundary conditions given by defect lines stretching from one end of a cylinder to the other, and so terminating on the boundaries. 
The corresponding fusion tree $\mathcal{V}^{(\varphi)}$ has an extra line coming from the defect line running into the bulk of the system, as illustrated in fig.~\ref{fig:PartitionCutTwisted}.
The heights on the boundary adjacent to the defect line must have a non-zero fusion product with $\varphi$.  
The corresponding square-lattice transfer matrix $T^{(\varphi)}$ acts downwards in figure \ref{fig:PartitionCutTwisted}, perpendicularly to the defect line, and is said to have a {\em twisted} boundary condition. 
A famous example is the antiperiodic boundary condition in the Ising model, which play a crucial role in the mapping to free fermions \cite{Lieb61}. 
Examples with $d_\varphi>1$ have been found by exploiting integrability \cite{Chui2001,Chui2002,Chui2003}, for example the $(1,s)$ defects in the ferromagnetic ABF models.

Partition function identities like (\ref{annulusfusion}) generalize to twisted boundary conditions. 
The simplest relation is given by inserting a closed loop.
The partition functions with and without the loop are related by \eqref{ZdZ} as always. 
We stretch the loop around the periodic direction, and then form a defect line running across the cylinder by fusing opposite sides of the loop together using \eqref{F0c} and gluing to the boundaries. In pictures,
\begin{align*}
d_a & \substack{\ket{A} \\ \annulusx \\ \ket{B} } 
= \substack{\ket{A}  \\ \annulusaloop  \\ \ket{B} }= \substack{ \ket{A}  \\ \annulusaloopprime \\  \ket{B}} =\sum_{\varphi} N_{aa}^{\varphi}\sqrt{\frac{d_\varphi}{d_a^2}}\,  \substack{ \ket{A} \\ \annulusaH \\ \ket{B} }
=\sum_{\varphi} N_{aa}^{\varphi}\sqrt{\frac{d_\varphi}{d_a^2}}\; \substack{ \mathcal{D}_a^{(\varphi)}|A\rangle \\ \annulusphi \\  \mathcal{D}_a^{(\varphi)} \ket{B}}\quad
\end{align*}
where have denoted by $\mcd_a^{(\varphi)}$ the defect creation operator acting on $\mathcal{V}^{(\varphi)}$. In an equation,
\begin{align}
	Z\big(\ket{A},\ket{B}\big) =\frac{1}{d_a} Z\Big(\mcd_a|A\rangle,\mcd_a|B\rangle\Big) + 
	\frac{1}{d_a}\sum_{\varphi\ne 0} N^\varphi_{aa} \sqrt{d_\varphi}\, Z^{(\varphi)}\mkern-2mu \Big(\mcd_a^{(\varphi)}|A\rangle,\mcd_a^{(\varphi)}|B\rangle\Big) \,.
\label{Zstretch}
\end{align}
We can generalize the defect creation operator further via
\begin{align}
\substack{  \ket{\mathcal{B}}\\ \bca}\quad \longrightarrow\quad \substack{ \ket{\mathcal{B}}\\ \bcb}\quad \equiv\quad  \substack{ \mathcal{D}^{(B)}_{RG} \ket{\mathcal{B}}\\ \bcc}
\end{align}
For $G\ne B$, $\mathcal{D}^{(B)}_{RG}$ resembles the boundary-condition-changing operators used in boundary CFT \cite{Cardy1989}. 
The operators here, however, are topological in that the triangle defects and hence the regions of distinct boundary conditions can be moved around at will without changing the partition function.

A more general identity relates twisted partition functions to untwisted ones. We reverse the logic leading to \eqref{Zstretch}, and start with boundary states $\mcd_a^{(\varphi)}|A\rangle$ and $\mcd_b^{(\varphi)}|B\rangle$. Pulling the defects off the boundary and doing an $F$-move leaves a line wrapped around the cycle with a bubble in it. The bubble can be removed using \eqref{bubbleremoval}, giving
\[
\substack{\mathcal{D}_a^{(\varphi)}|A\rangle \\ \annulusphi \\ \mathcal{D}_b^{(\varphi)} |B\rangle}=
\substack{\ket{A} \\ \AnnulusHa \\ \ket{B} }= \sum_{c} \sqrt{d_ad_bd_\varphi}\,
\Msixj{a& a& \varphi}{ b&b & c} \substack{ \ket{A}\\  \annulusz \\ \ket{B}} \,.
\]
For $\varphi=0$, this relation reduces to \eqref{annulusfusion}.
The partition functions therefore obey
\begin{align}
	Z^{(\varphi)}\mkern-2mu \Big(\mcd_a^{(\varphi)}|A\rangle,\mcd_b^{(\varphi)}|B\rangle\Big)
		= \sum_c \sqrt{d_ad_bd_\varphi}\, \Msixj{a& a& \varphi}{ b&b & c} Z\Big(\mcd_c|A\rangle,|B\rangle\Big) \,.
	\label{defectacross}
\end{align}
Using \eqref{defectacross} allows these twisted cylinder partition functions of this from to be expressed as sums over those without the twist.
As an example we apply \eqref{defectacross} to Ising with a spin-flip defect around the cylinder to get
\begin{align}
	Z^{(\psi)}\mkern-2mu \Big(\kett{\rm free},\kett{\rm free}\Big) = Z\big(\kett{+},\,\kett{+}\big) - Z\big(\kett{+},\,\kett{-}\big)
\end{align}
where $\kett{+}$ and $\kett{-}$ are fixed-up and fixed-down boundary conditions, $\kett{\rm free}$ are free boundary conditions, and we used the tetrahedral symbols \eqref{Isingtetra}. 
This twisted partition function is the difference of the same two fixed ones that give the untwisted one, \eqref{Zff} with $N=2$, generalizing the CFT expression \cite{Cardy1989} to any couplings. The extension to clock models and the spin-shift defects is fairly obvious, with the minus sign replaced by an $N$th roots of unity.

One important feature of twisted boundary conditions is that they do not break translation invariance. 
For twisted boundary conditions one needs to follow a translation by one site with a local unitary transformation in order to place the defect back into its original position.
This unitary transformation is given by an $F$-move. 
In the absence of the defect, translation is defined by \eqref{transdef}, so that $T\mct=\mct T$ when the Boltzmann weights are independent of position.
In the presence of a defect $\varphi$ between sites $j$ and $j+1$ as in figure \ref{fig:PartitionCutTwisted}, the matrix elements of the modified single site translation operator are given by
\begin{align}
	\big(\mct^{(\varphi)}\big)_{\{h\},\{h'\}} =  \sqrt{d_{h^{}_j} d_{h'_{j+1}}} 
	\Msixj{\varphi & h_j & h_{j+1}}{\upsilon & h'_{j+1} & h'_{j}} \;  \prod_{l\ne j} \delta_{h^{}_l h'_{l+1}}
\label{transmod}
\end{align}
so that as long as $A_Q(\chi)$ is independent of $Q$,
\begin{align}
T^{(\varphi)}\mct^{(\varphi)} =\mct^{(\varphi)} T^{(\varphi)}
\label{TTTT}
\end{align}
acting on $\mathcal{V}^{(\varphi)}$. 
This explicit form of the translation operator is very useful for doing momentum-resolved exact diagonalization.

\subsection{Scaling dimensions from Dehn twists}
\label{sec:twistedbc}

Twisted boundary conditions let us compute another set of universal quantities,  the eigenvalues of ``Dehn twists'' in the presence of a defect. Prosaically, a Dehn twist $\modT$ amounts to cutting open the cylinder around a cycle, twisting one side of the cut by $2\pi$, and then gluing the cylinder back together. 
In the presence of twisted boundary conditions it can mix different defect configurations. 
For example, the Dehn twist $\modT^{(\varphi)}$ in the presence of a defect $\varphi$ perpendicular to the cut gives
\begin{align}
\modT^{(\varphi)}: \; \TxGG\quad \rightarrow\quad \TGGx \,,
\label{Tphi}
\end{align}
in essence inserting a horizontal defect line.  

In a translation-invariant system, $\modT$ amounts to acting with the translation operator repeatedly to move the heights back to their original positions.  Without the twist, we have $\modT = (\mct)^L=1$ on the lattice, yielding the usual momentum quantization $e^{ikL}=1$ for the eigenvalues $e^{ik}$ of $\mct$. With twisted boundary conditions, the modified translation operator \eqref{transmod} commutes with the transfer matrix as displayed in \eqref{TTTT}. The translation operator still can be diagonalized in terms of its momentum eigenvalues, but the momentum eigenvalues and hence those of $\modT^{(\varphi)}$ are shifted. 

In a CFT, standard arguments relate the eigenvalues of the Dehn twist in the presence of twisted boundary conditions to the {\em conformal spin} $s_\varphi$ of a chiral operator.
Heuristically, one can think of this operator as creating the defect at the bottom of the cylinder, and then annihilating it at the top. In a CFT,  the energy difference between an excited state and the ground state is proportional to the sum of the right and left scaling dimensions $(\Delta_\varphi,\,\overline{\Delta}_\varphi)$ of the operator $\varphi$ creating the state: 
\begin{align}
E^{(\varphi)} - E_0 = \frac{2\pi}{L}\Big( \Delta_\varphi +\overline{\Delta}_\varphi\Big)  
\end{align}
where the fermi velocity (and $\hbar$) are set to be 1. 
In a gapless system, the right and left momenta are proportional to the right and left energies, so that 
\begin{align}
k_\varphi = \frac{2\pi}{L}\Big( \Delta_\varphi - \overline{\Delta}_\varphi\Big) \equiv \frac{2\pi}{L} s_\varphi  \,.
\label{kDelta}
\end{align}
assuming the ground state is chirally invariant.  We also identified the length in the continuum with the number of sites $L$ of the lattice, an assumption we discuss further below. 

When the continuum limit of a lattice model becomes a CFT, the Dehn-twist eigenvalues must be identical, as they are independent of system size. These eigenvalues thus provide another set of universal quantities computable exactly on the lattice. Letting $e^{i2\pi t_\varphi}$ be the eigenvalue of $\modT^{(\varphi)}$, using \eqref{kDelta} relates it to the conformal spin:
\begin{align}
s_\varphi = \frac{n}{2} + t_\varphi
\label{confspin}
\end{align}
for some integer $n$.
The reason for the half-integer ambiguity in \eqref{confspin} is explained in Part~I: When $\varphi$ is odd (as with the duality defect $X$ in parafermion models), one must act with $\modT^{(\varphi)}$ twice to leave the model invariant. When $\varphi$ is even or there is no grading, the ambiguity is up to an integer. Another subtlety is described in the discussion around \eqref{confspin2} below. 

In Part~I, we found $s_\sigma=\tfrac{1}{16}$ in Ising by explicit analysis of $(\modT^{(\sigma)})^2$. A similar computation gives the shifts for Fibonacci model in the presence of the $\tau$ defect, i.e.\ spin 1 in the ABF model at $k=3$. A basis of non-trivial $\tau$-defect configurations with twisted boundary conditions is 
\begin{align}
\label{FibTube}
\mathds{1}^{(\tau)} =  \TxGG \,, \qquad \quad \modT^{(\tau)} =   \TGGx \,, \qquad \quad \psi^{(\tau)} = \TGGG \,.
\end{align}
All other $\tau$-defect configurations with twisted boundary conditions can be related to these by doing $F$-moves. 
We take the Hilbert space in the horizontal direction so that multiplication of operators amounts to stacking the right-hand side of \eqref{Tphi} in the vertical direction. Using the $F$-moves \eqref{defectFib} to simplify yields then 
the non-trivial identities
\begin{align}
\big(\modT^{(\tau)}\big)^2&=\frac{1}{\phi} \mathds{1}^{(\tau)} + \frac{1}{\sqrt{\phi}}\psi^{(\tau)} \,,\cr
\modT^{(\tau)}\, \psi^{(\tau)} &= \frac{1}{\sqrt{\phi}}\mathds{1}^{(\tau)} - \frac{1}{\phi} \psi^{(\tau)} \,.
\end{align}
Putting the two together gives
\begin{align}
\label{FibSpins}
\big(\modT^{(\tau)}\big)^3=\mathds{1}^{(\tau)} +\frac{1}{\phi}\modT^{(\tau)}-\frac{1}{\phi}\big(\modT^{(\tau)}\big)^2 \quad\implies\ 
\big(\modT^{(\tau)}-\mathds{1}^{(\tau)}\big)\Big((\modT^{(\tau)})^2+\phi \modT^{(\tau)} +\mathds{1}^{(\tau)}\Big)=0 \,.
\end{align}
The eigenvalues of $\modT^{(\tau)}$ are therefore $1,e^{\pm i 4\pi/5}$. Using \eqref{confspin} gives shifts $t_\varphi=0,\pm\tfrac25 $ up to an integer ambiguity (here the defect implements a self-duality). These shifts do correspond to conformal spins: a dimension $2/5$ operator occurs in the three-state Potts CFT for the antiferromagnetic critical point, and a dimension 3/5 operator appears in the tricritical Ising CFT for  the ferromagnetic. 

The vector space $\mathcal{V^{(\varphi)}}$ can be split into the sectors with these eigenvalues by utilizing the tube category described in the next subsection \eqref{sec:tube}. As the Dehn twist commutes with the transfer matrix, these sectors can be thought of as having a conserved ``flux'', and the tube category allows this idea to be generalized as well. In fact, the tube category yields a general procedure allowing all eigenvalues $t_\varphi$ and hence the shifts to be computed, see Sec.~\ref{sec:tube} below.

We content ourself here with giving another nice example of a Dehn twist, by generalizing the Ising computation of Part~I to all the Potts and clock models. We consider the Dehn twist $\modT^{(X)}$ in the presence of the duality defect, and define 
\begin{align}
\psi_a=\psia
\end{align}
so that $\psi_0^{}=\mathds{1}^{(X)}$. 
Using $F$-moves from \eqref{FPotts} then gives
\begin{align}
\Big(\modT^{(X)}\big)^2&= \frac{1}{\sqrt{N}} \sum_{a=0}^{N-1} \psi_a \,,\qquad\quad
\psi_a\psi_b = \omega^{ab} \psi_{(a+b)\,{\rm mod}\,N} \,.
\label{PottsDehn}
\end{align}

Finding the eigenvalues of $\modT^{(X)}$ is an amusing exercise. The easiest way we found is to rewrite the $\psi_a$ in terms of projection operators (the tube idempotents) using
\begin{align}
\psi_a = \omega^{-a(a-1)/2}\big(\psi_1\big)^a \,,\qquad\quad
\big(\psi_1\big)^N  = (-1)^{(N-1)/2} \psi_0^{} \,,
\end{align}
as follows from \eqref{PottsDehn}. The projection operators for even $N$ are
\begin{align}
\Psi_r = 
\frac{1}{N}\sum_{a=1}^{N} \omega^{ar} \big(\omega^{\frac12}\psi_1\big)^a
\quad\implies\quad \big(\omega^{\frac12}\psi_1\big)^a= \sum_{r=1}^{N} \omega^{-ar} \Psi_r \,.
\label{projdef}
\end{align}
They obey $\Psi_r\Psi_{r'}=\delta_{rr'}\Psi_r$. Rewriting \eqref{PottsDehn} thus gives
\begin{align} 
\Big(\modT^{(X)} \Big)^2 
= \frac{1}{\sqrt{N}}  \sum_{r=0}^{N-1} \Psi_r\; \omega^{\frac12r^2}\,
\sum_{a=0}^{N-1} \omega^{-\frac12 (a+r)^2}\quad\hbox{ for even }N.
\label{Tsqsum}
\end{align}
For odd $N$, the projection operators are as in \eqref{projdef} without the $\omega^{-\frac12}$, and
\begin{align} 
\Big(\modT^{(X)}\Big)^2 = \frac{1}{\sqrt{N}}  \sum_{r=0}^{N-1} \Psi_r\; \omega^{\frac12 (r-\frac12)^2}\,\sum_{a=0}^{N-1} \omega^{-\frac12 (a+r-\frac12)^2} \quad\hbox{ for odd }N.
\label{Tsqsum2}
\end{align}
The sums over $a$  in \eqref{Tsqsum} and \eqref{Tsqsum2} are called quadratic Gauss sums, and are independent of $r$. They can be evaluated by using an analog of quadratic reciprocity \cite{Gelaki2009}, giving
\begin{align*}
\frac{1}{\sqrt{N}}\sum_{a=0}^{N-1} \omega^{-\frac12(a+r)^2} = e^{-i\frac\pi4}\ \hbox{ for even }N \,,\qquad
\frac{1}{\sqrt{N}}\sum_{a=0}^{N-1} \omega^{-\frac12(a+r-\frac12)^2} = e^{-i\frac\pi4}\ \hbox{ for odd }N \,,\qquad
\end{align*}
The square of the Dehn twist is therefore
\begin{align} 
\Big(\modT^{(X)}\Big)^2 = e^{-i\frac\pi4} \sum_{r=0}^{N-1} \alpha_r \Psi_r  \,,\qquad\quad\hbox{with }\alpha_r = 
\begin{cases}
\omega^{\frac{r^2}{2}}\qquad&\hbox{for even }N,\\
\omega^{\frac12 (r-\frac12)^2}\qquad&\hbox{for odd }N.
\end{cases}
\label{Tsq}
\end{align}

The projection operators in \eqref{Tsq} are complete and orthogonal, so the eigenvalues of $\modT^{(X)}$ must be
$\pm\sqrt{\alpha_r}\,e^{-i\pi/8}$. The nicest way to write out the resulting momentum shifts is in terms of an integer $m$ defined by $m = \tfrac{N}{2}-r$ for even $N$ and $m=\tfrac{N-1}{2}-r$ for odd $N$, yielding
\begin{align}
t_m=  \frac{N-1}{16} -\frac{m(N-m)}{4N} + \frac{n}{2} \,.
\label{paradim}
\end{align}
The index $m$ can be restricted to be non-negative and less than $N/2$, as other values result in at most half-integer shifts. Nowhere in this calculation did we fix the values $A_Q(\chi)$, so the values given in \eqref{paradim} constrain any field theory arising in a continuum limit of the $\mathbb{Z}_N$ clock model.

The conformal spins implied by \eqref{paradim} look somewhat unfamiliar in parafermion CFTs describing the continuum limits of the ferromagnetic clock model with $A_Q$ fine-tuned (for $N>3$) to \eqref{paraweights}. The reason is that the CFT spectrum with twisted boundary conditions contains non-integer-spin operators as well as  ``$C$-disorder'' operators  \cite{Zamolodchikov86} with dimensions
\begin{align}
\Delta_p = \overline{\Delta}_p = \frac{N-1}{16} -\frac{p(N-p)}{4(N+2)} \,,
\end{align}
for $p\le \tfrac{N}{2}$ a non-negative integer. The similarity to \eqref{paradim} is striking, but only $t_0$ can be identified with $\Delta_0$, coming from the chiral part of the $C$-disorder operator with $p=0$. Moreover, in any event only for $N=3$ is the critical behavior generically that of the parafermion CFT, as otherwise it is typically governed by a $c=1$ CFT \cite{Nienhuis1984}. 

The other eigenvalues in \eqref{paradim} therefore must arise in CFT from combining left and right sectors in unusual ways. This observation is in accord with the spins seen from extensive numerical simulations of clock chains with twisted boundary conditions \cite{Yu2017}. There is a catch, however. For the $N=3$ ferromagnetic chain described by the three-state Potts CFT in the continuum limit, CFT and numerics (table IX of \cite{Yu2017}) strongly suggest that twisted boundary conditions result in states with spins $\pm \tfrac{1}{8}$ and  $\pm \tfrac{1}{24}$ and half-integer shifts of these. These numbers are in perfect accord with the Dehn twist results \eqref{paradim} using \eqref{confspin}. For the antiferromagnetic $N=3$ case, however, the states in table X of \cite{Yu2017} have conformal spins $\pm \tfrac{1}{16}$ and $\pm \tfrac{1}{48}$, up to {\em quarter}-integer shifts. As our analysis is blind to the couplings, we obtain the same momentum shifts in both cases.  We account for the difference by noting that the unit cell of this antiferromagnet is two sites, so the effective length of the lattice model is $L_{\rm eff}=L/2$. If we use the effective length in the continuum results, the CFT formula \eqref{kDelta} and thus the correspondence to the lattice \eqref{confspin} are modified to
\begin{align} k_\varphi= \frac{2\pi}{L_{\rm eff}} s_\varphi \,,\qquad \implies \qquad
s_\varphi = \frac{L_{\rm eff}}{L} \Big( t_\varphi + \frac{n}{2}\Big) \,.
\label{confspin2}
\end{align}
giving the correct answer for both ferromagnet and antiferromagnet with $N=3$. Comparing the results of \cite{Yu2017} for mixed ferromagnet/antiferromagnets with $N=4$ and $N=5$ to \eqref{paradim} suggests that $L_{\rm eff}=L/4$ in these cases, whereas for $N=6$ no modification is needed.

\subsection{Tube category and the cylinder partition functions}
\label{sec:tube}

We have shown how diagonalizing the Dehn twist (in the basis of cylinder partition functions) allows us to extract universal quantities such as the momentum eigenvalues in the presence of twisted boundary conditions. 
In this section we describe how this can be formalized using the {\em tube category}.
The cylinder partition functions form representations of the tube category,
and the irreducible representations give an explicit definition of the `flux' sectors.

We begin by writing down a basis of topological defect configurations for the cylinder with twisted boundary conditions. 
Ultimately, any cylinder partition function can be reduced to one of the form $(T^{(R)})^N \circ Q_{BPR}^G$.
Where
\begin{align}
	Q_{BPR}^G \; \defineas \;\; \tubeDefect  \,.
	\label{twisted_symmetry}
\end{align}
The operator $Q_{BPR}^G$ provides a map from $\mathcal{V}^{(B)}$ to $\mathcal{V}^{(R)}$. 
It generalizes the dualities in the presence of a twisted boundary condition. 
When $B=R$, this operator commutes with the transfer matrices, 
while in general it provides linear relations between transfer matrices with different duality twisted boundary conditions. 
That is when $B\neq R$ in \eqref{twisted_symmetry} it allows us to understand how duality can ``change'' boundary conditions.
The representation theory then tells us what properties from a given $B$ twisted boundary condition carry over to an $R$ twisted boundary condition.
These facts are a simple consequence of the defect commutation relations \eqref{DefComm1} and \eqref{eq:trivalent_pentagon}:
\begin{align*}
T^{(R)} Q_{BPR}^G &= \TransferSymma
=\TransferSymmb \\
&=\TransferSymmc
 =\TransferSymmd \\
&=\TransferSymme 
=\TransferSymmf \\[7pt]
&=\ Q_{BPR}^G T^{(B)}
\end{align*}
and so
\begin{align}
\big(T^{(R)}\big)^N Q_{BPR}^G = Q_{BPR}^G \big(T^{(B)}\big)^N
\label{TXXT}
\end{align}
for any $N$. 
The relation \eqref{TXXT} applies for any Boltzmann weight built from the category. 
The relations \eqref{FibSpins}, \eqref{Tsq} for $\modT$ thus still apply without imposing translation invariance, so for example the off-critical staggered cases discussed in section \ref{sec:selfduality} still split into sectors with the same eigenvalues under the Dehn twist. The only reason to impose translation invariance is to be able to compare these eigenvalues with the conformal spins, as we did in section \ref{sec:twistedbc}.
Moreover we can use the representation theory of the algebra generated by $Q_{BPR}^G$ to decompose the transfer matrices, and the Hilbert spaces which they act on into sectors.
These sectors are defined by irreducible representations of the algebra generated by $Q_{BPR}^G$.  
Analyzing this action yields a host of linear identities between partition functions and their duals on the cylinder and the torus, generalizing those described for Ising in Part~I.

The algebra generated by $Q_{BPR}^G$ is called the {tube category}: $\tube(\mcc)$~\cite{ocneanu1994}.
The tube category is a $\mathbb{C}$-linear category whose objects are given by labeled circles,
\begin{align}
	\operatorname{obj}\tube(\mcc) = \Big\{\; \TubeObja \quad\raisebox{-1mm}{,}\quad \TubeObjb \quad\raisebox{-1mm}{,}\quad \TubeObja\oplus\TubeObjb \quad\raisebox{-1mm}{,}\quad \dots \;\Big\} \,,
\end{align}
where we have identified the left black dot with the right black dot, and $B, P, G \in \tube(\mcc)$.
Morphisms in the tube category are diagrams on tubes (equivalently cylinders/annuli):
For example,
\begin{align}
	\TubeMor \;,\, \TubeBPRG \; \in \operatorname{mor}\tube(\mcc) \,.
\end{align}
The collection of all possible objects (labeled boundaries) and morphisms (labeled tubes) is the tube category.
In practice, it is helpful to pick a representation of the tube category.
The following ``single strand'' representation is particularly convenient
\begin{align}
\label{rep1}
	\operatorname{Rep}_1 \tube(\mcc) &= \mathbb{C}\Bigg[ \TubeBPRG \Bigg].
\end{align}
Once we pick a representation such as the one above, $\tube(\mcc)$ is a finite-dimensional $C^*$ algebra with multiplication given by stacking the diagrams vertically, and denoted by $\circ$.
The basis in \eqref{rep1} constitutes a representation of $\operatorname{mor}\tube(\mcc)$ as any tube-diagram in $\operatorname{mor}\tube(\mcc)$ can be reduced to one in our representation by fusing together the strands on the boundary, and applying a finite number of $F$-moves. 
Any other representation of the tube category, when viewed as an algebra, will be Morita equivalent to $\operatorname{Rep}_1 \tube(\mcc)$ as algebras.
Labeling basis elements of this representation by $Q$, $Q'$ and $Q''$ gives
\begin{align}
Q \circ Q' = \sum_{Q''} M_{QQ'}^{Q''} Q'' \,,
\label{XXMX}
\end{align}
where $M_{QQ'}^{Q''}$ are the structure constants. 
In the basis given by the $Q_{BPR}^G$ from \eqref{twisted_symmetry}, the structure constants are found by $F$-moves, and are
\begin{align}
	M_{Q_{BPR}^G , Q_{B' P' R'}^{G'}}^{Q_{B''P''R''}^{G''}} &=
	\delta_{RR''}\delta_{B' B''}\delta_{B R'} \sqrt{d_{GG'G''PP'P''B}}
\Msixj{G & P & B}{G' & P' & P''} \Msixj{G & P & R}{P'' & G'' & G'} \Msixj{G'' & P'' & B'}{P' & G' & G} .
\end{align}
In a different basis for the tube category, for example the basis with two strands coming into and out of the tube, the multiplication would have looked quite a bit different. 
The spectrum of the transfer matrix, however, is independent of the choice of basis, as the irreducible representations of the tube category are independent of how it is represented.  For example, a collection of $F$-moves allow us to fuse two strands into one, and so map between the 1-strand basis and the 2-strand basis.

The algebra \eqref{XXMX} can be block diagonalized. The blocks are labeled by irreducible representations, while the diagonal elements of the blocks correspond to the minimal idempotents of the tube category. The collection of all minimal idempotents in a block is referred to as the isomorphism class of a given minimal idempotent.
Off-diagonal elements in each block are isomorphisms between minimal idempotents. 
The sum of minimal idempotents within a given block is called a minimal central idempotent. 
The minimal central idempotents of $\tube(\mcc)$ are in one-to-one correspondence with the simple objects of the Drinfeld center $\mcz(\mcc)$~\cite{Muger2003}.

Here we define a minimal idempotent in a way that is useful for computing them.
The complete set of minimal idempotents is denoted by $E_{R,i}$ and by definition can be expanded in terms of the $Q$ as
\begin{align} 
E_{R,i} = \sum_{P,G}C_{GPR}^{(i)} Q_{RPR}^G \,,
\end{align}
where $C_{GPR}^{(i)} \in \mathbb{C}$.
They are completely characterized by requiring they be orthogonal, minimal and complete.
Orthogonality means that
\begin{align}
E_{R,i} \circ E_{B,j} = \delta_{RB} \delta_{ij} E_{R,i} \,.
\label{idempconda}
\end{align}
The collection of idempotents is minimal if $E_{R,i}\circ \operatorname{mor}\tube(\mcc) \circ E_{B,j}$ is at most one dimensional, i.e.,
\begin{align}
	E_{R,i}\circ \operatorname{mor}\tube(\mcc) \circ E_{B,j} \cong  
\begin{cases} 
\mathbb{C} \quad \text{if $E_{R,i}$ is isomorphic to $E_{B,i}$}\\
0. \quad  \text{otherwise}\\
\end{cases}
\label{idempcondb}
\end{align}
In practice this means that the vector space $\{ E_{R,i}\circ  Q \circ E_{B,j} : Q \in \operatorname{mor}\tube(\mcc) \}$ is one dimensional if $E_{R,i}$ and $E_{B,j}$ are isomorphic (see \eqref{isomorphic} below), and equal to $0$ otherwise. 
If $\dim(E_{R,i}\circ \operatorname{mor}\tube(\mcc) \circ E_{B,j}) >1$ then one or both of the idempotents is not minimal.
Completeness of the idempotents means that for any $Q \in \operatorname{mor}\tube(\mcc)$ there exist morphisms $u_{R,i},v_{R,i} \in \operatorname{mor}\tube(\mcc)$ such that, 
\begin{align} 
Q = \sum_{R,i} u_{R,i} \circ E_{R,i} \circ v_{R,i} \,.
\end{align}
In particular, $\sum_{i}E_{R,i} = \operatorname{id}_R$.
We say minimal idempotents $E$ and $E'$ are isomomorphic when there exist morphisms $U$, $V \in \operatorname{mor}\tube(\mcc)$ such that,
\begin{align}
\label{isomorphic}
E = U \circ V \quad \text{and} \quad E' = V \circ U \,.
\end{align}

We now can make some universal statements about the spectrum of the transfer matrix. 
First, all the minimal idempotents commute with the transfer matrix and with each other. 
Using the above properties it follows that the spectrum of the transfer matrix $T_R$ can be decomposed as
\begin{align} 
	T^{(R)}  = \sum_i E_{R,i}\circ T^{(R)}  \circ  E_{R,i} = \sum_i T_{R,i} \,. 
\label{TETE}
\end{align} 
Correspondingly, the vector space $\mathcal{V}^{(R)}$ on which $T^{(R)}$ acts also can be split.
Letting $\mathcal{V}_{R,i}= \{ \ket{\psi} \in \mathcal{V}^{(R)} \; : \; E_{R,i} \ket{\psi} = \ket{\psi} \} $, completeness of the minimal idempotents guarantees that
\begin{align}
	\mathcal{V}^{(R)} \cong \bigoplus_{i} \mathcal{V}_{R,i} \,.
\end{align}
Every tube, including the minimal idempotents, also commutes with the Dehn twist $\modT^{(R)} = Q_{R  \mathds{1} R}^R$. Thus $ E_{R,i} \modT^{(R)} = \modT^{(R)} E_{R,i}$, and therefore $E_{R,i} \modT^{(R)} E_{R,i} = \modT^{(R)} E_{R,i}$. 
Since the Dehn twist is unitary, we have $E_{R,i} \modT^{(R)} E_{R,i} = \theta_{R,i} E_{R,i} $ for some complex phase $\theta_{R,i}$. 
Hence in the basis of minimal idempotents
\begin{align}
	\modT^{(R)} = \sum_{i} \theta_{R,i}\, \operatorname{id}_{R,i}
\end{align}
where $\operatorname{id}_{R,i}$ is the identity operator on $\mathcal{V}_{R,i}$.

As pointed out above, the isomorphism classes of minimal idempotents of $\tube(\mcc)$ are in one-to-one correspondence with the simple objects of $\mathcal{Z}(\mcc)$.
Thus by simply knowing the fusion category $\mathcal{C}$ described by the topological defects, 
and the Drinfeld center $\mathcal{Z}(\mathcal{C})$ we can compute many properties about the annulus partition functions. 
For example, when $\mathcal{C}$ is a modular tensor category, $\mathcal{Z}(\mathcal{C}) \cong \mathcal{C} \times \mathcal{C}^*$ and so the momentum eigenvalues for object $(x,y) \in \mathcal{C} \times \mathcal{C}^*$ take the form $k= h_x-h_y + n$, where $n \in \mathbb{Z}$ and $\theta_{x/y} = e^{2\pi i h_{x/y}}$ is the topological spin.
Lets illustrate this identification explicitly with the Fibonacci model. 
The diagrams in \eqref{FibTube} are the generators of the Fibonacci tube category in the sector where a $\tau$ strand terminates on the boundaries at the top and bottom of the cylinder. The minimal idempotents of the tube category are eigenstates of $\modT$. 
Because the Fibonacci category is modular, $\mathcal{Z}(\mathcal{C}) \cong \mathcal{C} \times \mathcal{C}^*$ here, and in particular the simple objects are labeled $(\mathds{1}, \mathds{1})$, $(\tau, \mathds{1})$, $(\mathds{1}, \tau)$, and $(\tau, \tau)$.
The topological spins of those simple objects are given by, $1, e^{i 4 \pi /5}, e^{-i 4 \pi /5}$ and $1$ respectively. 
The tube idempotents labeled $(\tau, \mathds{1})$, $(\mathds{1}, \tau)$, and $(\tau, \tau)$ are the irreducible representations of the algebra generated by \eqref{FibTube}, and therefore also give rise to the topological spins computed using \eqref{FibSpins}.  
This correspondence explains the somewhat counterintuitive result that even in the presence of twisted boundary conditions, there remains a sector with no momentum shift in the Fibonacci theory.

We now hammer out the tube category for the case when $\mcc$ is the Fibonacci theory, which we denote $\mathrm{Fib}$.
The tube category can be split up based on its boundary conditions. 
Tubes which have the same boundary condition on top and bottom are called endomorphisms, and the algebra generated by all such endomorphisms is called the endomorphism algebra. 
The Fibonacci fusion category has two endomorphism algebras, one for each simple object:
\begin{align} \begin{split}
	\operatorname{End}(\mathds{1}) &= \mathbb{C} \Bigg[\; \Txxx\;,\ \TGZG \;\Bigg]\ =\ \mathbb{C} \big[\mathds{1},\, \mcd_\tau\big] \,,
\\	\operatorname{End}(\tau) &= \mathbb{C}\Bigg[\; \TxGG\,,\ \TGGx\;,\ \TGGG \;\Bigg]\ =\ \mathbb{C} \Big[\mathds{1}^{(\tau)},\,\modT^{(\tau)},\,\psi^{(\tau)}\Big] \,.
\end{split} \end{align}
Each endomorphism algebra can be decomposed into minimal idempotents independently.
In this case both endomorphism algebras are commutative, and so we have as many minimal idempotents as we do diagrams, i.e., $\operatorname{End}(\mathds{1}) \cong \mathbb{C} \oplus \mathbb{C}$ and $\operatorname{End}(\tau) \cong \mathbb{C} \oplus \mathbb{C}\oplus \mathbb{C}$.
A simple direct computation gives the minimal idempotents for $\operatorname{End}(\mathds{1})$ as
\begin{subequations} \begin{align}
	(\mathds{1},\mathds{1})_{\mathds{1}} &=\frac{1}{{1+ \phi^2}}\, \Txxx \;+\; \frac{\phi}{{1+ \phi^2}}\, \TGZG \;,
\\	(\tau ,\tau)_{\mathds{1}} &=\frac{\phi^2}{{1+ \phi^2}}\, \Txxx \;-\; \frac{\phi}{ {1+ \phi^2}}\, \TGZG \;.
\end{align}
The notation $(a,b)$ is chosen so that the idempotent can be easily identified with the simple object in the Drinfeld center, and the subscript denotes the boundary condition.
Both of these minimal idempotents have eigenvalue $+1$ under the Dehn twist operator $\modT^{(\mathds{1})}$.
The remaining minimal idempotents are given by\footnote{In practice, it is helpful to diagonalize the Dehn twist $\modT^{(G)}$ in the tube basis in order to quickly and efficiently find many of the idempotents.
	For the Fibonacci theory the eigenstates of $\modT^{(\tau)}$ completely determine the minimal idempotents with boundary condition $\tau$.} 
\begin{align}
	(\tau, \mathds{1})_\tau &= \frac{1}{1+\phi^2}\,
		\Bigg(\; \TxGG \;+\; e^{4\pi i/5}\, \TGGx \;+\; \sqrt{\phi}\, e^{-3\pi i/5}\, \TGGG \;\Bigg) \,,
\\	(\mathds{1}, \tau)_\tau &= \frac{1}{1+\phi^2}\,
	\Bigg(\; \TxGG \;+\; e^{-4\pi i/5}\, \TGGx \;+\; \sqrt{\phi}\, e^{3\pi i/5}\, \TGGG \;\Bigg) \,,
		\\	 (\tau, \tau)_\tau & = \frac{1}{1+\phi^2}
		\Bigg( \phi\, \TxGG \;+\; \phi\, \TGGx \;+\; \frac{1}{\sqrt{\phi}}\, \TGGG \;\Bigg) \,.
\end{align} \end{subequations}
The Dehn twist $\modT^{(\tau)}$ eigenvalues are given by $\theta_{(\tau,\mathds{1})_\tau}=\theta_{\tau} = e^{6 \pi i/5}$, $\theta_{(\mathds{1},\tau)_\tau} = \theta_{\tau}^*= e^{-6 \pi i/5}$, and $\theta_{(\tau,\tau)_\mathds{1}} = \theta_\mathds{1} = 1$ respectively.
Notice that the $(\tau, \tau)$ idempotent has support on both $\operatorname{End}(\mathds{1})$ and $\operatorname{End}(\tau)$. 
This reflects the fact that the full tube algebra, which includes all diagrams on the tube is non-Abelian. 
It is a consequence of the fact that some tube diagrams connect tubes with the $\tau$ boundary condition to tubes with the $\mathds{1}$ boundary condition.

The tube category thus lets us reproduce the Dehn-twist results in an elegant fashion. We can go further and use \eqref{isomorphic} to construct isomorphisms mapping between idempotents, i.e., find $U$ and $V$ where 
\begin{align}
	(\tau, \tau)_\mathds{1} = U \circ V \quad\quad \text{and} \quad\quad (\tau, \tau)_\tau = V \circ U \,.
\end{align}
The only remaining tube diagrams are off-diagonal and provide exactly the maps we are looking for: 
\begin{align}
	\operatorname{mor}(\mathds{1} \to \tau ) = \mathbb{C} \Bigg[\; \TTadb \;\Bigg] \,,\qquad\quad
	\operatorname{mor}(\tau \to \mathds{1} ) = \mathbb{C} \Bigg[\; \TTada \;\Bigg] \,.
\end{align}
The notation means the right-hand side is the collection of morphisms in the tube category which take a circle with one marked point labeled by $\mathds{1}$ to a circle with one marked point labeled by $\tau$.
These morphism spaces are one-dimensional and so up to a constant we have, 
\begin{align} 
	U \propto \TTada \quad\quad \text{ and} \quad\quad V \propto \TTadb \;.
\end{align}
An immediate consequence of these isomorphisms is an identity for the partition functions found from from gluing the ends of the cylinder into a torus. Namely, the partition function found from gluing the $(\tau, \tau)_\mathds{1}$ idempotent into a torus is identical to the partition function found from gluing the $(\tau, \tau)_\tau$ idempotent into a torus. 
In a sequel we will revisit the consequences for toroidal partition functions in more detail.



\section{Conclusion}
\label{sec:conclusion}

We have developed a comprehensive approach to finding and exploiting topological defects in two-dimensional classical lattice models, including their quantum-spin-chain limits. This approach requires that the models be defined using a fusion category, with the transfer matrix written in terms of projection operators best defined using fusion diagrams. Such models include some of the best-known models of lattice statistical mechanics, including the Ising, Potts and self-dual eight-vertex models. The category setup allows our knowledge of lattice topological defects to be expanded significantly, making straightforward the derivation of properties such as their behavior under branching and fusing.  Linear identities between partition functions with different defect configurations follow easily.

These identities have a number of elegant applications.
They give a very useful implementation of the venerable Kramers-Wannier duality, allowing us in Part~I to find new results for the even more venerable Ising model. Here we derived the analogous results for the Potts, clock and eight-vertex models, and showed how to generalize duality itself to a far-wider class of models. 
We computed universal quantities directly on the lattice using the defects, including the eigenvalues of Dehn twists and ratios of $g$-factors.
If the limit of the lattice model is a conformal field theory, then these results provide strong and exact constraints on what that CFT may be. 
Another consequence is the existence of exact degeneracies in the ground states and low-lying spectrum of non-critical models. 
Whereas the continuum description of such models is by an integrable gapped quantum field theory, the lattice models we analyze typically are not integrable. Self-duality provides a direct way of understanding why these degeneracies occur.

We showed that the lattice partition functions can be viewed as boundary conditions to the Turaev-Viro-Barrett-Westbury partition function, and gave a natural interpretation of the topological defects as elements in the Drinfeld center $\mcz(\mcc)$.
This was further corroborated by the computation of the Dehn twist eigenvalues which coincided with the topological spins of simple objects in $\mcz(\mcc)$.

Many generalizations await. 
One we have already worked out in detail is the implementation of the partition-function identities on the torus. 
We further find that when the continuum limit is a conformal field theory that we can provide an explicit map between the lattice partition function on the torus and the conformal field theory partition function for an arbitrary topological defect configuration on the lattice.
If the continuum limit is a CFT, we show how an arbitrary topological defect configuration maps onto characters of the Virasoro and extended algebras.

Another is to define (non-local) operators by terminating a topological defect somewhere in the bulk of the system. 
These terminations are naturally labeled by simple objects in the Drinfeld center. 
Indeed, we can view the termination as a boundary with the topology of a circle, and so has a natural action from the tube category.
We can study the behavior of the operator under 2$\pi$ twists, so as to find the lattice version of the topological spin. 
As with the Dehn twist eigenvalues, the lattice values are independent of system size, and so if the continuum limit is a CFT, they must match the conformal spins of the corresponding operators. 
We thus find a host of generalizations of fermion and parafermion operators, beyond the conserved currents described in \cite{Fendley2020}. 
In particular, here it is possible to analyze generally terminations that change the $\upsilon$ used on an edge. 
Moreover, the braiding coming from the Drinfeld center allows the results of  \cite{Fendley2020} to be implemented solely using fusion category data, so that the braiding used there is not required.

Even though our construction gives a host of lattice topological defects and corresponding dualities, it is not exhaustive. 
A very interesting generalization is to construct topological defects separating Morita-equivalent fusion categories.
A pair of fusion categories are Morita equivalent if there exists an invertible bimodule between them. 
Consequently they have the same Drinfeld center and so one expects their lattice partition functions to have similar properties. 
Indeed, one can use the invertible bimodule to construct a lattice topological defect between any two Morita equivalent fusion categories, thus generalizing the dualities studied in this paper.
We note that Morita-equivalent fusion categories may look wildly different, for example $\text{Rep}(G)$ and $\text{Vec}_G$ are Morita equivalent but have different objects, fusion rules, etc.
Consequently, the dualities give precise relations between partition functions which have very different microscopic degrees of freedom.

Another generalization of our construction gives orbifold defects. 
Orbifold defects again relate partition functions with different microscopic degrees of freedom, 
but unlike the Morita-equivalence dualities, orbifold defects are non-invertible.
These defects allow one to extend an ancient result \cite{Fendley1989} relating partition functions of seemingly distinct lattice models via an analog of the orbifold construction of conformal field theory. 
We illustrate the idea by considering one of the old examples, where the ABF models with $k$ even are related to the $D_n$ series of models  \cite{Pasquier1986} with $n=2+\tfrac{k}{2}$.
The $D_n$ Dynkin diagram gives the adjacency graph of the latter, replacing the $\mathcal{A}_{k+1}$ one from \eqref{Akadjacency}. For example, the $D_4$ model has the incidence diagram of the three-state Potts model, associated with the $\mathbb{Z}_3$ Tambara-Yamagami fusion category.
Since the relation of Boltzmann weights of the $D_n$ model with the corresponding $\mathcal{A}_{2n-3}$ model is a linear one \cite{Fendley1989}, it is natural to expect a topological defect separating the region of the two models.
Indeed, with some work the explicit form of the orbifold defects can be extracted from \cite{Fendley1989}.
In the $D_3=\mathcal{A}_3$ Ising case, such a defect is precisely the KW duality we described above. 
However, for higher $n$, our construction requires generalization. 
As with their continuum counterparts \cite{Frohlich2009}, there exists a topological ``orbifold'' defect that implements the lattice orbifold. 
Similarly to the Morita-equivalence defects, these orbifold defects can be parametrized by Frobenius algebra objects in $\mcc$.

\paragraph{Pre-acknowledgments} This work was done almost entirely in 2015. Part I appeared in January 2016, and we gave multiple talks describing the main results, e.g.\ \cite{Aasen2017,Fendley2018}. The only more recent results were a few small applications, mainly the Dehn-twist calculation in the clock models.  Over the five years we (fairly inexplicably) took to write up these results, a number of nice papers have appeared with some overlap. A partial list includes
\cite{Buican2017,freed2018,Vanhove2018,Belletete2018,thorngren2019,Wenjie2019,Chang_2019,lootens2019,Belletete2020}.

\paragraph{Acknowledgments} It is a pleasure to thank Jason Alicea, Kevin Walker, and Dominic Williamson for stimulating discussions.
DA was supported by a postdoctoral fellowship from the Gordon
and Betty Moore Foundation, under the EPiQS initiative, Grant GBMF4304.
PF was supported by EPSRC grants EP/S020527/1 and EP/N01930X (PF), while 
RM was supported by NSF DMR-1848336.

\clearpage
\appendix

\section{\texorpdfstring{$\mathcal{A}_{k+1}$}{A[k+1]} models}
\label{app:TetraSym}

In this section, we describe in detail the $\mathcal{A}_{k+1}$ category, explicitly writing out its tetrahedral symbols. This category underlies the ABF height models, one of the chief examples of this paper.
The simple objects in $\mathcal{A}_{k+1}$ are labelled by $0, \tfrac12, 1, \tfrac32, \dots, \tfrac{k}{2}$.
The quantum dimensions of these objects are
\begin{align}
	d_h = \frac{ \sin\frac{\pi(2h+1)}{k+2} }{ \sin\frac{\pi}{k+2} } \,.
\end{align}
The fusion rules are
\begin{align}
	N^a_{bc} \;=\; \bigg\{ \begin{tabular}{l @{\qquad} l}
		1	&	$a+b \geq c$,\; $b+c \geq a$,\; $c+a \geq b$,\; $a+b+c \in \mathbb{Z}$,\; and $a+b+c \leq k$ \,,
	\\[0.2ex]	0	&	otherwise \,.
	\end{tabular}
\end{align}

\subsection{Tetrahedral symbols}

Let $0 \leq a,b,c,d,e,f \leq \frac{k}{2}$ be elements of the $\mathcal{A}_{k+1}$ fusion category.
One choice of the tetrahedral symbols are~\cite{kauffman1993}
\begin{align}
	\Msixj{a & b & c}{d & e & f}
   \;&=\;
	\Wsixj{a & b & c}{d & e & f} (-1)^p
	\label{eq:def_TetraSym} \,,
\end{align}
where
\begin{align}
	p &\defineas \frac{ 3(a+b+c+d+e+f)^2 - (a+d)^2 - (b+e)^2 - (c+f)^2 }{2} \,.
\end{align}
and $\{\!\begin{smallmatrix} *&*&* \\ *&*&* \end{smallmatrix}\!\}_q$ are the $q$-deformed Wigner-6j symbols (cf.\ App.~\ref{app:Wigner6j}).
Thus, the tetrahedral symbols are identical to the Wigner-6j symbols up to some signs.

\paragraph{Limiting cases.}
When one of the element is zero, the symbol reduces to
\begin{align}
	\Msixj{\varphi & a & b}{0 & b & a} = \frac{N_{ab}^\varphi}{\sqrt{d_a d_b}} \,.
\end{align}
(Contrast with Eq.~\eqref{eq:Wigner6j_limit} for the Wigner-6j symbols.)
The expressions are also much simpler when one of the elements is $\frac12$ \cite{Lienert1992}:
\begin{subequations} \label{tetrahalf} \begin{align}
	\Msixj{\varphi & a &b}{\frac12 & b-\frac12 & a+\frac12} &= N_{ab}^\varphi N_{a+\frac12,b-\frac12}^\varphi \times (-1)^{a+b-\varphi} \sqrt{\frac{d_{\frac{\varphi+a-b}{2}} d_{\frac{\varphi+b-a-1}{2}}}{d_a d_b d_{a+\frac12} d_{b-\frac12}}} \,,
\\	\Msixj{\varphi & a &b}{\frac12 & b-\frac12 & a-\frac12} &= N_{ab}^\varphi N_{a-\frac12,b-\frac12}^\varphi \times \sqrt{\frac{d_{\frac{a+b+\varphi}{2}} d_{\frac{a+b-\varphi-1}{2}}}{d_a d_b d_{a-\frac12} d_{b-\frac12}}} \,.
\end{align}\end{subequations}

\subsection{\texorpdfstring{Wigner-6j symbols}{Wigner-6j symbols}}
\label{app:Wigner6j}
The $q$-deformed Wigner-6j symbols may be computed via the Racah formula \cite{Reshetikhin1988}.
\begin{align}\begin{split}
	\Wsixj{j_1 & j_2 & j_3}{j_4 & j_5 & j_6}
	&= \Delta(j_1,j_2,j_3) \, \Delta(j_1,j_5,j_6) \, \Delta(j_4,j_2,j_6) \, \Delta(j_4,j_5,j_3)
	\\	&\quad
		\times \sum_z \begin{pmatrix}
				\frac{ (-1)^z \; [z+1]! }{ [z-j_1-j_2-j_3]! \ [z-j_1-j_5-j_6]! \ [z-j_4-j_2-j_6]! \ [z-j_4-j_5-j_3]! }
			\\[1ex]	\times
				\frac{1}{ [j_1+j_2+j_4+j_5-z]! \ [j_2+j_3+j_5+j_6-z]! \ [j_3+j_1+j_6+j_4-z]! }
			\end{pmatrix} ,
	\label{eq:def_W6j}
\end{split}\end{align}
where
\begin{align}\begin{split}
	q &\defineas \exp\tfrac{2 \pi i}{k+2} \,,
\\	[m] &\defineas \frac{q^{m/2} - q^{-m/2}}{q^{1/2} - q^{-1/2}}
		= \frac{\sin \frac{m\pi}{k+2}}{\sin \frac{\pi}{k+2}} \,,
\\	[n]! &\defineas \left\{ \! \begin{array}{cl}
			\prod_{m=1}^n [m]  &  n > 0 \,,
		\\	1  &  n = 0 \,,
		\\	\infty  &  n < 0 \,,
		\end{array} \right.
\\	\Delta(j_1, j_2, j_3) &\defineas \left\{ \! \begin{array}{cl}
		\sqrt{\dfrac{ [j_1+j_2-j_3]! \ [j_3+j_1-j_2]! \ [j_2+j_3-j_1]! }{ [j_1+j_2+j_3+1]! }}	&	N^{j_1}_{j_2 j_3} = 1 \,, \\[3ex]
		0	&	\text{otherwise} \,, \end{array} \right.
\end{split}\end{align}
The summation $\sum_z$ is over integers $z$ such that the summand is well-behaved (i.e., no terms being zero or infinity).
Since $[k+2] = 0$, thus $[n]!$ vanishes for $n \geq k+2$.
Also note that $[n]$ yields all the quantum dimensions as $d_x = [2x+1]$.

The Wigner-6j symbols is invariant under the same permutations as those of the tetrahedral symbols, i.e., Eqs.~\eqref{eq:tetrasym_permute_cols}, \eqref{eq:tetrasym_swap_rows}.
However, they have different limiting cases.
\begin{align}
	\Wsixj{a & a & 0}{b & b & c} = \frac{N^a_{bc} (-1)^{a+b+c}}{\sqrt{d_a d_b}} .
	\label{eq:Wigner6j_limit}
\end{align}
We also have
\begin{align}
	\Wsixj{a & b & c}{\frac{k}{2}-d & \frac{k}{2}-e & \frac{k}{2}-f}
		&= (-1)^{k+a+b+c}
		\Wsixj{a & b & c}{d & e & f} .
\end{align}
It follows that
\begin{align}
	\renewcommand{\arraystretch}{1.1}
	\Wsixj{\frac{k}{2}-a & \frac{k}{2}-b & c}{\frac{k}{2}-d & \frac{k}{2}-e & \quad f\quad}
		&= (-1)^{k+a+b+d+e}
		\Wsixj{a & b & c}{d & e & f} .
\end{align}
The associated $F$-symbols are defined as
\begin{align}
	\big[F_e^{abd}\big]_{c,f} = (-1)^{a+b+d+e} \Wsixj{a&b&c}{d&e&f} \sqrt{d_c d_f} .
\end{align}
Hence, the Racah-Elliot-5 relation also appears different.
\begin{align}
	\sum_x d_x (-1)^{S}
		\Wsixj{a & b & x}{c & d & g} \Wsixj{c & d & x}{e & f & h} \Wsixj{e & f & x}{b & a & j}
	&= \Wsixj{g & h & j}{e & a & d} \Wsixj{g & h & j}{f & b & c} ,
\end{align}
with the sign determined by $S = a+b+c+d+e+f+g+h+j+x$.

\comment{
\section{Properties of tetrahedral symbols}
\label{app:tetra}

\paragraph{Orthogonality relations.} The tetrahedral symbols satisfies the \emph{Racah-Elliot} orthgonality relations.
\\
Racah-Elliot-3:
\begin{align}
	\sum_x d_x \Msixj{a & b & m}{c & d & x} \Msixj{a & b & n}{c & d & x}
	&= \frac{\delta_{mn}}{d_m} N^m_{ab} N^m_{cd} .
	\label{eq:RE3}
\end{align}
Racah-Elliot-5:
\begin{align}
	\sum_x d_x \Msixj{a & b & x}{c & d & g} \Msixj{d & c & x}{f & e & h} \Msixj{e & f & x}{b & a & j}
	&= \Msixj{g & h & j}{e & a & d} \Msixj{g & h & j}{f & b & c} .
	\label{eq:RE5}
\end{align}
Or equivalently
\begin{align}
	\sum_x d_x \Msixj{\beta & \gamma & g}{\rho & \nu & x} \Msixj{\gamma & \alpha & h}{\mu & \rho & x} \Msixj{\alpha & \beta & j}{\nu & \mu & x}
	&= \Msixj{g & h & j}{\alpha & \beta & \gamma} \Msixj{g & h & j}{\mu & \nu & \rho} .
\end{align}
\paragraph{$F$-symbols.} The $F$-symbols for the equivalent tensor category are
\begin{align}
	\big[ F_d^{abc} \big]_{xy} &= \sqrt{d_{xy}} \Msixj{a & b & x}{c & d & y} .
\end{align}
Thus Racah-Elliot-3 implies unitarity of the tensor category.

\begin{align}
	\sum_x
		\big[ F_a^{bcd} \big]_{g,x} \big[ F_j^{bxe} \big]_{a,f} \big[ F_f^{cde} \big]_{x,h}
	=	\big[ F_j^{gde} \big]_{a,h} \big[ F_j^{bch} \big]_{g,f} \;.
\end{align}
Racah-Elliot-4 is equivalent to the \emph{hexagon equation} while Racah-Elliot-5 is equivalent to the \emph{pentagon equation}.

$R$-symbols.
\begin{eqnarray}
R_c^{ab} &=& e^{i\pi (h_c-h_a-h_b)}\\
h_a &=& a^2-\frac{a(a+1)}{k+2}
\end{eqnarray}
satisfy Bonderson 2.78

}

\section{6j symbols for fusion categories with non-self-conjugate objects}
\label{app:notselfdual}

A $\mathbb{C}$-linear, semi-simple, rigid monoidal categories is really just a a fancy name for a fusion category.
And since everyone already knows what a spherical, unitary fusion category is, we just have to explain the remaining qualifiers.

The point of this section is to speculate on cases where $x \neq \bar{x}$ for some $x$.
It is also important to distinguish between upper/lower indices in the fusion symbol
\begin{align}
	N^{c}_{ab} = N^{b}_{c\bar{a}} = N^{\bar{a}}_{b\bar{c}} = N^{\bar{c}}_{\bar{a}\bar{b}} = N^{\bar{b}}_{\bar{c}a} = N^{a}_{\bar{b}c} .
\end{align}
(Note $N^\ast_{\ast\ast}$ is invariant under exchanging the two lower indicies: $N^c_{ab} = N^c_{ba}$.)
For each $a$, its quantum dimension $d_a$ is the dominant eigenvalue of the matrix $N^\ast_{a\ast}$.
They quantum dimensions satisfy
\begin{align}
	d_0 &= 1 \,,
&	d_a &\geq 1 \,,
&	d_{\bar{a}} &= d_a \,,
&	\sum_x N^x_{ab} d_x &= d_a d_b \,.
\end{align}
We require the theory to be \emph{multiplicity-free}, that is, $N^c_{ab} \in \{0, 1\}$ be no larger than 1; thus no additional degrees of freedom lie at the graph vertices.
The $A$ and $B$-symbols (in absence of fusion multiplicities) are defined as
\begin{subequations}\begin{align}
	A^{ab}_c & \;\defineas\; \sqrt{\frac{d_ad_b}{d_c}} \big[ F_b^{\bar{a}ab} \big]^\ast_{0,c} \;,
\\	B^{ab}_c & \;\defineas\; \sqrt{\frac{d_ad_b}{d_c}} \big[ F_c^{c\bar{b}b} \big]_{a,0} \;.
\end{align}\end{subequations}
Note that the Frobenius-Schur indicators $\varkappa_a = A_0^{a\bar{a}} = B_a^{0a}$ are special cases of these symbols.

We establish the criteria for which the tetrahedral tetrahedral are defined.
First we must fix the gauge, i.e., choose a set of basis vectors $\psi_c^{ab} \in V_c^{ab}$, which in turns fixes the $F$-symbols.
\begin{align}
	\fbox{\text{The tetrahedral symbols are defined if $N_{ab}^c = 1 \Longrightarrow A_c^{ab} = B_c^{ab} = +1$ for all $a,b,c$.}}
\end{align}
Note that the symbols are gauge-dependent.  Hence some categories may admit one or multiple tetrahedral symbols, while some may admit none.
(For example, we require that all Frobenius-Schur indicator must be $+1$, which rules out $\SU2_k$, since $\varkappa_x = (-1)^{2x}$ in the theory.)

\subsection*{Properties of tetrahedral symbols.}

First we define the tetrahedral symbols
\begin{align}
	\Msixj{a & b & c}{d & e & f}
	& \;\defineas\;
	\frac{1}{\sqrt{d_{abcdef}}} \tetradiagram .
\end{align}
Recall that
\begin{align}
	\big[ F_d^{abc} \big]_{x,y} = \Braket{ \FmoveRsmall | \FmoveLsmall } = \frac{1}{\sqrt{d_{abcd}}} \FmoveC \;.
	\label{eq:def_Tetra22}
\end{align}
Therefore, the tetrahedral symbols are related to the $F$-symbols via
\begin{align}
	\big[ F_d^{abc} \big]_{x,y} &= \sqrt{d_{xy}} \Msixj{a & b & \bar{x}}{c & d & y} .
\end{align}

\paragraph{Properties.}
Each vertex imposes a constraint of the fusion rules:
\begin{align}
	N^{\bar{a}}_{bc} N^d_{ec} N^e_{fa} N^f_{db} = 0
	\;\Rightarrow\; \Msixj{a & b & c}{d & e & f} = 0 \,.
\end{align}
The symbols are invariant under cyclic permutation of the columns.
\begin{align}
	\Msixj{a & b & c}{d & e & f} = \Msixj{b & c & a}{e & f & d} = \Msixj{c & a & b}{f & d & e} .
\end{align}
When any two columns are interchanged, we take the dual of the top row and the conjugate the entire symbol.
\begin{align}
\label{colexchange}
	\begin{array}{ccccccc @{} l}
		\Msixj{a & b & c}{d & e & f}
	&	=	&	\Msixj{\bar{a} & \bar{c} & \bar{b}}{d & f & e}^\ast
	&	=	&	\Msixj{\bar{c} & \bar{b} & \bar{a}}{f & e & d}^\ast
	&	=	&	\Msixj{\bar{b} & \bar{a} & \bar{c}}{e & d & f}^\ast
	&	.
	\\[-1.6ex]
	&&	\hspace{3ex} \underbracket{\hspace{4ex}}_{\text{exchange}} \hspace{0.7ex}
	&&	\underbracket{\hspace{7ex}}_{\text{exchange}} \hspace{0.7ex}
	&&	\underbracket{\hspace{4ex}}_{\text{exchange}} \hspace{4ex}
	\end{array}
\end{align}
We may also exchange the rows among any two columns, at the cost of dualizing some entries.
\begin{align}
	\begin{array}{ccccccc @{} l}
		\Msixj{a & b & c}{d & e & f}
	&	=	&	\Msixj{\bar{a} & e & \bar{f}}{\bar{d} & b & \bar{c}}
	&	=	&	\Msixj{\bar{d} & \bar{b} & f}{\bar{a} & \bar{e} & c}
	&	=	&	\Msixj{d & \bar{e} & \bar{c}}{a & \bar{b} & \bar{f}}
	&	.
	\\[-1.6ex]
	&&	\hspace{3ex}\underbracket{\hspace{4ex}}_{\text{swap rows}}
	&&	\underbracket{\hspace{7ex}}_{\text{swap rows}}
	&&	\underbracket{\hspace{4ex}}_{\text{swap rows}}\hspace{3ex}
	\end{array}
\end{align}
Finally, we also have the limiting case
\begin{align}
	\Msixj{\bar{a} & a & 0}{b & b & c}
	= \Msixj{a & b & \bar{c}}{\bar{b} & a & 0}
	= \frac{N^c_{ab}}{\sqrt{d_ad_b}} \,.
\end{align}

\paragraph{Orthogonality relations.}
The unitarity of the $F$-symbols imply
\begin{align}
	\sum_y d_y \Msixj{a & b & x}{c & d & y} \Msixj{a & b & x'}{c & d & y}^\ast
	= \frac{ \delta_{xx'} N^{\bar{x}}_{ab} N^{x}_{c\bar{d}} }{d_x} \,,
\end{align}
analogous to Racah-Elliot-3.

For the analogous equation to Racah-Elliot-5, consider the following diagram.
\begin{align}
	\circlespentagon
\end{align}
After evaluating the diagram two different ways,%
	\footnote{%
		The left hand side comes from applying three $F$-moves to eliminate $g$, $h$, and $j$,
			resulting in coefficients $\big[F_\gamma^{\rho\bar\nu\beta}\big]_{\bar{g},x}$, etc.
		A useful identity for the right hand side is $\vertexabcC = \sqrt{d_{def}} \Msixj{a & b & c}{d & e & f}^\ast \vertexabc$.}
	eliminating the common factor $\sqrt{d_{\alpha\beta\gamma ghj \mu\nu\rho}}$ from both sides, we get the pentagon equation:
\begin{align}
	\sum_x d_x
		\Msixj{\rho & \bar\nu & g}{\beta & \gamma & x}
		\Msixj{\mu & \bar\rho & h}{\gamma & \alpha & x}
		\Msixj{\nu & \bar\mu & j}{\alpha & \beta & x}
	=	\Msixj{g & h & j}{\alpha & \beta & \gamma}
		\Msixj{g & h & j}{\mu & \nu & \rho}^\ast .
	\label{eq:super_pentagon}
\end{align}

\paragraph{Other relations.}
\begin{align}
	\sum_f d_f \Msixj{a & \bar{a} & c}{b & \bar{b} & f} &= \delta_{c,0} \sqrt{d_{ab}} \,,
\\	\sum_{x,\mu,\nu,\rho} d_x d_\mu d_\nu d_\rho \Msixj{\rho & \bar\nu & g}{\beta & \gamma & x} 
		\Msixj{\mu & \bar\rho & h}{\gamma & \alpha & x}
		\Msixj{\nu & \bar\mu & j}{\alpha & \beta & x}
		\Msixj{g & h & j}{\mu & \nu & \rho}
	&=	\mathscr{D}^2 \Msixj{g & h & j}{\alpha & \beta & \gamma} .
\end{align}

\section{Tambara-Yamagami fusion categories and tetrahedral symbols}
\label{app:Potts}

In this appendix we review the Tambara-Yamagami fusion categories and write down their 6j-symbols~\cite{Tambara1998}.

Let $A$ be an Abelian group. 
The fusion category has one object for every group with quantum dimension equal to 1, and a single non-Abelian object $X$ with quantum dimensions $d_X = \sqrt{|A|}$, where $|A|$ is the number of group elements in $A$.
The identity element $\I$ corresponds to $0 \in A$.
The fusion rules are given by,
\begin{align}
	a \otimes b = (a{+}b) \,, \qquad a \otimes X = X \otimes a = X \,, \qquad X \otimes X = \; \bigoplus_{a \in A} a \,,
\end{align}
where $a,b$ are elements of $A$ and $a+b$ denotes group composition in $A$.
We also have that $\bar{X} = X$ and $\bar{a} = -a$. 
The fusion category is $\mathbb{Z}_2$ graded with $\mcc_0 = \{a \in A \}$ and $\mcc_1 = \{ X \}$.

The 6j-symbols depend on a symmetric non-degenerate bi-character $\chi$,
That is, $\chi: A \times A \rightarrow \mathbb{C}^\times$ with $\chi(a+b,c) = \chi(a,c)\chi(b,c)$, $\chi(a,b+c) = \chi(a,b)\chi(a,c)$, $\chi(a,b) = \chi(b,a)$, and $\det \chi \neq 0$. 
We also work in a gauge where $\chi(0,a) = \chi(a,0) = 1$, where 0 is the identity element of $A$.
The 6j-symbols are given by 
\begin{align}
\label{ATY6j}
	\Msixj{b{-}c & c{-}a & a{-}b}{a & b & c} &= 1 \,,
&	\Msixj{X & X & a}{X & X & b} &= \frac{ \chi(a,b)}{d_X} \,,
&	\Msixj{a & b & c}{X & X & X} &= \frac{N_{bc}^{\bar{a}}}{\sqrt{d_X}} \,.
\end{align}
The remaining 6j-symbols can be inferred from symmetry.

\subsection*{Example: \texorpdfstring{$\mathbb{Z}_3$}{Z3} Tambara-Yamagami}

The simple objects are $0$, $1$, $2$, and $X$.
The non-trivial fusion symbols are
\begin{align}
	N^0_{XX} = N^1_{XX} = N^2_{XX} = 1 \,,
	\quad	N^X_{X1} = N^X_{2X} = 1 \,,
	\quad	N^0_{12} = 1 \,,
	\quad	N^1_{22} = N^2_{11} = 1 \,.
\end{align}
Therefore, $\bar{1} = 2$ and $\bar{X} = X$.
The quantum dimensions are
\begin{align}
	d_0 = d_1 = d_2 = 1 \,,
	\quad	d_X = \sqrt{3} \,.
\end{align}
The only non-Abelian anyon is $X$, and its $F$-symbol is given by
\begin{align}
	\big[ F_X^{XXX} \big]_{a,b} = \frac{1}{\sqrt{3}}
			\begin{pmatrix} 1 & 1 & 1 \\ 1 & \omega & \omega^2 \\ 1 & \omega^2 & \omega \end{pmatrix}_{a,b}
		&=	\frac{1}{\sqrt{3}} \omega^{ab} \,,
	&	\text{where } \omega &= e^{\frac{2\pi i}{3}} \,.
\end{align}
Therefore, the tetrahedral symbols are given by
\begin{align}
	\Msixj{X & X & a}{X & X & b} &= \frac{\omega^{-ab}}{\sqrt{3}} ,
&	\Msixj{1 & 1 & 1}{X & X & X} &= \Msixj{2 & 2 & 2}{X & X & X} = \frac{1}{\sqrt[4]{3}} .
\end{align}
The remaining symbols may be inferred from symmetry.

\phantomsection
\addcontentsline{toc}{section}{References}

\setlength{\bibsep}{4.5pt plus 0.3ex}

\bibliography{references}
\bibliographystyle{apsrev4-1}

\clearpage
\appendix

\end{document}